\documentclass[10pt,a4paper]{article}
\usepackage{graphicx}
\usepackage{subfigure}
\usepackage{amsmath,amsthm}
\usepackage{amssymb,amsfonts}
\usepackage{threeparttable}
\usepackage{multirow}
\usepackage{booktabs}
\usepackage{authblk}
\usepackage{url}
\usepackage{fdsymbol}
\usepackage{hyperref}
\usepackage{tablefootnote}
\usepackage{textgreek}

\title{Evaluating Independence and Conditional Independence Measures}
\author{Jian Ma\thanks{Email: majian@hitachi.cn}}
\affil{Hitachi China Research Laboratory}

\begin{document}

\maketitle

\begin{abstract}
\noindent
Independence and Conditional Independence (CI) are two fundamental concepts in probability and statistics, which can be applied to solve many central problems of statistical inference. There are many existing independence and CI measures defined from diverse principles and concepts. In this paper, the 16 independence measures and 16 CI measures were reviewed and then evaluated with simulated and real data. For the independence measures, eight simulated data were generating from normal distribution, normal and Archimedean copula functions to compare the measures in bivariate or multivariate, linear or nonlinear settings. Two UCI dataset, including the heart disease data and the wine quality data, were used to test the power of the independence measures in real conditions. For the CI measures, two simulated data with normal distribution and Gumbel copula, and one real data (the Beijing air data) were utilized to test the CI measures in prespecified linear or nonlinear setting and real scenario. From the experimental results, we found that most of the measures work well on the simulated data by presenting the right monotonicity of the simulations. However, the independence and CI measures were differentiated on much complex real data respectively and only a few can be considered as working well with reference to domain knowledge. We also found that the measures tend to be separated into groups based on the similarity of the behaviors of them in each setting and in general. According to the experiments, we recommend CE as a good choice for both independence and CI measure. This is also due to its rigorous distribution-free definition and consistent nonparametric estimator.

\end{abstract}

\section{Introduction}
Independence and Conditional Independence (CI) are the fundamental concepts in probability and statistics. They can be applied to solve many central problems in the fields, such as graphical models \cite{Lauritzen1996}, variable selection \cite{Fan2008,Ma2021}, signal processing \cite{oja2000independent}, causal discovery \cite{ma2021estimating}, among others. Dawid delivered an unified conceptual framework upon them \cite{Dawid1979} for the theory of statistical inference. 

Given random variables $X,Y,Z$ and their joint probability density function $P(X,Y)$ and marginals $P(X),P(Y)$, independence of $(X,Y)$ is defined as
\begin{equation}
	P(X,Y) = P(X)P(Y).
\end{equation}
CI of $(X,Y)$ given $Z$ can be defined with conditional probability functions as
\begin{equation}
	P(X,Y|Z) = P(X|Z)P(Y|Z).
\end{equation}
With these definitions, many measures of independence and CI has been defined previously. Linear correlation as a measure of independence dates back to the early data of statistics, and partial correlation \cite{Baba2004} is its counterpart for CI. They both make implicit Gaussian assumptions, which limit their applications to only linear relationships. For decades, many works have been contributed to search an ideal measure for independence that can tackle nonlinearity and non-Gaussianity. 

In the middle of 20th century, copula theory was developed for a unified theory of representation of (in)dependence relationships \cite{nelsen2007introduction}. At its core is Sklar theorem which states that any joint probability function can be represented as a so-called copula function associated with marginals. Copula theory provides a powerful tool for studying independence.

Today, there exists many different types of measures for independence and CI. Pearson's $r$ is one of the most well-known measures for independence \cite{Pearson1896} and is also well-known for applicable to only linear correlations. Kernel method in machine learning \cite{Muandet2017} is proposed to develop nonlinear measures defined in reproducing kernel Hilbert spaces (RKHSs), which results in the Hilbert-Schmidt Independence Criterion (HSIC) for independence \cite{Gretton2007,Pfister2018} and Kernel-based CI test \cite{zhang2011uai}. HSIC can be considered as a nonlinear generalization of covariance in RKHSs. There are also other generalizations of the covariance/correlation concepts with techniques, such as distance correlation \cite{Szekely2007,Szekely2009} by utilizing characteristics functions. 

Mutual Information (MI) is a bivariate measure of independence proposed in information theory \cite{infobook}. It is distribution-free and hence can be applied to any cases of independence, whether linear or nonlinear. Conditional MI (CMI) is another measure in information theory defined for measuring CI.  

Many measures for independence are bivariate by definition, such as Pearson's $r$ and MI. There are many works for developing multivariate version of these measures. For example, dHSIC is a multivariate extension of HSIC. In information theory, many multivariate MI were proposed, such as total correlation \cite{Watanabe1960}, co-information \cite{Bell2003}, multi-information \cite{Studeny1998}. 

Copula theory provides a powerful tool for developing independence measures. In fact, many basic measures, like Kendall's $\tau$ and Spearman's $\rho$ can be represented by copula function \cite{nelsen2007introduction}. Inspired by this, Schweizer and Wolff proposed a copula based measures also \cite{Schweizer1981}.

In 2008, Ma and Sun \cite{Ma2011} proposed an independence measure, called Copula Entropy (CE), that is defined as a copula function based Shannon entropy. They proved that CE is equivalent to negative MI, and therefore built a bridge between copula theory and information theory. They also proposed a representation of transfer entropy, which is essentially CMI, with only CE \cite{ma2021estimating}. The rank-based nonparametric estimators for CE and TE were also proposed.

Today, there are already many other measures for independence and CI defined with different mathematical concepts in the literature. Some measures are proved to be interrelated with each other, such as HSIC and dCor \cite{Sejdinovic2013}. For the reviews of an incomplete list of measures, please refer to \cite{Tjoestheim2022} for independence and \cite{Li2020} for CI. 

So here comes question: which one is an ideal measure? To answer this, Renyi \cite{Renyi1959} proposed a set of seven axioms for an ideal measure. He then identified maximal correlation as the one that satisfies all the axioms. When developing their copula based measure, Schweizer and Wolff's revised three of the seven Renyi's axioms, which made them weaker than before \cite{Schweizer1981,Schweizer1991}. Joe also proposed a generalization of Renyi's axioms for multivariate dependence \cite{Joe1989}. For CI measure, Dawid also proposed a set of three properties in \cite{Dawid1979}. All this axioms state the properties that an ideal measure of independence should have in theory.

In practice, empirical comparison between measures is also important. Since the measures were developed from different statistical principle and concepts and work at different scales and intervals, one may wonder how they behave on the same data generating from diverse model assumptions. This comparison can be done with simulated or real data. By simulated data, one can check the behaviors of the measures in prespecified settings, such as linear relationships governed by normal distribution, or more complex conditions. By real data, one can investigate the true power of measures in much complex real situations provided that the true relationships of independence or CI are known in advance. Previously, Josse and Holmes compared three independence measures with both simulated and real data \cite{Josse2016}. Li and Fan compared eight CI measures with simulated data \cite{Li2020}. Considering the abundant works on measures of independence and CI, their comparison lists are far from complete.

In this paper, we want to evaluate more independence and CI measures with simulated and real data. 16 independence measures and 16 CI measures defined with different mathematical concepts will be included in these comparisons, including just mentioned the basic correlation/partial correlation, kernel-based measures, distance correlation and its variants, CE based ones, and many others (see section \ref{sec:measures}). Simulation experiments will be designed for evaluating them in different conditions: Gaussian or non-Gaussian, bivariate or multivariate, linear or non-linear. Three real dataset from the UCI machine learning repository \cite{asuncion2007uci} are used for testing the true power of these measures. Through comparisons, we try to answer three questions: how each measure behaves in several prespecified settings; what are the relationships between these measures according to their behaviors; and which one among them is preferred to the others in practice.

This paper is organized as follows: section \ref{sec:measures} introduce the measures of independence and CI to be evaluated, section \ref{sec:exp} present experiments, experimental results will be given in section \ref{sec:results}, followed by discussions in section \ref{sec:discussion}, section \ref{sec:conclusions} concludes the paper.

\section{Measures}
\label{sec:measures}
\subsection{Independence Measures}
\label{sec:indep}
\subsubsection{Copula Entropy}
With copula theory\cite{nelsen2007introduction}, Ma and Sun \cite{Ma2011} defined a new mathematical concept, named Copula Entropy, as the Shannon entropy of copula function.

Let $\mathbf{X}$ be random variables with marginals $\mathbf{u}$ and copula density function $c$. The CE of $\mathbf{X}$ is defined as
\begin{equation}
	H_c(\mathbf{x})=-\int_{\mathbf{u}}{c(\mathbf{u})\log c(\mathbf{u})d\mathbf{u}}.
	\label{eq:ce}
\end{equation}	

They also proved that CE is equivalent to MI in information theory \cite{infobook}. CE has several ideal properties, such as multivariate, symmetric, invariant to monotonic transformation, non-positive (0 iff independent), and equivalent to correlation coefficient in Gaussian cases. It is a perfect measure for statistical independence. 

Ma and Sun \cite{Ma2011} also proposed a non-parametric method for estimating CE, which composes of two simple steps: 1) estimating empirical cdf; and 2) estimation CE from the estimated empirical cdf. In the first step, the rank statistic is used to derive empirical cdf; in the second step, the famous KSG method \cite{kraskov2004} for estimating entropy is suggested. The proposed estimation method is rank-based and to estimate the entropy of rank statistic essentially. 

\subsubsection{Kendall's $\tau$}
Kendall's $\tau$ \cite{Kendall1938} (Ktau) is one of the well-known non-parametric statistic for bivariate independence. It is a kind of rank correlation between two random variables $X,Y$. 

Let $s$ be the sign function, then Kendall's $\tau$ is defined as
\begin{equation}
	\tau = \frac{1}{n^2}\sum_{i,j=1}^{n}{s(x_i-x_j)s(y_i-y_j)},
\end{equation}
where $n$ is sample size. The $\tau$ such defined is the difference between the probability of concordance and discordance of all the sample pairs.

\subsubsection{Hoeffding's D}
Hoeffding \cite{Hoeffding1948} proposed a non-parametric test for bivariate independence. The hypothesis tests the equality of joint and marginal probability densities, as follows:
\begin{equation}
	F_{XY}=F_X F_Y,
	\label{eq:fxy}
\end{equation}
where $X,Y$ are two random variables, and $F_XY$ and $F_X,F_Y$ are joint and marginal probability densities respectively. Hoeffding defined for the test the functional $D$ as
\begin{equation}
D(x,y) =\int{\{F_{XY}(x,y)-F_X(x)F_Y(y)\}^2d F_{XY}(x,y)}.
\label{eq:hoeffding}
\end{equation} 
Then Hoeffding's D statistic can be derived from empirical probability densities accordingly.

\subsubsection{Bergsma-Dassios's $\tau^*$}
Inspired by Kendall's $\tau$, Bergsma and Dassios \cite{Bergsma2014} defined a new statistic $\tau^*$ (BDtau) for testing bivariate independence. It extends $\tau$ by defined a statistic from the concordance and discordance of all the two sample pairs, instead of a sample pairs. 

Define the sign functions $a$ as
\begin{equation}
a(z_1,z_2,z_3,z_4) = sign(|z_1-z_2|+|z_3-z_4|-|z_1-z_3|-|z_2-z_4|),
\end{equation}
Then the $\tau^*$ for two random variables $X,Y$ is defined as
\begin{equation}
	\tau^*(X,Y)=\frac{1}{n^4}\sum_{i,j,k,l=1}^{n}{a(x_i,x_j,x_k,x_l)a(y_i,y_j,y_k,y_l)}.
\end{equation}

\subsubsection{HHG}
Heller et al. \cite{heller2016consistent} proposed a partition-based statistics for bivariate independence by testing the hypothesis of distribution equality. It is based on Hoeffding's $D$ statistic \eqref{eq:hoeffding}. For a given partition of samples, the HHG statistic is an aggregation over all partitions on the summation of Pearson' score or the maximization of the likelihood ratio score of each cell. The Pearson's score (HHG.chisq) or the likelihood ratio (HHG.lr) score are derived from the observed and expected sample counts in a cell, as follows:
\begin{equation}
	\{\frac{(o_C-e_C)^2}{e_C},o_C\log \frac{o_C}{e_C}\},
\end{equation}
where $o_C,e_C$ are the observed and expected counts. The HHG statistic is distribution free since it is based on ranks.

\subsubsection{Ball Correlation}
Pan et al. \cite{Pan2020} proposed a dependence measure, call Ball covariance, with theory of probability measure. The Ball covariance is defined as the average distance between joint Borel probability measure and product of marginal Borel probability measures to test the counterpart of Hoeffding's independence hypothesis \eqref{eq:fxy} in Banach space. The statistic of Ball Covariance can be represented as 
\begin{equation}
	BCov(x_1,\ldots,x_K)=\frac{1}{N^2}\sum_{i,j=1}^{N}{(P_{ij}^{\mathbf{x}}-\prod_{k=1}^{K}{P_{ij}^{x_k}})^2},
\end{equation}
where $N$ is the sample size, and $P_{ij}^{\mathbf{x}}$ and $P_{ij}^{x_k}$ are empirical joint and marginal Borel probability measure respectively. It holds that the Ball covariance of two random variables is equal to zero iff they are independent. 

\subsubsection{BET}
Zhang \cite{Zhang2019} proposed a framework for independence test, called binary expansion testing (BET), through approximating bivariate copula with multiscale binary expansion. He defined a cross interaction odds ratio (CIOR) $\lambda$ on the binary partitions for quantifying bivariate independence, which is analogous to the odd ratio defined on the contingency table. They introduced a test statistic for testing bivariate independence based on the observation that the symmetry of $\lambda$ can be deployed for transforming BET into a multiple testing problem.

\subsubsection{Quantification of Asymmetric Dependence $\zeta$}
The dependence between two random variables is asymmetric if one is the function of the other. Trutschnig \cite{Trutschnig2011} proposed a measure $\zeta$ (QAD) for asymmetric dependence. He defined a metric $D_1$ between copula function $A,B$ with the so called Markov kernels $K$ 
\begin{equation}
	D_1(A,B)=\int_{I}\int_{I} |K_A(x,[0,y])-K_B(x,[0,y])|d\lambda(x)d\lambda(Y)
\end{equation}
and therefore derived a metric space $(C,D_1)$ which is complete and separable. He then defined a dependence measure $\zeta$ as the $D_1$-distance from copula $C$ to product copula $\Pi$
\begin{equation}
	\zeta(C,\Pi) = D_1(C,\Pi).
\end{equation}
The measure $\zeta \in [0,1]$ can characterize both independence and completely dependence: $\zeta(C_0)=0$ if $C_0$ is for independence cases and $\zeta(C_1)=1$ if $C_1$ is for the cases where two random variables corresponding to $C_1$ has functional relationships. More importantly, $\zeta$ is asymmetric for two variables with functional relationships.

\subsubsection{Dependence Coefficient}
When random variable $X,Y$ are independent, the following holds
\begin{equation}
	E(Y|X)=E(Y).
	\label{eq:eyx}
\end{equation}
This equation can be used to design dependence measures. Chatterjee \cite{Chatterjee2021} proposed a simple correlation coefficient (CODEC) base on \eqref{eq:eyx}. Given ${(X_i,Y_i)}$ as the realization of $X,Y$, such that $X_1\leq \cdots \leq X_n$, and $r_i$ be the rank of $Y_i$, the new coefficient they proposed is defined as
\begin{equation}
	\xi_n(X,Y)=1-\frac{3\sum_{i=1}^{n-1}{|r_{i+1}-r_i|}}{n^2-1}.
\end{equation}
The proposed statistic is non-parametric and equal 0 iff independent and 1 iff one is a measurable function of the other.

\subsubsection{Product Copula based Measure}
Genest et al. \cite{Genest2019} proposed a model-free independence test based on empirical copula. The statistic (mixed) is defined as
\begin{equation}
	||n^{\frac{1}{2}}(C_n-\Pi)||_2^2=\int_{I^d}n\{C_n-\Pi\}^2d\mathbf{u},
\end{equation}
where $C_n$ is empirical copula, and $\Pi$ is product copula. They constructed a estimator that estimates empirical copula with empirical checkerboard copula.

Erdely \cite{Erdely2017} proposed a subcopula-based dependence measure (subcop) which is defined as follows:
\begin{equation}
	d(S)=\sup_{Dom S}\{S-\Pi\}- \sup_{Dom S}\{\Pi-S\},
\end{equation}
where $S$ is bivariate subcopula. He proposed a estimator for $d(S)$ with empirical subcopula.

\subsubsection{Distance Correlation}
Distance Correlation (dCor) is a nonlinear generalization of traditional correlation concept proposed by  Sz\'ekely, et al \cite{Szekely2007,Szekely2009}. It generalizes bivariate second-order correlation to multivariate nonlinear cases via distance covariance. dCor between random vectors $X$ and $Y$ is defined as
\begin{equation}
\mathbf{dCor}(X,Y) = \frac{\nu^2(X,Y)}{\sqrt{\nu^2(X)\nu^2(Y)}},
\end{equation}
where $\nu^2(X,Y)$ is distance covariance defined with characteristic function $f$ as
\begin{equation}
\nu^2(X,Y;w) = \|f_{X,Y}(t,s) - f_X(t)f_Y(s)\|_w^2.
\label{eq:dcov}
\end{equation}
Here, $\|\cdot\|_w$ is the norm in the weighted $L_2$ function space defined with positive weight function $w(\cdot,\cdot)$ \cite{Szekely2007,Szekely2009}. dCor characterizes independence: $\mathbf{dCor}(X,Y) \geq 0$, and $\mathbf{dCor}(X,Y) = 0$ if and only if $X,Y$ are independent.

Jin and Matteson \cite{Jin2017} generalized the above bivariate distance correlation to more general measure of multivariate independence. They introduced two dCov based measures: one is defined with the characteristic function $f$ of multiple variables and the product of the $f_i$ of a variable as
\begin{equation}
\nu^2(\mathbf{X})=|| f_{\mathbf{X}}(\mathbf{t})-\prod_{i}{f_i(t_i)} ||_w^2,
\end{equation}
and the other is defined as the aggregation of bivariate distance correlation. 

Shao and Zhang \cite{Shao2014} proposed another variants of distance correlation, called martingale difference correlation (MDC), which is essentially to test the hypothesis \eqref{eq:eyx}. They presented the definition of MDC similar to dCor by extending the definition of distance covariance to martingale difference divergence (MDD) as
\begin{equation}
MDD(Y|X) = ||\frac{1}{2}\Delta_t^2(f_{X,Y}(t,0)-f_X(t)f_Y(0))||^2,
\end{equation}
where $\Delta_t^2$ is for second order derivative on $t$.

\subsubsection{Hilbert-Schmidt Independence Criteria}
Hilbert-Schmidt Independence Criterion (HSIC) is another widely studied independence measure \cite{Gretton2007} and it has multivariate version -- d-variable HSIC (dHSIC) \cite{Pfister2018}. dHSIC defines a nonlinear dependence measure in Reproducing Kernel Hilbert Spaces (RKHSs) with kernel function, as follows:
\begin{equation}
\mathbf{dHSIC}(P(X_1,\cdots,X_d)) = \|\Pi(P(X_1)\otimes,\cdots,\otimes P(X_d))-\Pi(P(X_1,\cdots,X_d))\|,
\end{equation}
where $\Pi$ is kernel mean embedding function, and $\otimes$ is tensor products of kernels. dHSIC can be considered as the distance in RKHS between the embeddings of joint distribution and margins. dHSIC also characterizes independence: $\mathbf{dHSIC}(P(X_1,\ldots,X_d))=0$ if and only if $X_1,\ldots, X_d$ are independent. 

\subsubsection{Nonlinear Nonparametric Statistic}
Voile and Nawrocki \cite{Viole2012} proposed a nonlinear correlation measure based on partial moments as a substitute of Pearson correlation coefficient, called Nonlinear Nonparametric Statistic (NNS) $\rho_{NNS}$, which is defined as ordered aggregations of co-partial moments. They showed that when Pearson correlation coefficient is $-1,0,1$ in linear correlation cases, $\rho_{NNS}=-1,0,1$ correspondingly and that $\rho_{NNS}=1$ when there is a nonlinear functional relationship between two random variables.

\subsection{Conditional Independence Measures}
\label{sec:ci}
\subsubsection{Copula Entropy}
CE has also theoretical relationship with CI. Ma \cite{ma2021estimating} proved that Transfer Entropy (TE) can be represented with only CE. Since TE is essentially conditional MI, an information-theoretical measure of CI, we can also measure CI with only CE as the proposition below.

Given random variables $X,Y,Z$, the measure $H_{ci}$ of CI between $X,Y$ given $Z$ can be measured as follows:
\begin{equation}
	H_{ci}(x,y,z)=H_c(x,z)+H_c(y,z)-H_c(x,y,z).
\label{eq:ci}
\end{equation}
Please refer to \cite{ma2021estimating} for the proof of this proposition.

With the CE estimation method, Ma also proposed a non-parametric method for estimating TE or testing CI by estimating the 3 CE terms according to \eqref{eq:ci} \cite{ma2021estimating}.

\subsubsection{Distance Correlation}
Wang et al. \cite{wang2015conditional} proposed a new measure for testing CI, called Conditional Distance Correlation (CDC), which is a generalization of distance correlation to CI. It is defined by replacing the characteristic functions in \eqref{eq:dcov} with conditional characteristic functions as
\begin{equation}
	\nu^2(X,Y|Z;w)=||f_{X,Y|Z}(t,s)-f_{X|Z}(t)f_{Y|Z}(s)||_w^2.
\end{equation}
Accordingly, they defined a simple statistic with the symmetric random kernel function.

Another generalization of distance correlation to CI is the partial Martingale Difference Correlation (CMDM) presented by Part et al. \cite{Park2015}, which is the extension of MDC for CI test.

\subsubsection{COnditional DEpendent Coefficient (CODEC)}
The equation \eqref{eq:eyx} for mutual independence can be easily extended to the case for CI. Given random variables $X,Y,Z$, the following holds
\begin{equation}
	E(Y|X,Z)=E(Y|Z),
	\label{eq:exzy}
\end{equation}
if $X$ is conditionally independent of $Y$ given $Z$. According to this, Azadkia and Chatterjee \cite{Azadkia2021} proposed a measure of CI which generalized the measure of unconditional independence they proposed in \cite{Chatterjee2021}. Given random variables $X,Y,Z$, the proposed measure for CI of $X$ and $Y$ given $Z$ is defined as
\begin{equation}
	T(X,Y|Z)=\frac{\int{E(Var(P(Y\geq t|X,Z)|Z)d\mu(t)}}{\int{E(Var(1_{\{Y\geq t\}}|Z))d\mu(t)}}.
	\label{eq:codec}
\end{equation}
They also proposed an estimator for the above measure based on rank. Given $(X_i,Y_i,Z_i)$ as the realization of random variables $X,Y,Z$. Let $N(i)$ be the index of $Z_j$ which is the nearest neighbor of $Z_i$ with respect to the Euclidean metric. Let $M(i)$ be the index of $(X_j,Z_j)$ which is the nearest neighbor of $(X_i,Z_i)$. Let $r_i$ be the rank of $Y_i$. Then the non-parametric estimator for the measure \eqref{eq:codec} is 
\begin{equation}
	T_n(Y,X|Z)=\frac{\sum_{i=1}^{n}(min\{r_i,r_{M(i)}\}-min\{r_i-r_{N(i)}\})}{\sum_{i=1}^{n}{(r_i-min\{r_i,r_{N(i)}\})}}.
\end{equation}

\subsubsection{Kernel Partial Correlation Coefficient}
Huang et al. \cite{Huang2020} proposed to test the hypothesis \eqref{eq:exzy} with kernel tricks. The measure they proposed, called kernel partial correlation (KPC) coefficient, is define with kernel function $k(\cdot,\cdot)$ as
\begin{equation}
\rho^2(Y,X|Z)=\frac{E[{MMD}^2(P(Y|XZ),P(Y|Z))]}{E[{MMD}^2(\delta_Y,P(Y|X))]},
\label{eq:kpc}
\end{equation}
where $MMD$ is maximum mean discrepancy which is the distance in RKHSs associated with $k(\cdot,\cdot)$. The measure defined in \eqref{eq:kpc} can be considered as the counterpart of the CODEC \eqref{eq:codec} in RKHSs.

\subsubsection{Prediction based Tests}
Chalupka et al. \cite{Chalupka2018} proposed a method, called Fast Conditional Independence Test (FCIT) by testing the hypothesis \eqref{eq:exzy} with predictions. The idea behind FCIT is that if X is conditionally dependent of Y given Z, then predicting Y from X and Z should be more accurate than predicting Y with only X. They implemented the method with decision trees. The prediction accuracy was measured with mean squared error.

Burkart and Kir\'aly \cite{Burkart2017} also proposed a similar method, called predictive conditional independence test (PCIT), to test the hypothesis \eqref{eq:exzy} with prediction models.

Sen et al. \cite{Sen2017} proposed another CI test (CCIT) based on prediction. In it, they test the following hypothesis for CI of $X,Y$ given $Z$:
\begin{equation}
	p_{X,Y,Z}(x,y,z)=p_{X|Z}(x|z)p_{Y|Z}(y|z)p_Z(z),
	\label{eq:ccit}
\end{equation}
where $p_{X,Y,Z}(x,y,z)$ is the joint distribution of $X,Y,Z$. If $X$ is conditionally independent of $Y$ given $Z$, then \eqref{eq:ccit} holds. They transformed the test into a regression problem. Given a sample from $(X,Y,Z)$. First, separate the sample into two parts. Then simulate a sample from a part of original sample that is close to conditional independent distribution $p(X,Y|Z) = p_{X|Z}(x|z)p_{Y|Z}(y|z)p_Z(z)$ with the nearest-neighbor bootstrap method. Next, combine the simulated sample with another part of original sample as a data for binary classification problem. If the result of classification is close to random guess, then we accept the hypothesis of CI, or reject it otherwise.

\subsubsection{Conditional Mutual Information Estimators}
Conditional Mutual Information (CMI) is a measure for CI in information theory. Runge \cite{Runge2017} proposed a nearest neighbor estimator for CMI (CMI1). The proposed estimator is based on
\begin{equation}
	I_{X,Y|Z}=H_{XZ}+H_{YZ}-H_Z-H_{XYZ},
	\label{eq:ixyz}
\end{equation}
where $H$ denotes the Shannon entropy. Using the nearest neighbor estimator of $H$, he proposed the CMI estimate to be
\begin{equation}
	\hat{I}_{X,Y|Z}=\psi(k)+\frac{1}{n}\sum_{i=1}^{n}[\psi(k_{Z,i})-\psi(k_{XZ,i})-\psi(k_{YZ,i})],
	\label{eq:eixyz}
\end{equation}
where $\psi(\cdot)$ is the Gamma function, and $k$ is the number of nearest neighbors in the balls of $(X,Y,Z)$ centered at the sample $i$.

Mesner and Shalizi \cite{Mesner2021} proposed a variant of the estimator of \eqref{eq:ixyz} (CMI2) for the mixed discrete and continuous variables cases, which presents more accurate estimate for such cases than the estimator \eqref{eq:eixyz}.

\subsubsection{Partial Correlation}
Partial correlation (pcor) is one of the widely used measures for CI in practice. Given random variable $X_1,X_2$ and $\mathbf{Y}\in R^p$, and the partial covariance matrix is defined as
\begin{align}
	\Sigma_{\mathbf{X}|\mathbf{Y}} &=\Sigma_{\mathbf{X}\mathbf{X}}-\Sigma_{\mathbf{X}\mathbf{Y}}\Sigma_{\mathbf{Y}\mathbf{Y}}^{-1}\Sigma_{\mathbf{Y}\mathbf{X}}\\
	&= \left[
	\begin{array}{cc}
	\theta_{11} & \theta_{12}\\
	\theta_{21} & \theta_{22}\\
	\end{array}
	\right]	
\end{align}
where $\Sigma$ is the covariance matrix of $(\mathbf{X},\mathbf{Y})$. Then partial correlation  $\rho_{\mathbf{X}|\mathbf{Y}}$ is defined as \cite{Baba2004}
\begin{equation}
	\rho_{\mathbf{X}|\mathbf{Y}}=\frac{\theta_{12}}{\sqrt{\theta_{11}\theta_{22}}}.
\end{equation}
It has been proved that $X_1,X_2$ are conditionally independent given $\mathbf{Y}$ iff the joint distribution is multivariate normal distribution and the partial correlation  $\rho_{\mathbf{X}|\mathbf{Y}} = 0$.

The partial correlation can be computed with the correlation between the residuals of regression models. Given the following regression functions $X_1=\beta_1 Y + \epsilon_1$ and $X_2=\beta_2 Y + \epsilon_2$, then the partial correlation $\rho_{\mathbf{X}|\mathbf{Y}}=cor(\epsilon_1,\epsilon_2)$, where $cor(\cdot,\cdot)$ is correlation function. 

Shah and Peters \cite{Shah2020} proposed a statistic for CI test, called Generalised Covariance Measure (GCM), as a normalized covariance between the residuals of regression functions. Scheidegger et al. \cite{Scheidegger2021} proposed a generalised GCM (wGCM) by a weight function on the covariance of the residuals.

\subsubsection{Kernel based Conditional Independence}
The kernel-base CI test (KCIT) proposed in \cite{zhang2011uai} is based on the idea, called kernel mean embedding, that test partial correlation by transforming distributions into RKHS with kernel functions. They proved that CI in RKHSs is equivalent to the covariance of residuals of nonlinear regression in RKHSs $E(\tilde{f}\tilde{g})=0$. They proposed a statistic based on the covariance of residuals of kernel ridge regression as
\begin{equation}
	T_{CI}=\frac{1}{n}Tr(\tilde{K}_{XZ|Z}\tilde{K}_{Y|Z}),
\end{equation}
where $\tilde{K}$ is the centralized kernel matrix for residuals.

KCIT cannot be used with large dataset because of high computational cost. Strobl et al. \cite{Strobl2019} proposed two approximation of KCIT, called Randomized Conditional Independence Test (RCIT) and Randomized conditional Correlation Test (RCoT), with random Fourier features. Both approximations have linear sample complexity.

\subsubsection{Partial Copula based Test}
Given random variables $X,Y,Z$, the partial copula is the joint distribution $(U_X,U_Y)$ with conditional distributions as marginal functions 
\begin{align}
	U_X = F_{X|Z}(X|Z), & & U_Y = F_{Y|Z}(Y|Z).
\end{align}
Petersen and Hansen \cite{Petersen2021} proposed to test the CI of $X,Y$ given $Z$ by the independence of the residuals $U_X,U_Y$. They proposed a method (pcop) for such test that first estimates partial copula with quantile regression and then tests the independence of the residuals with a generalized measure of correlation.

\section{Experiments}
\label{sec:exp}
\subsection{General settings}
\subsubsection{Data and Experiments}
We compared the independence and CI measures with both simulated data and real data. For the independence measures, eight groups of data were simulated from the known joint distributions and two real data (heart disease data and wine quality data) were used. For the CI measures, a simulated data and a real data (Beijing air data) were used. All the real data are from the UCI machine learning repository \cite{asuncion2007uci}.

In the simulation experiments, we first generated the data from bivariate or multivariate normal distributions and copula functions with nonlinear marginals. Then the bivariate or multivariate measures were estimated from the simulated data. In the real data experiments, the measures between attributes were estimated directly from the data.

As introduced in section \ref{sec:measures}, these measures of independence and CI are designed with different concepts. We are interested in the relationships between them on the conditions of different types of dependence. Since these measures work at different scales and intervals, we cannot make comparisons between the estimation results of them directly. There is also no standard reference for the estimations of each measure, except correlation/partial correlation in normal cases and CE of which the analytical value can be derived if probability distribution function is known. To compare these measures, we calculated the correlations between the estimations of these measures and then clustered the measures based on the calculated correlations with hierarchical clustering \cite{Murtagh2014}. By this, we will derive clusters of the measures and check the (dis)similarities of the measures in different conditions.

\subsubsection{Implementations}
To evaluate the measures of independence and CI in section \ref{sec:measures}, we conducted experiments with the implementations of those measures (as listed in Table \ref{tb:impl}). All the \texttt{R} and \texttt{Python} packages are available on the CRAN and PyPI, except the \textsf{parCopCITest} for the partial copula based independence measure which is available on the Github \footnote{\url{https://github.com/lassepetersen/partial-copula-CI-test}}. For all the implementations of the measures, the default parameters of the implementation functions were used. All the implementations for the estimators of the measures, except of Ktau, Hoeff, BDtau and CMI1, are offical from the authors of the measures.

\begin{table}
	\centering
	\caption{The implementations of the measures.}
	\begin{tabular}{l|c|c|c|c}
		\toprule
		\textbf{Package}&Independence&CI&Language&Version\\
		\midrule
		\textsf{copent}&CE&TE/CI&\texttt{R}&0.2\\
		\textsf{stats}&Ktau&&\texttt{R}&4.1.3\\		
		\textsf{energy}&dCor&&\texttt{R}&1.7-10\\
		\textsf{dHSIC}&dHSIC&&\texttt{R}&2.1\\
		\textsf{HHG}&HHG&&\texttt{R}&2.3.4\\
		\textsf{independence}&Hoeff,BDtau&&\texttt{R}&1.0.1\\
		\textsf{Ball}&Ball&&\texttt{R}&1.3.12\\
		\textsf{qad}&QAD&&\texttt{R}&1.0.1\\
		\textsf{BET}&BET&&\texttt{R}&0.4.2\\
		\textsf{MixedIndTests}&Mixed&&\texttt{R}&0.8.0\\
		\textsf{subcopem2D}&subcopula&&\texttt{R}&1.3\\
		\textsf{EDMeasure}&MDM&CMDM&\texttt{R}&1.2.0\\
		\textsf{FOCI}&CODEC&CODEC&\texttt{R}&0.1.3\\
		\textsf{NNS}&NNS&&\texttt{R}&0.8.61\\			
		\textsf{RCIT}&&RCoT&\texttt{R}&0.1.0\\		
		\textsf{cdcsis}&&CDC&\texttt{R}&2.0.3\\		
		\textsf{GeneralisedCovarianceMeasure}&&GCM&\texttt{R}&0.2.0\\		
		\textsf{weightedGCM}&&wGCM&\texttt{R}&0.1.0\\		
		\textsf{KPC}&&KPC&\texttt{R}&0.1.1\\		
		\textsf{ppcor}&&pcor&\texttt{R}&1.1\\		
		\textsf{parCopCITest}&&pcop&\texttt{R}&-\tablefootnote{The version committed to Github at Jan 19, 2021.}\\		
		\textsf{causallearn}&&KCI&\texttt{Python}&0.1.2.2\\
		\textsf{pycit}&&CMI1&\texttt{Python}&0.0.7\\		
		\textsf{knncmi}&&CMI2&\texttt{Python}&0.0.1\\		
		\textsf{fcit}&&FCIT&\texttt{Python}&1.2.0\\		
		\textsf{CCIT}&&CCIT&\texttt{Python}&0.4\\		
		\textsf{pcit}&&PCIT&\texttt{Python}&1.2.2\\		
		\bottomrule
	\end{tabular}
	\label{tb:impl}
\end{table}

In simulation experiments, the \texttt{R} package \textsf{mnormt} was used for generating the data of normal distributions, and the \texttt{R} package \textsf{copula} and the \texttt{Python} package \textsf{pycop} were used for generating data from normal or Archimedean copula functions. The functionality implemented in the \texttt{R} package \textsf{corrplot} was used for hierarchical clustering.

\subsection{Independence Measures}
\subsubsection{Simulated data}
We did eight simulation experiments to evaluate the independence measuere on different types of dependence: Gaussian or non-Gaussian, bivariate or multivariate, linear or nonlinear. The normal distribution will be used to simulate linear dependence and the normal and Archimedean copula will be used to simulate nonlinear dependence. For multivariate dependence, we considered two cases: joint dependence and dependence between random vectors. For joint dependence, trivariate normal distribution and trivariate Archimedean copula are used; For the latter, a quadvariate normal distribution is used to simulate the dependence between two bivariate random vectors. We let the control parameters ($\rho$ or $\alpha$) increase monotonically so as to make the dependence strength increasing accordingly.

\paragraph{Bivariate normal distribution}
In the first experiment, we simulated a group of bivariate normal distribution. The covariance $\rho$ of two random variables was set from 0 to 0.9 by step 0.1 and the sample size is 800. With this setting, we test the measure's ability of measuring the basic linear correlation and check whether the measures is a strictly increasing function of $|\rho|$ in Gaussian cases.

\paragraph{Bivariate normal copula}
We also test the nonlinear dependence with bivariate normal copula. In this case, the dependence structure is same with normal distribution but the dependence relationship is nonlinear due to nonlinear marginals. In simulation, the covariance parameter $\rho$ of bivariate normal copula was set from 0 to 0.9 by step 0.1, and the marginals are normal distribution with 0 mean and standard deviation as 2 and exponential distribution with rate 2. The sample size is 800.

\paragraph{Bivariate Archimedean copula} To test the measure's ability of measuring nonlinear independence relationships, we did three simulation experiments with bivariate Archimedean copula functions. Three common types of Archimedean copula function were considered, including Clayton copula, Gumbel copula, and Frank copula \cite{nelsen2007introduction}. The definitions of these copula are as follows:
\begin{equation}
	Clayton_{\alpha}(u,v)=max\left([u^{\alpha} + v^{\alpha} -1]^{-\frac{1}{\alpha}},0\right),
\end{equation}
where $ \alpha \in (-1,\infty)\setminus \{0\}$. By definition, Clayton copula is an Archimedean copula that has asymmetric dependence with large tail.
\begin{equation}
	Gumbel_{\alpha}(u,v)=\exp \left\lbrace  -[(-\ln u)^{\alpha}+(-\ln v)^{\alpha}]^{\frac{1}{\alpha}} \right\rbrace ,
\end{equation}
where $\alpha \in [1,\infty]$. Gumbel copula represents asymmetric dependence.
\begin{equation}
	Frank_{\alpha}(u,v)=-\frac{1}{\alpha}\ln\left(1+\frac{(e^{-\alpha u}-1)(e^{-\alpha v}-1)}{e^{-\alpha}-1}\right),
\end{equation}
where $\alpha \in (-\infty,\infty) \setminus \{0\}$. Frank copula is symmetric copula.

In simulations, the $\alpha$ of these three copulas are all set from 1 to 10, which means increasing dependence between two random variables. The two marginals of the copulas are normal distribution with 0 mean and standard deviation as 2, and exponential distribution with rate as 2. The sample size is 800 in all simulations.

\paragraph{Multivariate normal distribution}
Six measures in section \ref{sec:indep}, including CE, BET, dHSIC, NNS, mixed, and subcop, can measure multivariate dependence. We test their ability of multivariate dependence with a very basic case -- trivariate normal distribution. The covariance matrix of the trivariate normal distribution has only one parameter $\rho$ as
\begin{equation}
\left[ 
\begin{array}{ccc}
1&\rho&\rho\\
\rho&1&\rho\\
\rho&\rho&1\\
\end{array}\right].
\end{equation}
In simulation, the mean of the trivariate normal distribution is $\mathbf{0}$ and the $rho$ is set from 0 to 0.9 by step 0.1. The sample size is 800. The dependence between three variables were estimated from the simulated data.

\paragraph{Multivariate Archimedean copula}
We also test these measures' ability of measuring multivariate nonlinear dependence with trivariate Archimedean copula. The 3-dimensional Gumbel copula was considered in simulation. The $\alpha$ was also set from 1 to 10 and the three marginals are normal distribution with 0 mean and standard deviation as 2, and two exponential distributions with rate as 0.5 and 2 respectively. The sample size is 800. The dependence between three variables were estimated from the simulated data.

\paragraph{Multivariate normal distribution for dependence between random vectors}
Ten measures in section \ref{sec:indep}, including CE, HHG, Ball, BET, QAD, dCor, MDM, dHSIC, and NNS, can measure dependence between random vectors. For CE, the dependence between random vectors $\mathbf{X},\mathbf{Y}$ can be easily derived as follows:
\begin{equation}
	H_c(\mathbf{x};\mathbf{y}) = H_c(\mathbf{x},\mathbf{y})-H_c(\mathbf{x})-H_c(\mathbf{y}).
\end{equation}

We evaluated their ability with quadvariate normal distribution associated with four random variables $(X_1,X_2,X_3,X_4)$. The four variables are separated into two groups $(X_1,X_2)$ and $(X_3,X_4)$ with the below covariance matrix 
\begin{equation}
	\left[
	\begin{array}{cccc}
	1&\rho_{12}&\rho&\rho\\
	\rho_{12}&1&\rho&\rho\\
	\rho&\rho&1&\rho_{34}\\
	\rho&\rho&\rho_{34}&1\\
	\end{array}
	\right],
\end{equation}
where $\rho_{12}$ and $\rho_{34}$ are the covariance for $(X_1,X_2)$ and $(X_3,X_4)$ respectively, and $\rho$ is for the dependence between two groups of variables. By changing the value of $\rho$, we can simulate different measure strength between two groups of random variables. In simulation, the mean of the quadvariate normal distribution is $\mathbf{0}$ and $\rho_{12}=0.8,\rho_{34}=0.75$. The $\rho$ was set from 0 to 0.8 by step 0.1. The sample size is 800. The dependence between $(X_1,X_2)$ and $(X_3,X_4)$ were estimated with the above ten measures from the simulated data.

\subsubsection{Real Data}
We evaluate the independence measures with two real data from the UCI machine learning repository: the heart disease data and the wine quality data.

\paragraph{Heart disease data}
The UCI heart disease data \cite{asuncion2007uci} concerns heart disease diagnosis from biomedical measurements. It was collected from four locations worldwide, including Cleveland, Budapest, California, and Zurich. This dataset contains 899 samples, each with 76 attributes, including diagnosis and other biomedical attributes. Among them, 13 attributes were recommended by professionals to be valuable for diagnosis.

Previously, we have used this data to study the variable selection problem that is to select the biomedical attributes related to diagnosis in the dataset \cite{Ma2021}. Three measures, including CE, dCor and dHSIC were compared on this problem. In this study, we will use the same setting to compare all the 16 independence measures. The measures of the dependence between diagnosis and other attributes are estimated from the data and the attributes above a threshold are considered to be selected as associated with diagnosis. Since the measures work at different scale, we cannot compare the estimation results directly. To be fair, a common threshold for all the measures is chosen as the dependence between diagnosis and fbs (\#16) of each measures after balancing the tradeoff between true positive and false positive of selections. The number of the recommended attributes such selected by each measure will be compared. 

\paragraph{Wine quality data}
The UCI wine quality data \cite{asuncion2007uci} contains two datasets about the red and white variants of the Portuguese "Vinhe Verde" wine. It has 12 attributes, including 11 physicochemical and 1 sensory attributes. The aim is to study the relationship between physicochemical and sensory attributes. 

Even though evaluating the wine quality is somewhat subjective, there are still some common sense in the oenological theory \cite{Cortez2009}. For example, an increase in the alcohol tends to relate to high quality. The citric acid and residual sugar levels tends to make white wine with a good balance between freshness and sweet taste. The volatile acidity is negatively related to wine quality.

In the experiment, the measures of the dependence between physicochemical and sensory attributes will be estimated from the data on white wine. The estimation results will be interpreted with reference to the common sense in the oenological theory.

\subsection{Conditional Independence Measures}
\subsubsection{Simulated data}
\paragraph{Trivariate normal distribution}
To evaluate the CI measure, we did a simulation with trivariate normal distribution on random variables $X,Y,Z$. The covariance matrix of these variables is as follows:
\begin{equation}
	\left[
	\begin{array}{ccc}
	1&\rho_{xy}&\rho_{xz}\\
	\rho_{xy}&1&\rho_{yz}\\
	\rho_{xz}&\rho_{yz}&1\\
	\end{array}
	\right],
	\label{eq:cov3}
\end{equation}
where $\rho_{xy},\rho_{xz},\rho_{yz}$ are the covariance between random variables respectively. In simulation, the mean of the variables is $\mathbf{0}$ and the $\rho_{xy}=0.7,\rho_{yz}=0.6$, and the $\rho_{xz}$ was set from 0 to 0.9 by step 0.1. In this way, the conditional dependence between $(X,Y)$ given $Z$ are decreasing monotonically as $\rho_{xz}$ decreases. The sample size is 800 as before.

In the experiment for wGCM, the XGBoost algorithm was chosen for residual regression and the parameter 'beta' for weight function is set as 0.7.

\paragraph{Trivariate normal copula}
To evaluate the CI measures on nonlinear cases, we simulate a group of data from trivariate normal copula. The covariance matrix of the normal copula is the same as \eqref{eq:cov3} and again $\rho_{xy}=0.7,\rho_{yz}=0.6$, and the $\rho_{xz}$ was set from 0 to 0.9 by step 0.1. The three marginals associated with the normal copula are a normal distribution with 0 mean and standard deviation as 2, and two exponential distributions with rate as 0.5 and 2 respectively. The sample size is 800. The parameter setting for wGCM is the same as the above experiments.

\subsubsection{Real data}
\paragraph{Beijing air data}
The UCI Beijing air data \cite{asuncion2007uci} was used for evaluating the CI measures. It is about air pollution at Beijing. This hourly data set contains the PM2.5 data of US Embassy in Beijing. Meanwhile, meteorological data from Beijing Capital International Airport are also included.

Meteorological factors in data include dew point, temperature, pressure, cumulated wind speed, combined wind direction, cumulated hours of snow, cumulated hours of rain. The pressure factors are analyzed in our experiments. The data was collected hourly from Jan. 1st, 2010 to Dec. 31st, 2014, which results in 43824 samples with missing values. To avoid tackling missing values, only the data from April 2nd, 2010 to May 14th, 2010 were used in our experiments, which contains 1000 samples without missing values. 

Previously, we have used the data for studying the method for estimating transfer entropy which is essentially CMI \cite{ma2021estimating}. In that work, we proposed a method for estimating transfer entropy via CE and compared the method with CDC and KCI. 

In this study, the CI measures of the CI between the factor and the future value of PM2.5 given the historical value of PM2.5 were estimated from the data. The time lags range from 1 hour to 24 hours. Due to space limits, we studied only the pressure factor here.

\section{Results}
\label{sec:results}
\subsection{Independence Measures}
\subsubsection{Simulated data}
\begin{figure}
	\subfigure[CE]{\includegraphics[width=0.245\linewidth]{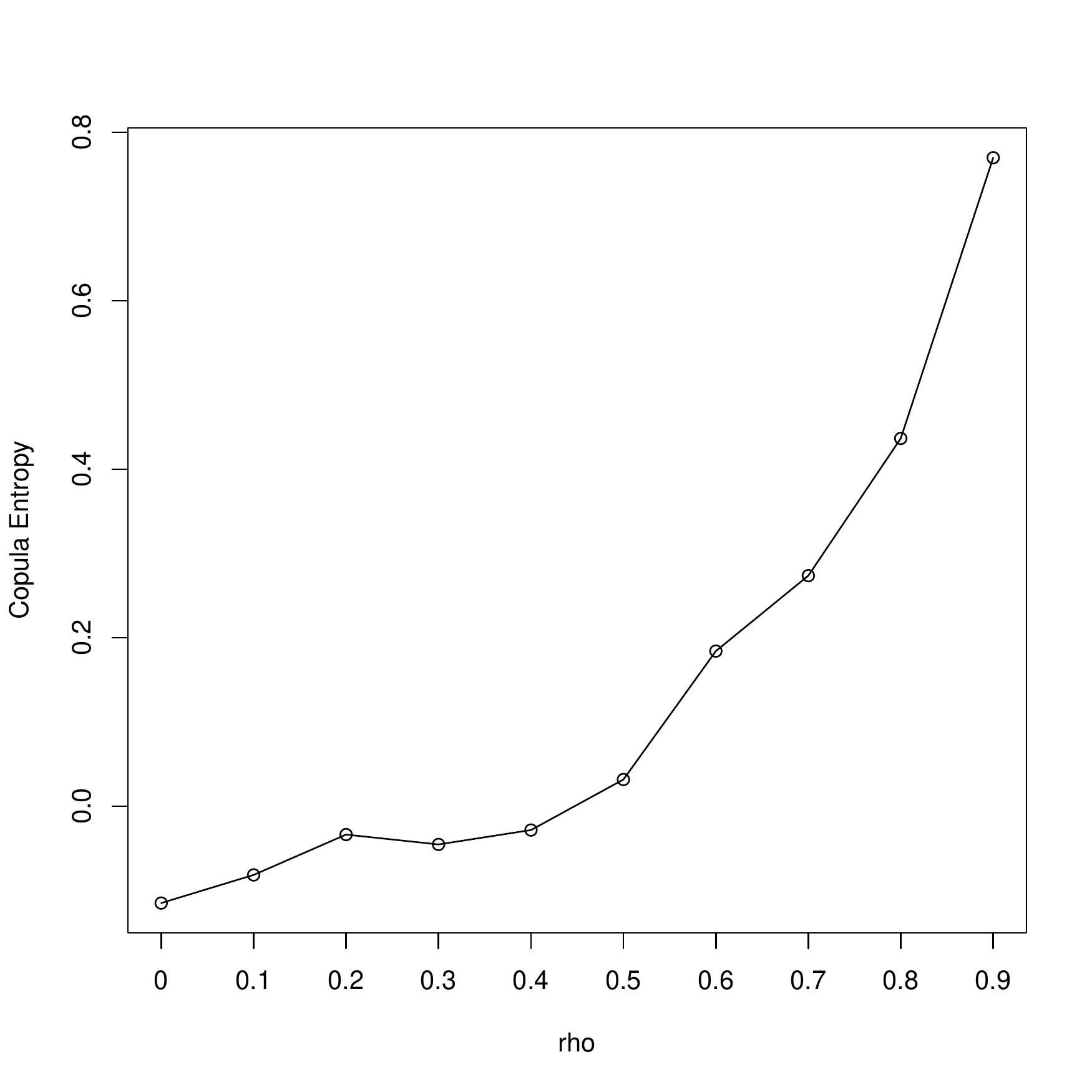}}
	\subfigure[Ktau]{\includegraphics[width=0.245\linewidth]{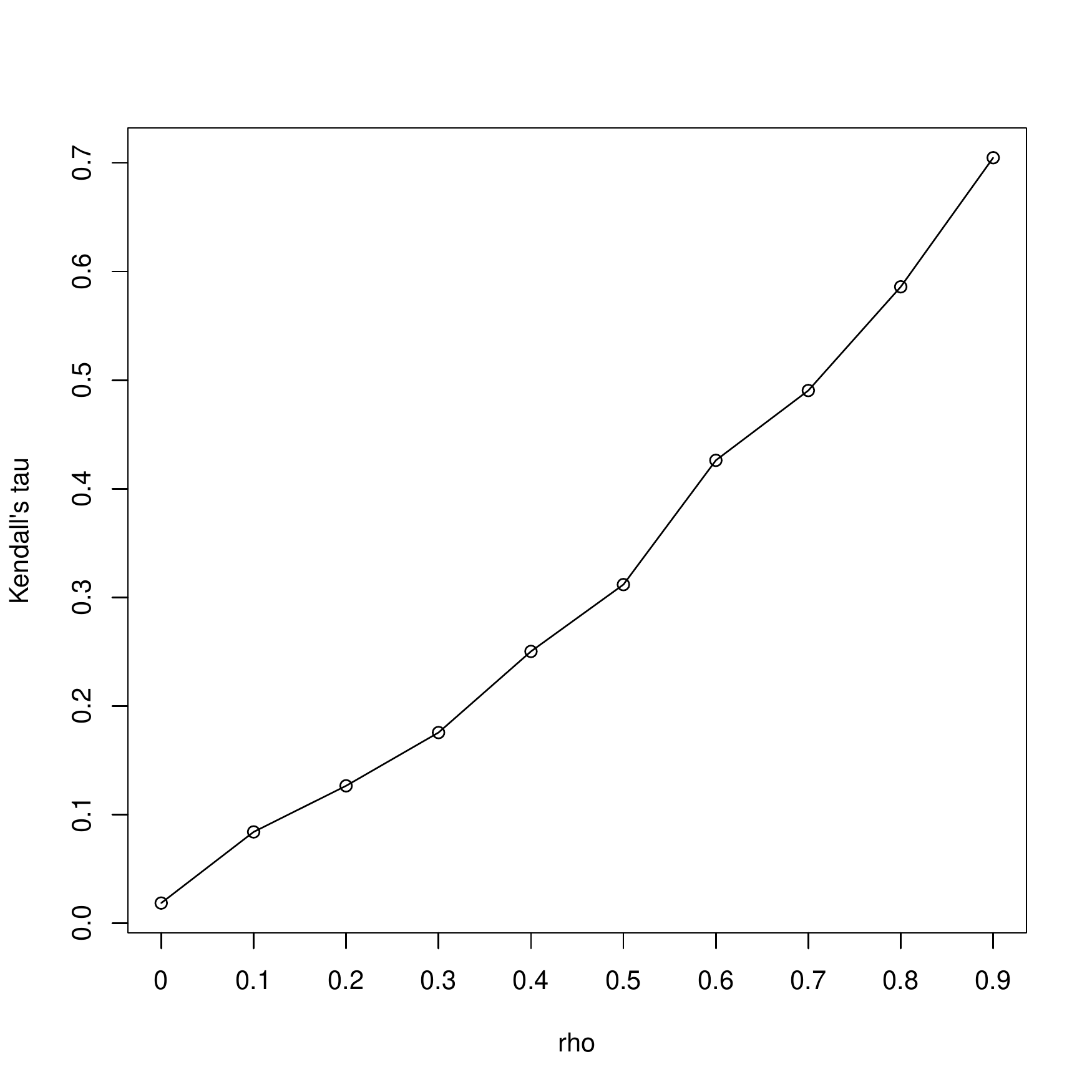}}
	\subfigure[Hoeff]{\includegraphics[width=0.245\linewidth]{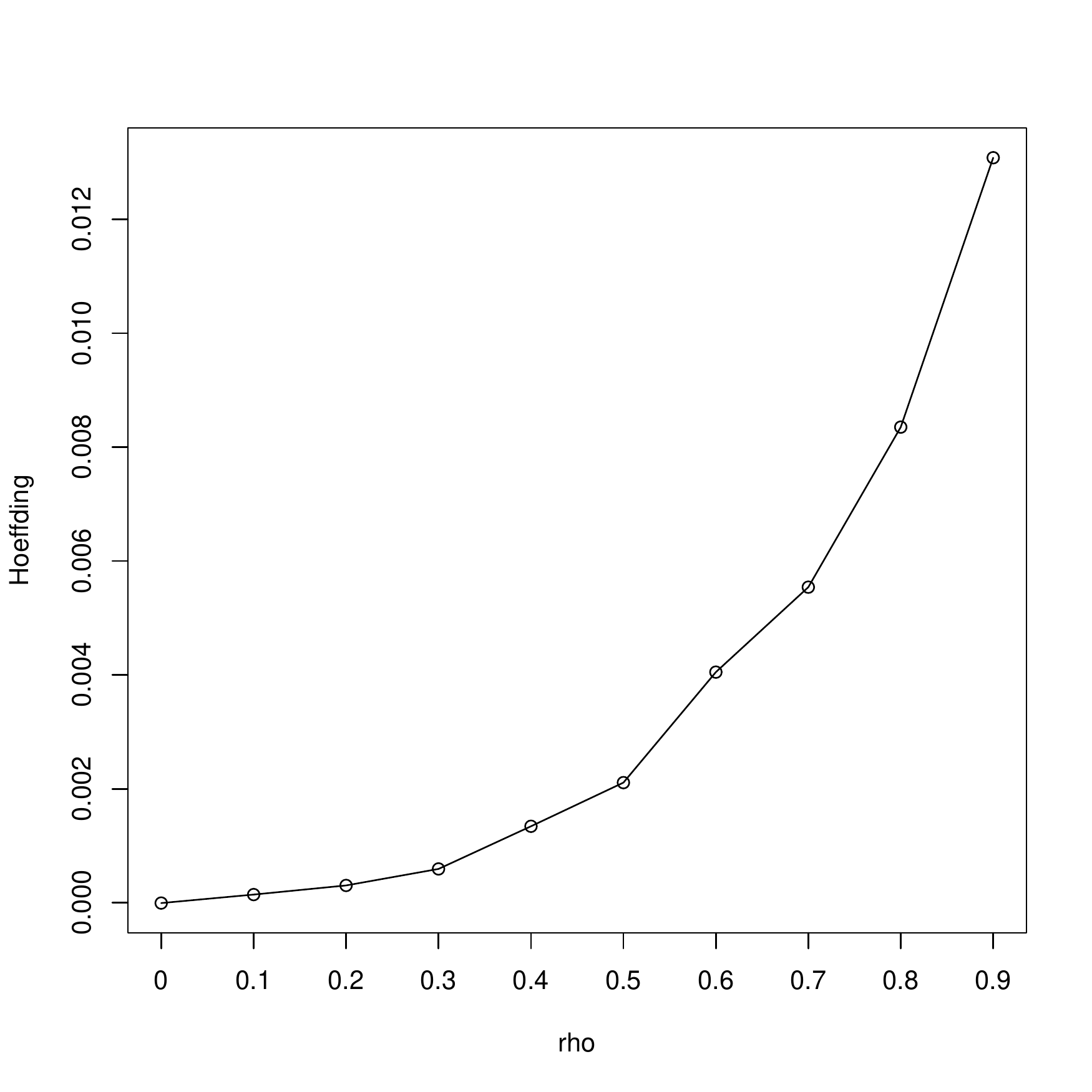}}
	\subfigure[BDtau]{\includegraphics[width=0.245\linewidth]{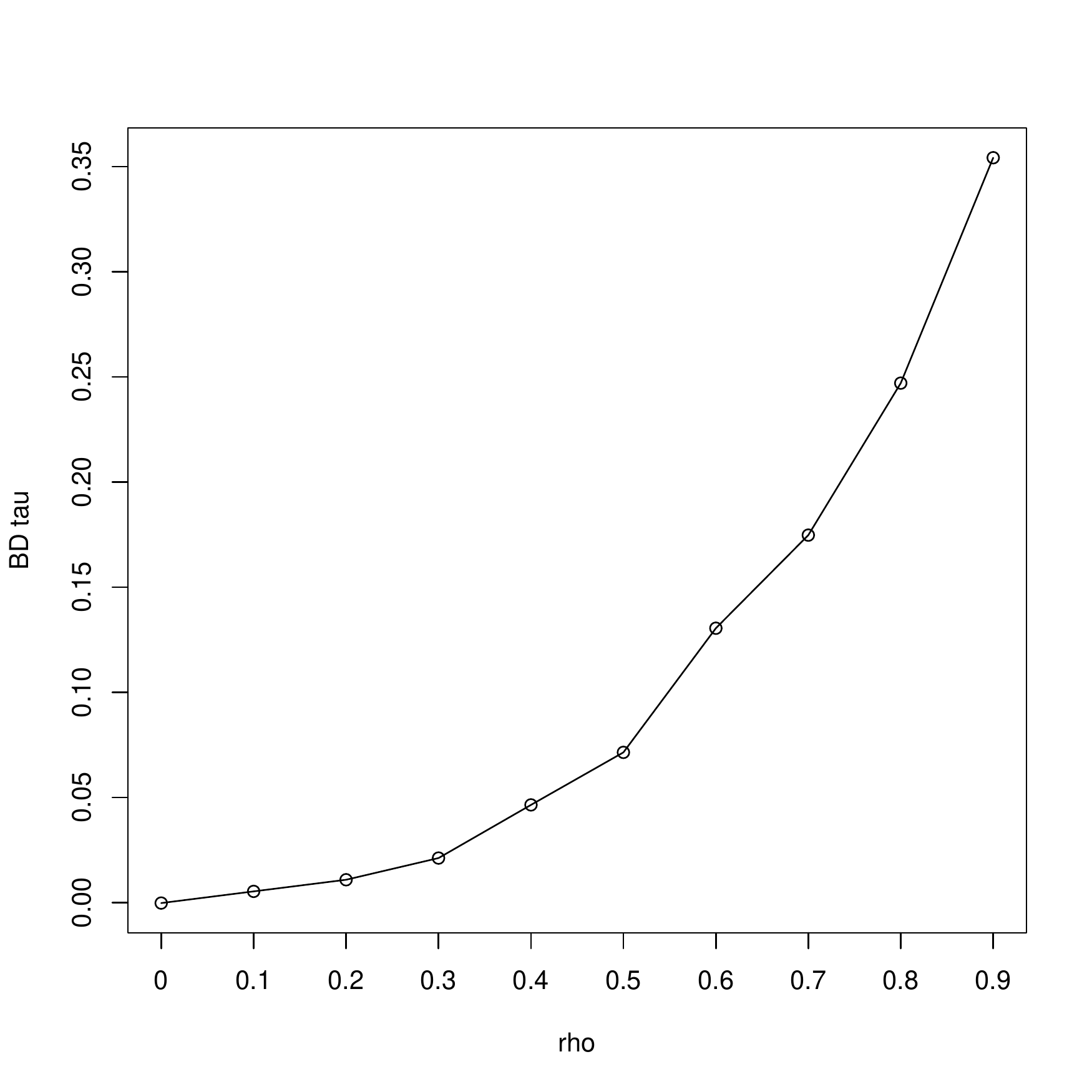}}
	\subfigure[HHG.chisq]{\includegraphics[width=0.245\linewidth]{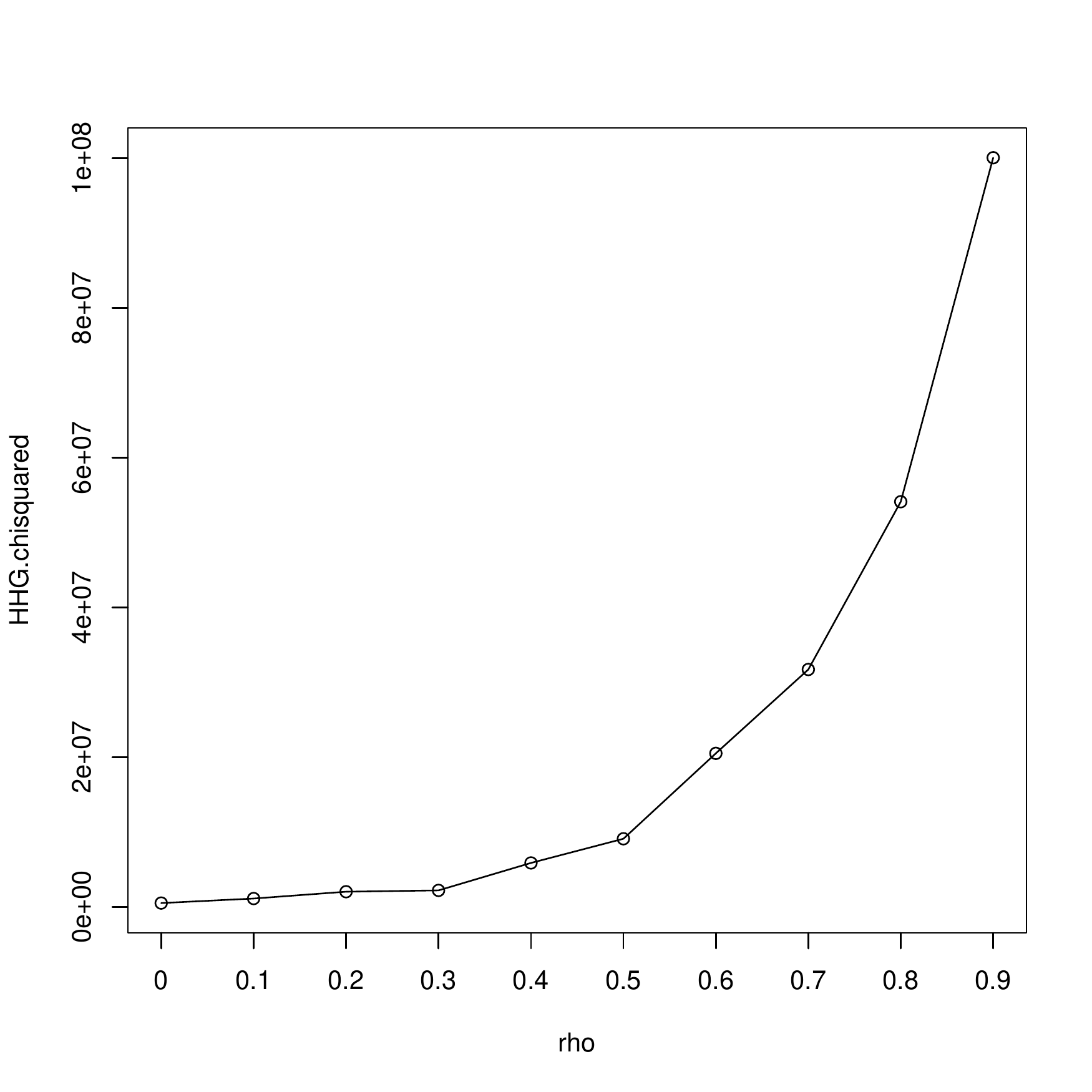}}
	\subfigure[HHG.lr]{\includegraphics[width=0.245\linewidth]{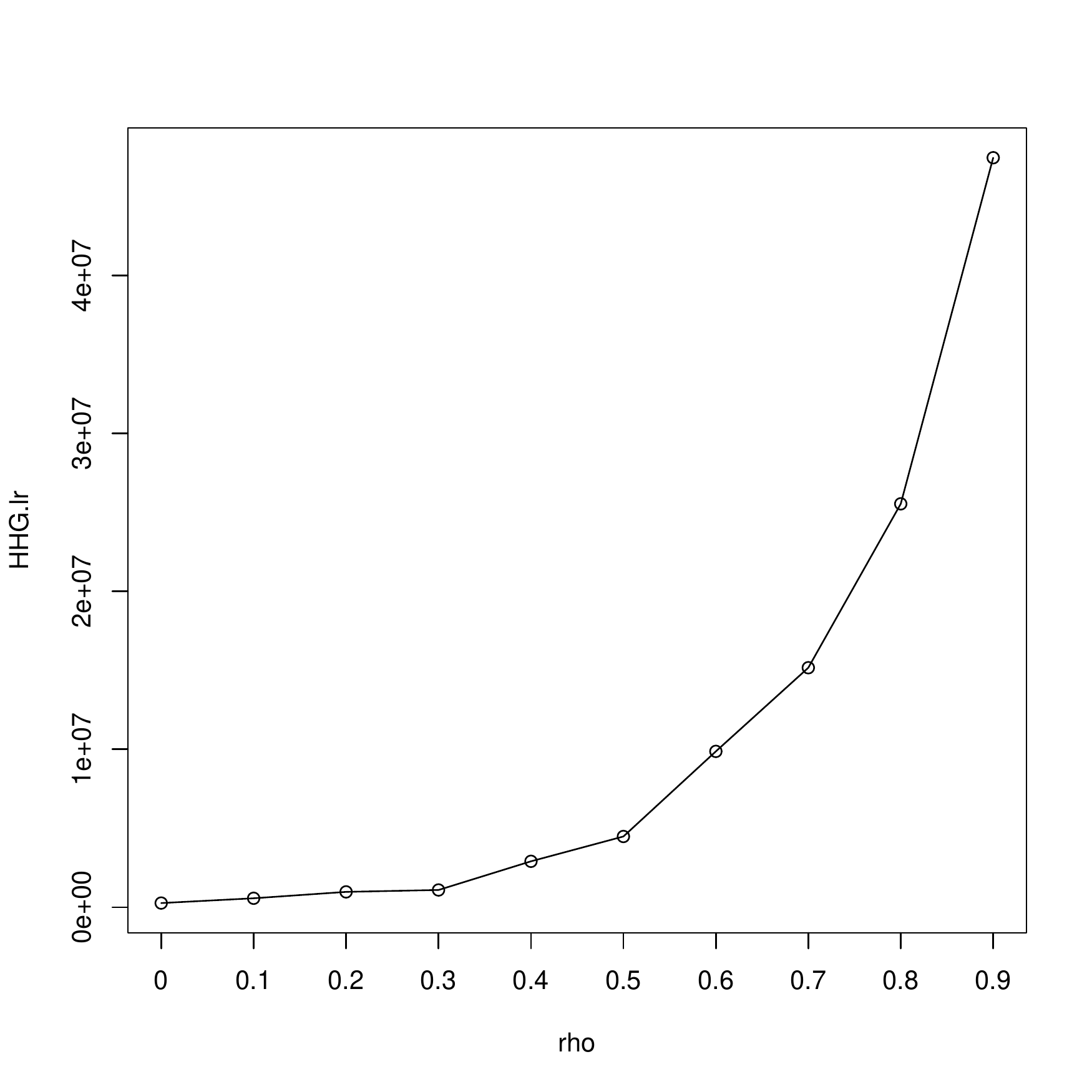}}
	\subfigure[Ball]{\includegraphics[width=0.245\linewidth]{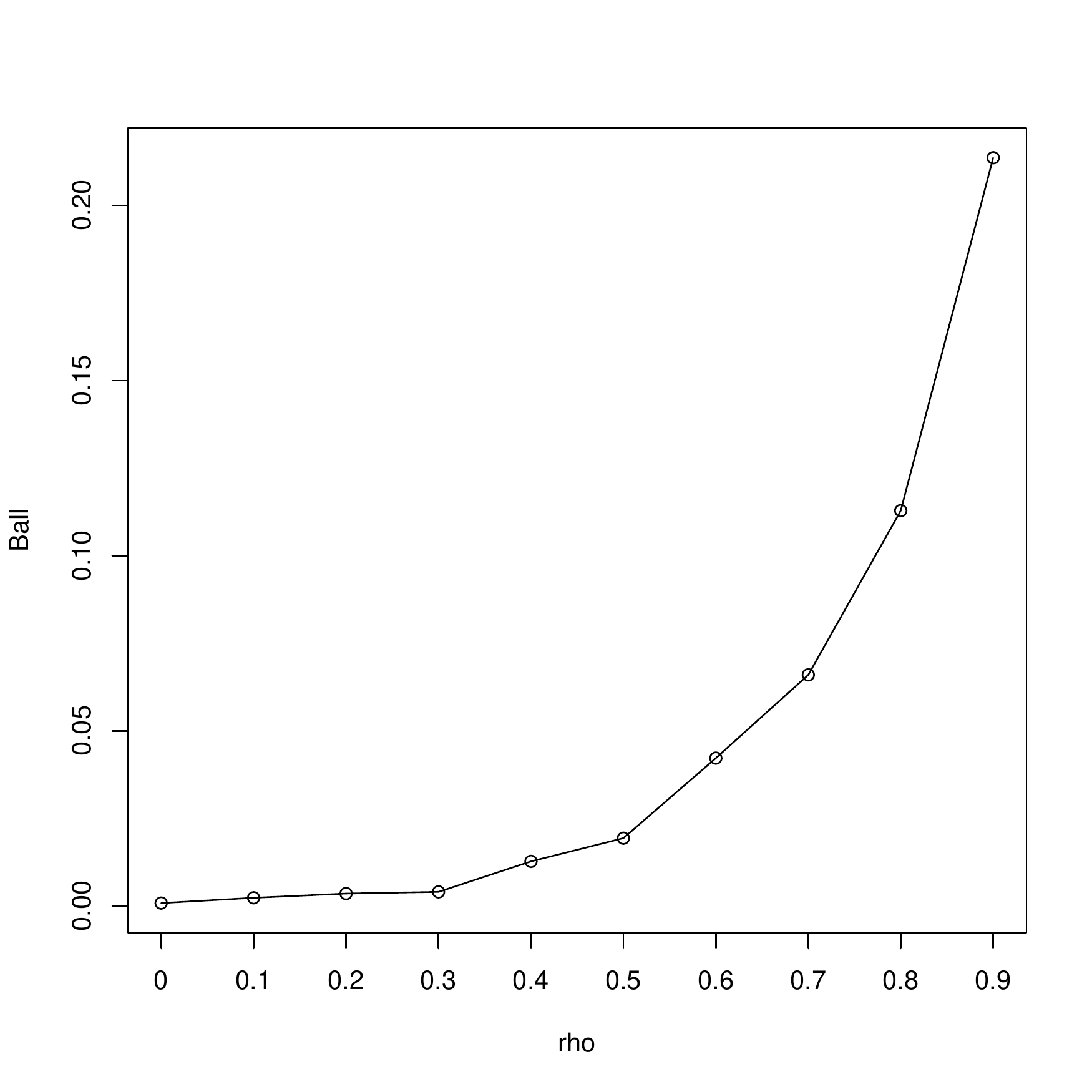}}
	\subfigure[BET]{\includegraphics[width=0.245\linewidth]{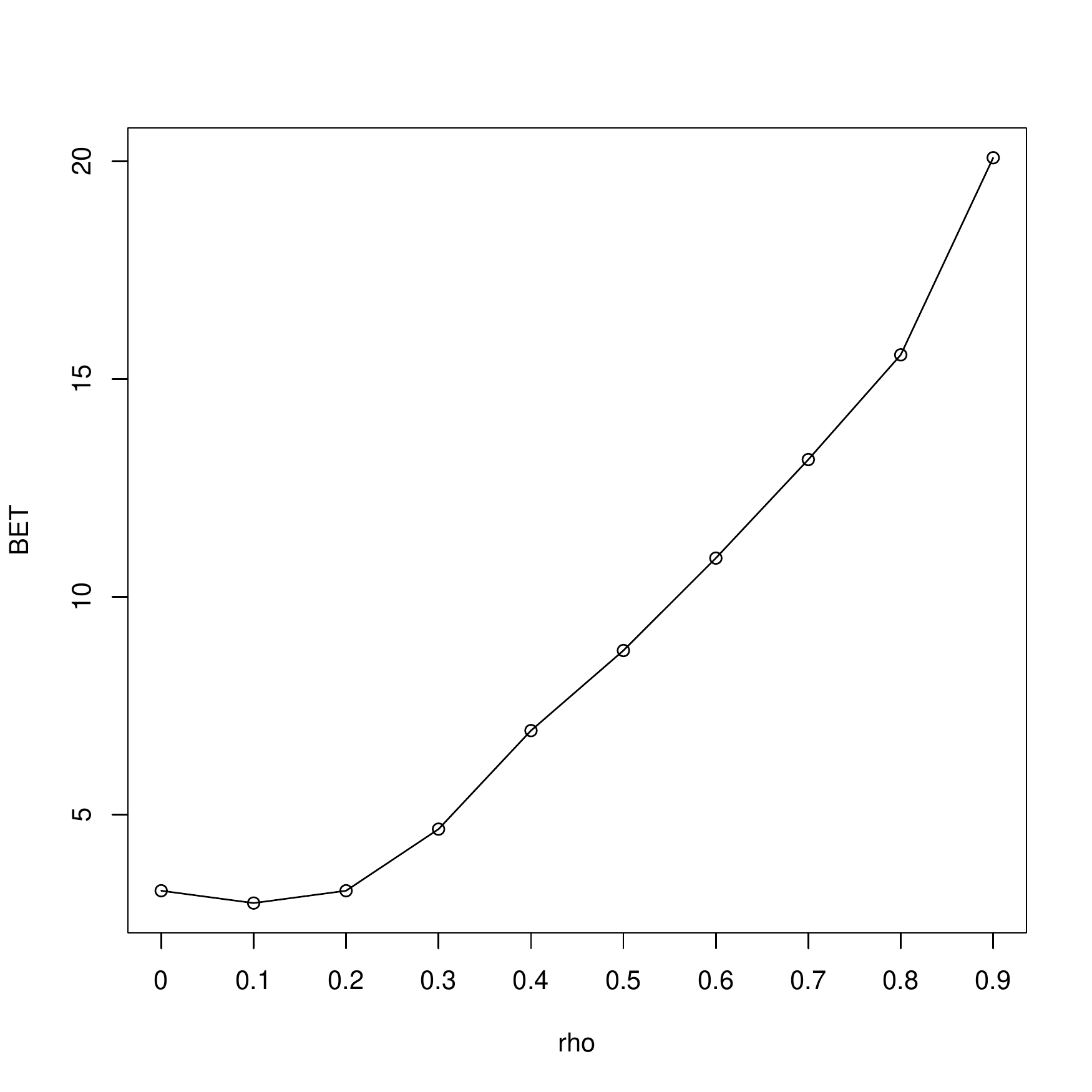}}
	\subfigure[QAD]{\includegraphics[width=0.245\linewidth]{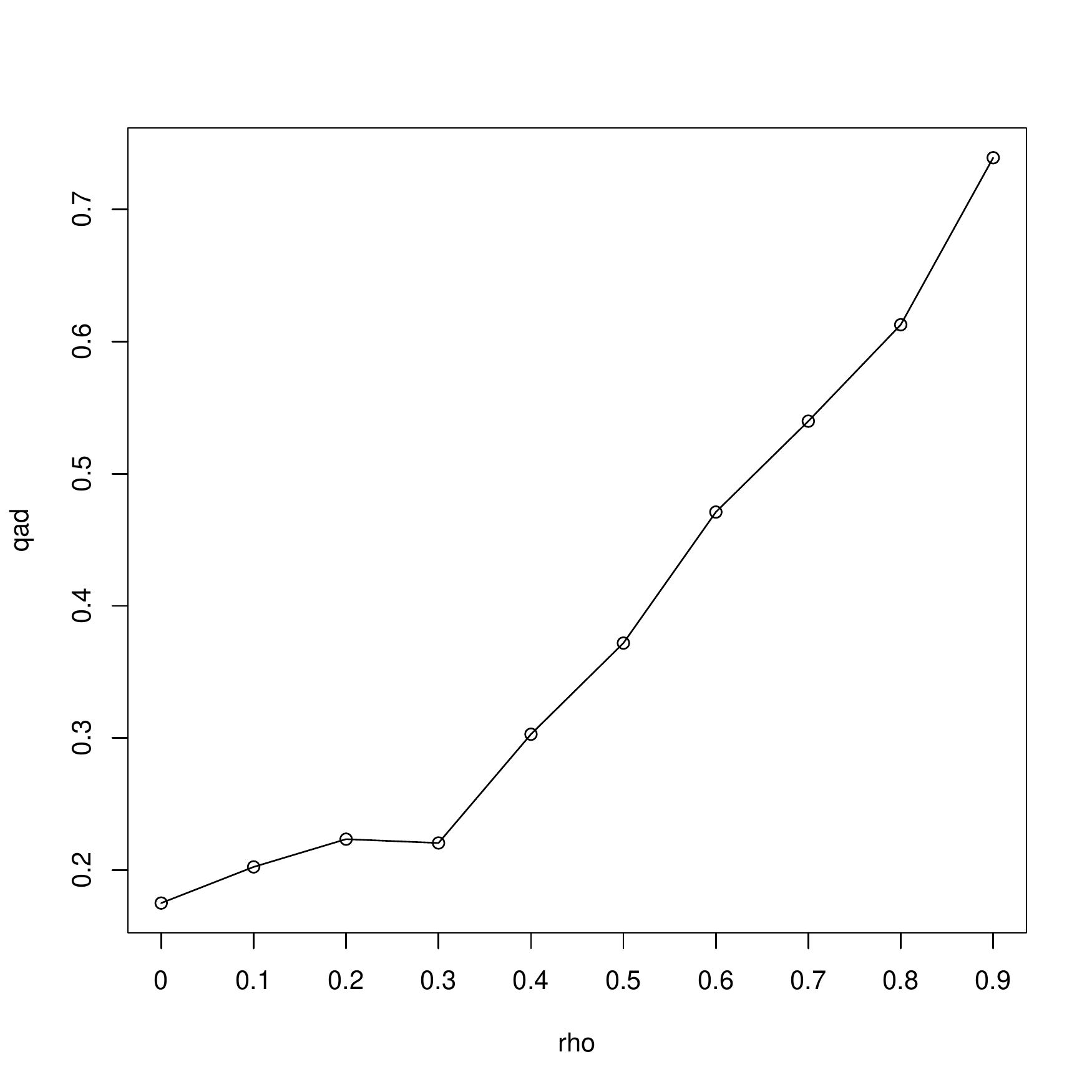}}
	\subfigure[mixed]{\includegraphics[width=0.245\linewidth]{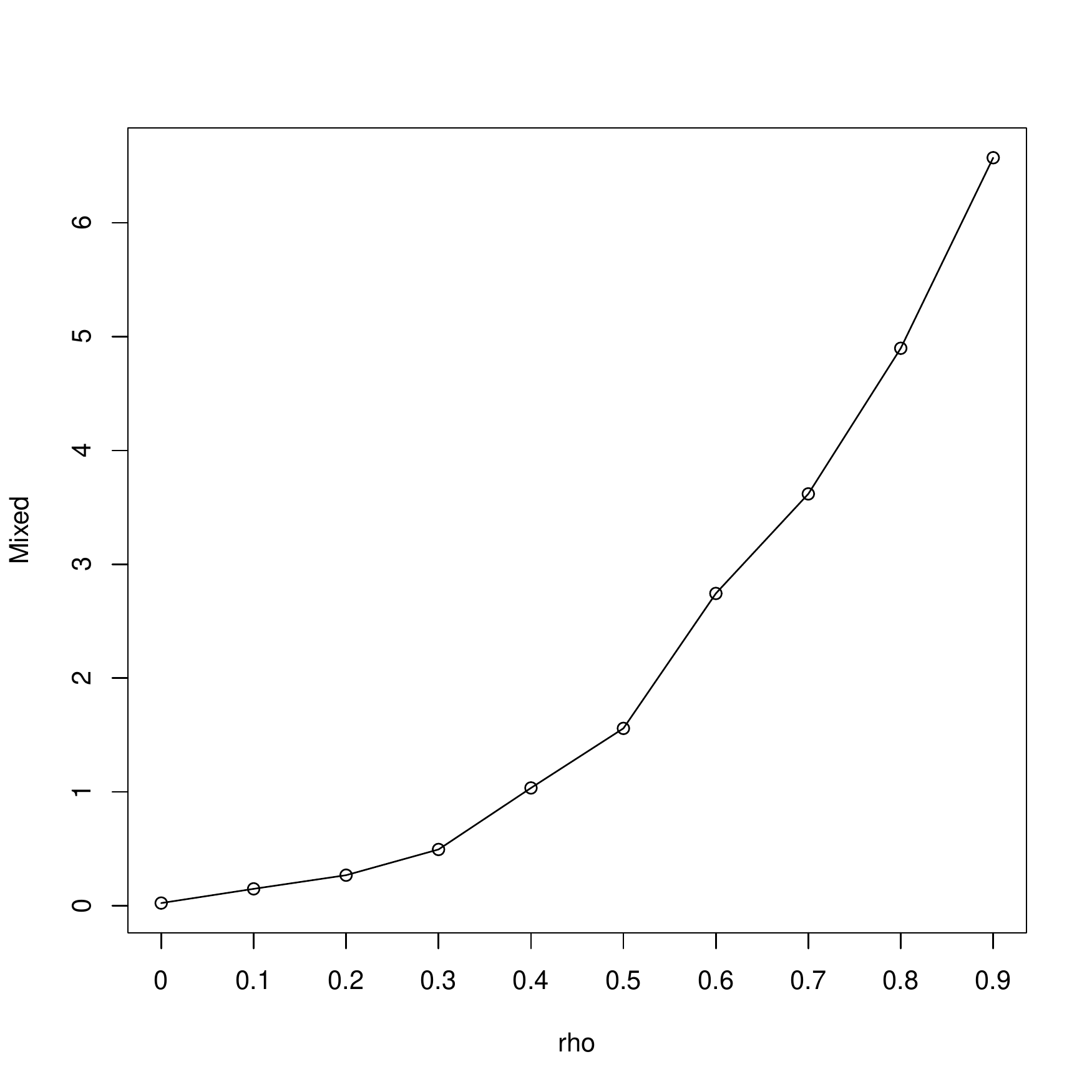}}
	\subfigure[CODEC]{\includegraphics[width=0.245\linewidth]{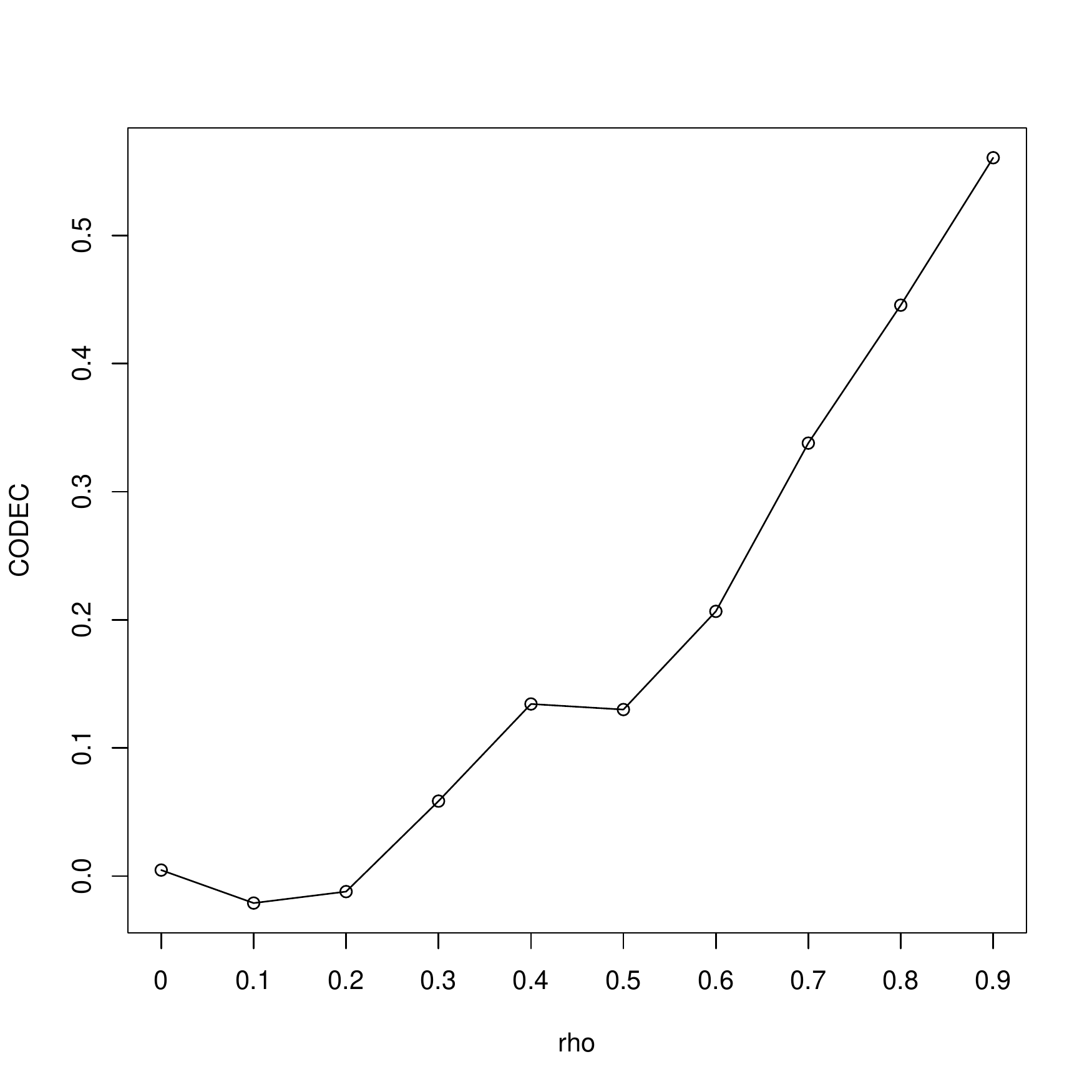}}
	\subfigure[subcop]{\includegraphics[width=0.245\linewidth]{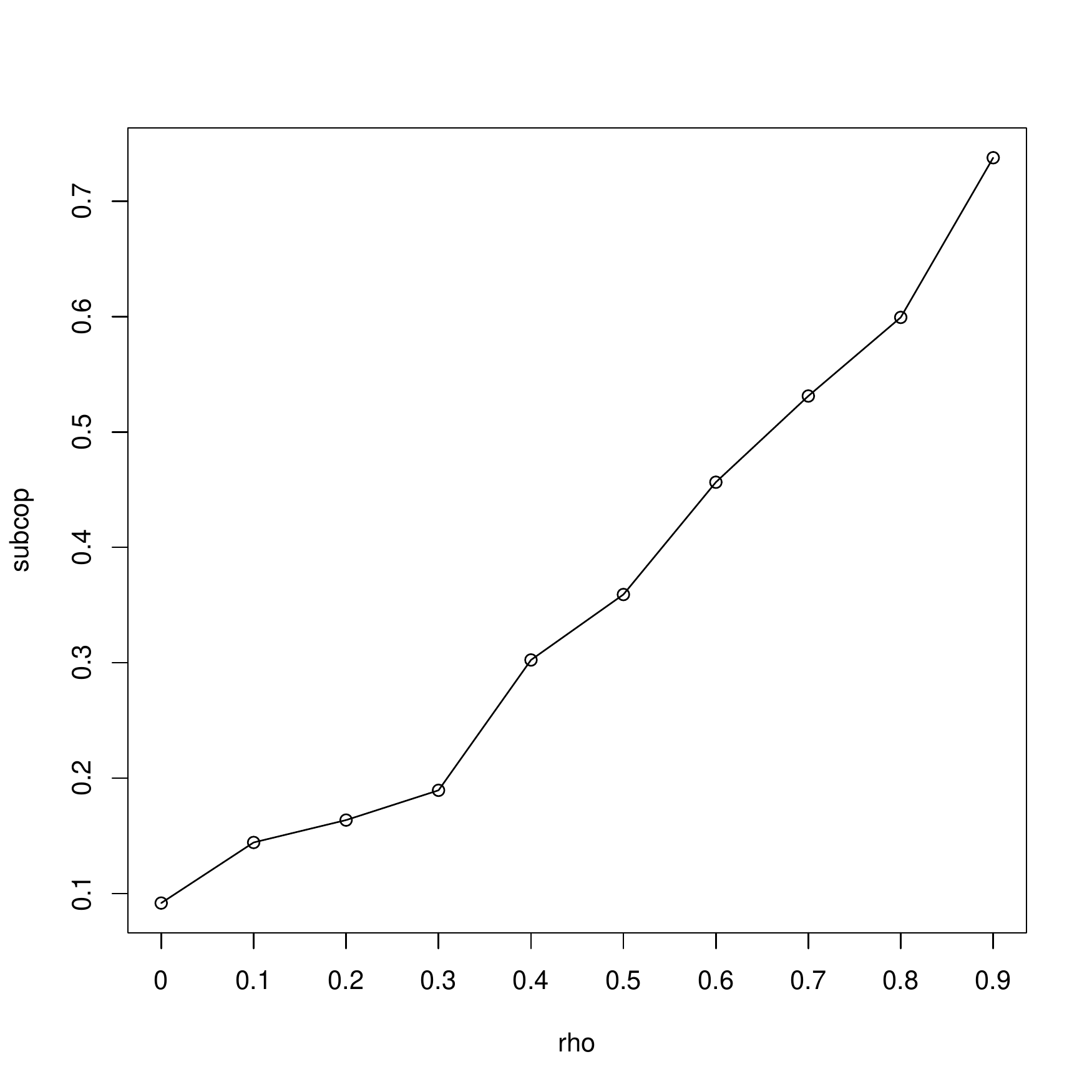}}
	\subfigure[dCor]{\includegraphics[width=0.245\linewidth]{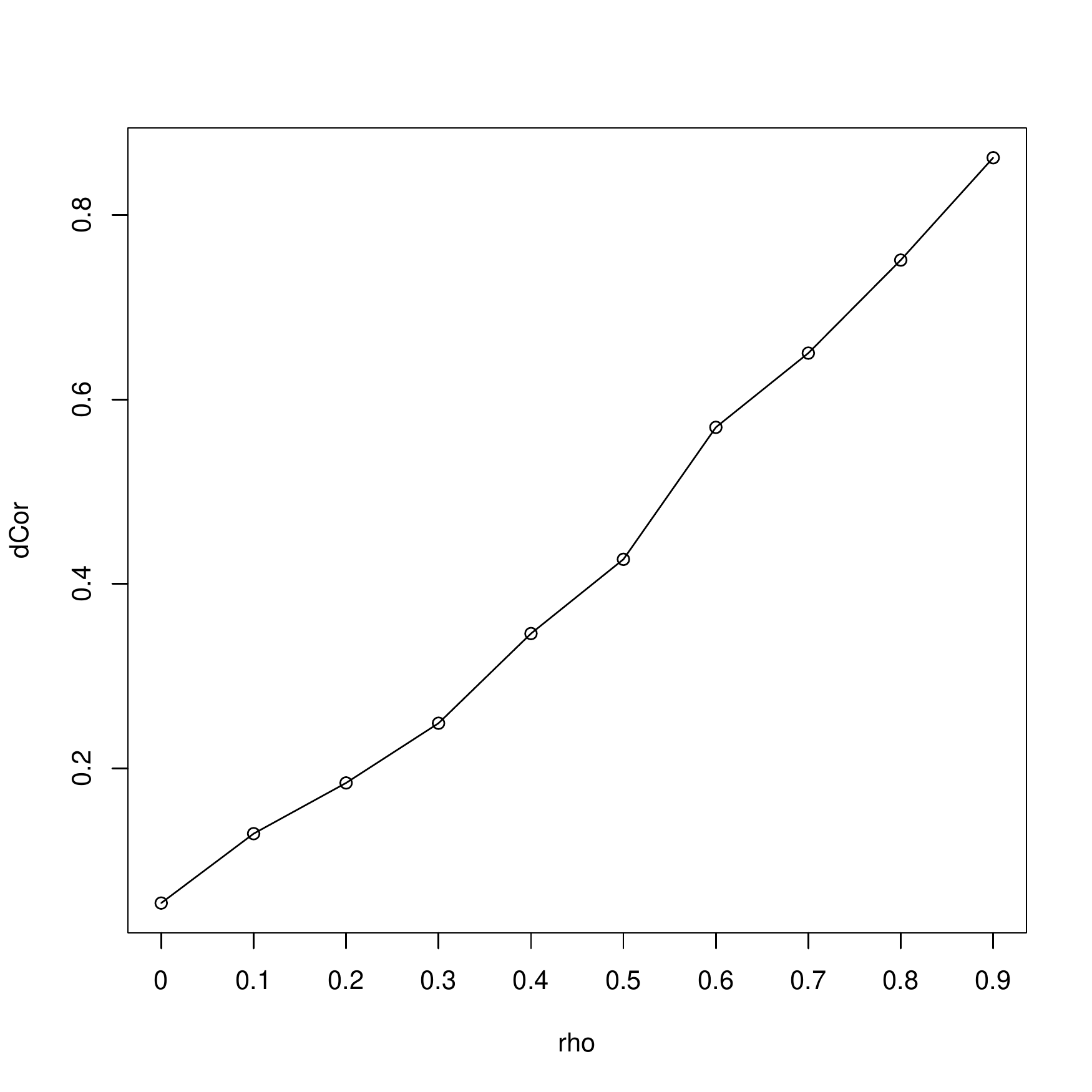}}
	\subfigure[mdm]{\includegraphics[width=0.245\linewidth]{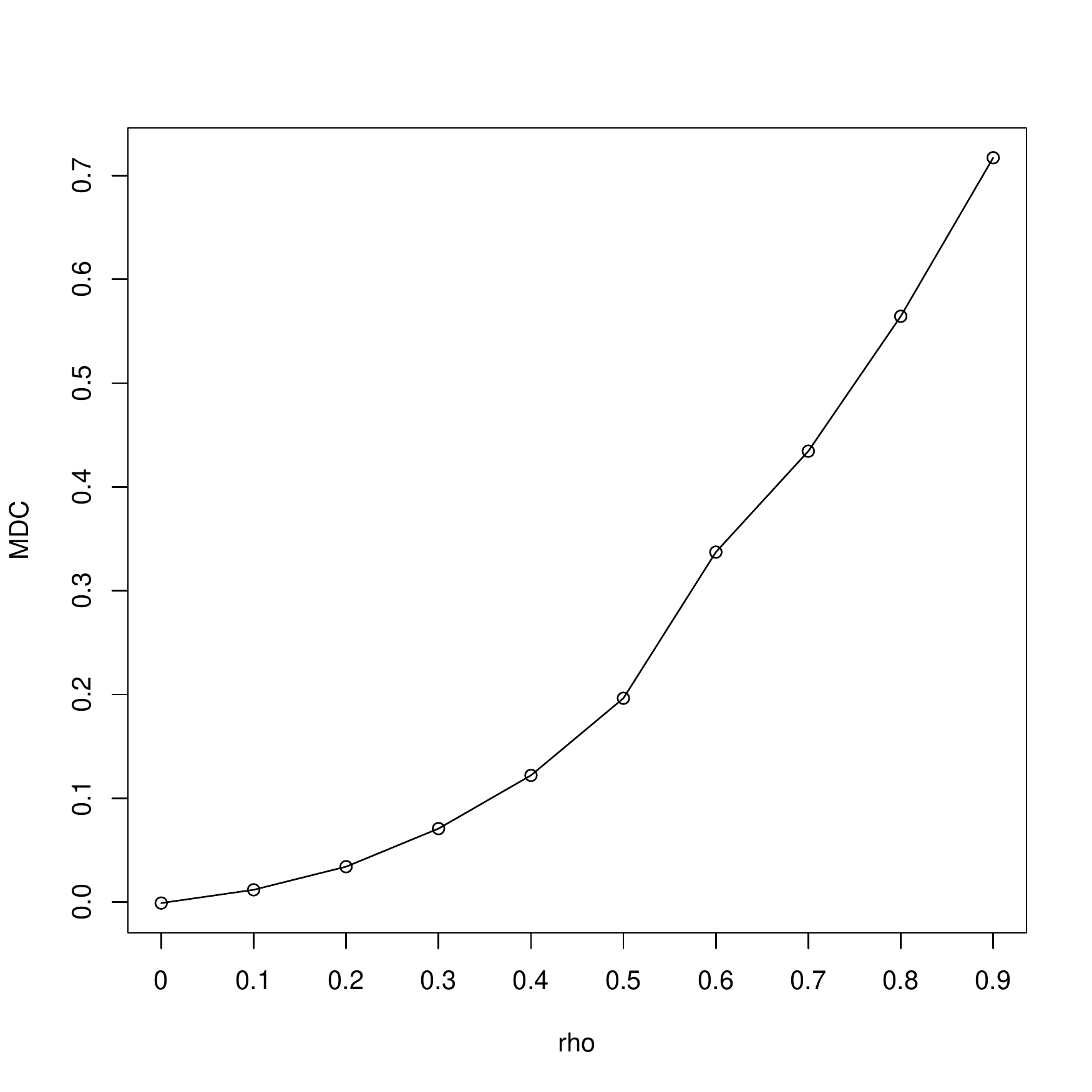}}
	\subfigure[dHSIC]{\includegraphics[width=0.245\linewidth]{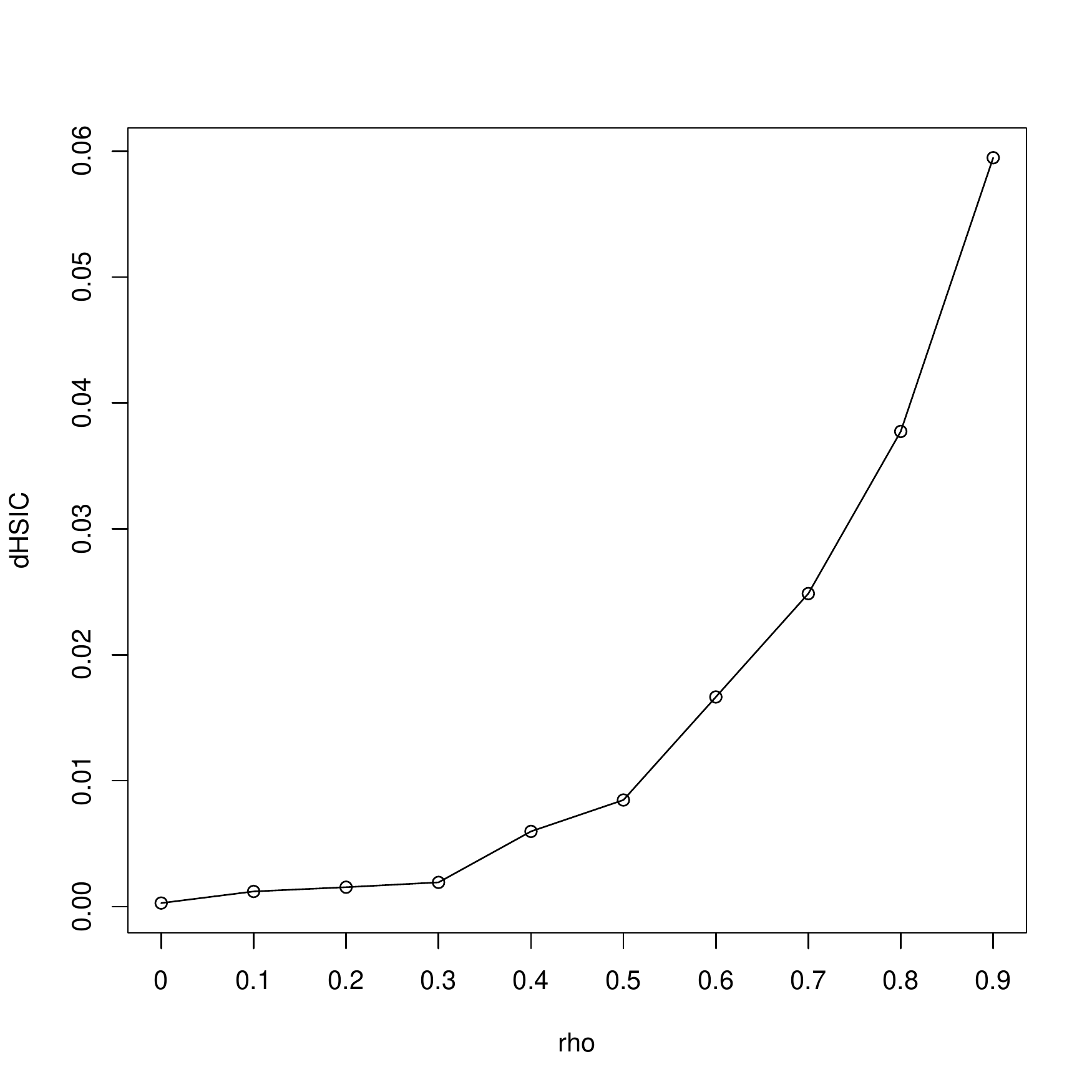}}
	\subfigure[NNS]{\includegraphics[width=0.245\linewidth]{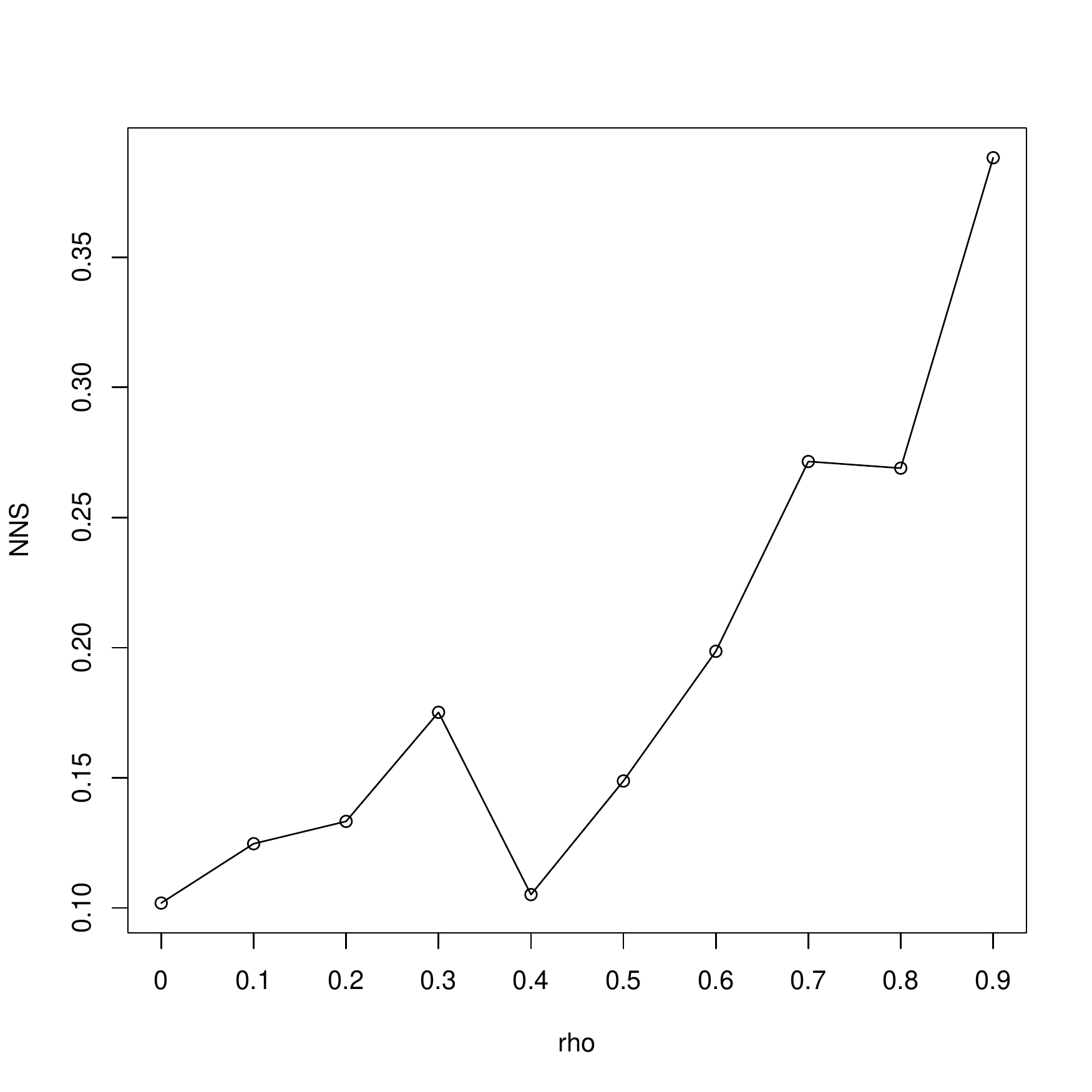}}
	\caption{Estimation of the independence measures from the simulated data of the bivariate normal distribution.}
	\label{fig:binormal}
\end{figure}

\begin{figure}
	\centering
	\includegraphics[width=0.9\linewidth]{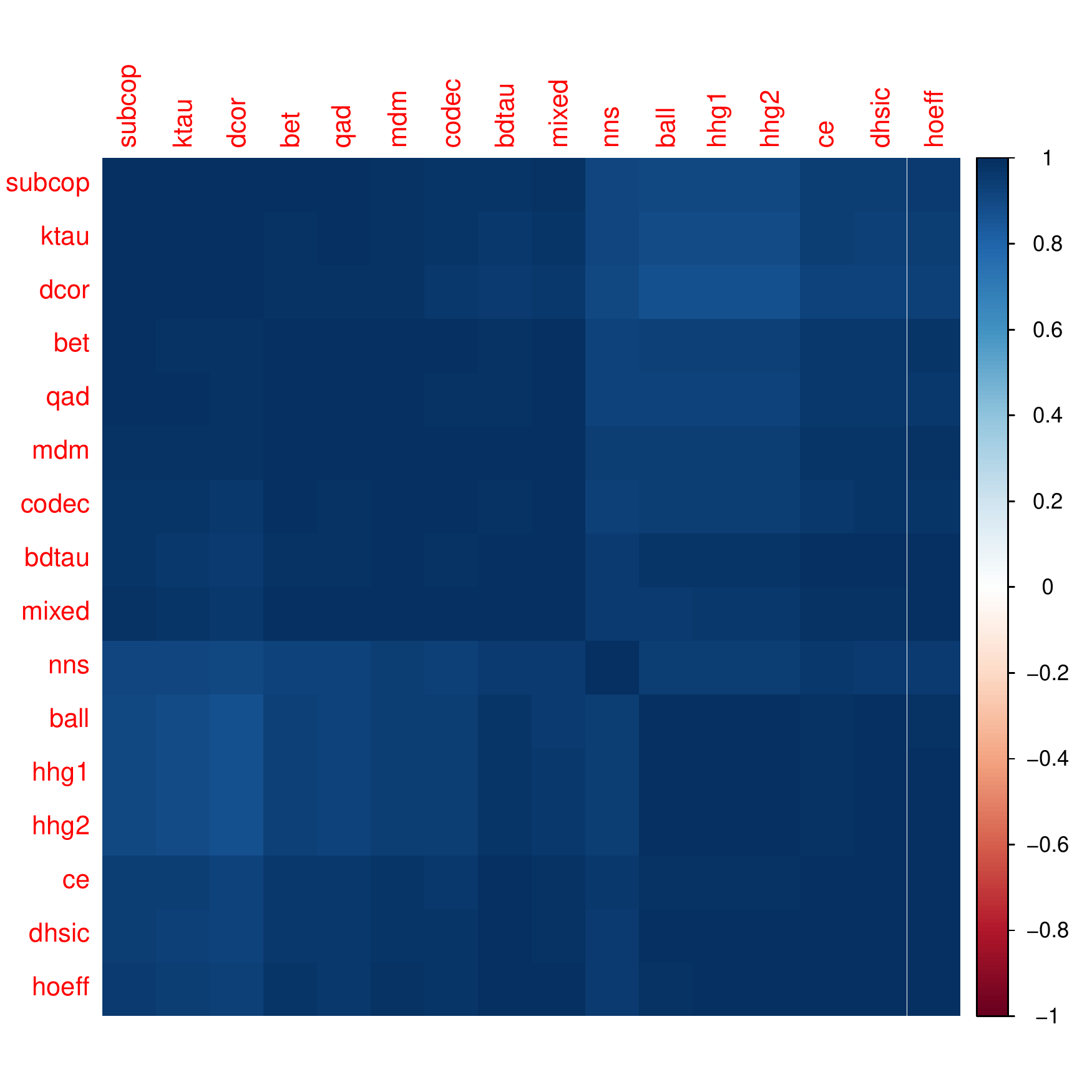}
	\caption{Correlation matrix of the independence measures estimated from the simulated data of the bivariate normal distribution.}
	\label{fig:binormalcm}
\end{figure}

\paragraph{Bivariate normal distribution}
For the first experiment, the estimation of the independence measures from the bivariate normal distribution is shown in Figure \ref{fig:binormal}. The correlation matrix of the estimation of the measures is shown in Figure \ref{fig:binormalcm}. It can be learned from these two figures that the estimation results can be roughly seperated into 5 groups: \{CE, dHSIC, hoeff\},\{Ball,HHG\}, \{BET, QAD, MDM, CODEC, mixed\},\{subcop, Ktau, dCor\},and \{NNS\}. As the $\rho$ increases, the measures in the first and second groups increase exponentially while the the measures in the third and fourth group increase approximately linearlly.

\begin{figure}
	\subfigure[CE]{\includegraphics[width=0.245\linewidth]{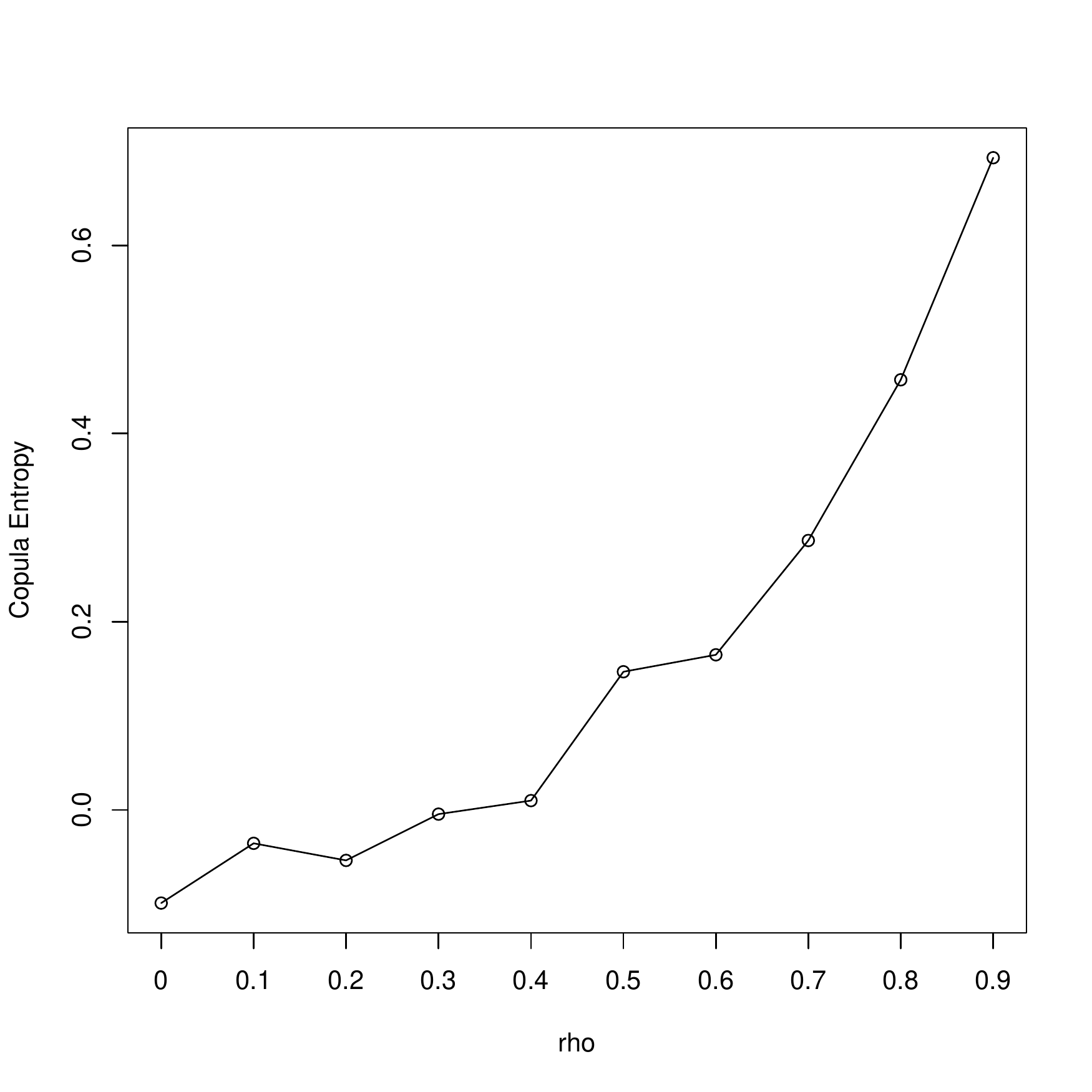}}
	\subfigure[Ktau]{\includegraphics[width=0.245\linewidth]{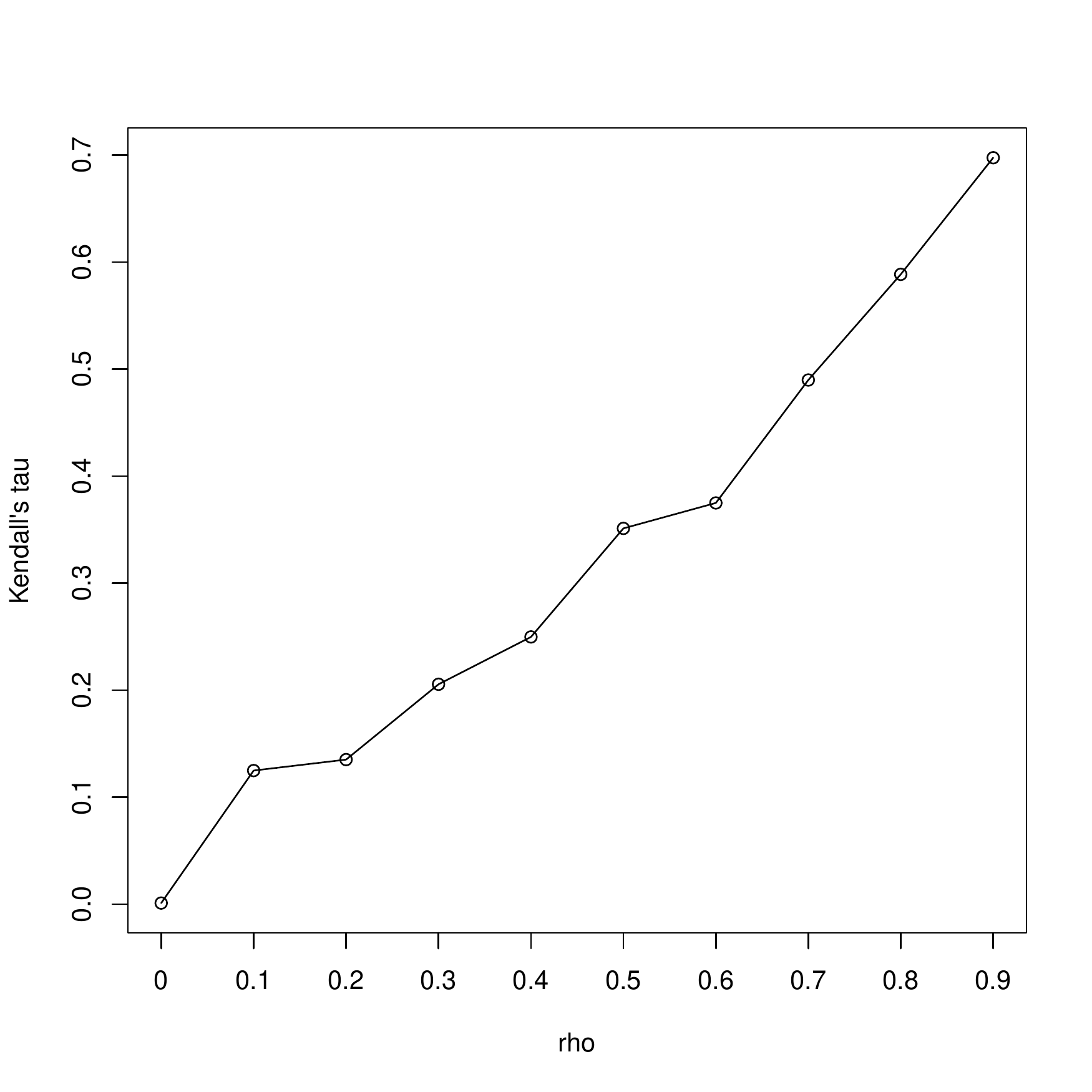}}
	\subfigure[Hoeff]{\includegraphics[width=0.245\linewidth]{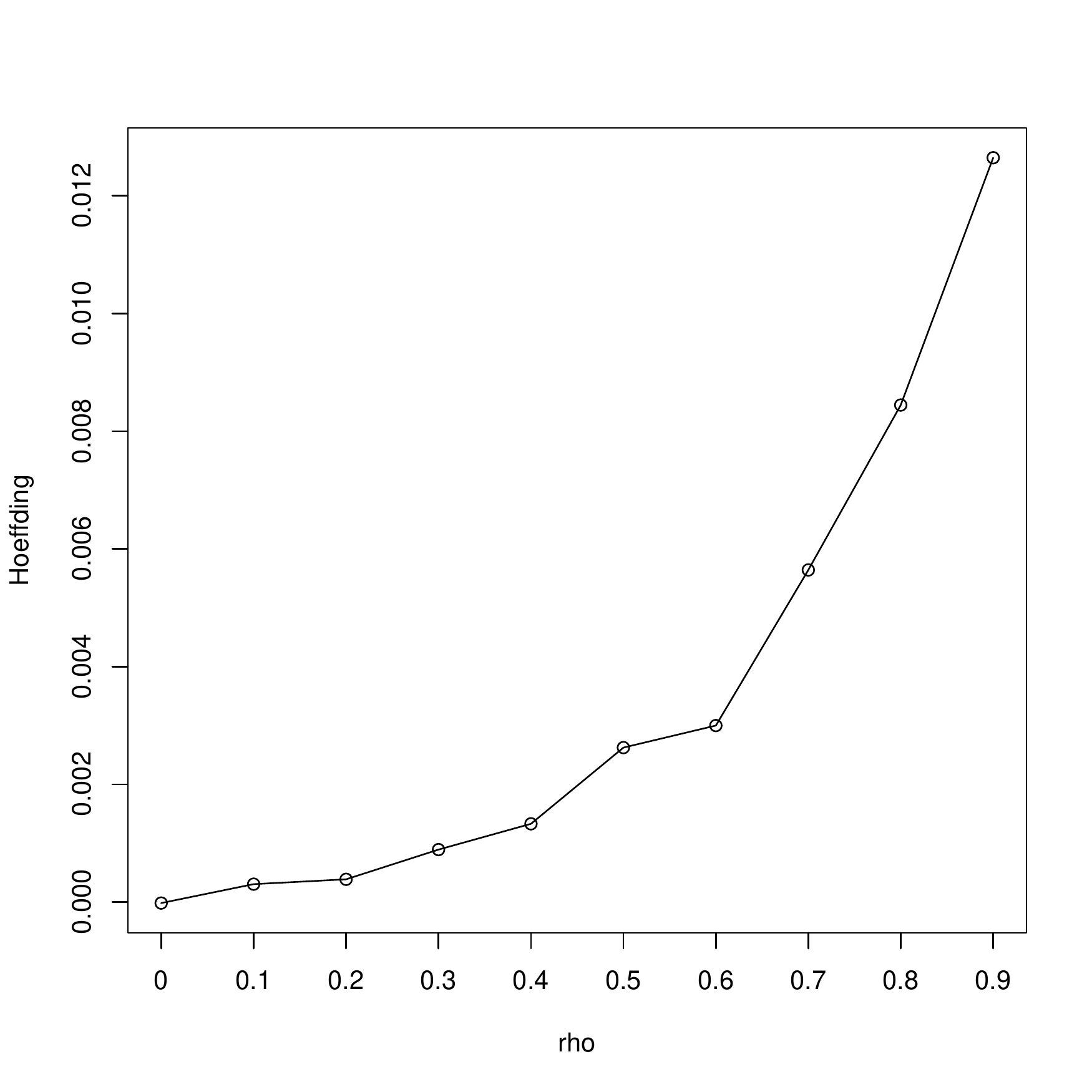}}
	\subfigure[BDtau]{\includegraphics[width=0.245\linewidth]{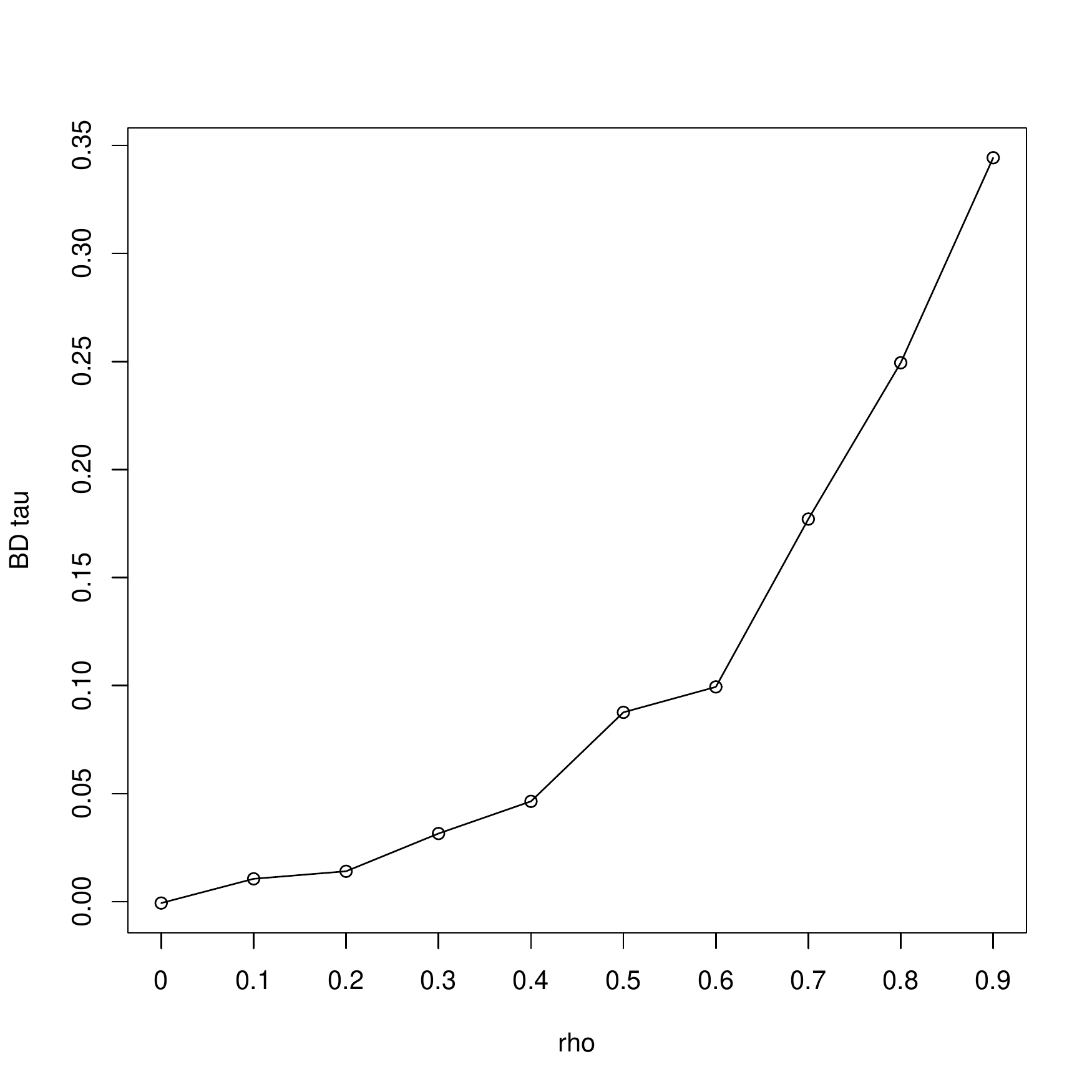}}
	\subfigure[HHG.chisq]{\includegraphics[width=0.245\linewidth]{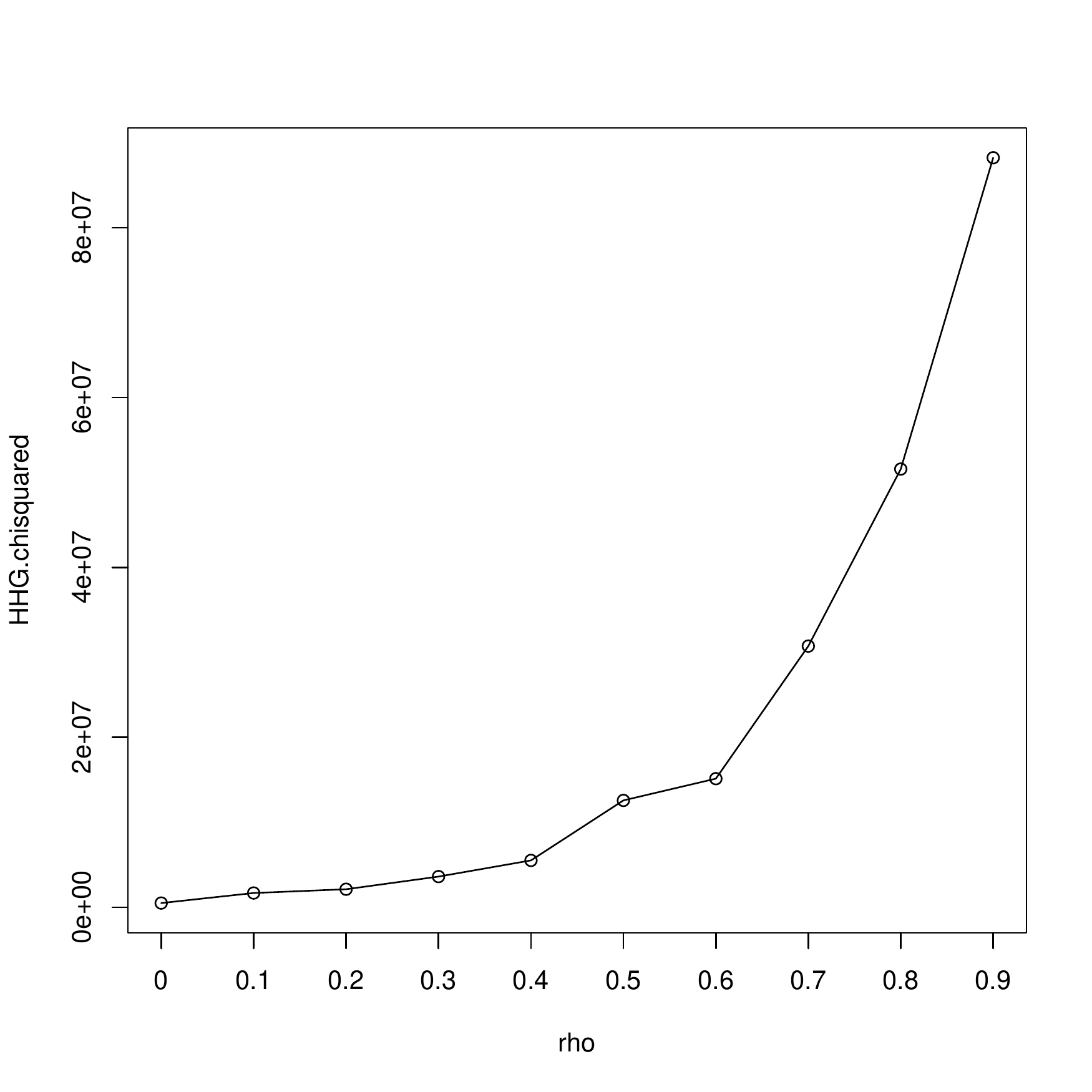}}
	\subfigure[HHG.lr]{\includegraphics[width=0.245\linewidth]{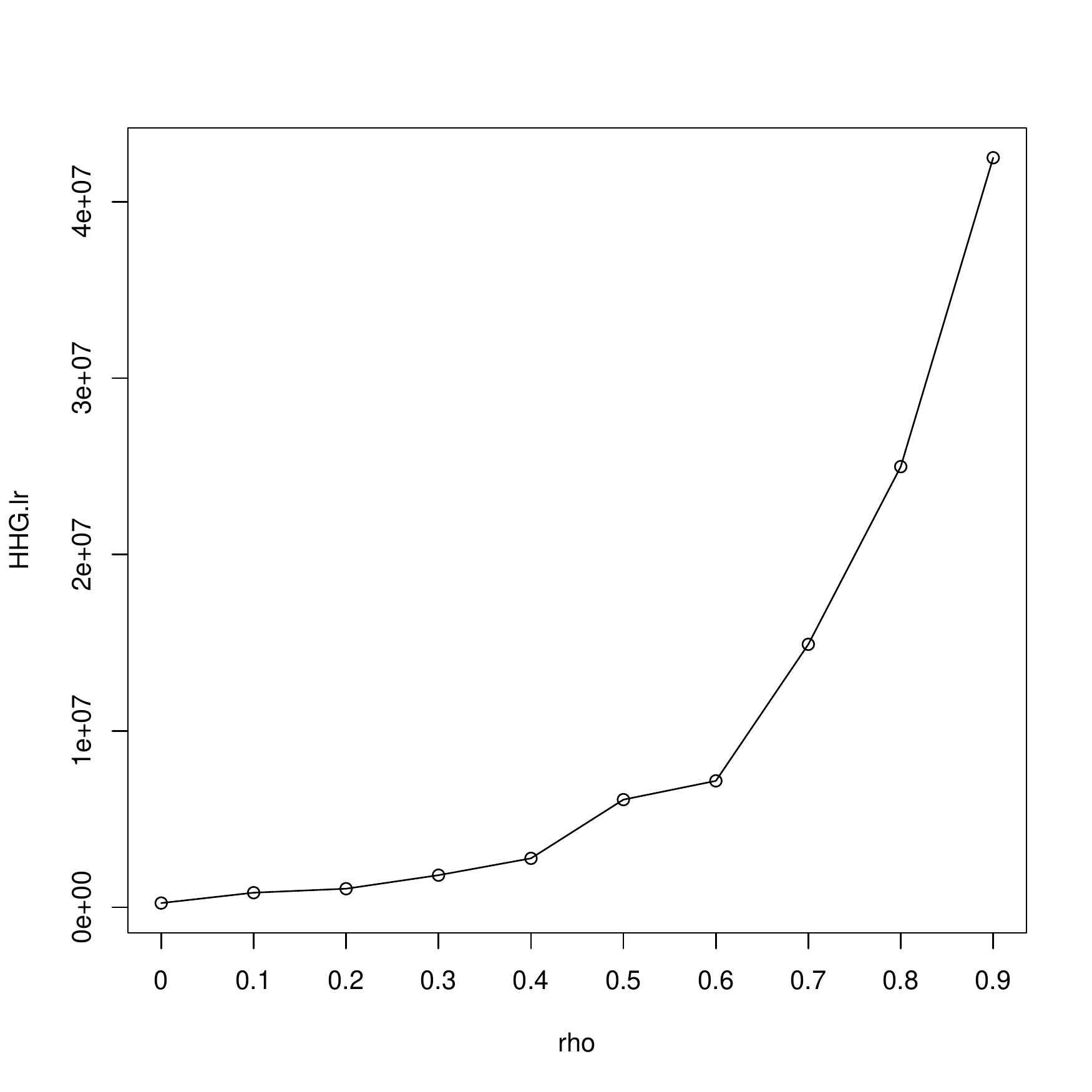}}
	\subfigure[Ball]{\includegraphics[width=0.245\linewidth]{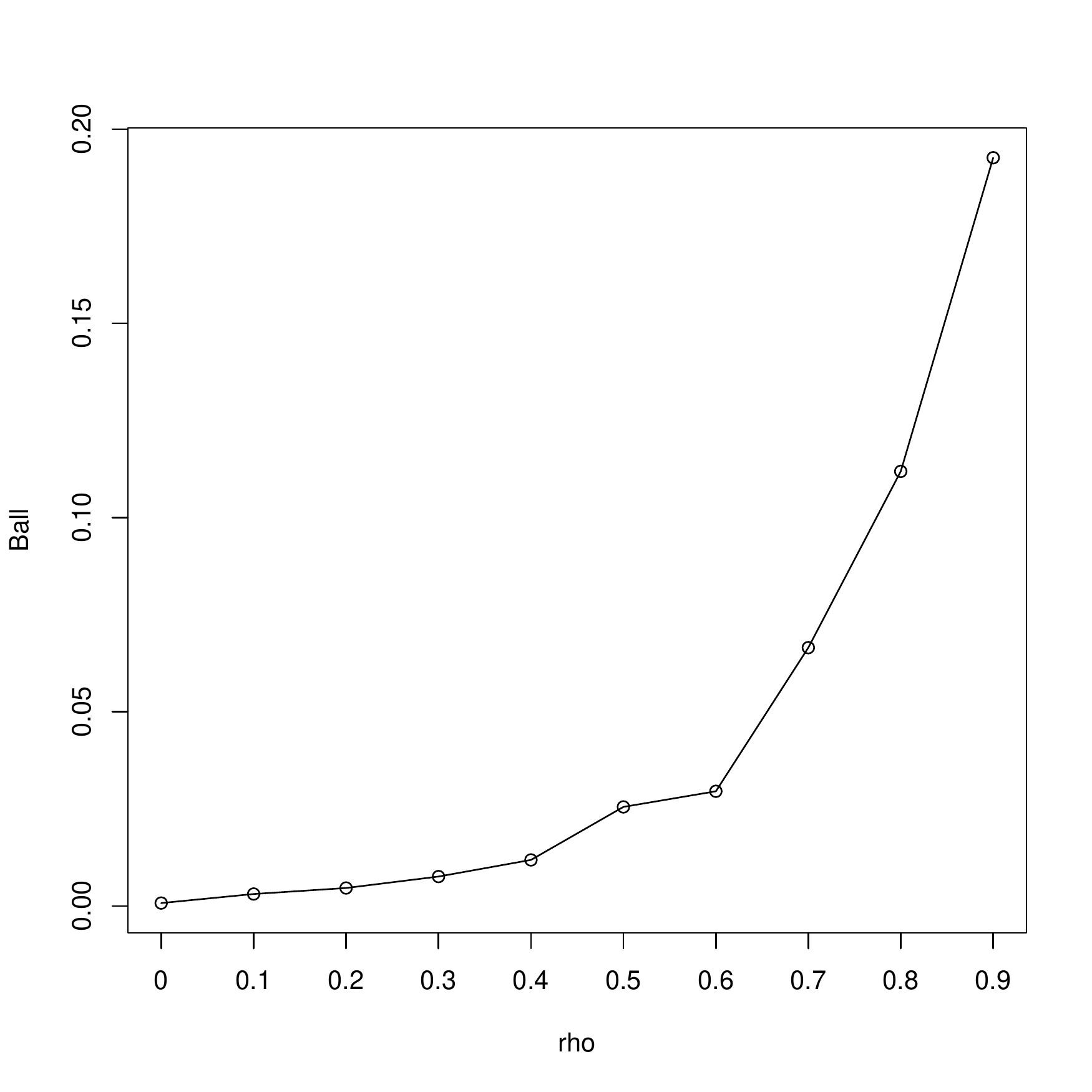}}
	\subfigure[BET]{\includegraphics[width=0.245\linewidth]{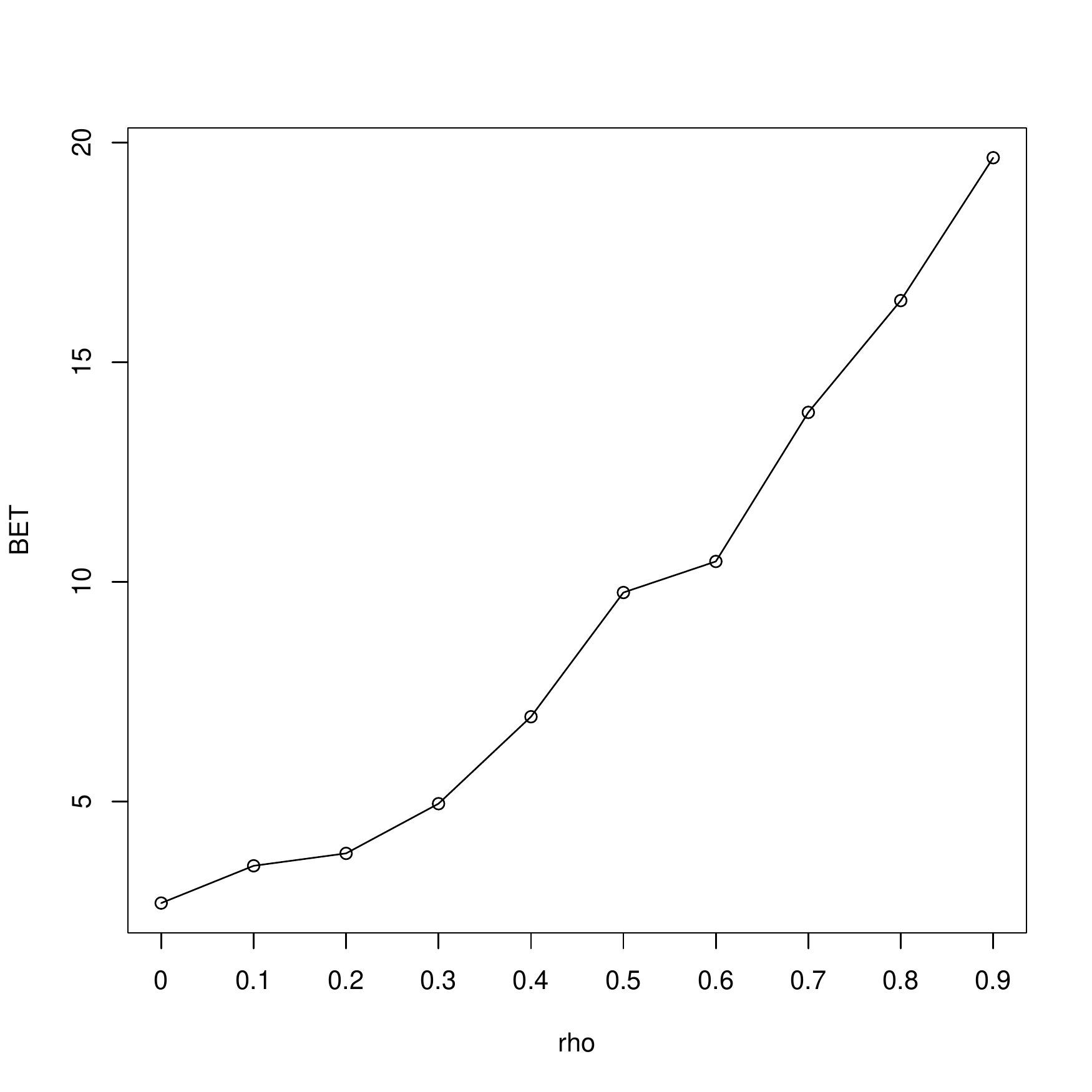}}
	\subfigure[QAD]{\includegraphics[width=0.245\linewidth]{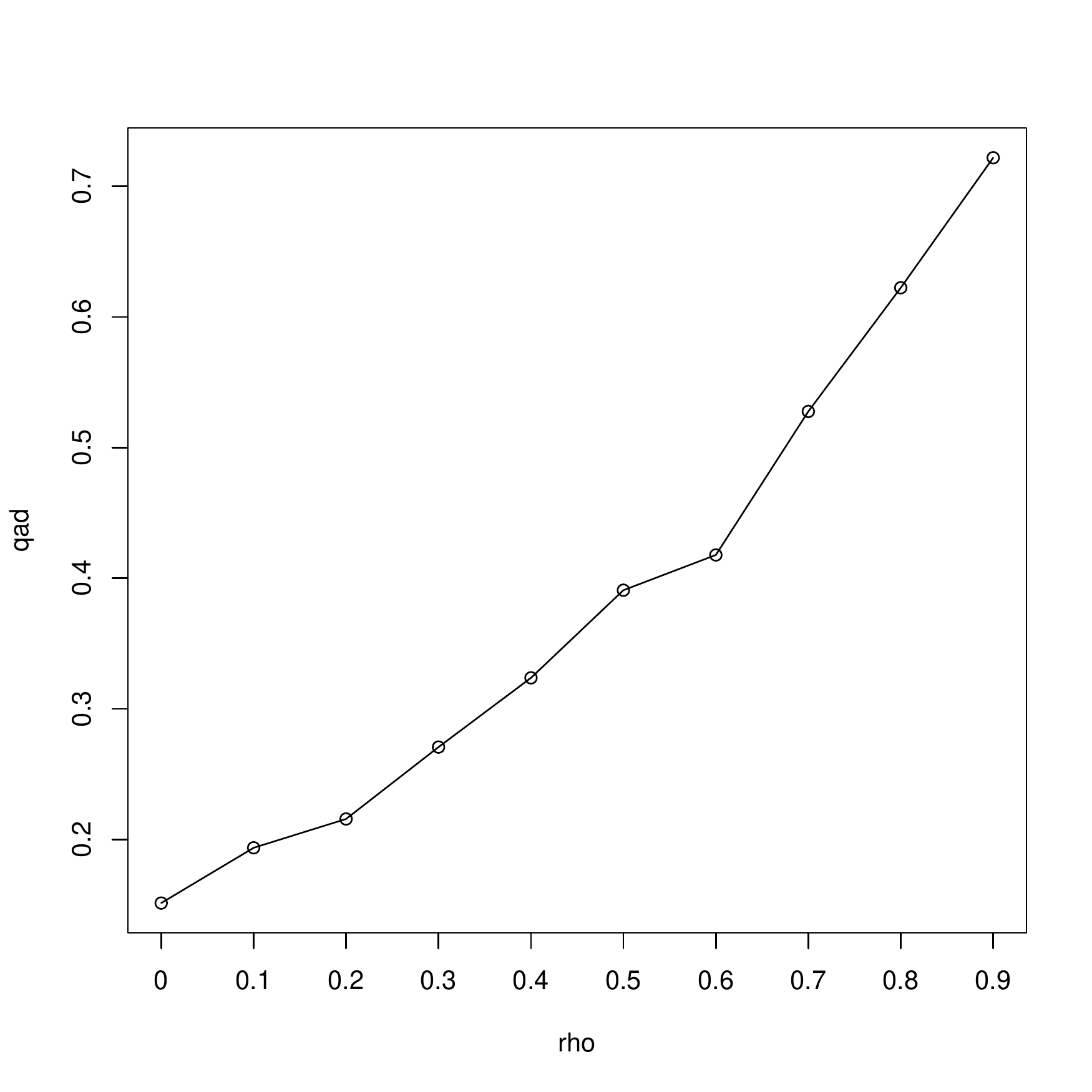}}
	\subfigure[mixed]{\includegraphics[width=0.245\linewidth]{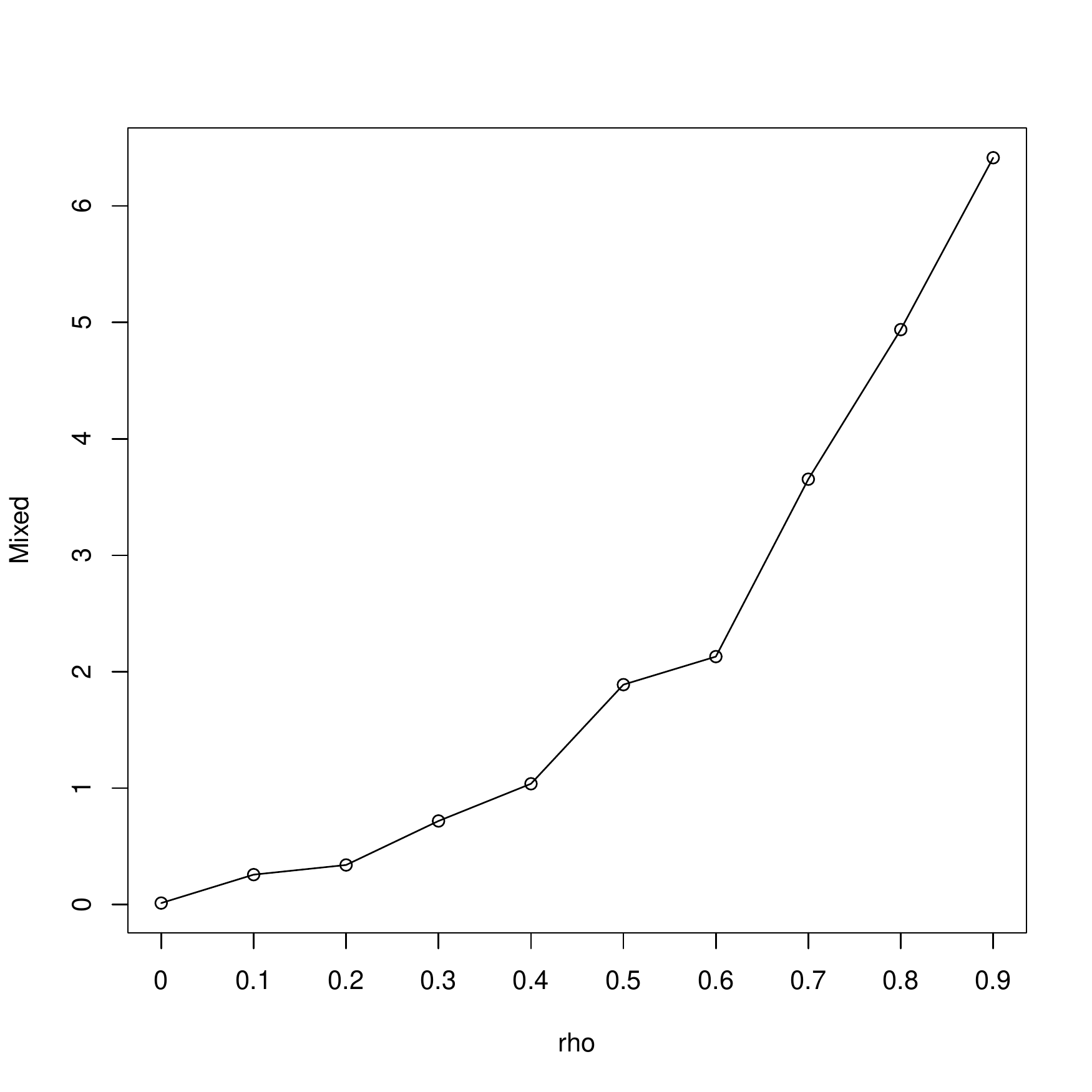}}
	\subfigure[CODEC]{\includegraphics[width=0.245\linewidth]{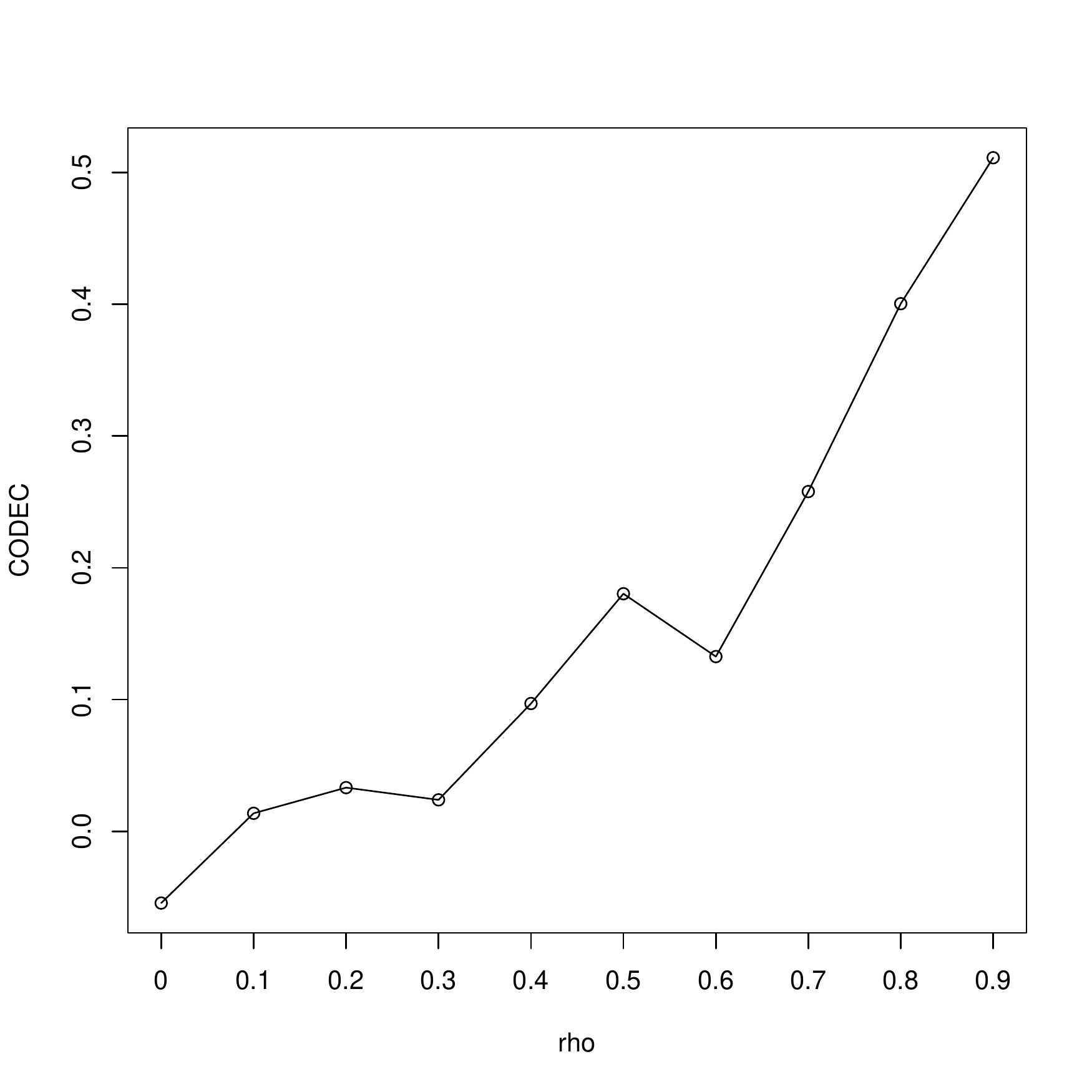}}
	\subfigure[subcop]{\includegraphics[width=0.245\linewidth]{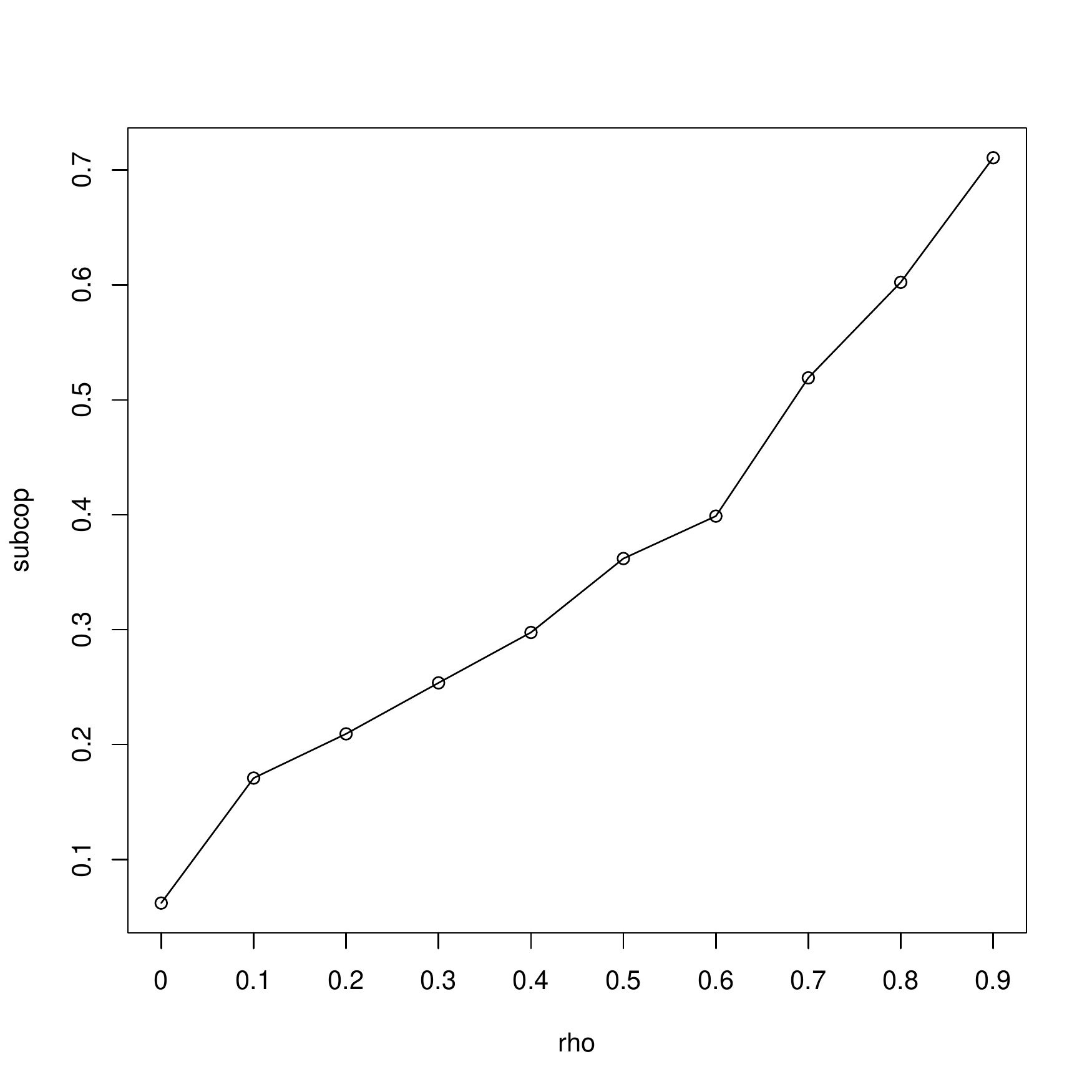}}
	\subfigure[dCor]{\includegraphics[width=0.245\linewidth]{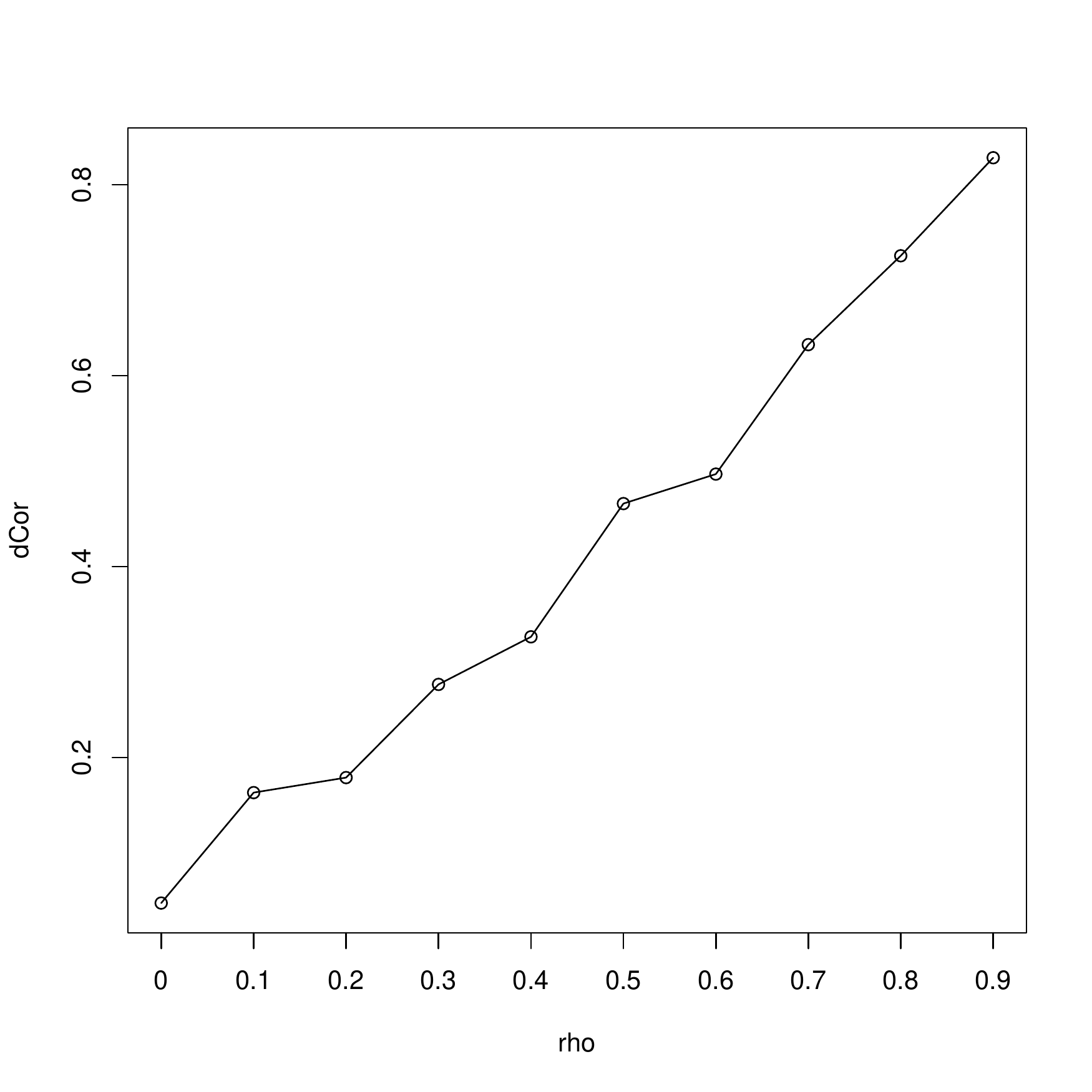}}
	\subfigure[mdm]{\includegraphics[width=0.245\linewidth]{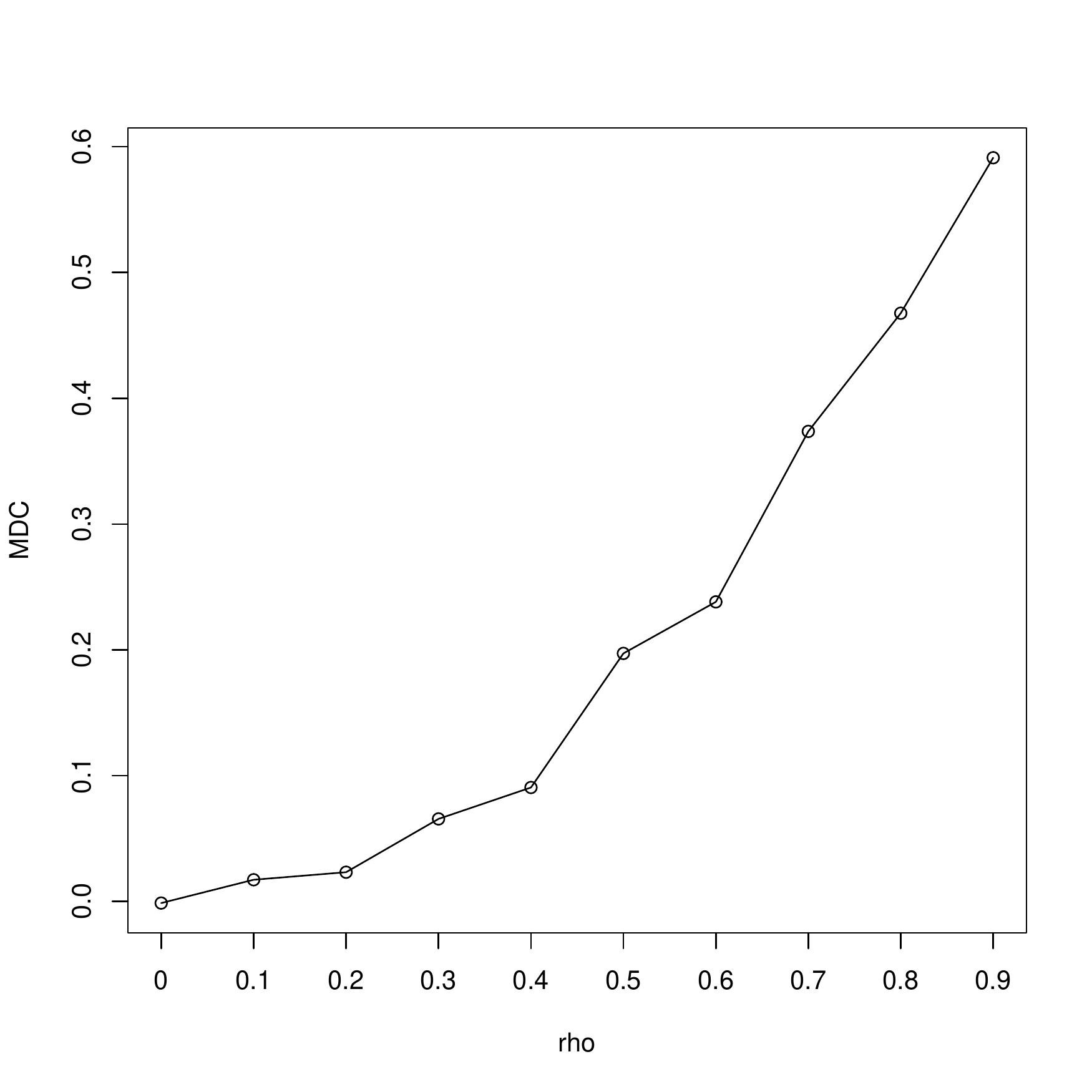}}
	\subfigure[dHSIC]{\includegraphics[width=0.245\linewidth]{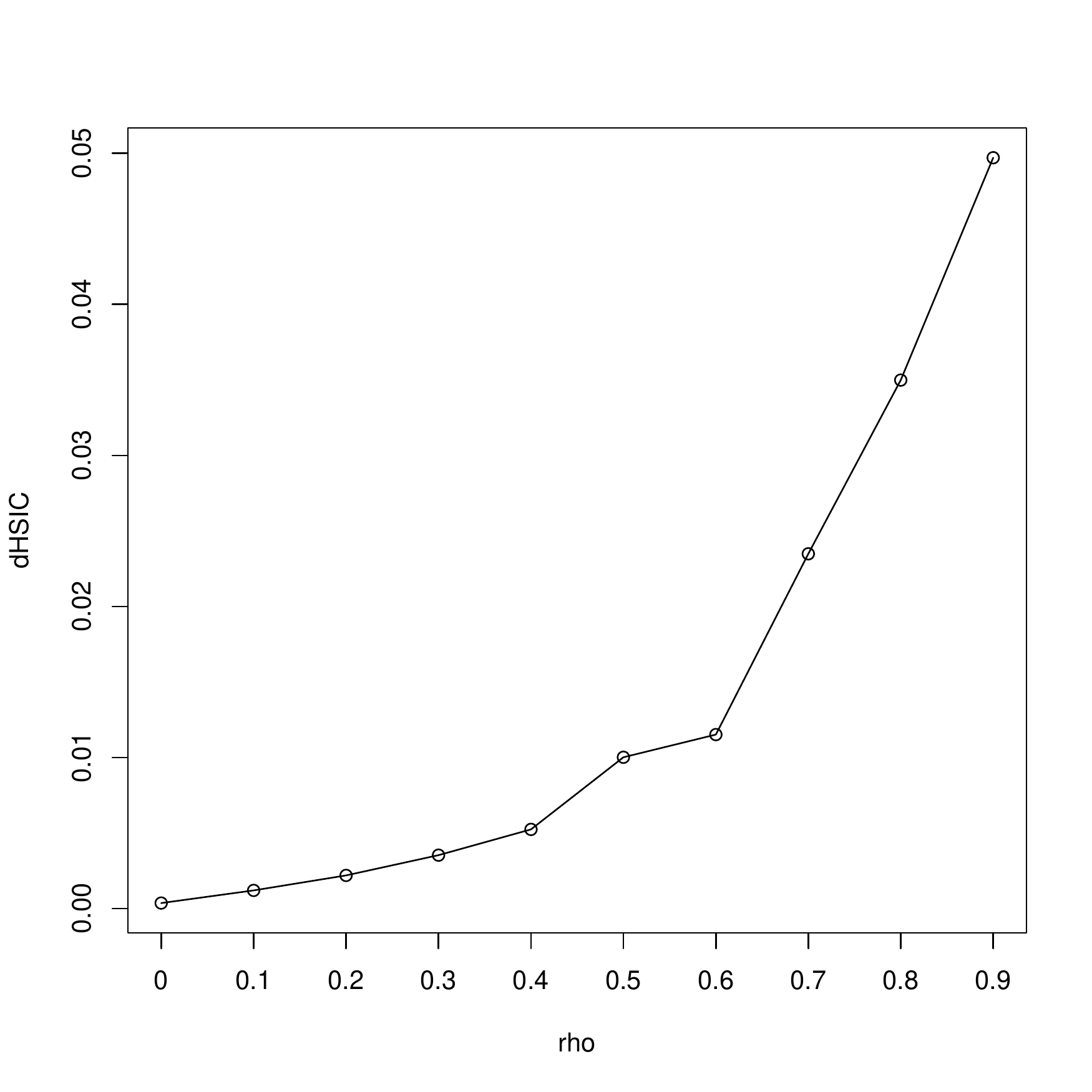}}
	\subfigure[NNS]{\includegraphics[width=0.245\linewidth]{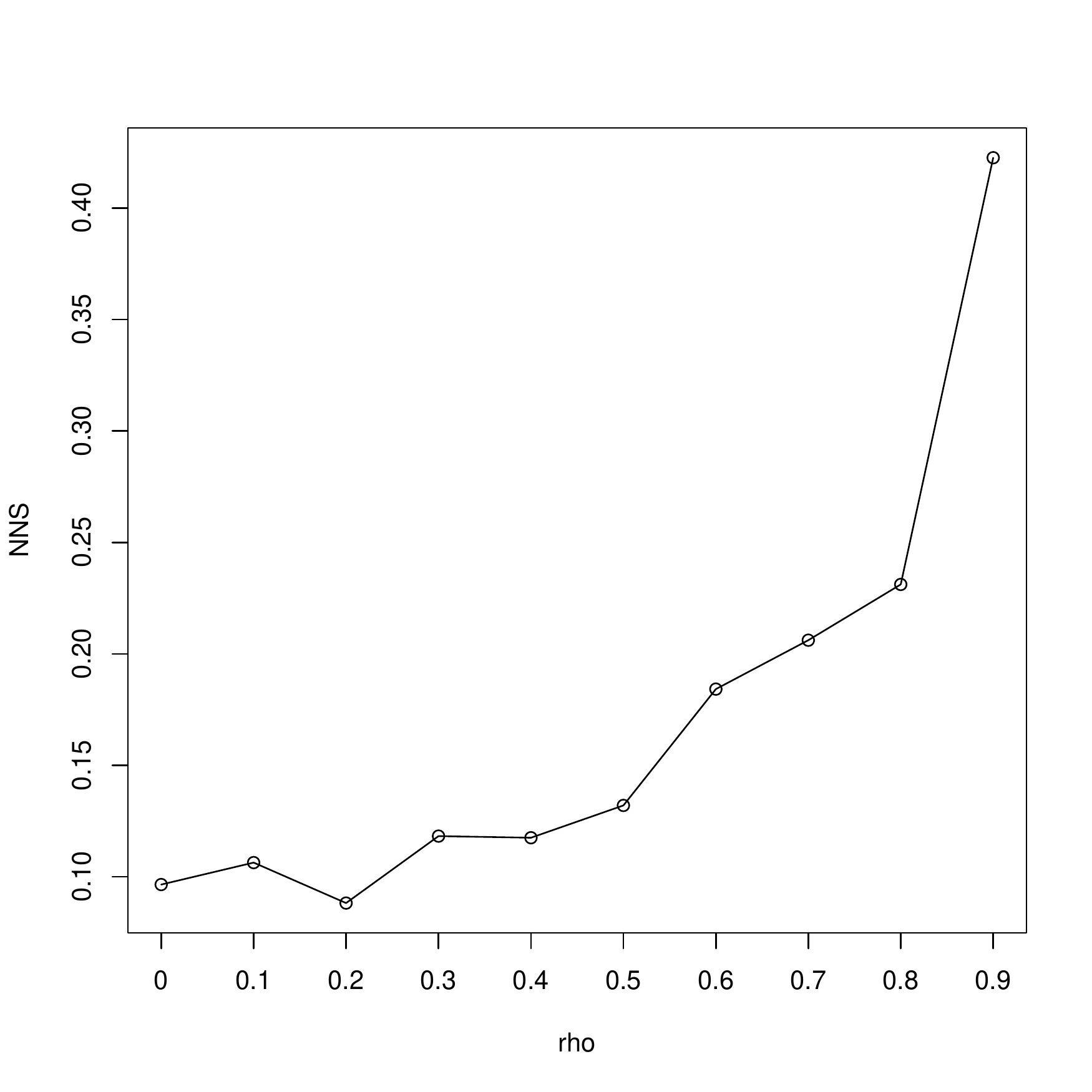}}
	\caption{Estimation of the independence measures from the simulated data of the bivariate normal copula.}
	\label{fig:binormalcop}
\end{figure}

\begin{figure}
	\centering
	\includegraphics[width=0.9\linewidth]{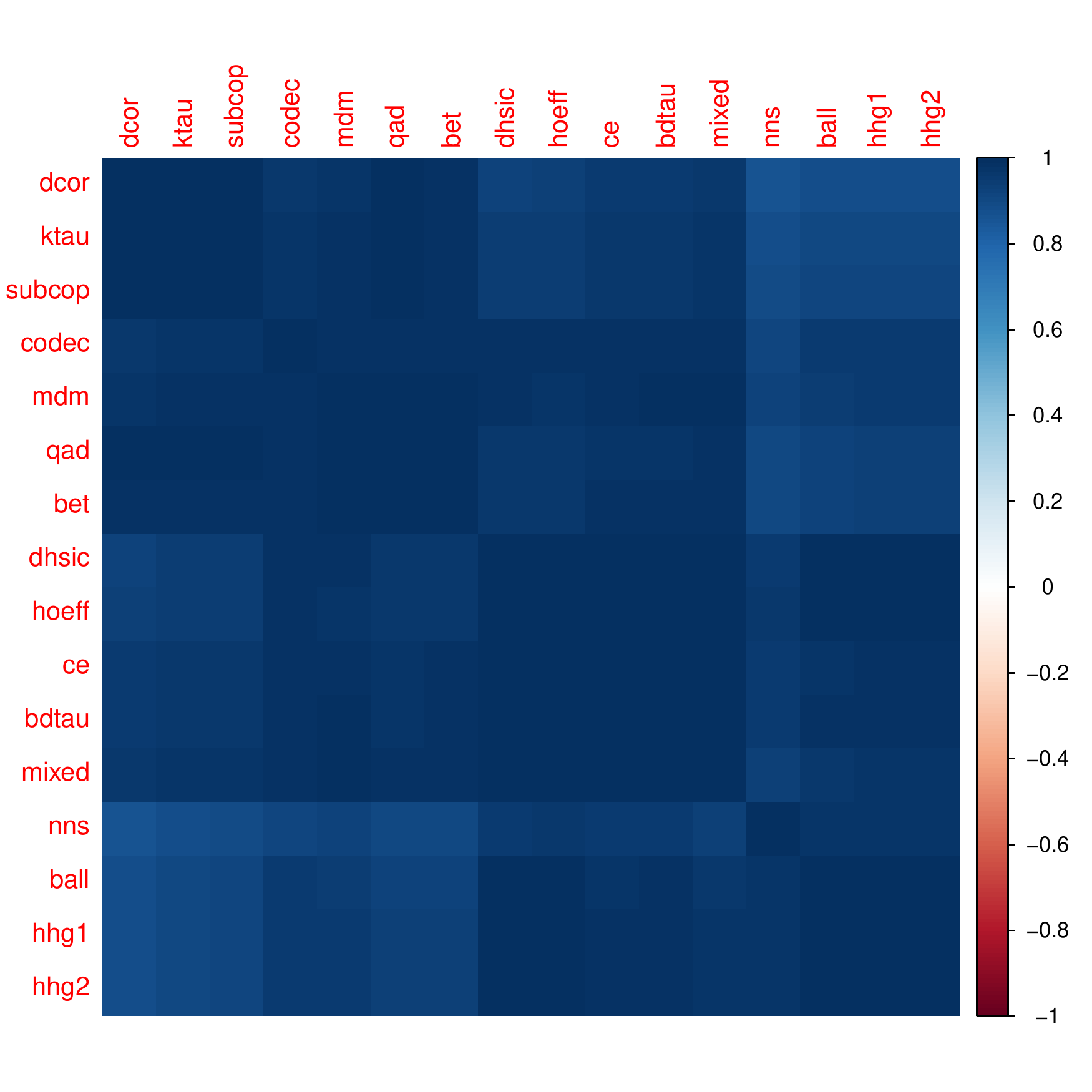}
	\caption{Correlation matrix of the independence measures estimated from the simulated data of the bivariate normal copula.}
	\label{fig:binormalcopcm}
\end{figure}

\paragraph{Bivariate normal copula}
For the second experiment, the estimation of the independence measures from the bivariate normal copula is shown in Figure \ref{fig:binormalcop} and the correlation matrix of the estimation of the measures is shown in Figure \ref{fig:binormalcopcm}. It can be learned that there are about 5 groups of the measures according to the correlations: \{dCor, Ktau, subcop\}, \{CODEC, MDM, QAD, BET\}, \{dHSIC, Hoeff, CE, BDtau, mixed\}, \{Ball, HHG\}, and \{NNS\}. As the $\rho$ increases, the measures in the first and second groups increases linearly while those in the other groups increases exponentially.

\begin{figure}
	\subfigure[CE]{\includegraphics[width=0.245\linewidth]{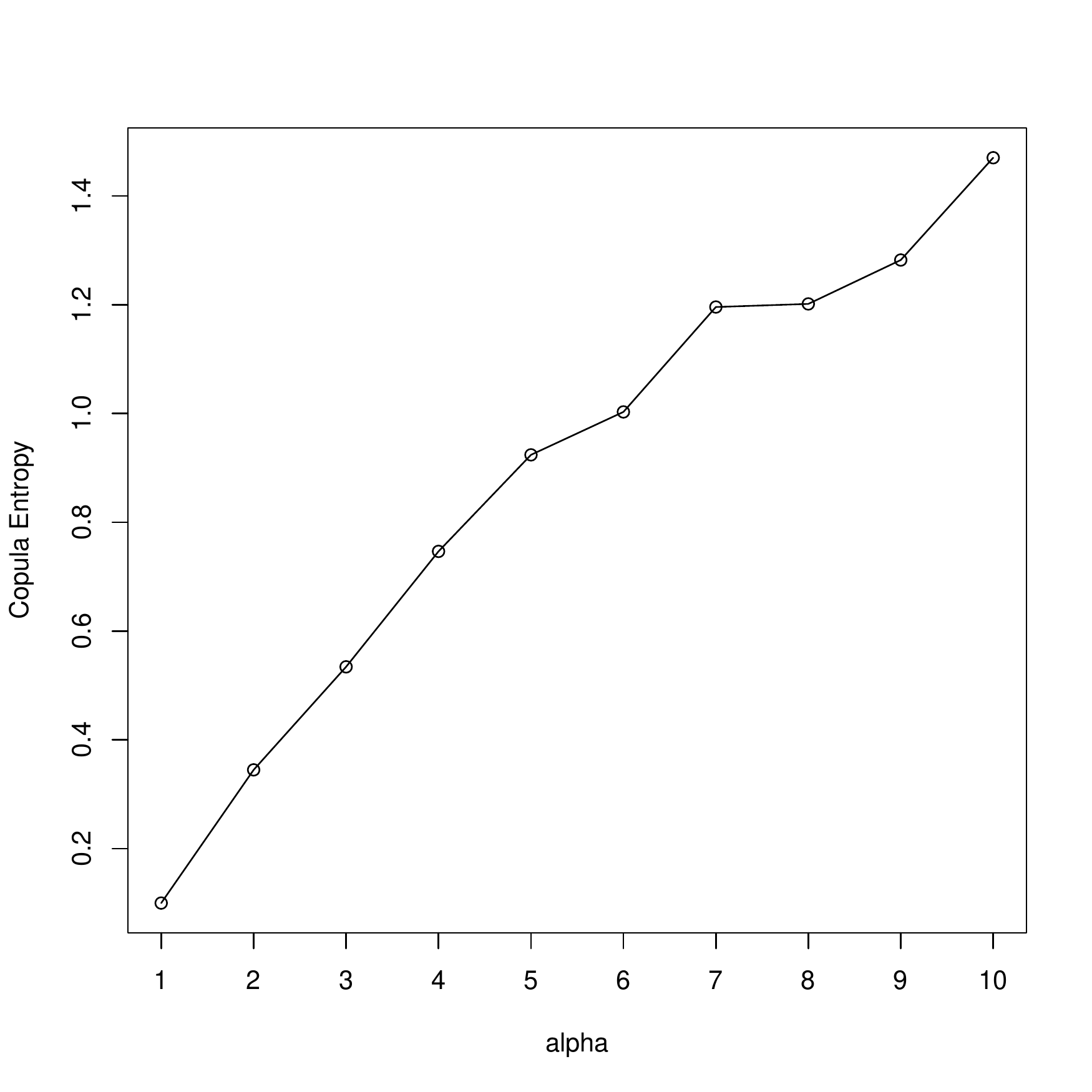}}
	\subfigure[Ktau]{\includegraphics[width=0.245\linewidth]{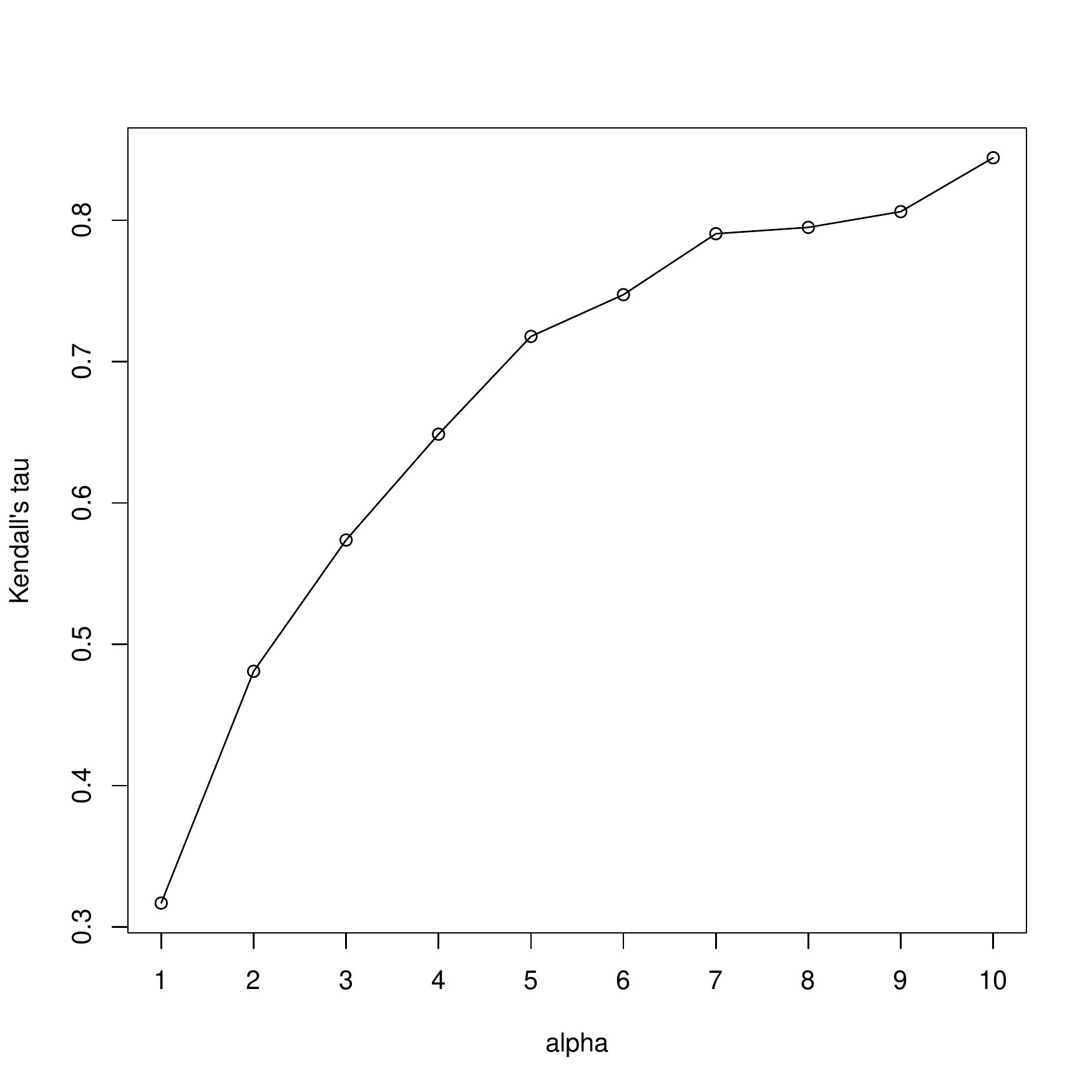}}
	\subfigure[Hoeff]{\includegraphics[width=0.245\linewidth]{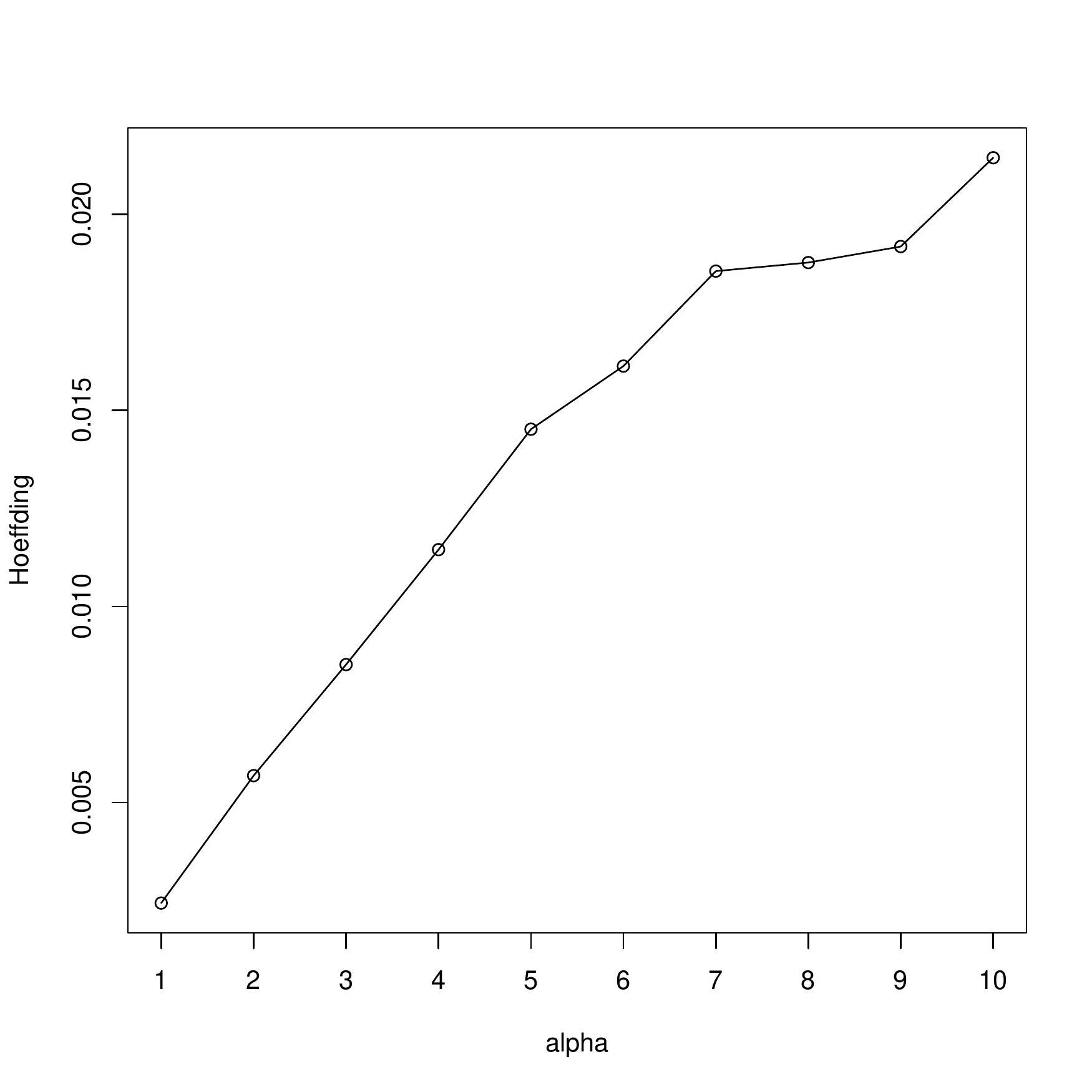}}
	\subfigure[BDtau]{\includegraphics[width=0.245\linewidth]{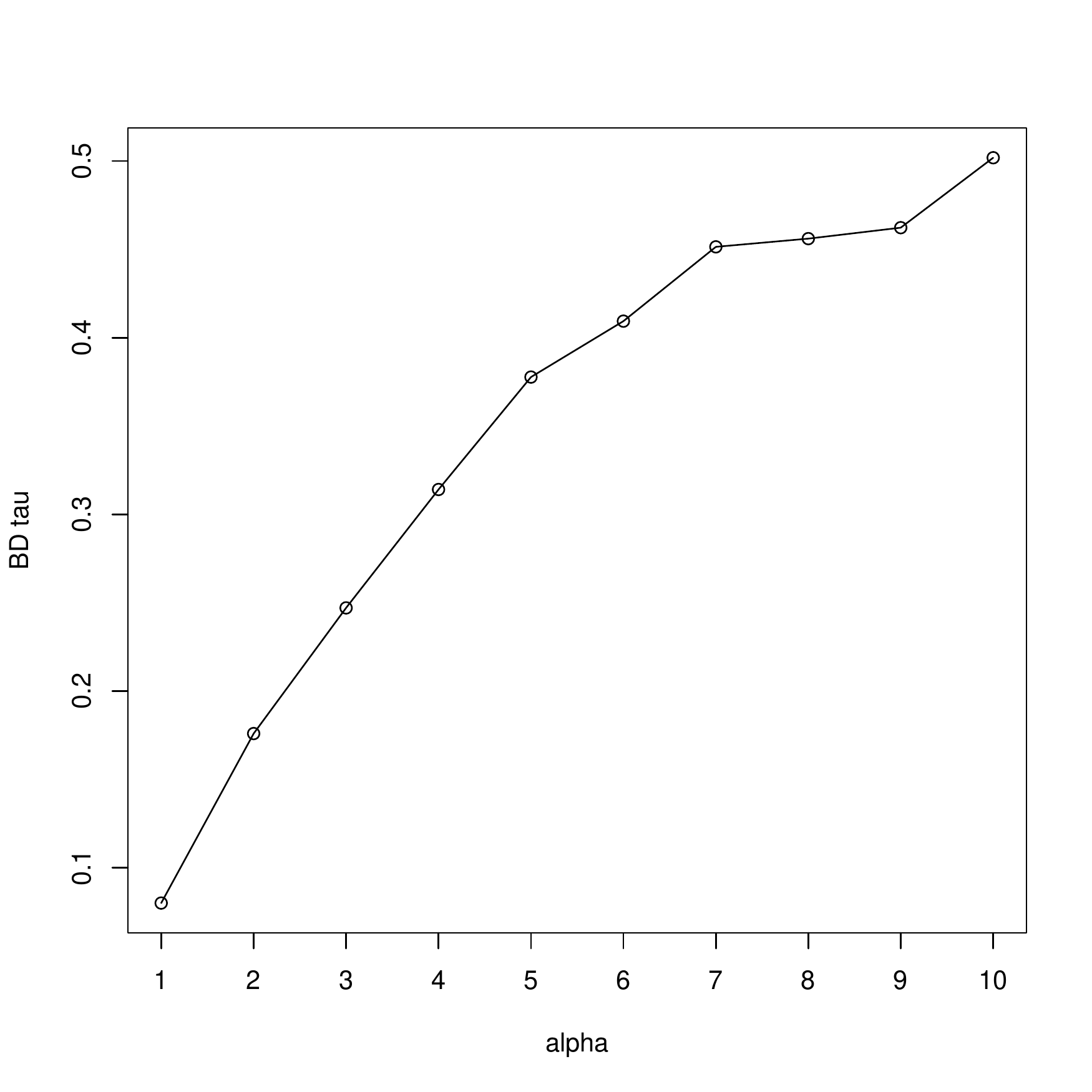}}
	\subfigure[HHG.chisq]{\includegraphics[width=0.245\linewidth]{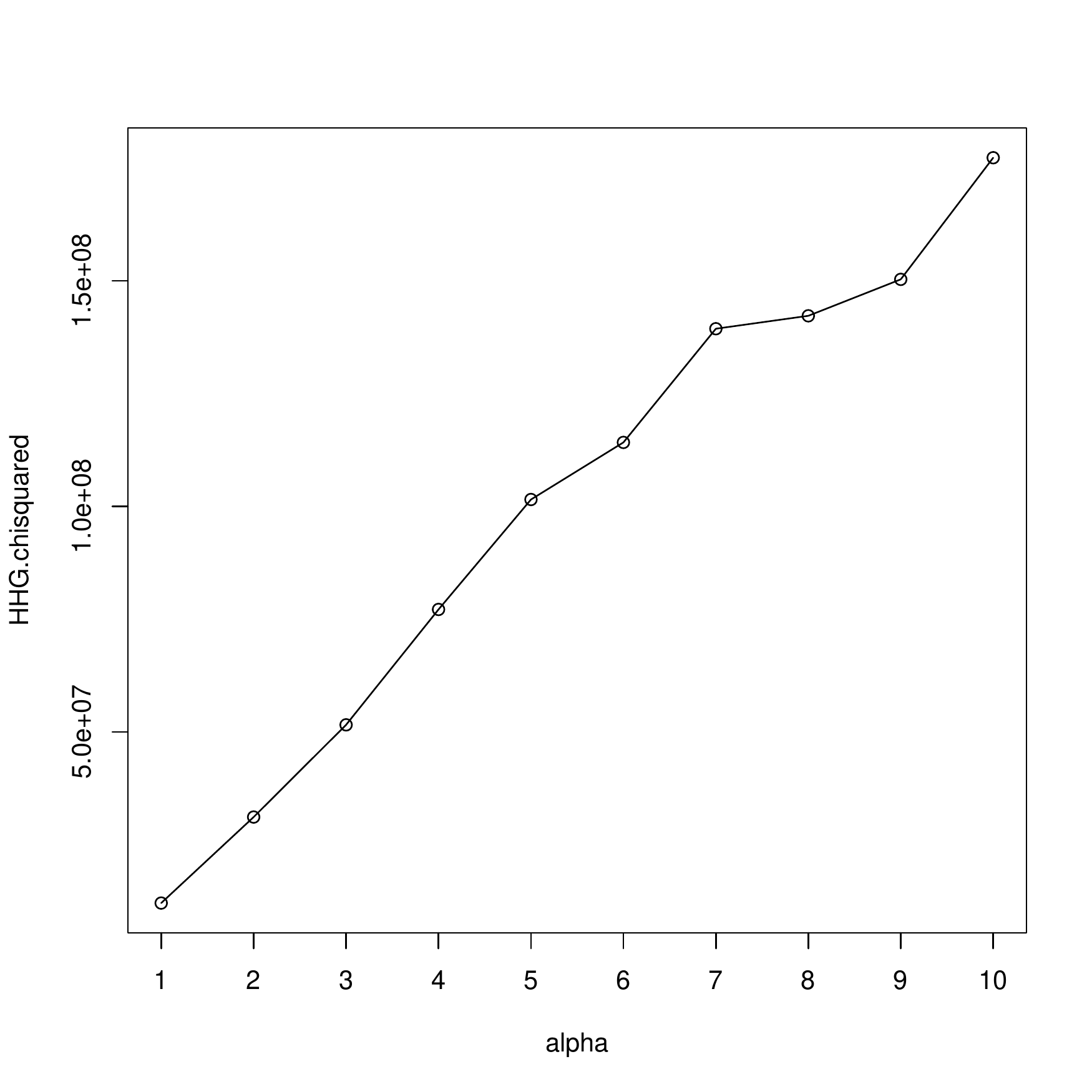}}
	\subfigure[HHG.lr]{\includegraphics[width=0.245\linewidth]{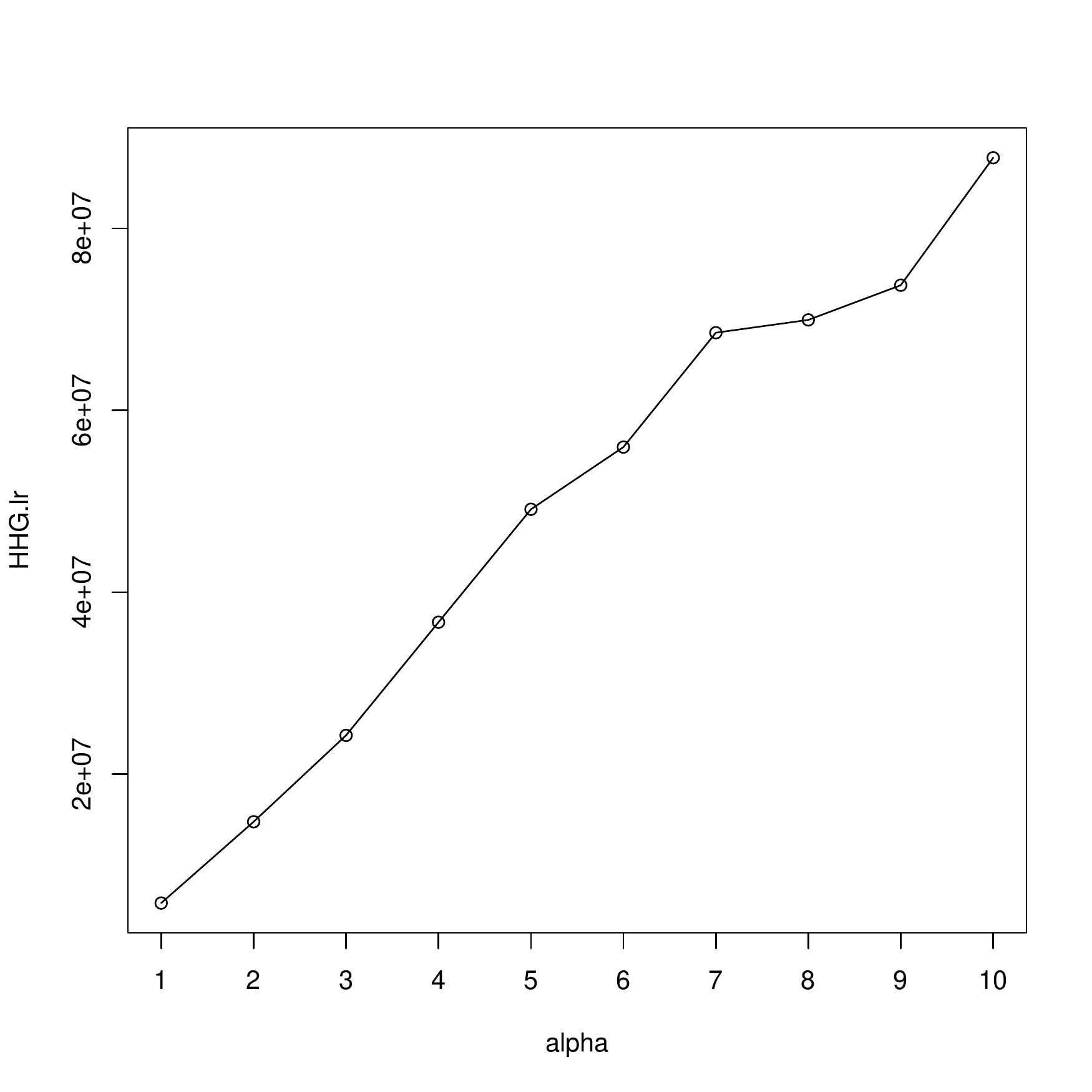}}
	\subfigure[Ball]{\includegraphics[width=0.245\linewidth]{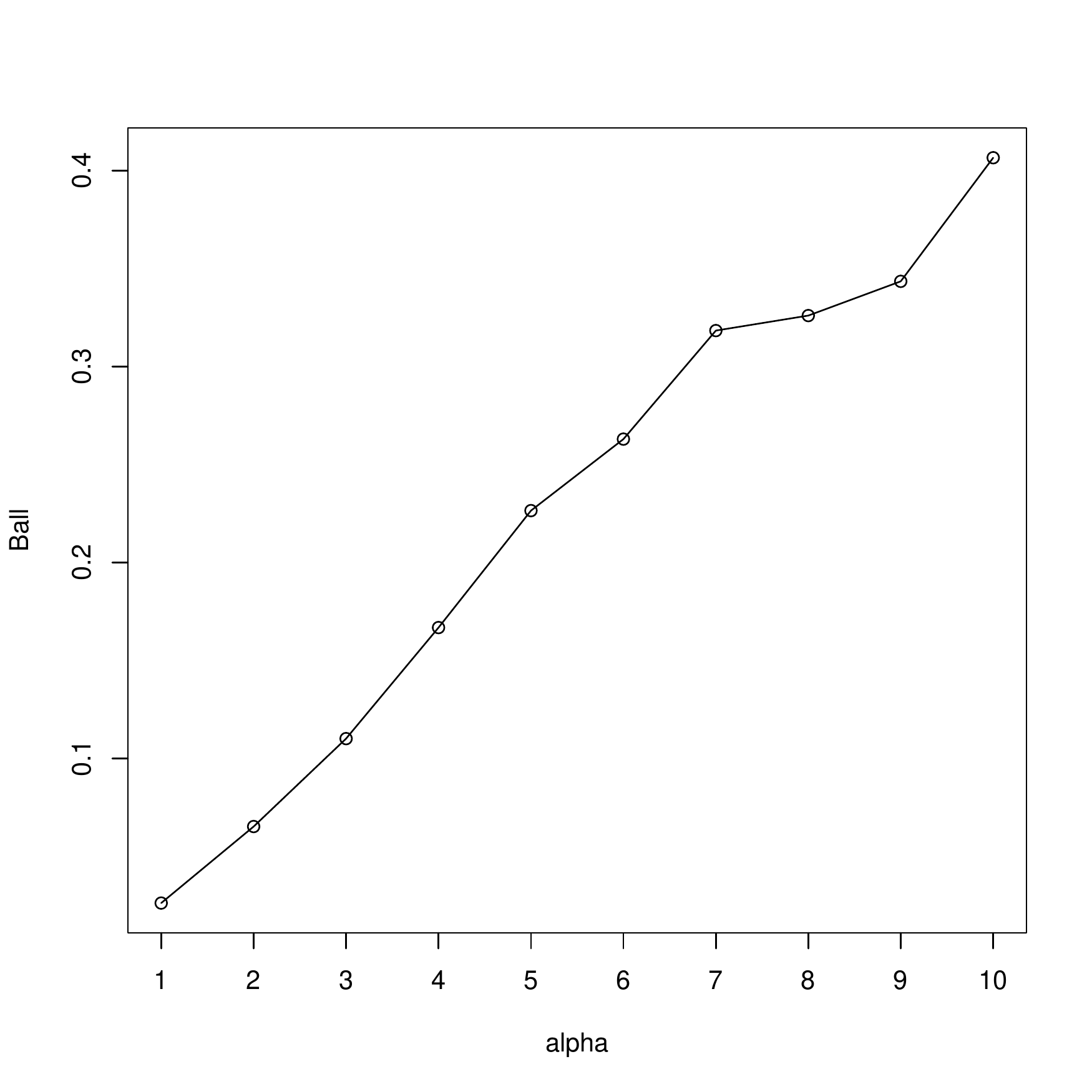}}
	\subfigure[BET]{\includegraphics[width=0.245\linewidth]{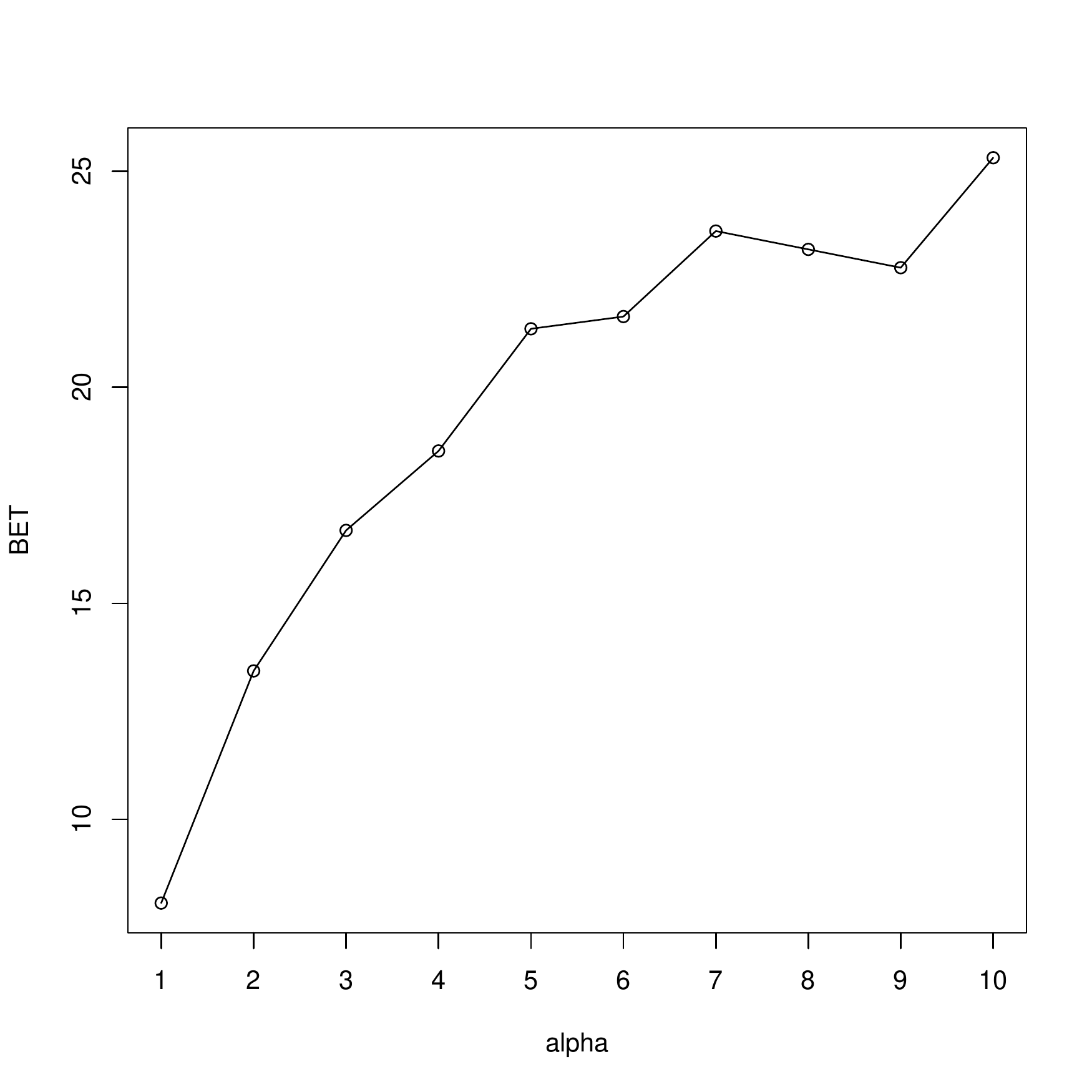}}
	\subfigure[QAD]{\includegraphics[width=0.245\linewidth]{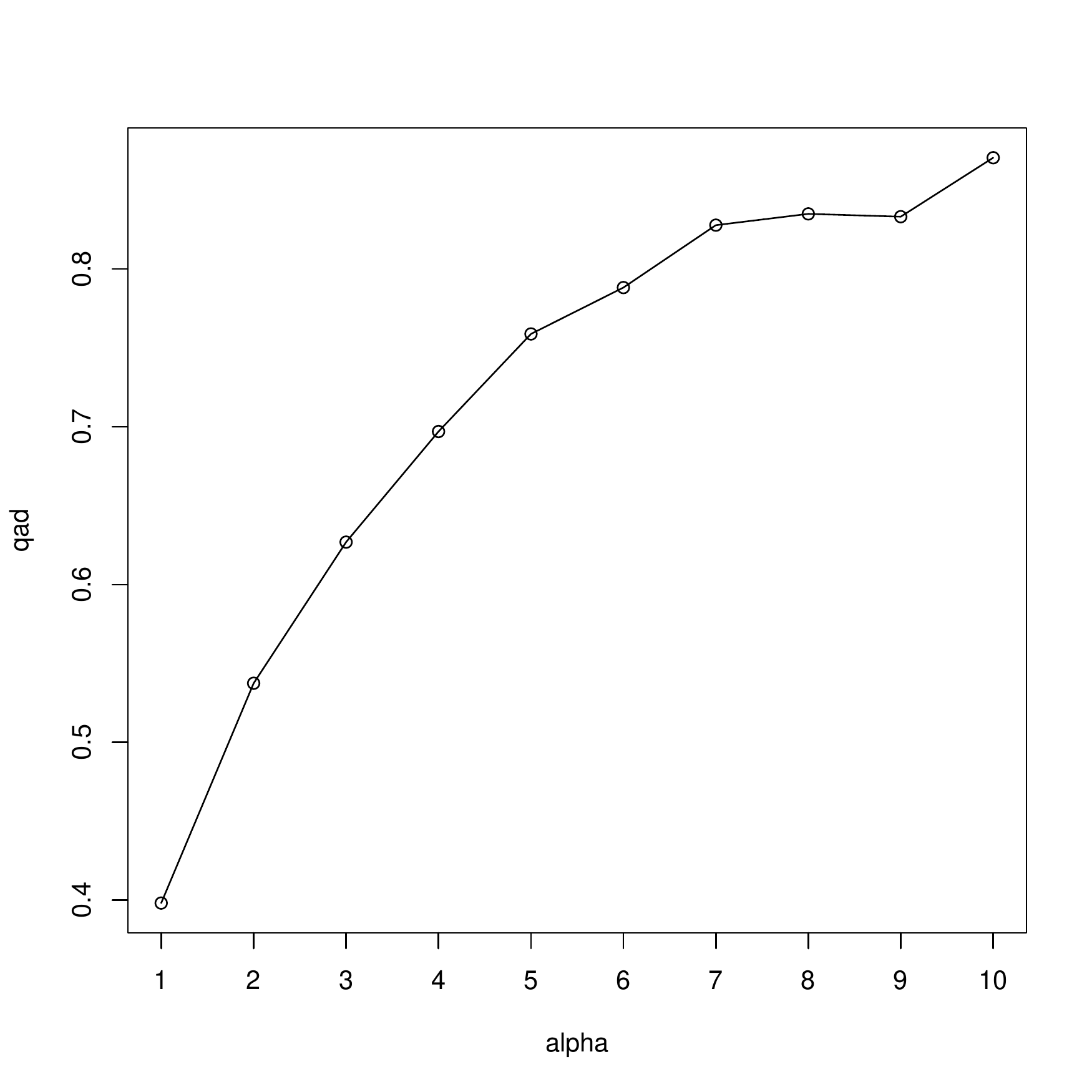}}
	\subfigure[mixed]{\includegraphics[width=0.245\linewidth]{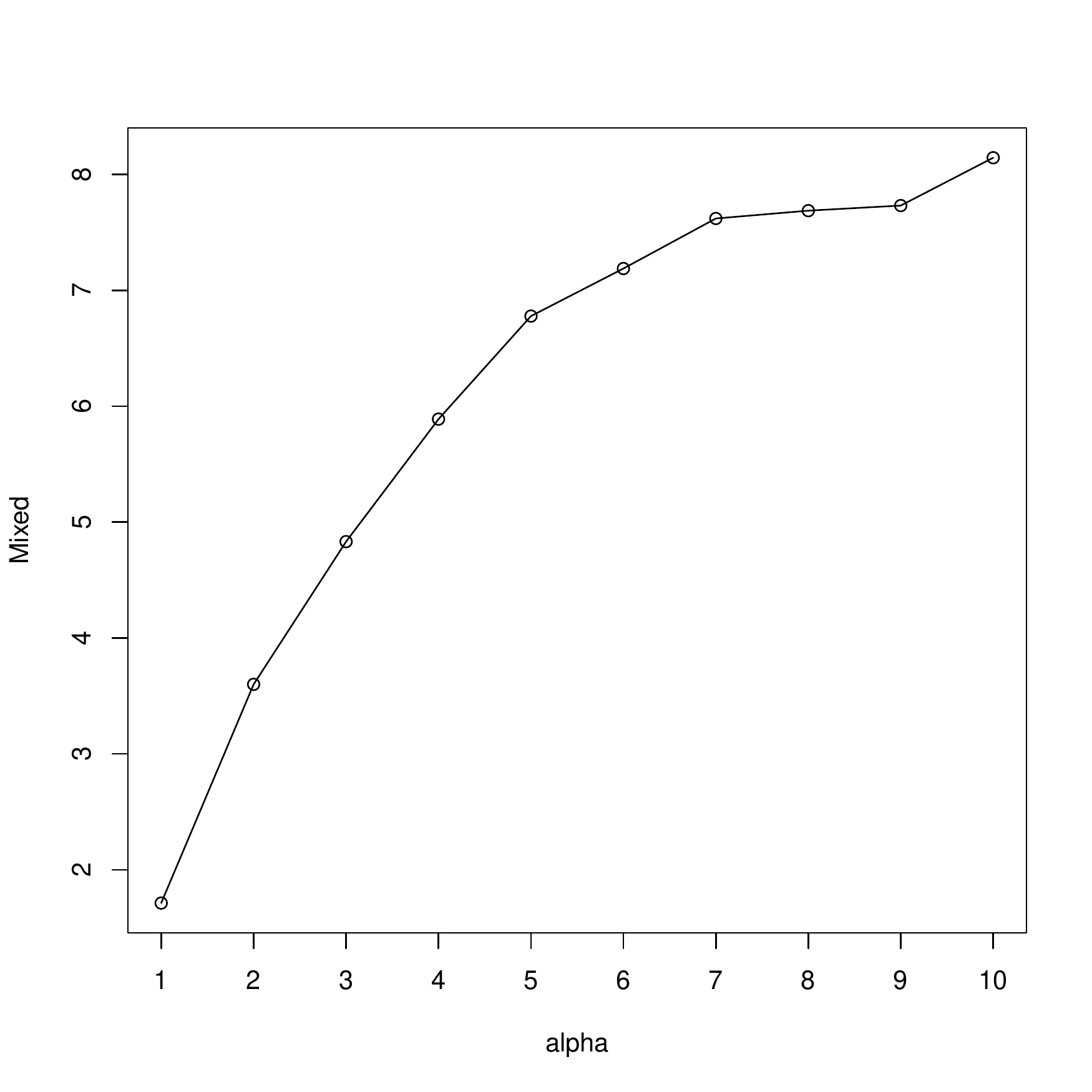}}
	\subfigure[CODEC]{\includegraphics[width=0.245\linewidth]{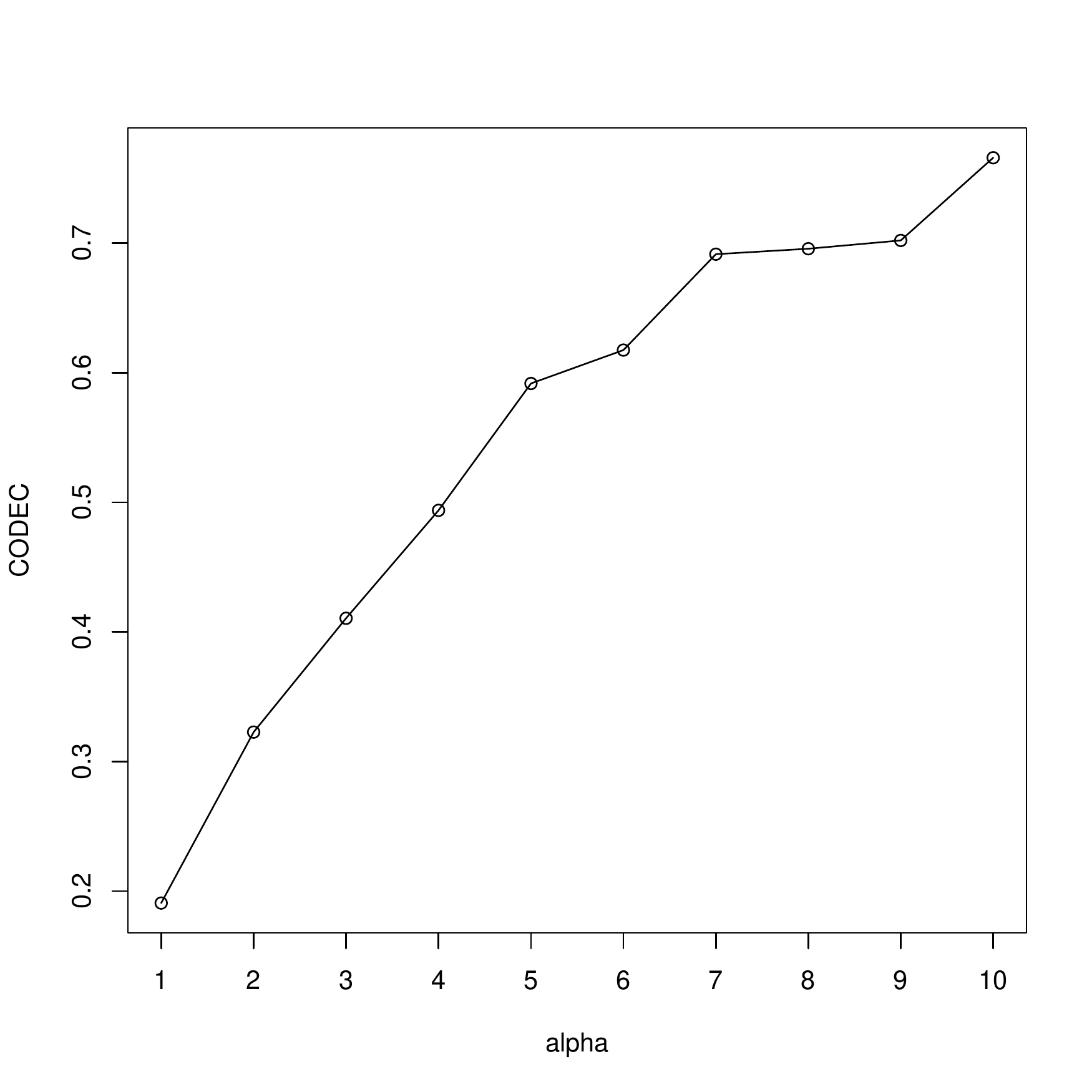}}
	\subfigure[subcop]{\includegraphics[width=0.245\linewidth]{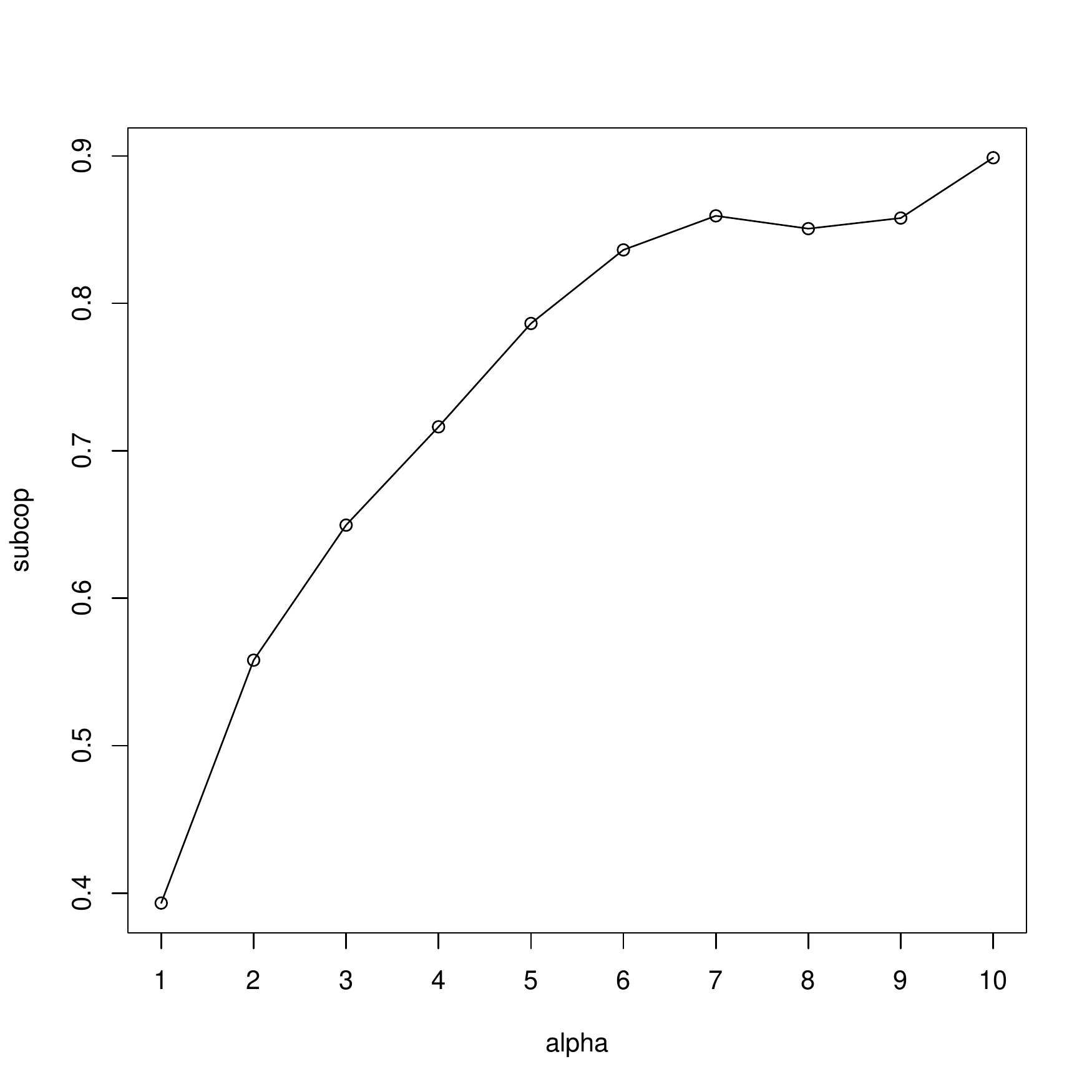}}
	\subfigure[dCor]{\includegraphics[width=0.245\linewidth]{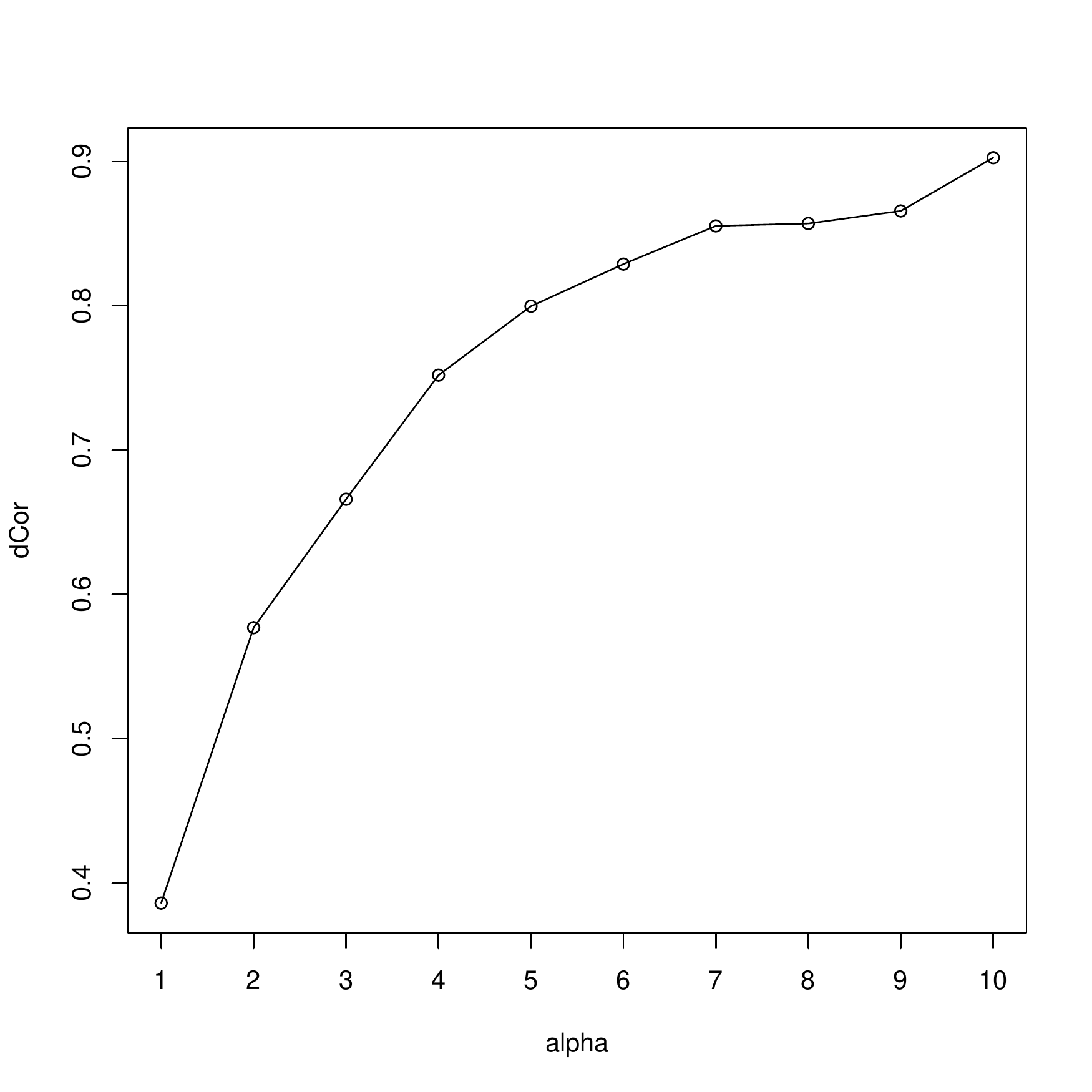}}
	\subfigure[mdm]{\includegraphics[width=0.245\linewidth]{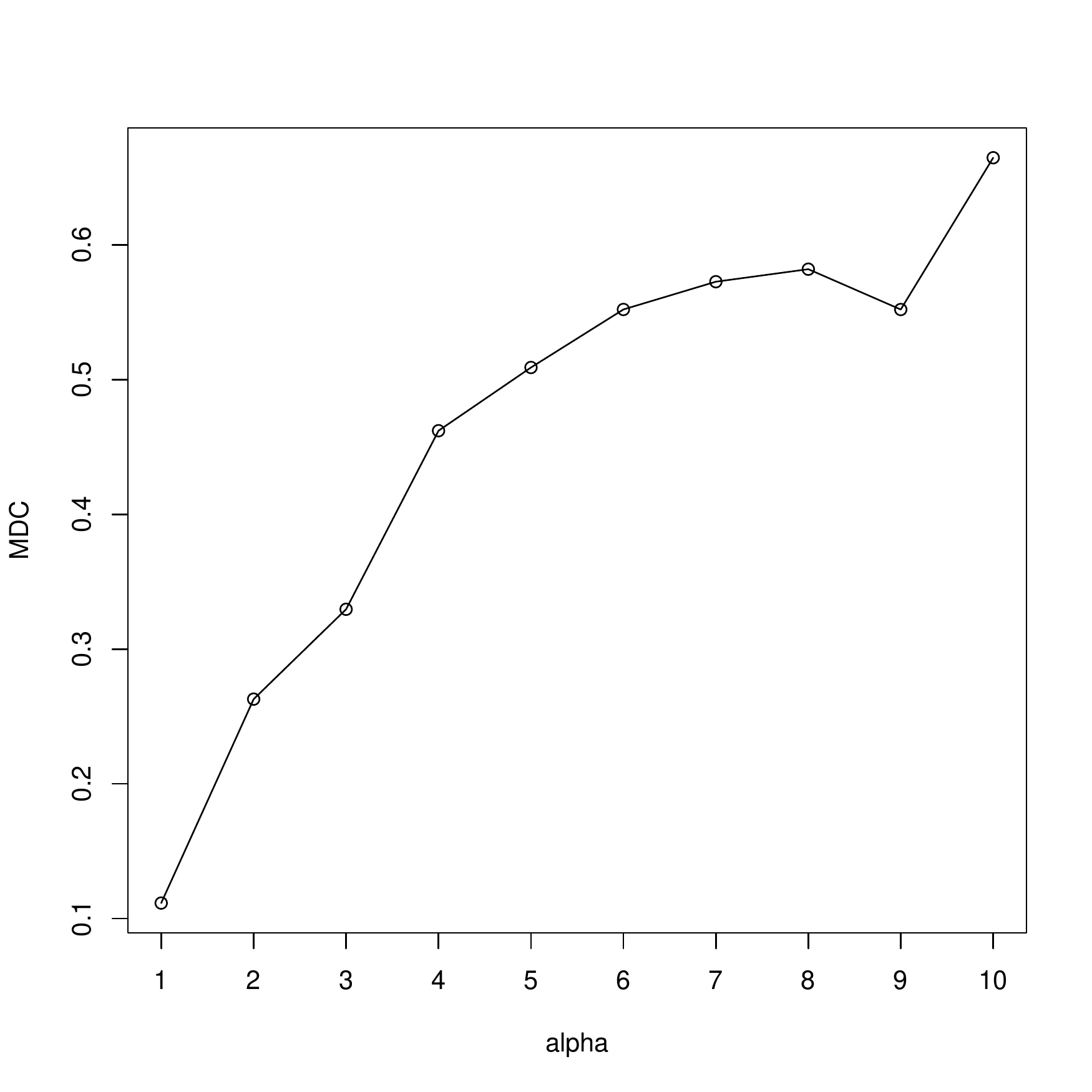}}
	\subfigure[dHSIC]{\includegraphics[width=0.245\linewidth]{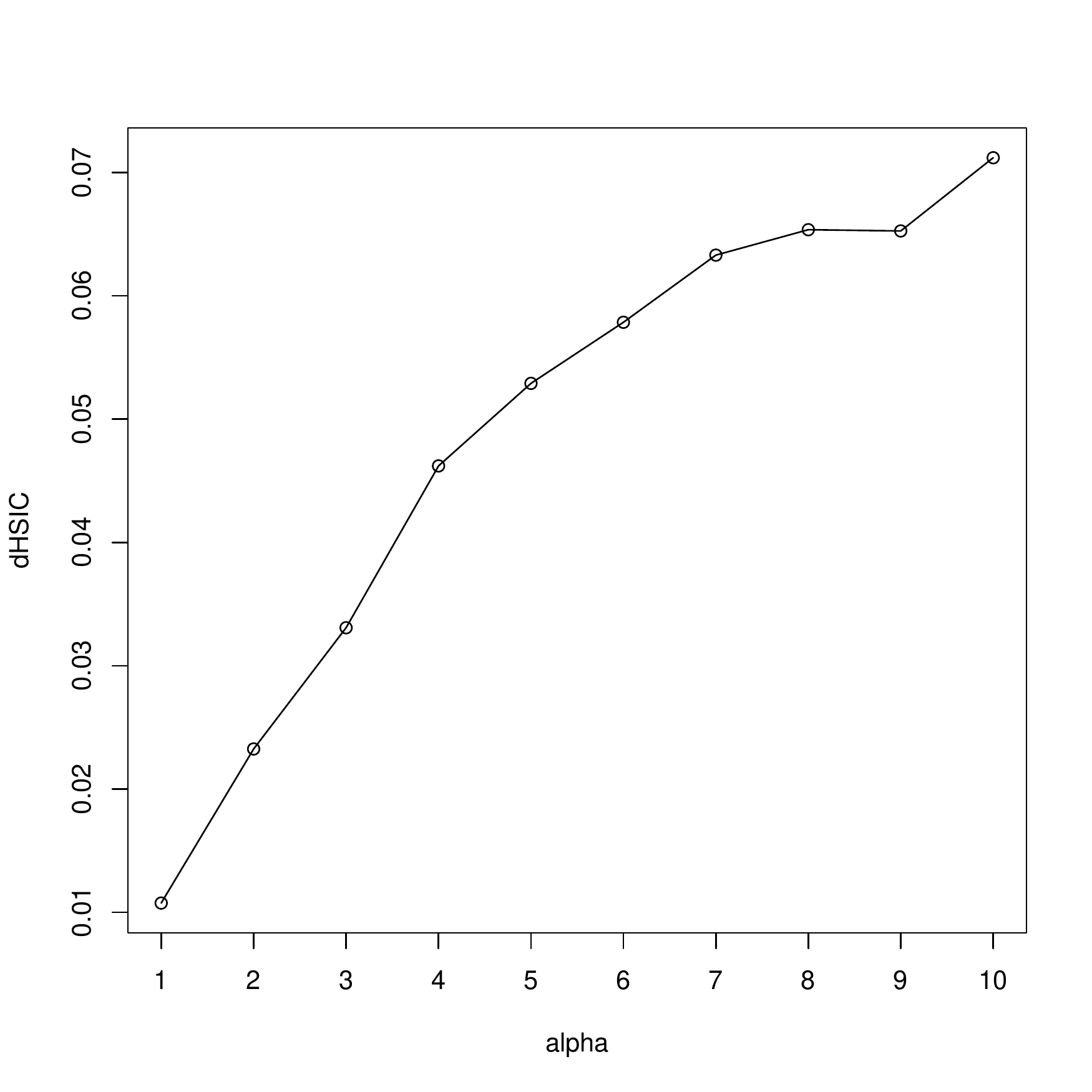}}
	\subfigure[NNS]{\includegraphics[width=0.245\linewidth]{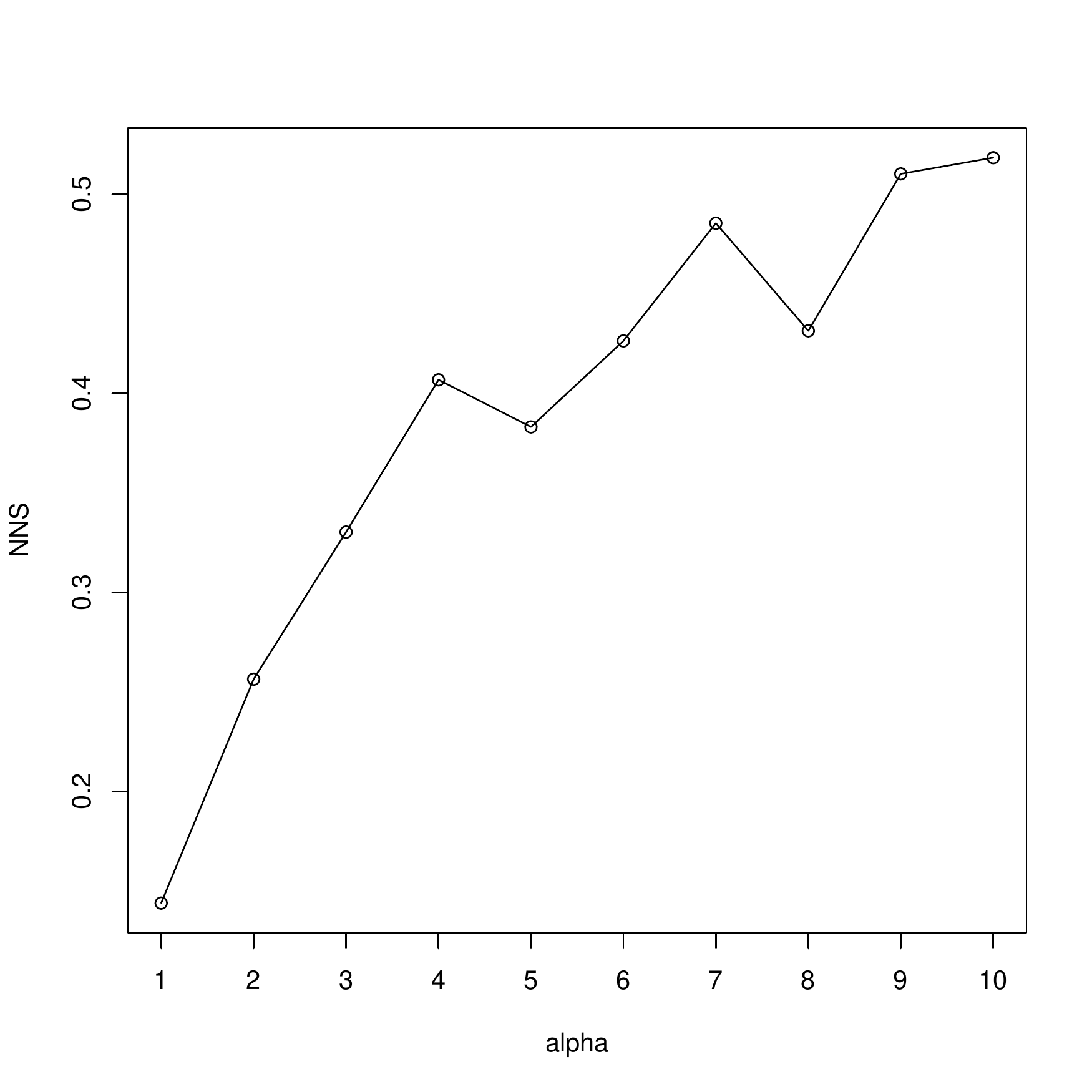}}
	\caption{Estimation of the independence measures from the simulated data of the bivariate Clayton copula.}
	\label{fig:biclayton}
\end{figure}

\begin{figure}
	\centering
	\includegraphics[width=0.9\linewidth]{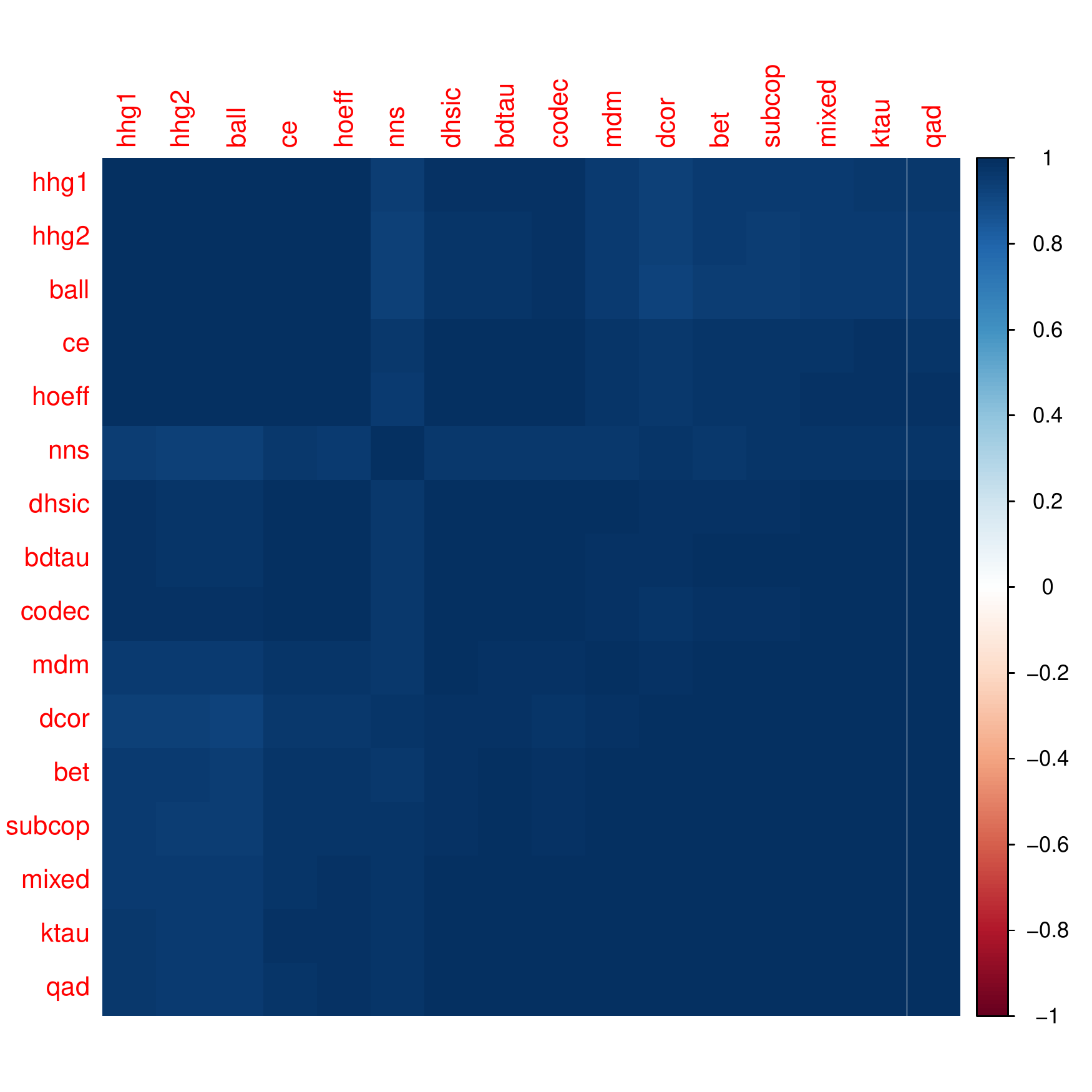}
	\caption{Correlation matrix of the independence measures estimated from the simulated data of the bivariate Clayton copula.}
	\label{fig:biclaytoncm}
\end{figure}

\paragraph{Bivariate Archimedean copula}
For the third experiment, the estimation of the independence measures from the bivariate Clayton copula is shown in Figure \ref{fig:biclayton} and the correlation matrix of the estimation of the measures is shown in Figure \ref{fig:biclaytoncm}. It can be learned that there are about 4 groups: \{CE, Ball, HHG, Hoeff\}, \{dHSIC, BDtau, CODEC\}, \{NNS\}, and all the others. As the $\alpha$ increases, the measures in the first group increase roughly linearly while the other measures increase roughly nonlinearly.

\begin{figure}
	\subfigure[CE]{\includegraphics[width=0.245\linewidth]{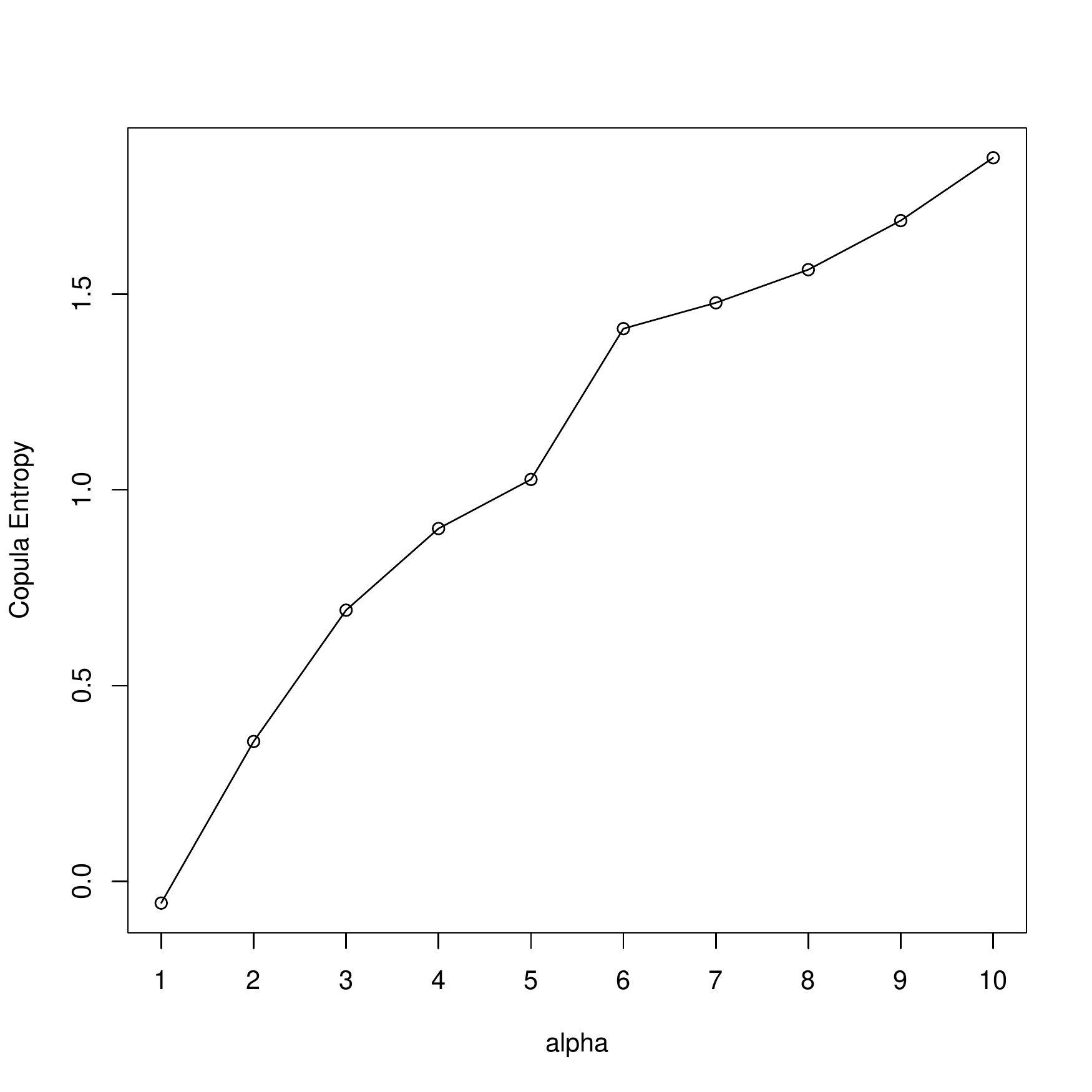}}
	\subfigure[Ktau]{\includegraphics[width=0.245\linewidth]{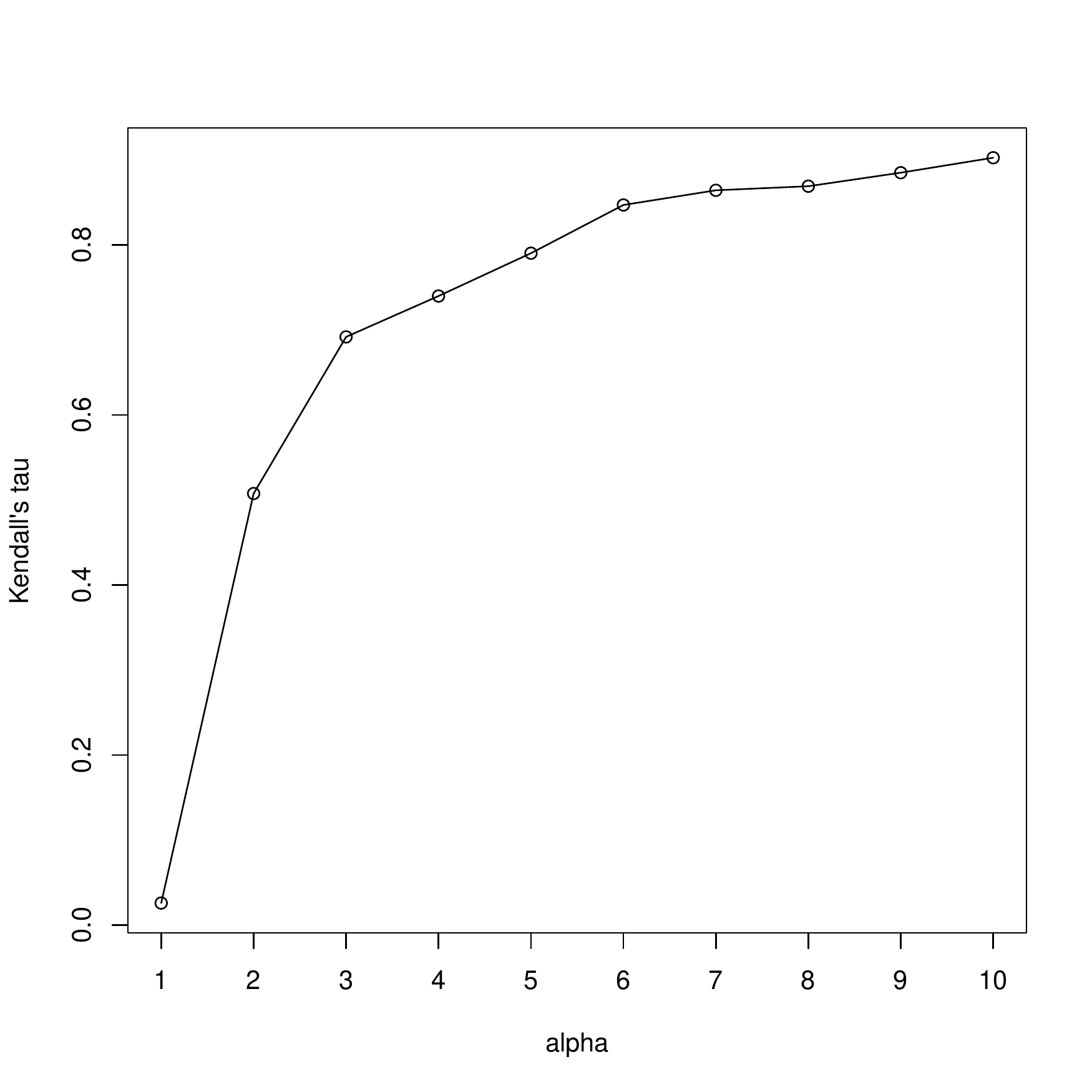}}
	\subfigure[Hoeff]{\includegraphics[width=0.245\linewidth]{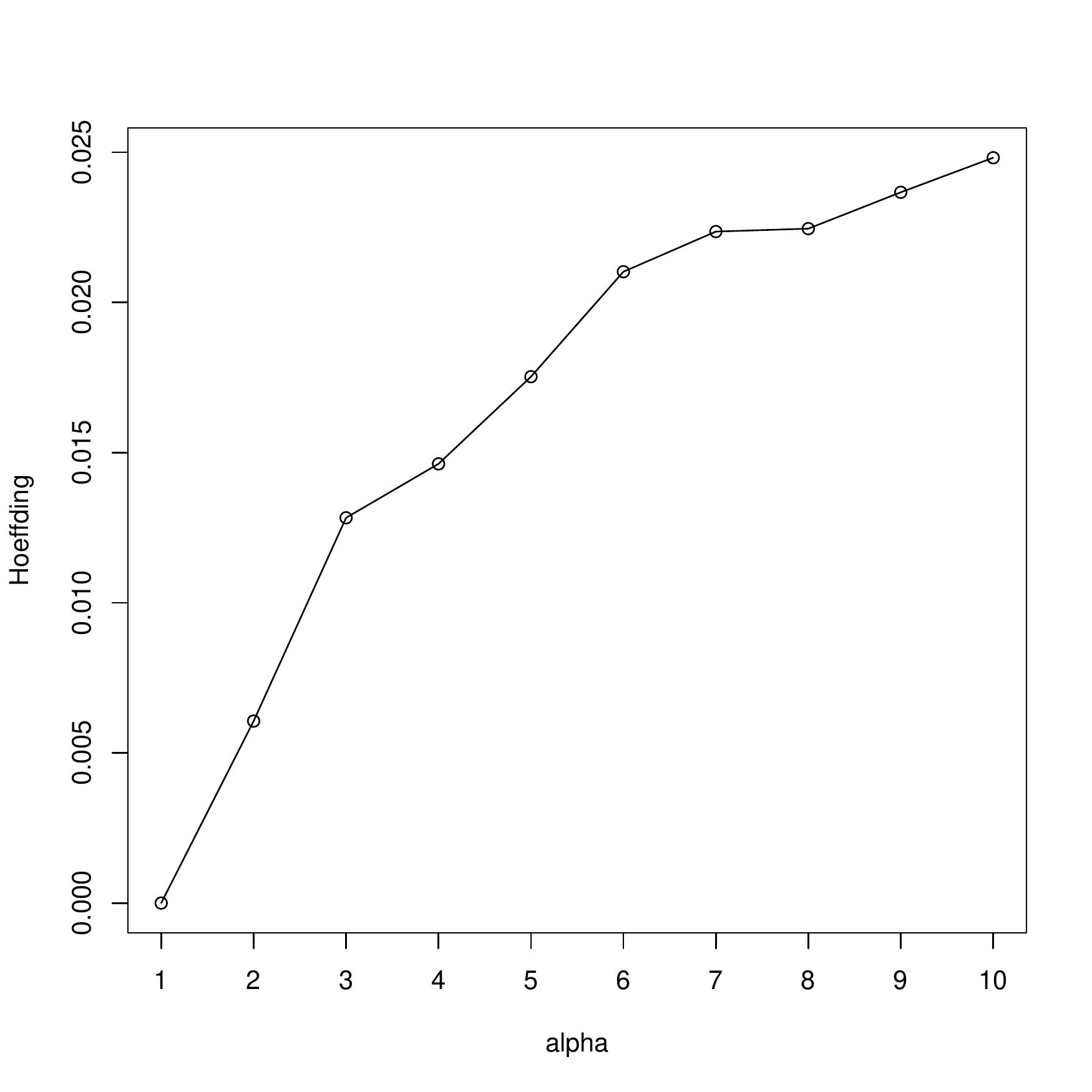}}
	\subfigure[BDtau]{\includegraphics[width=0.245\linewidth]{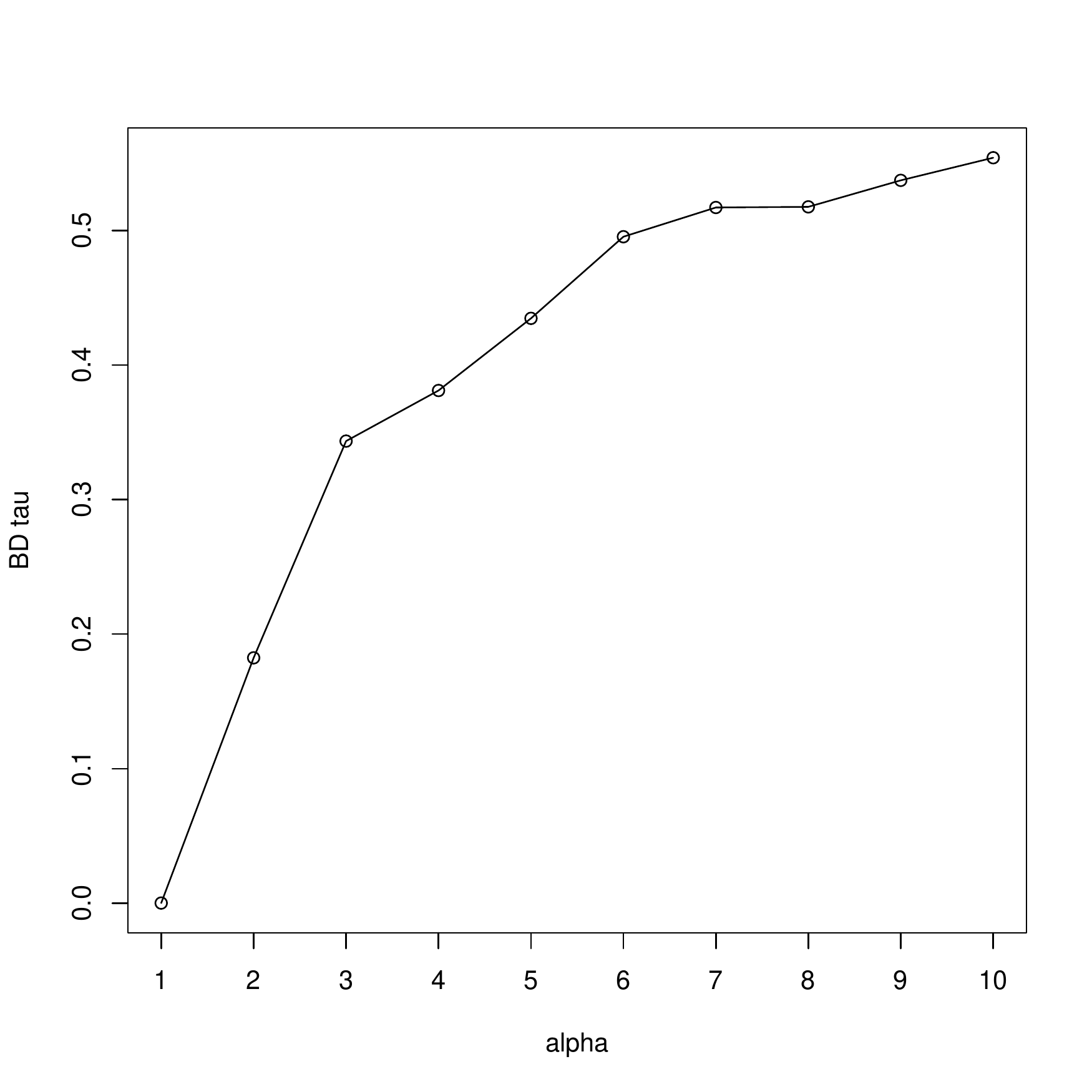}}
	\subfigure[HHG.chisq]{\includegraphics[width=0.245\linewidth]{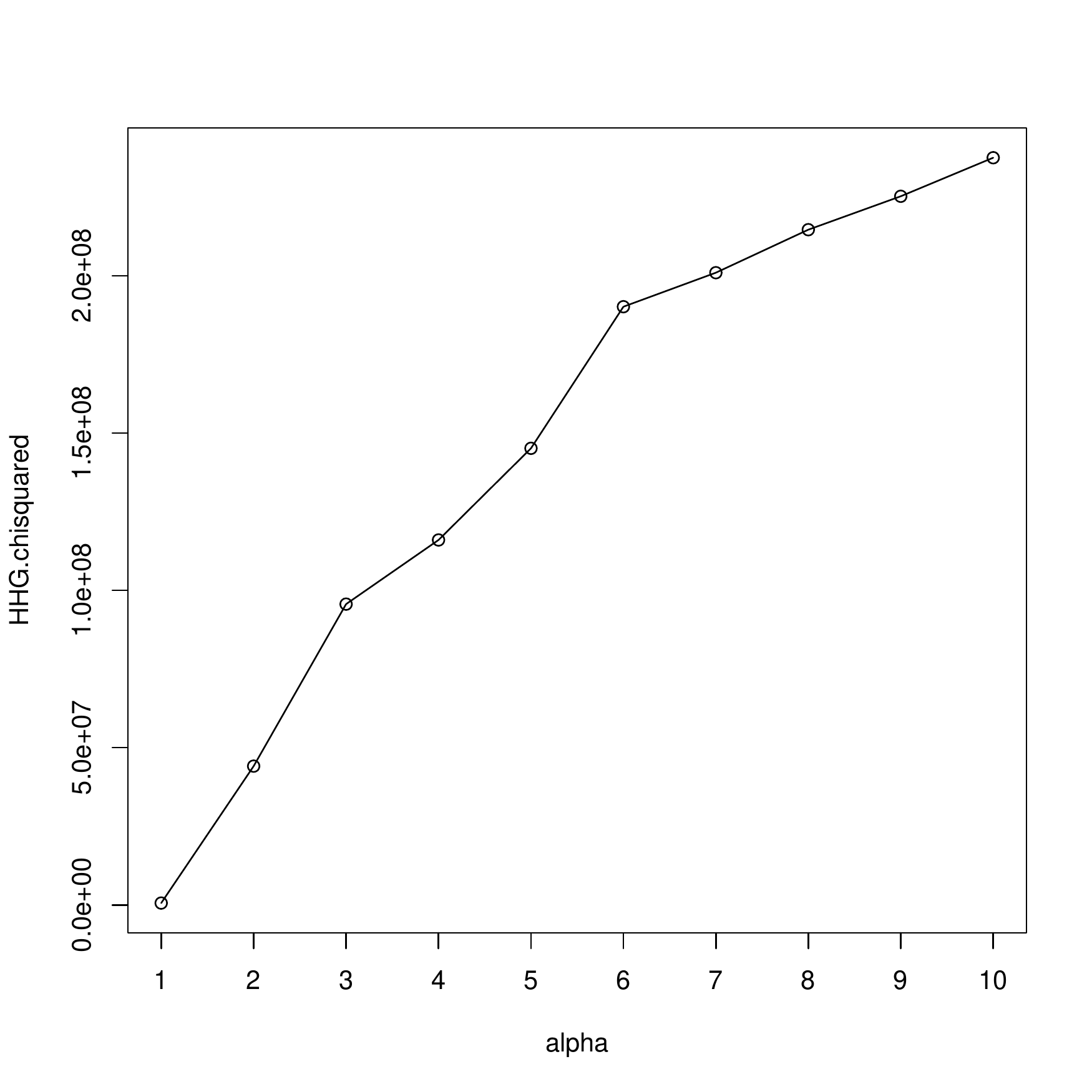}}
	\subfigure[HHG.lr]{\includegraphics[width=0.245\linewidth]{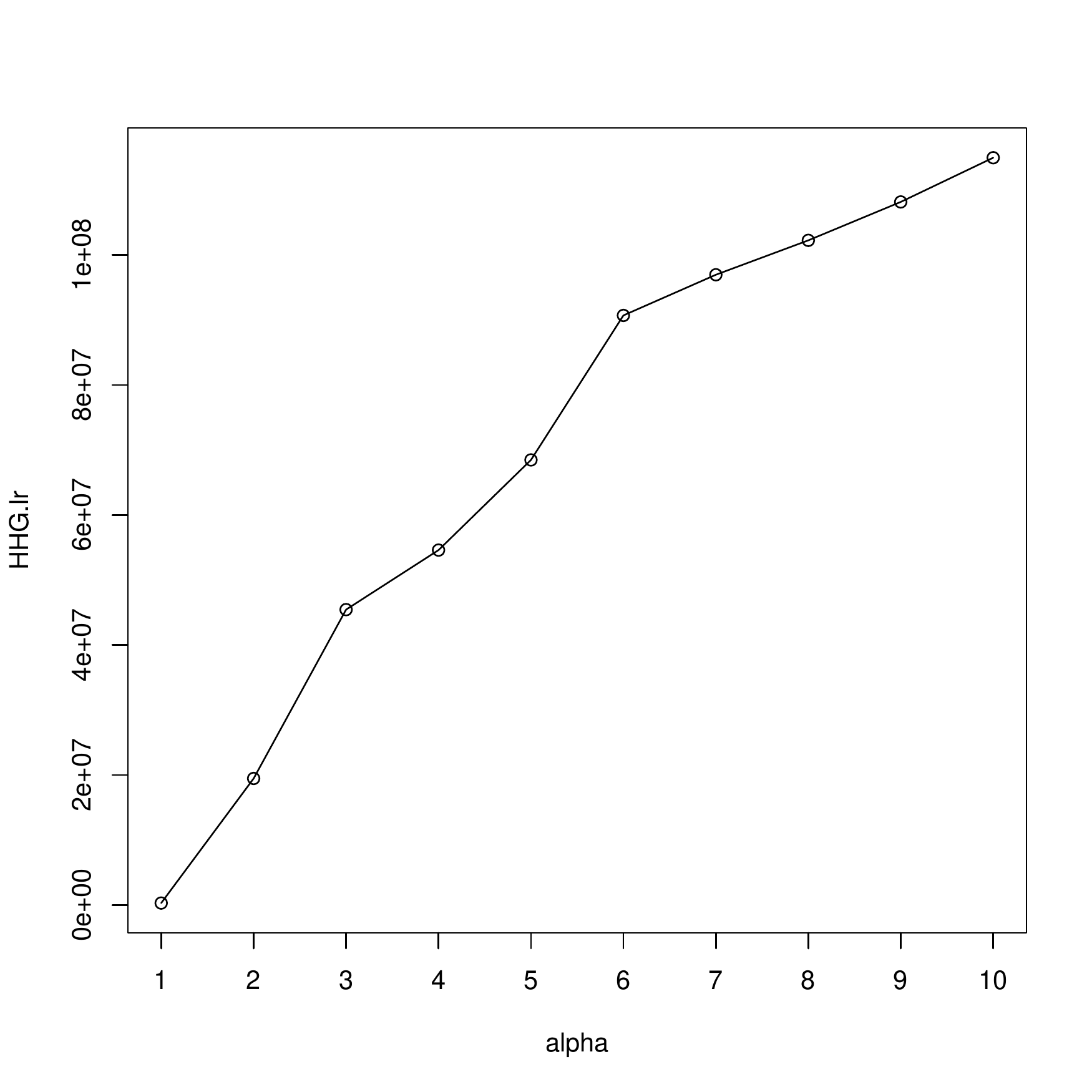}}
	\subfigure[Ball]{\includegraphics[width=0.245\linewidth]{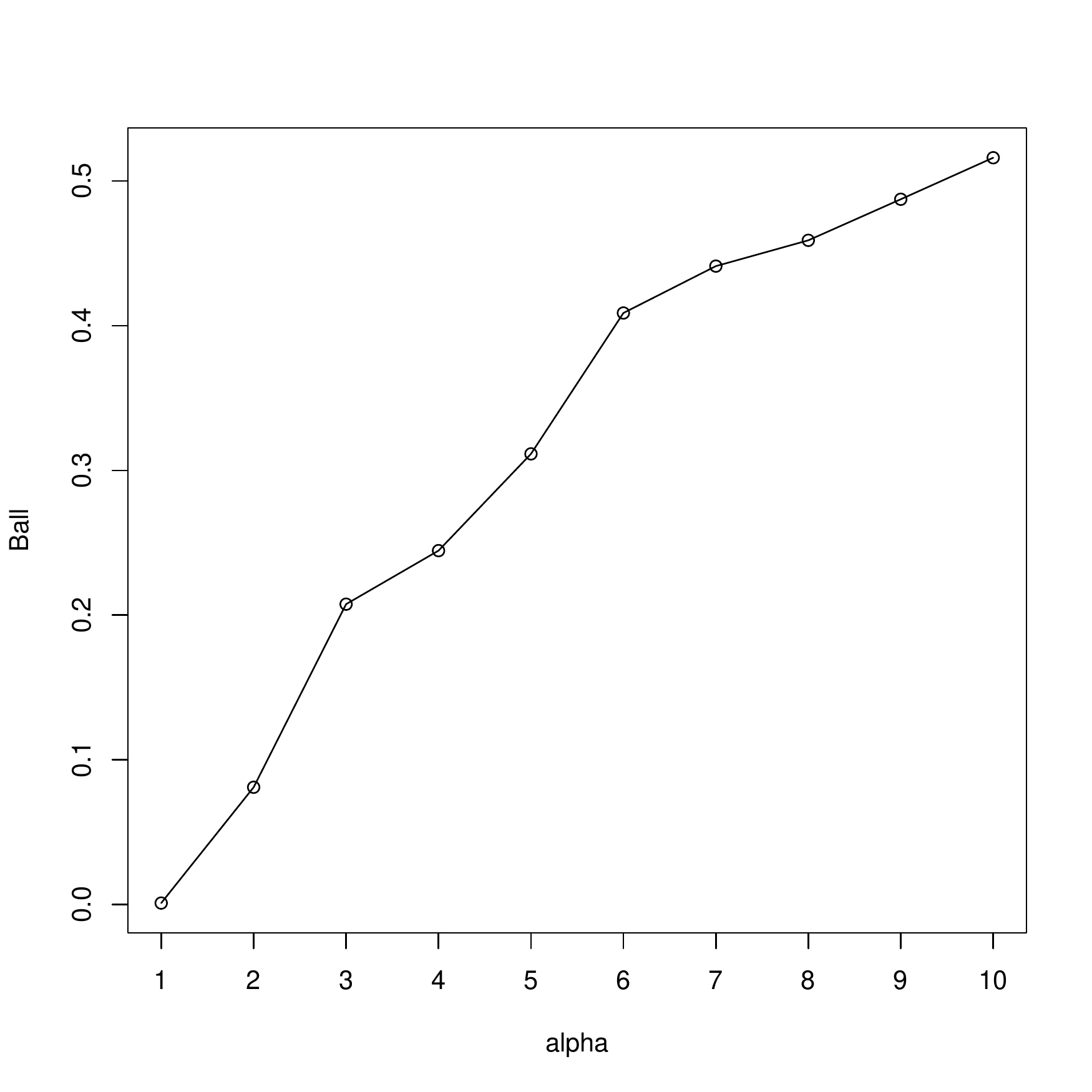}}
	\subfigure[BET]{\includegraphics[width=0.245\linewidth]{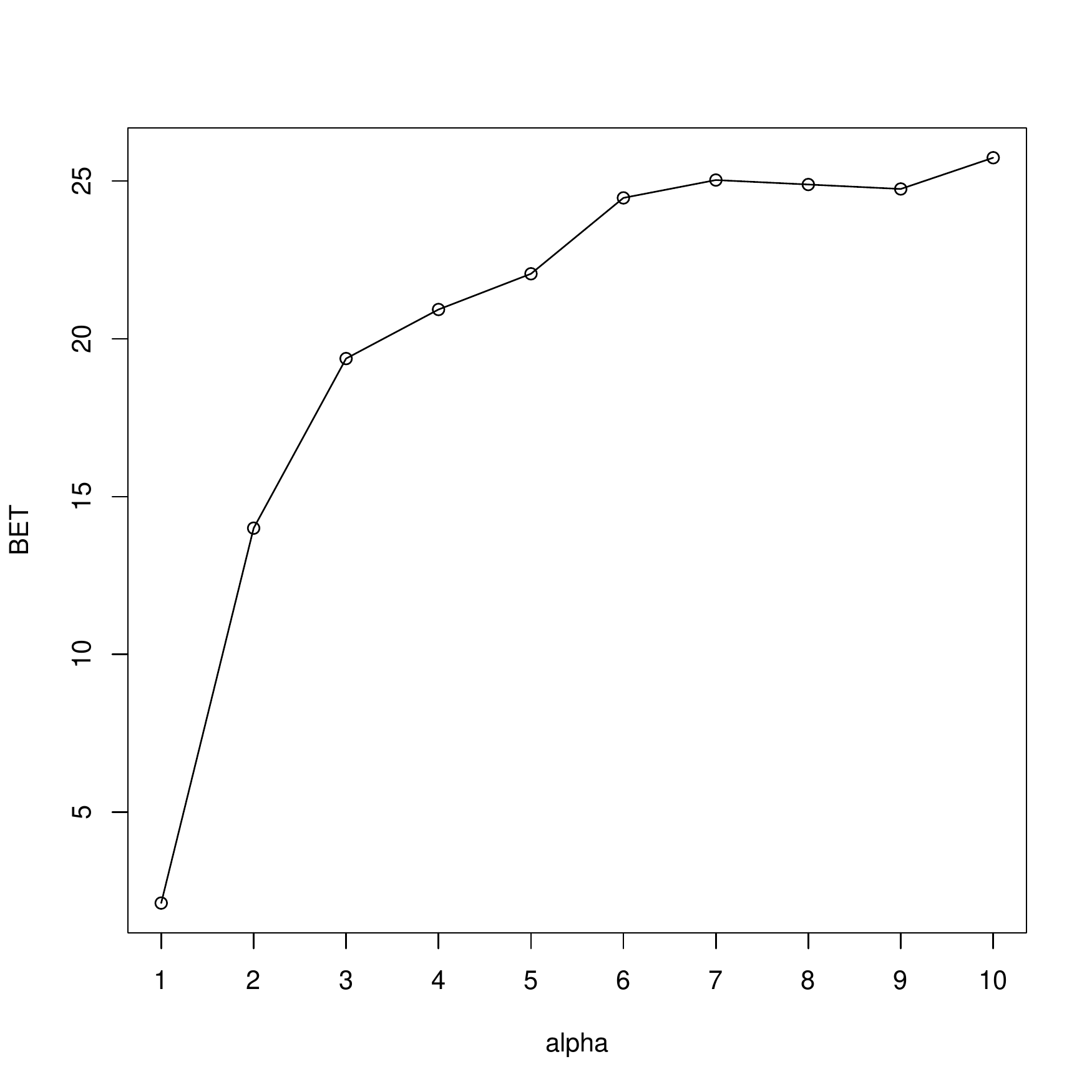}}
	\subfigure[QAD]{\includegraphics[width=0.245\linewidth]{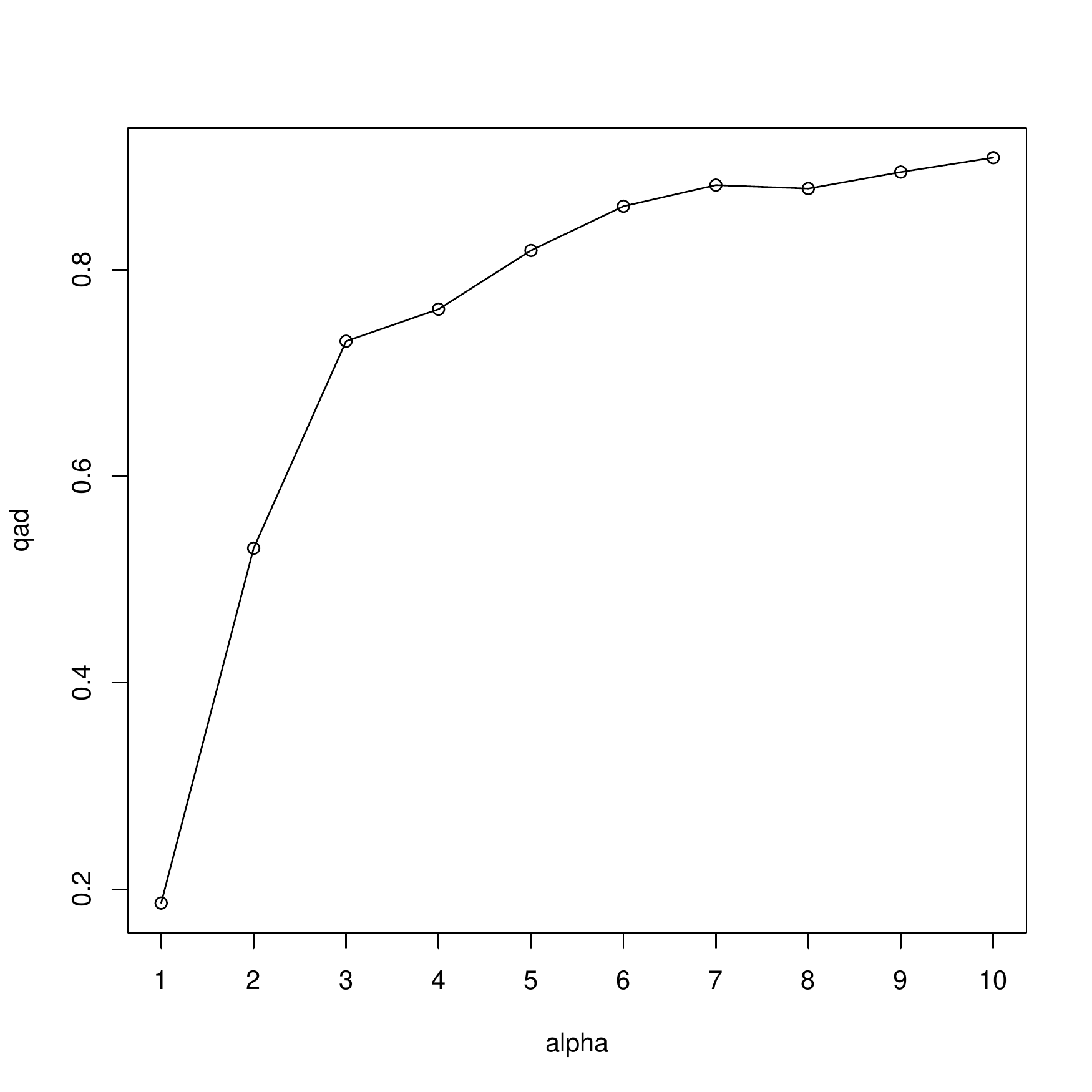}}
	\subfigure[mixed]{\includegraphics[width=0.245\linewidth]{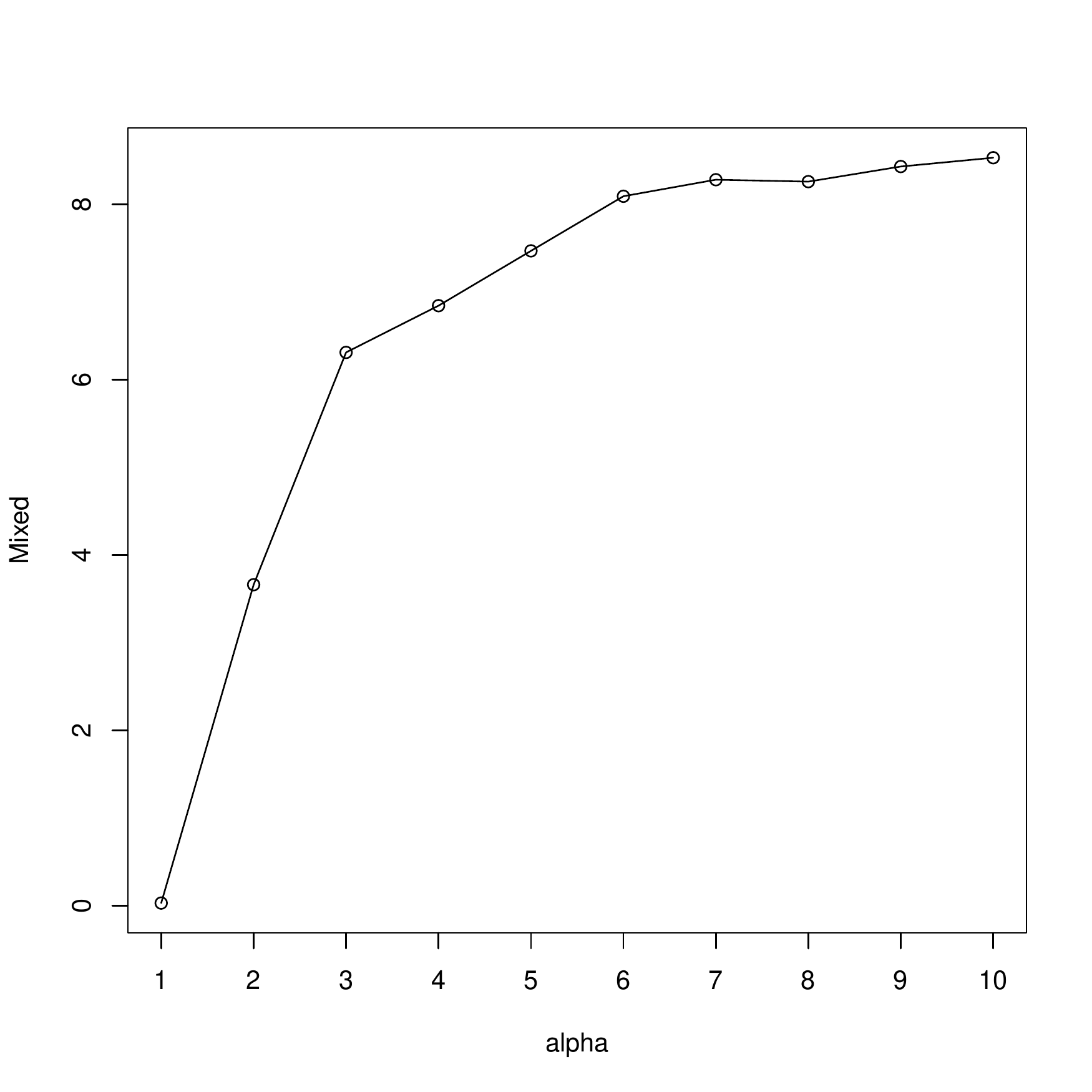}}
	\subfigure[CODEC]{\includegraphics[width=0.245\linewidth]{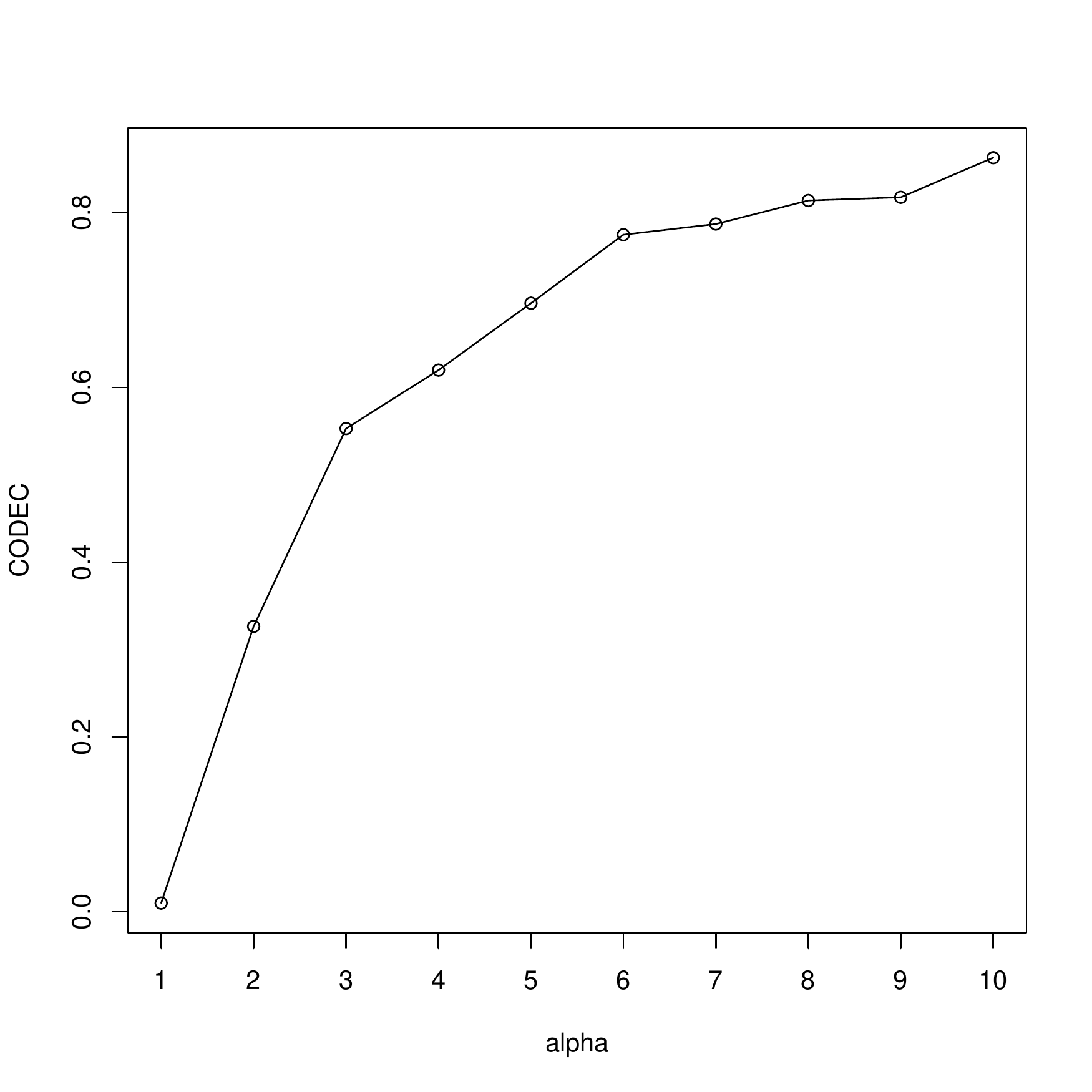}}
	\subfigure[subcop]{\includegraphics[width=0.245\linewidth]{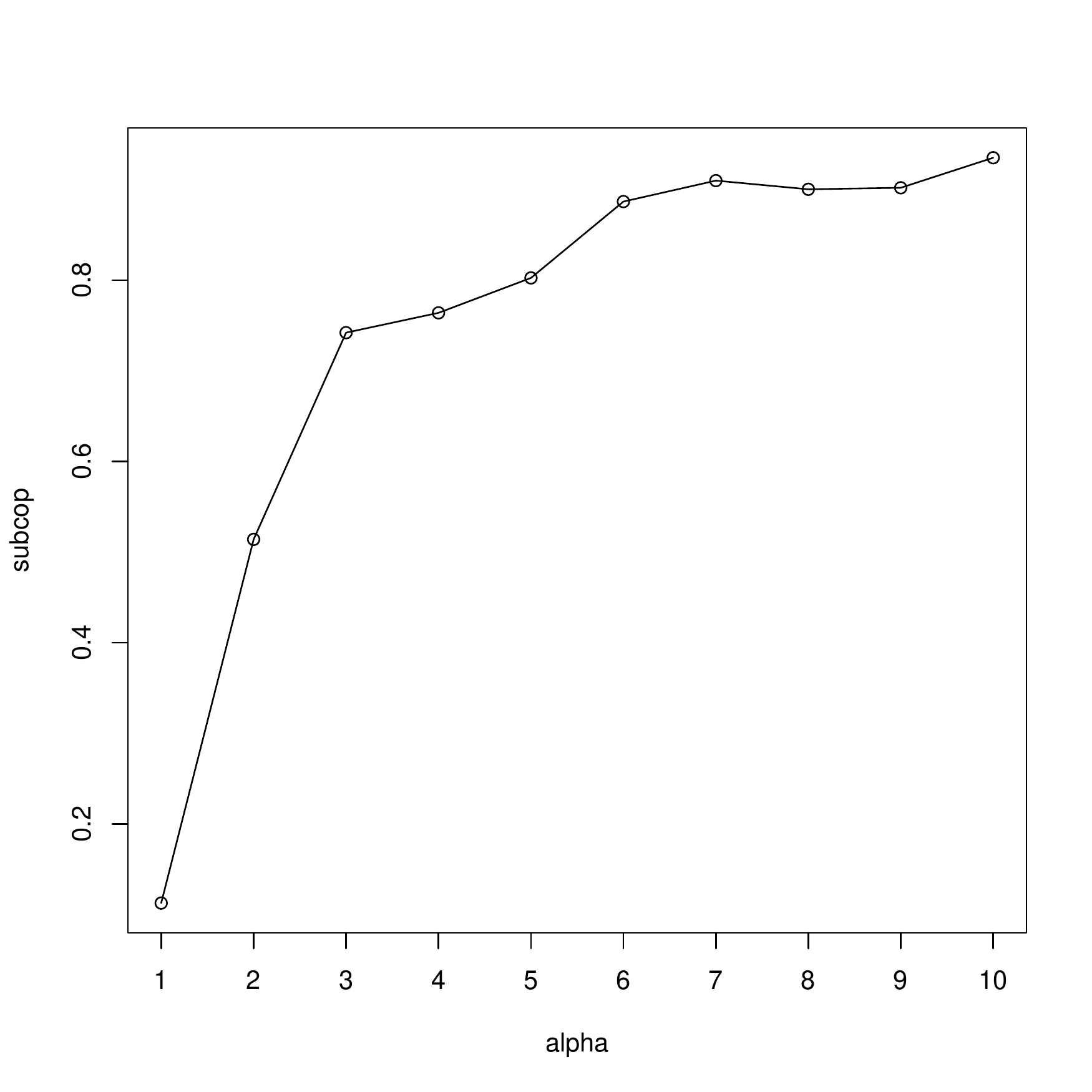}}
	\subfigure[dCor]{\includegraphics[width=0.245\linewidth]{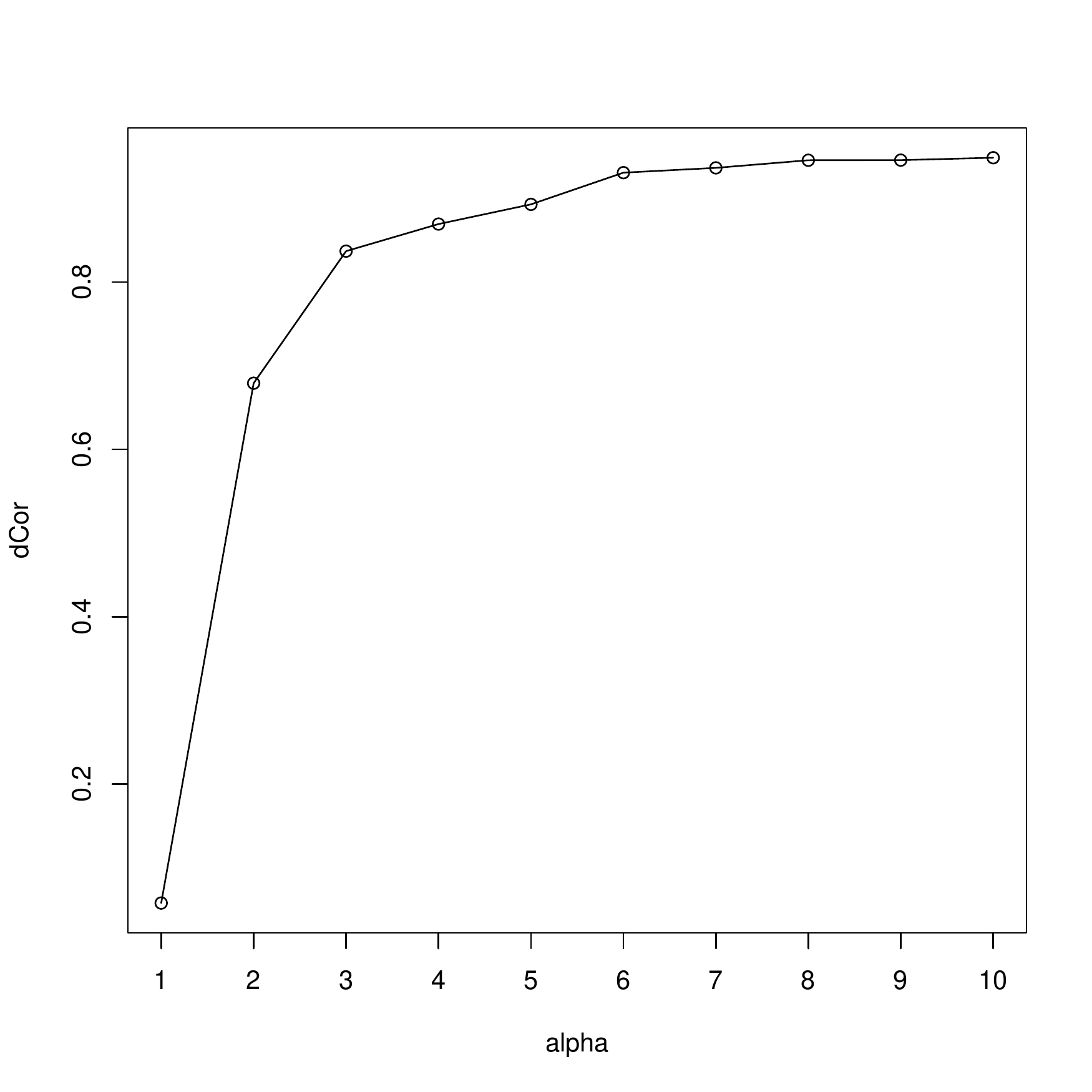}}
	\subfigure[mdm]{\includegraphics[width=0.245\linewidth]{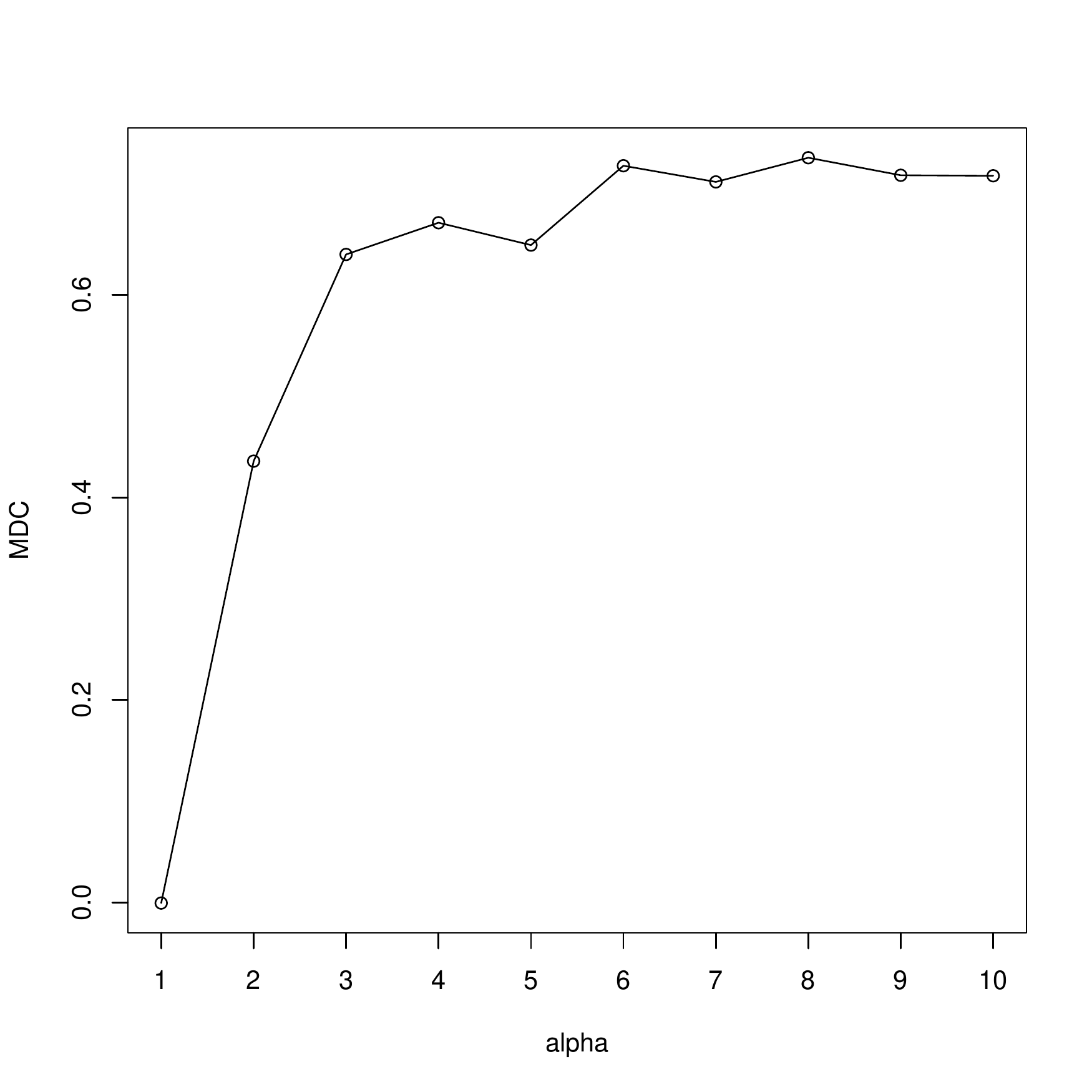}}
	\subfigure[dHSIC]{\includegraphics[width=0.245\linewidth]{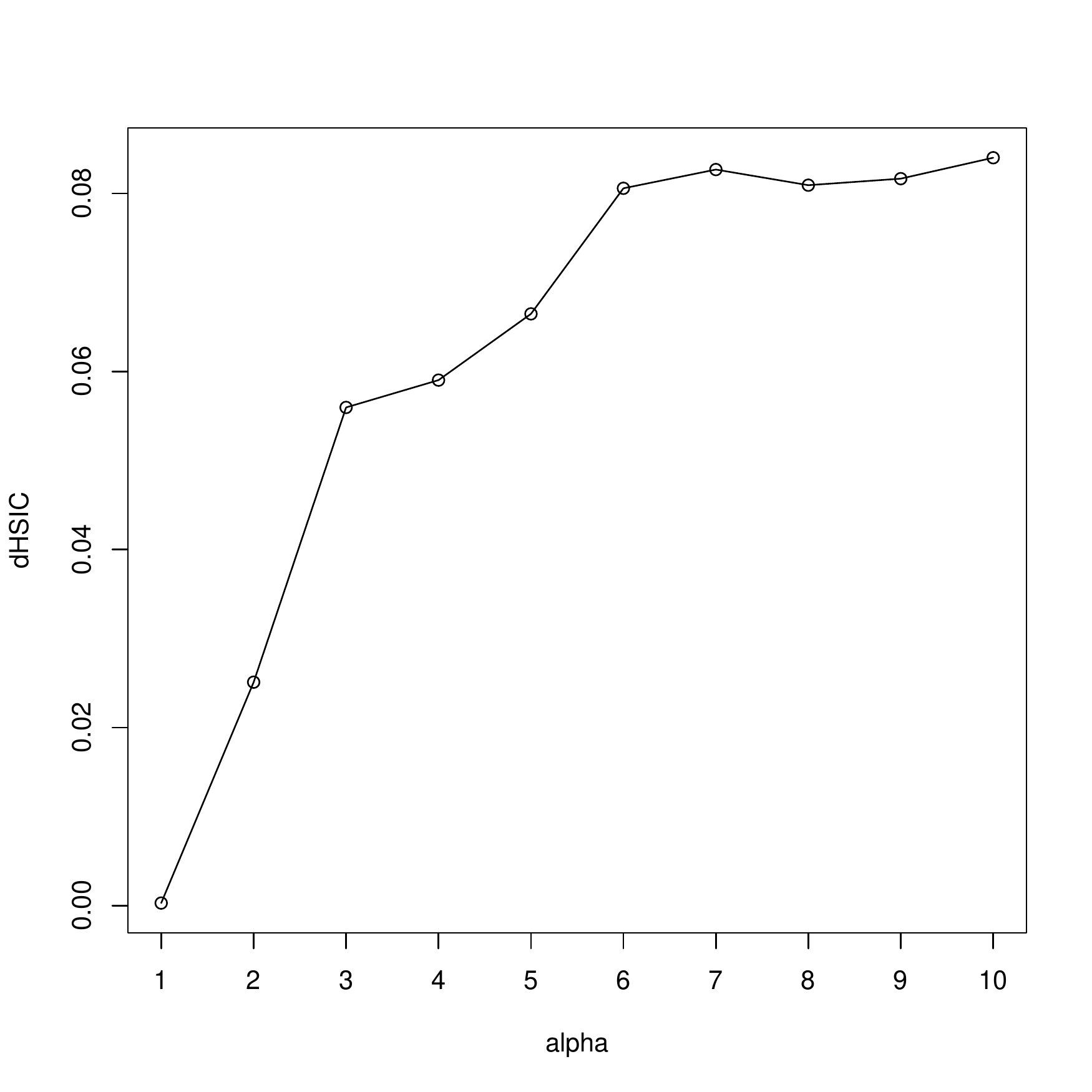}}
	\subfigure[NNS]{\includegraphics[width=0.245\linewidth]{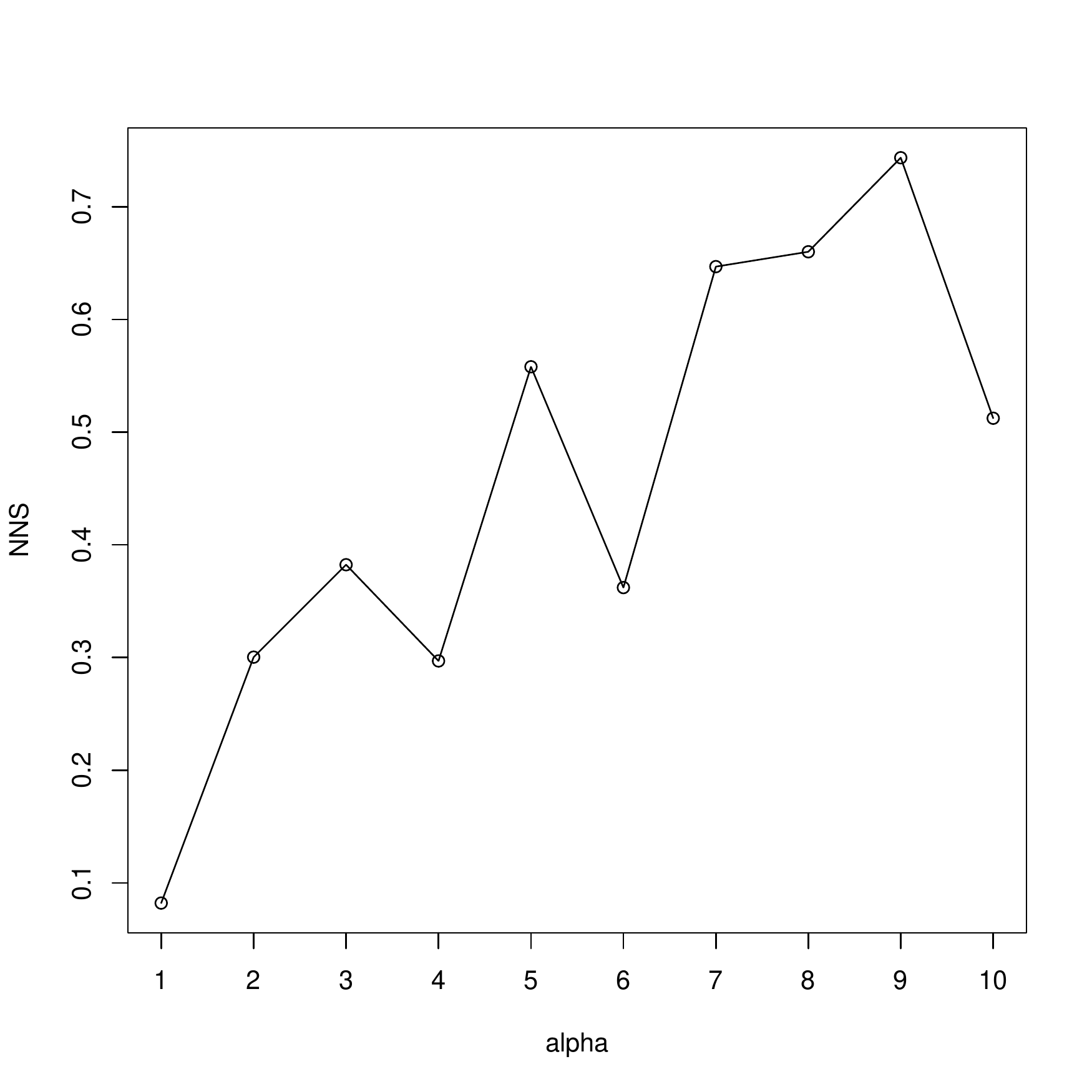}}
	\caption{Estimation of the independence measures from the simulated data of the bivariate Gumbel copula.}
	\label{fig:bigumbel}
\end{figure}

\begin{figure}
	\centering
	\includegraphics[width=0.9\linewidth]{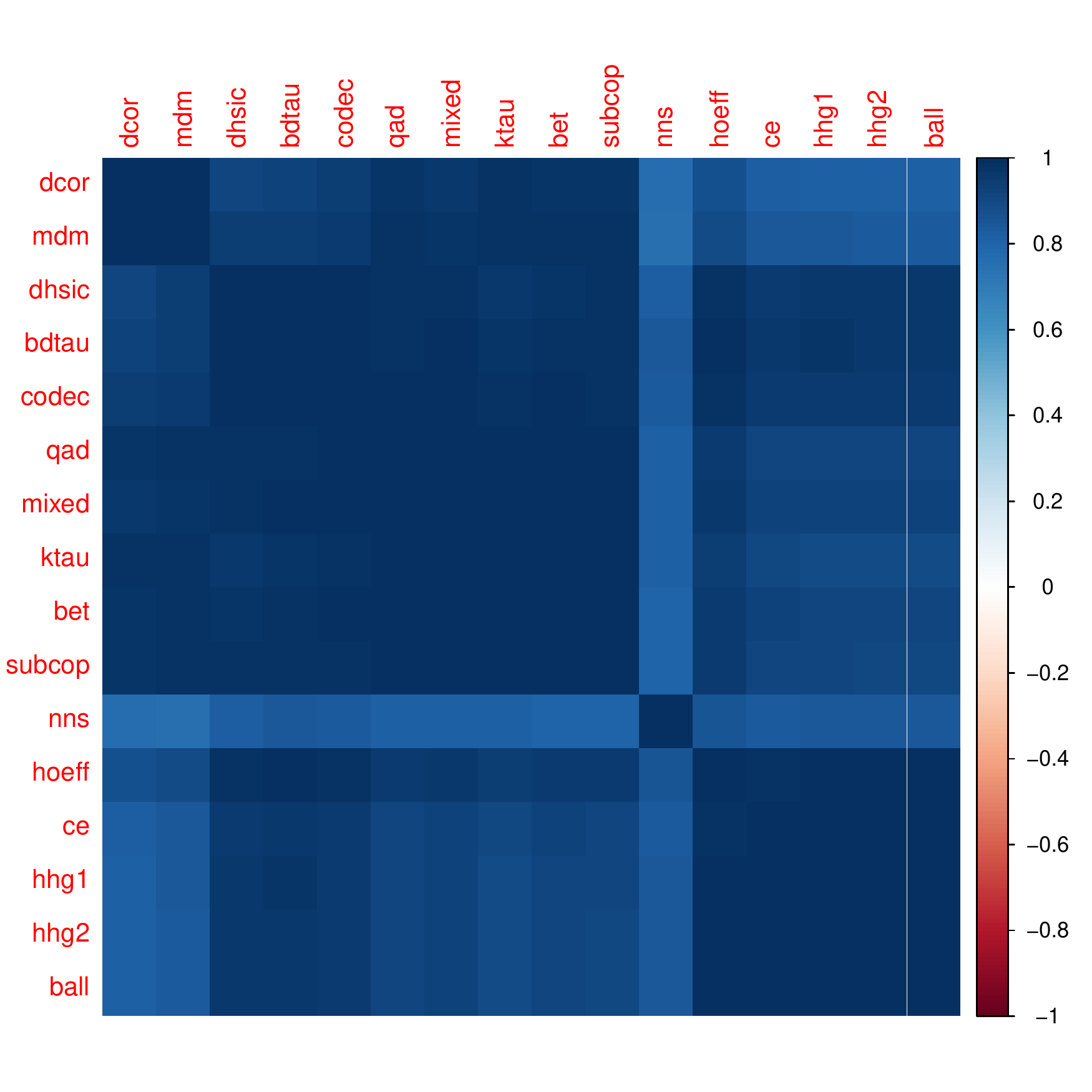}
	\caption{Correlation matrix of the independence measures estimated from the simulated data of the bivariate Gumble copula.}
	\label{fig:bigumbelcm}
\end{figure}

For the fourth experiment, the estimation of the independence measures from the bivariate Gumbel copula is shown in Figure \ref{fig:bigumbel} and the correlation matrix of the estimation of the measures is shown in Figure \ref{fig:bigumbelcm}. According the correlation matrix, the measures can be seperated into four groups: \{dCor, MDM\}, \{dHSIC, BDtau, CODEC, QAD, mixed, Ktau, BET, subcop\}, \{NNS\}, and all the others.

\begin{figure}
	\subfigure[CE]{\includegraphics[width=0.245\linewidth]{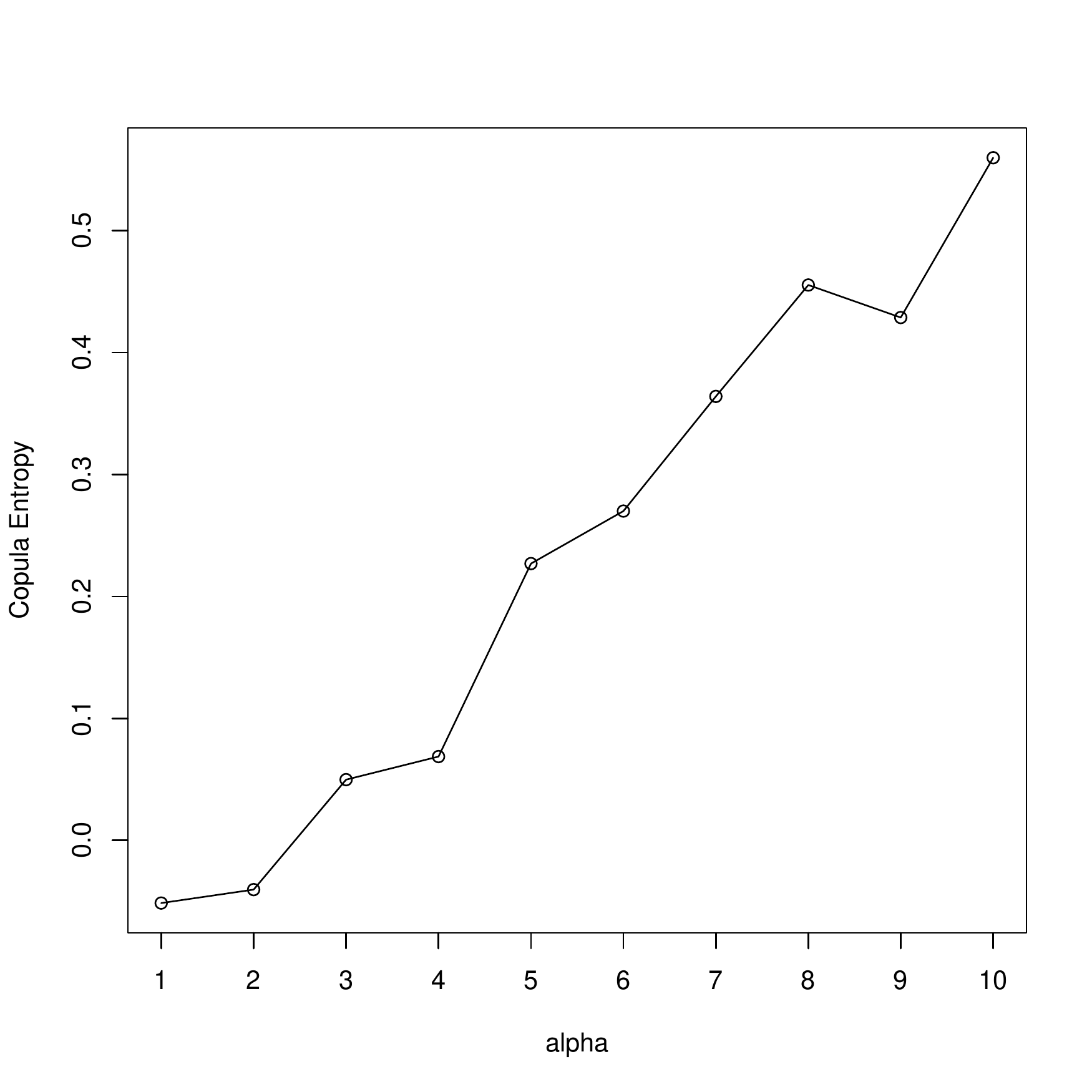}}
	\subfigure[Ktau]{\includegraphics[width=0.245\linewidth]{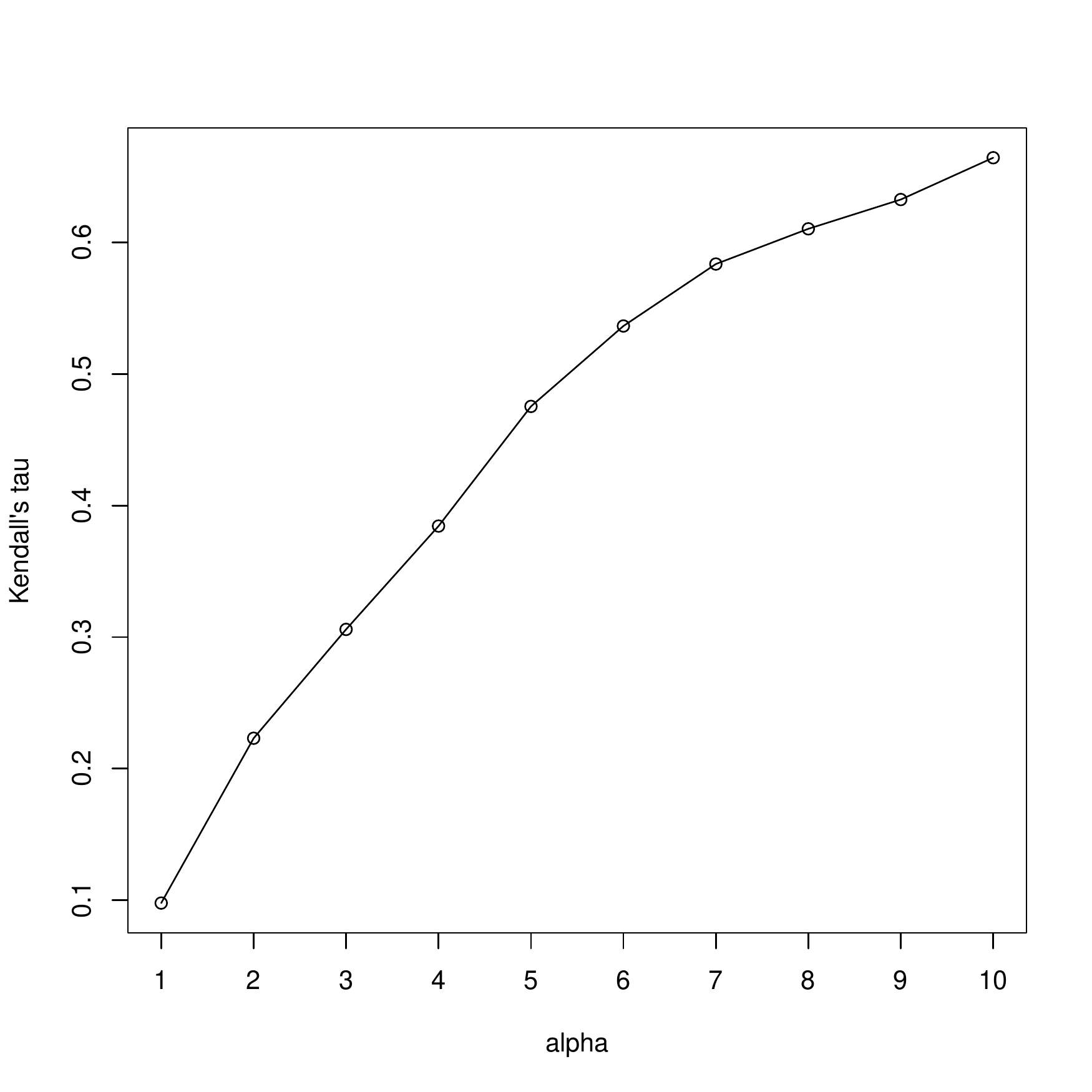}}
	\subfigure[Hoeff]{\includegraphics[width=0.245\linewidth]{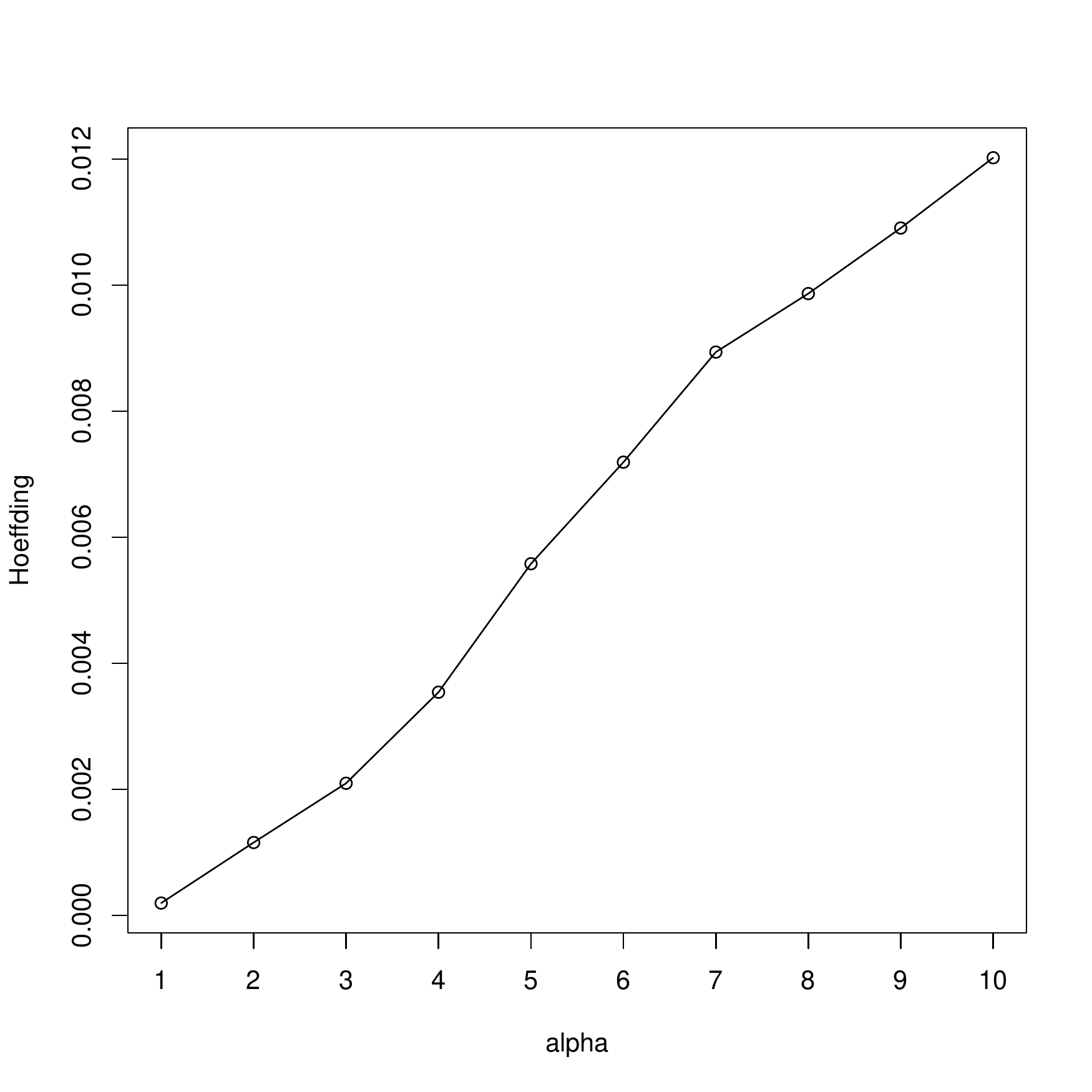}}
	\subfigure[BDtau]{\includegraphics[width=0.245\linewidth]{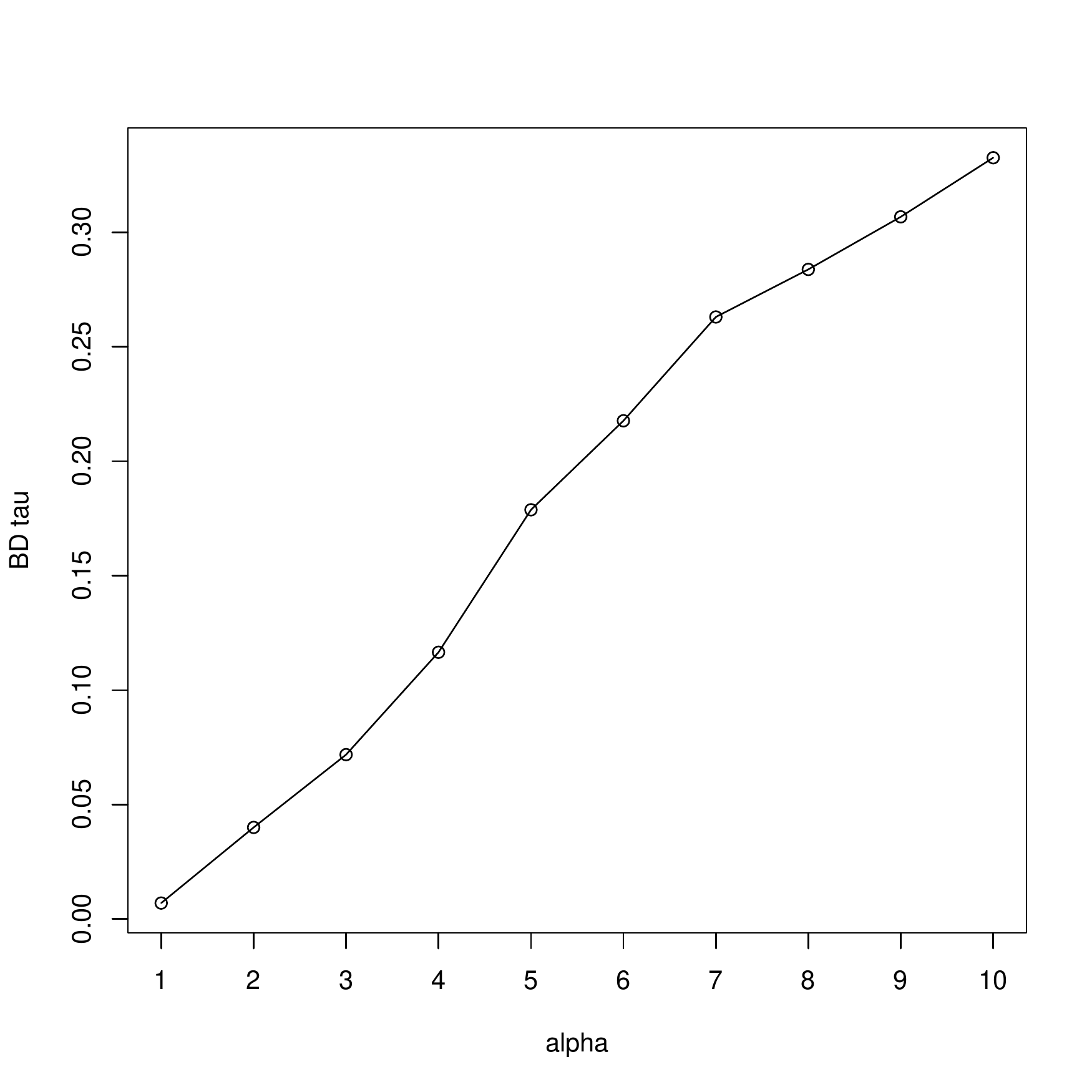}}
	\subfigure[HHG.chisq]{\includegraphics[width=0.245\linewidth]{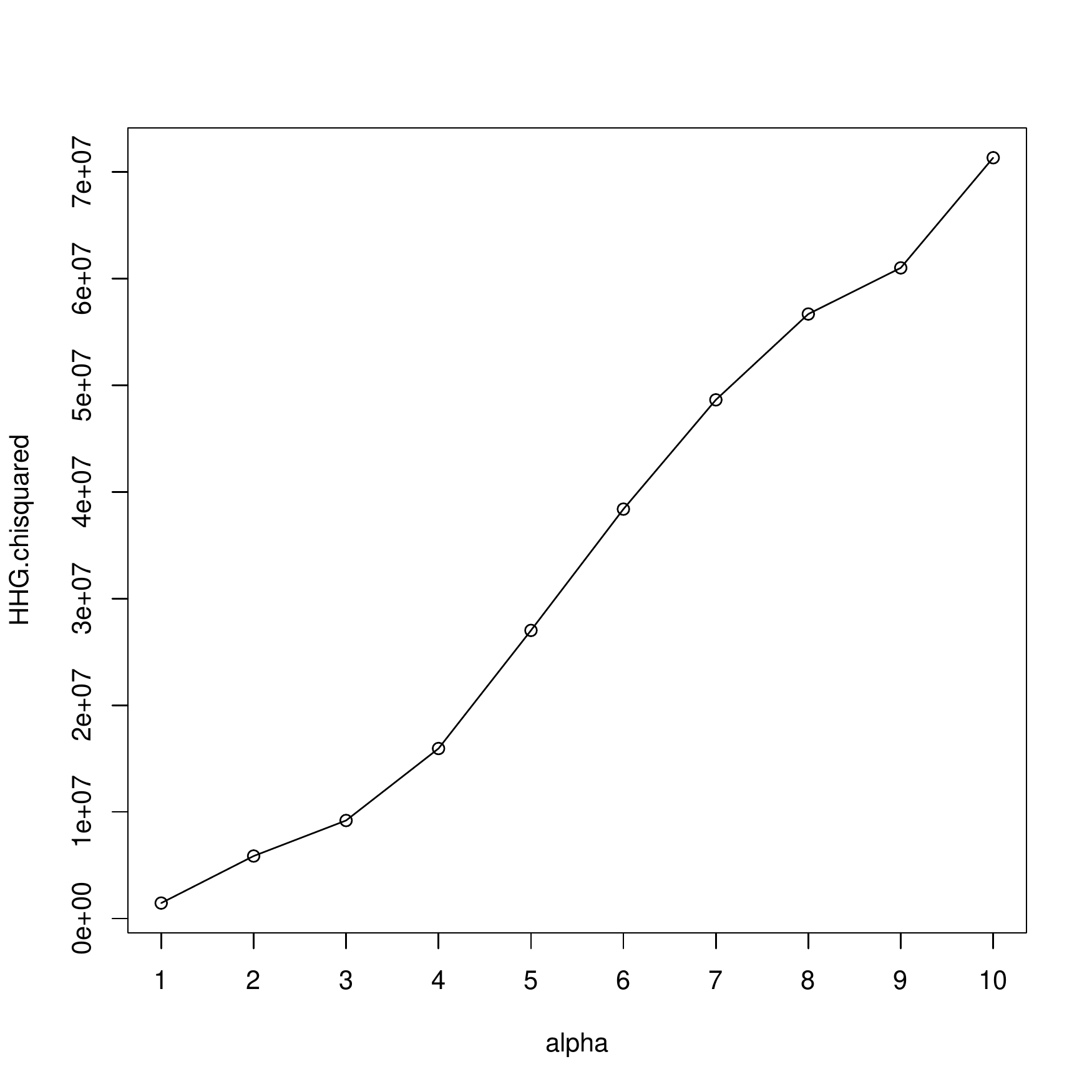}}
	\subfigure[HHG.lr]{\includegraphics[width=0.245\linewidth]{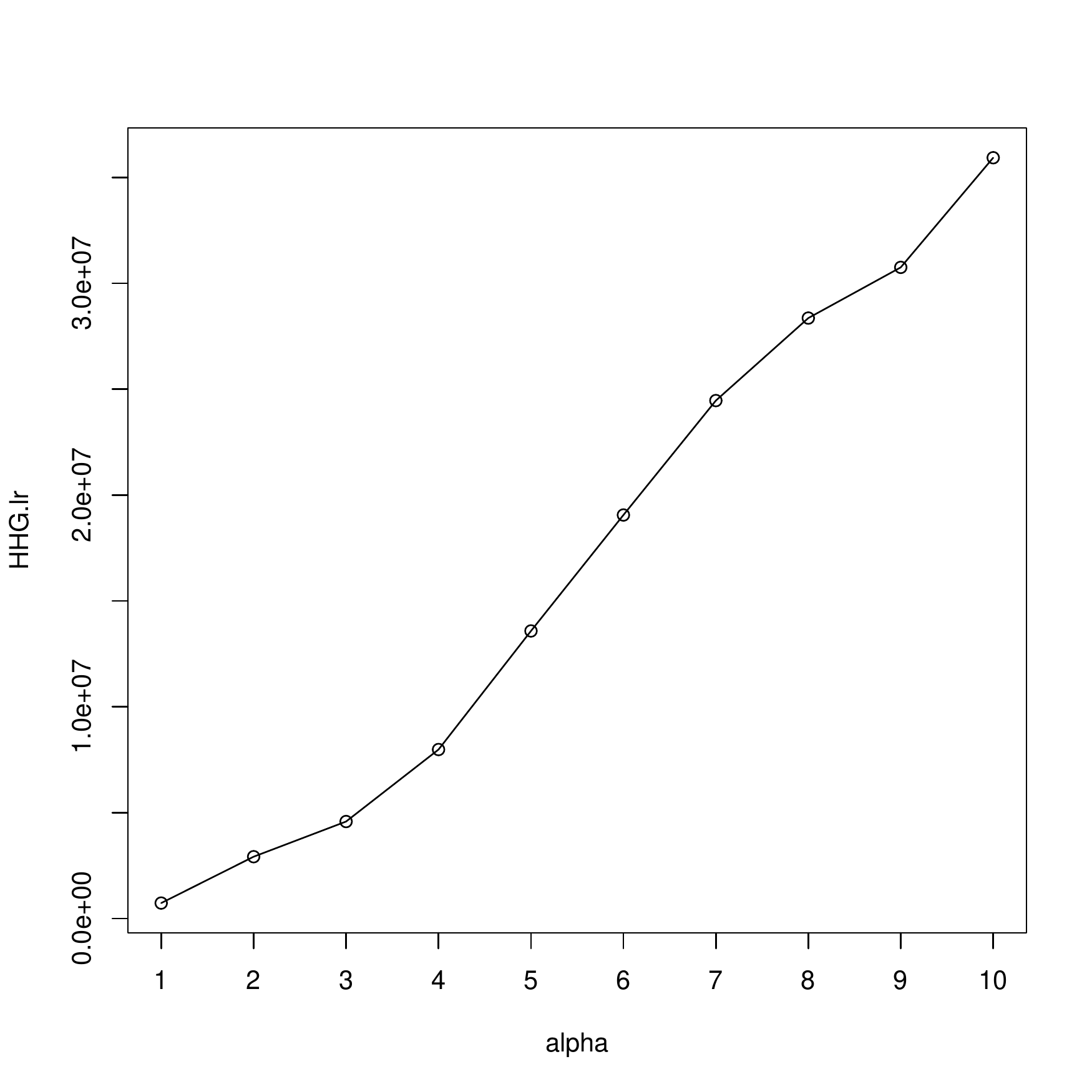}}
	\subfigure[Ball]{\includegraphics[width=0.245\linewidth]{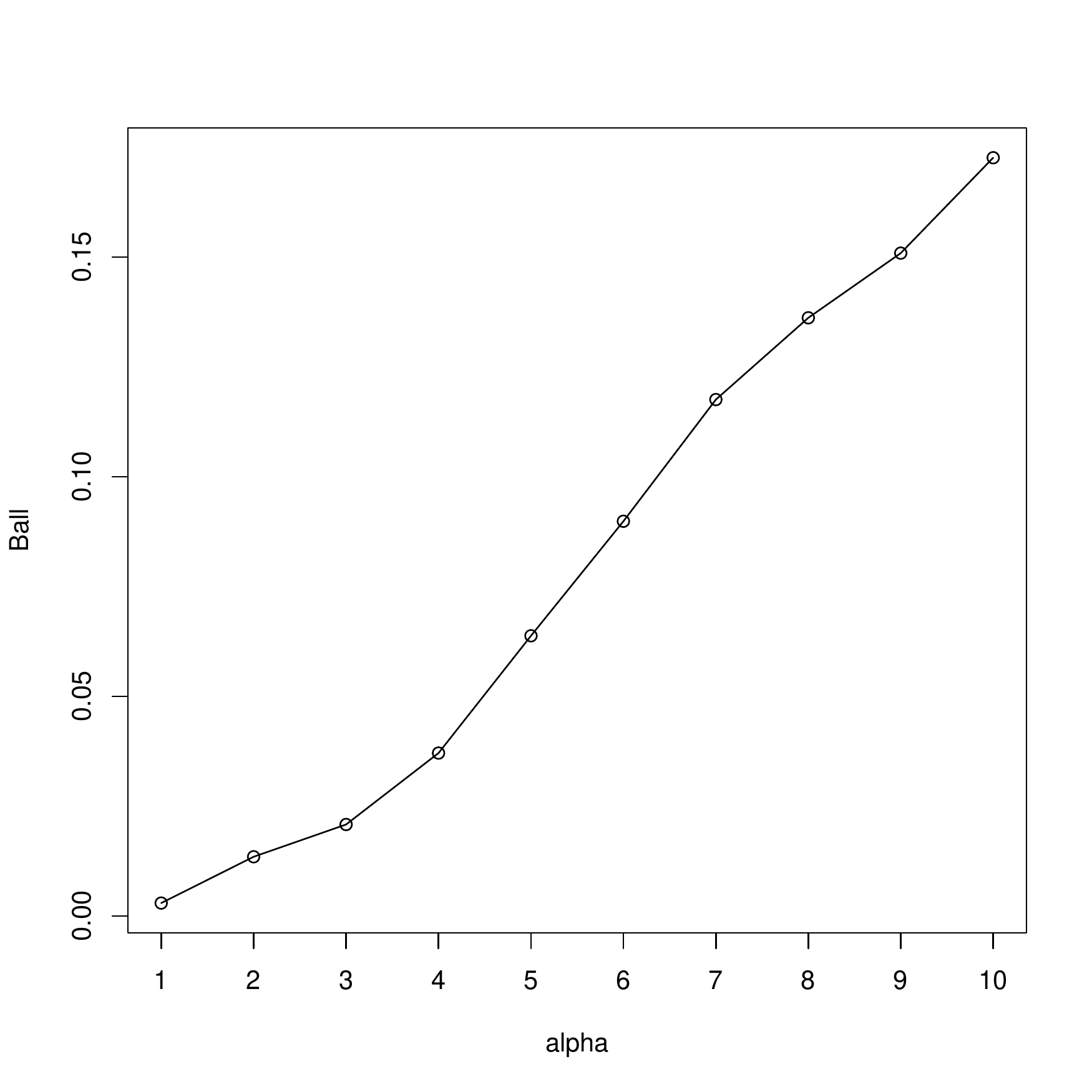}}
	\subfigure[BET]{\includegraphics[width=0.245\linewidth]{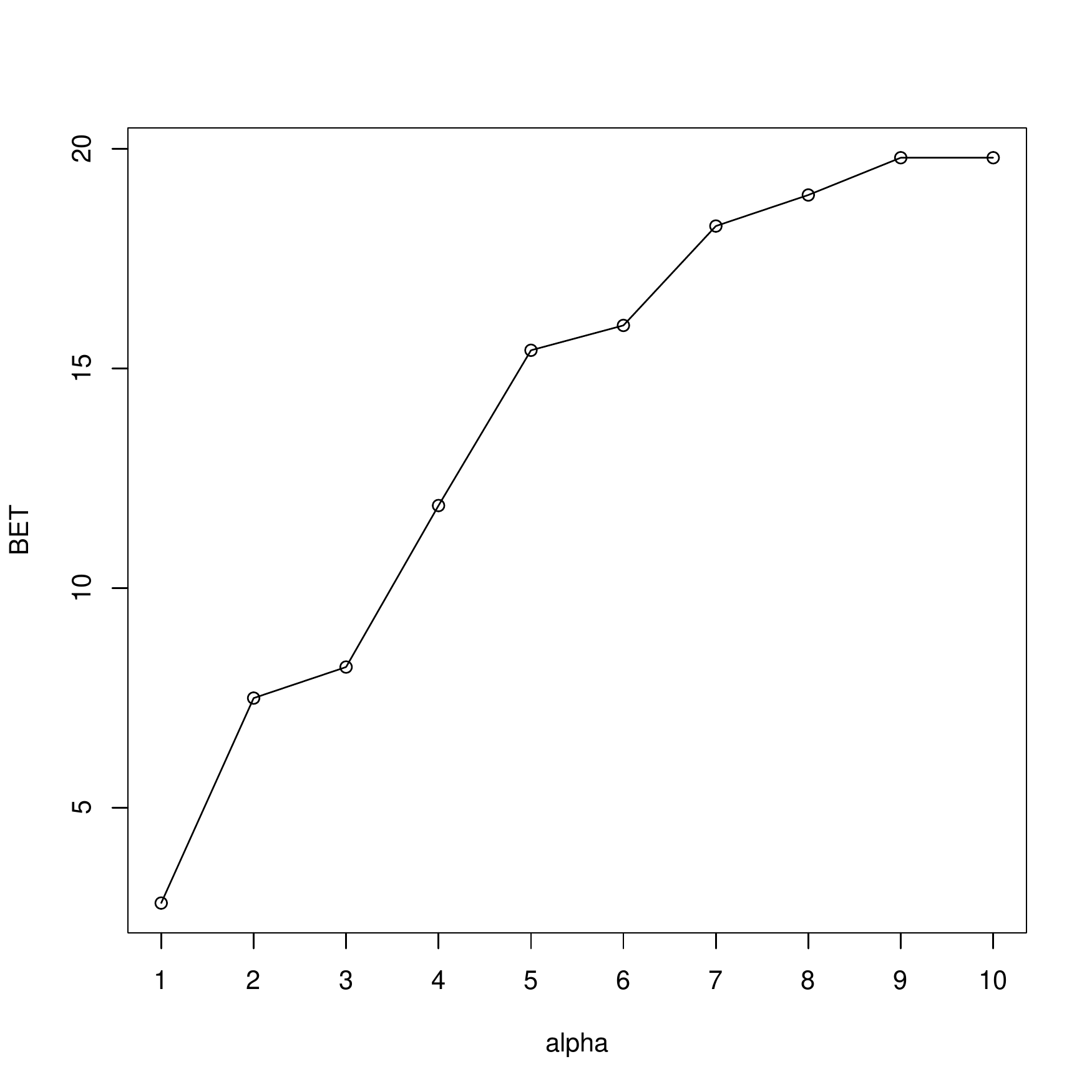}}
	\subfigure[QAD]{\includegraphics[width=0.245\linewidth]{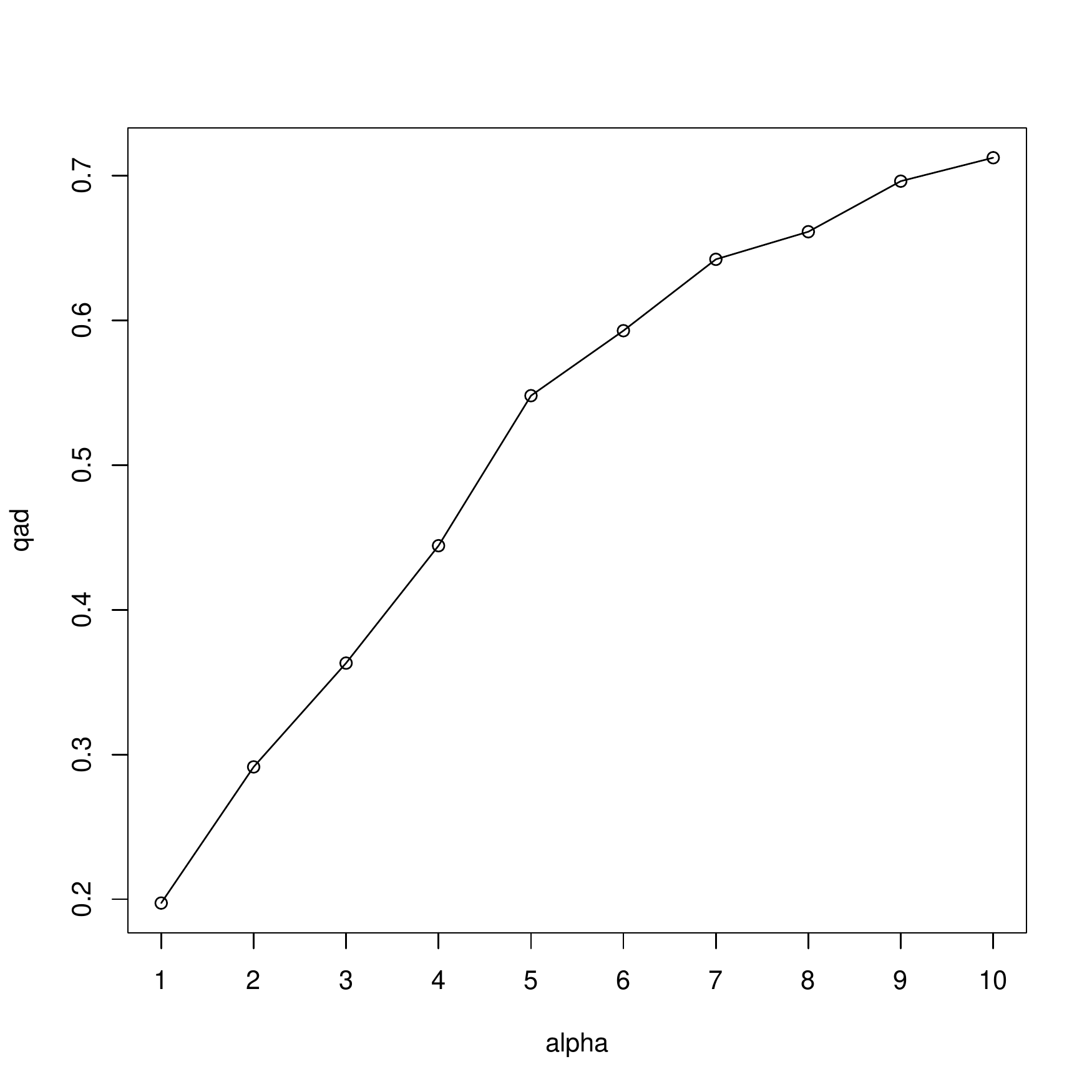}}
	\subfigure[mixed]{\includegraphics[width=0.245\linewidth]{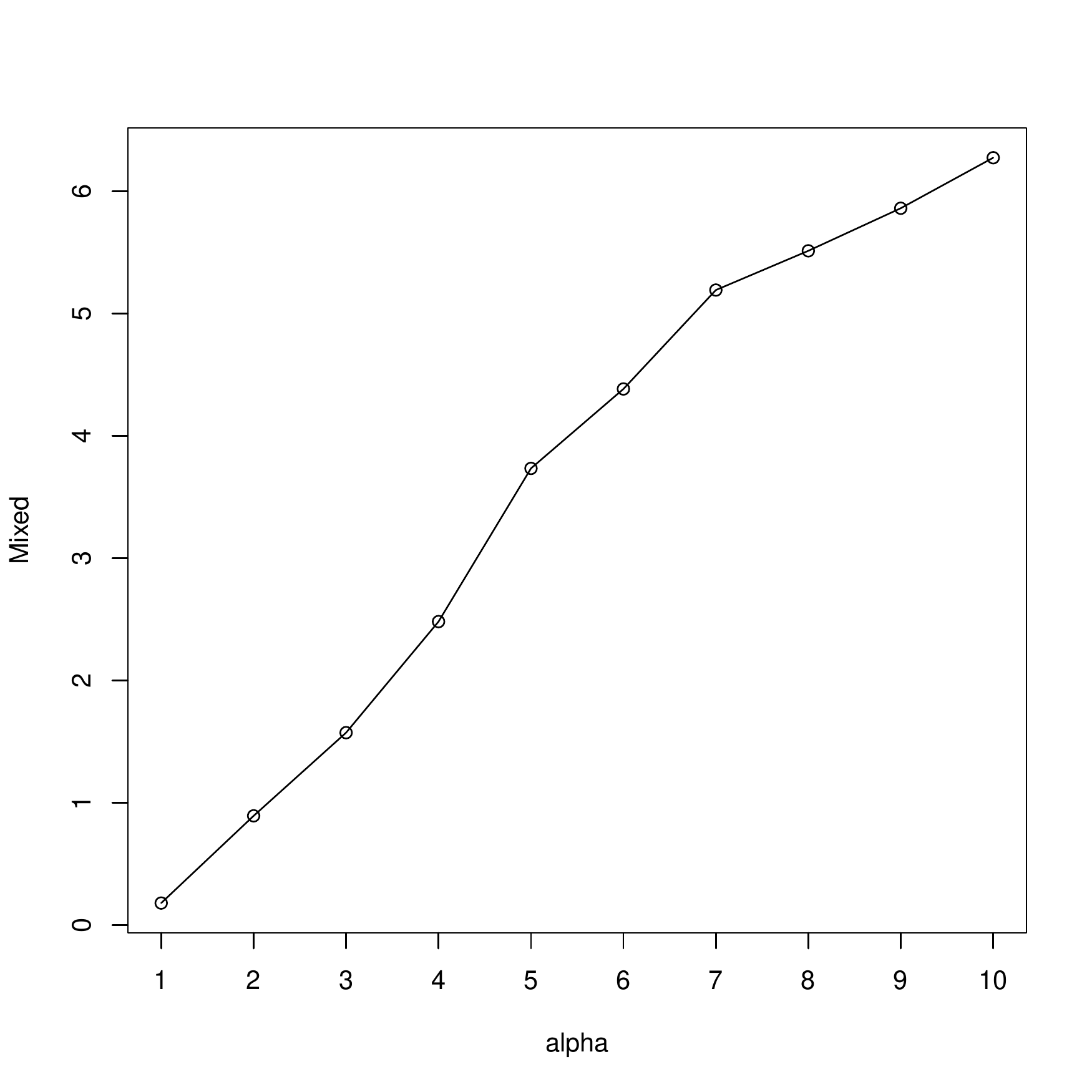}}
	\subfigure[CODEC]{\includegraphics[width=0.245\linewidth]{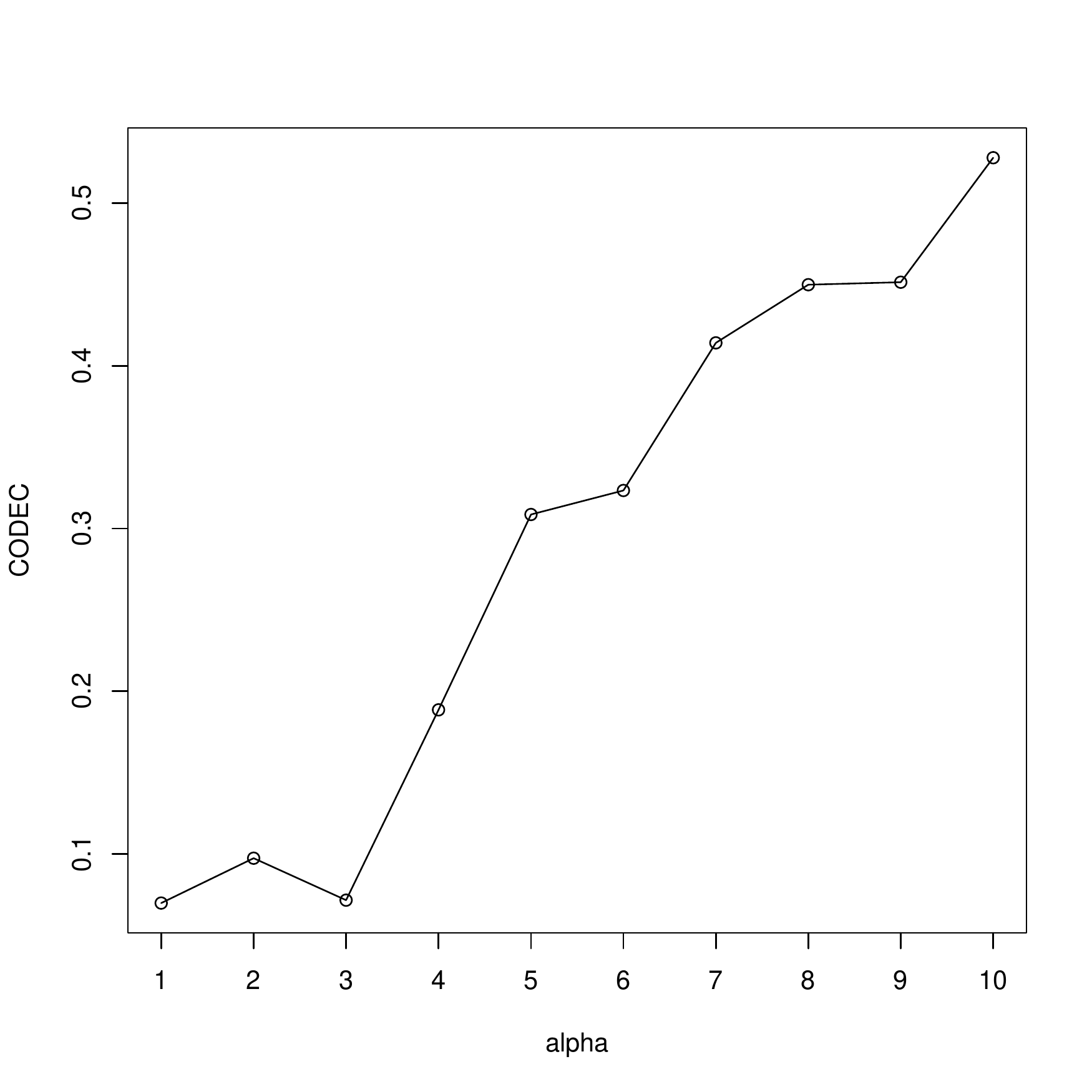}}
	\subfigure[subcop]{\includegraphics[width=0.245\linewidth]{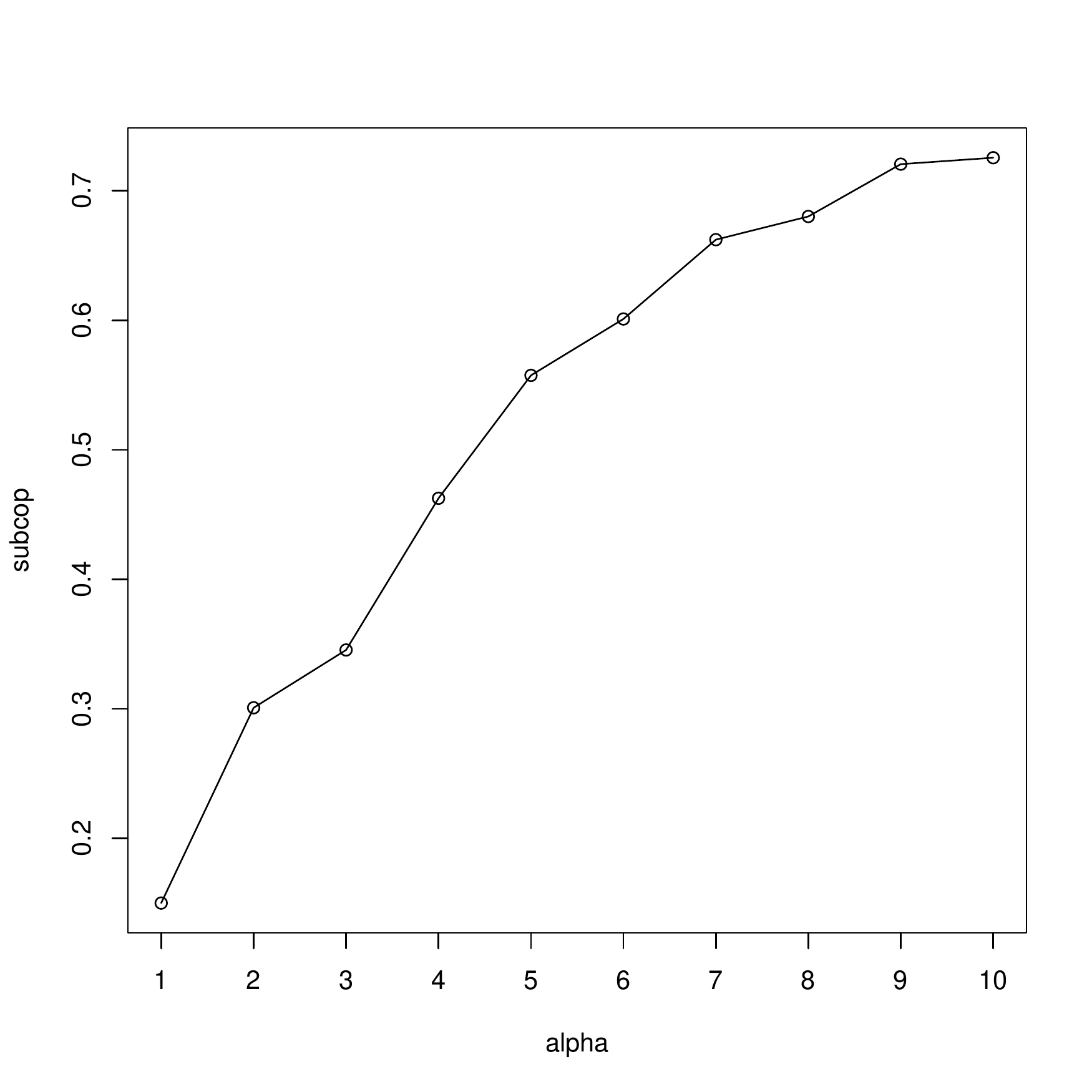}}
	\subfigure[dCor]{\includegraphics[width=0.245\linewidth]{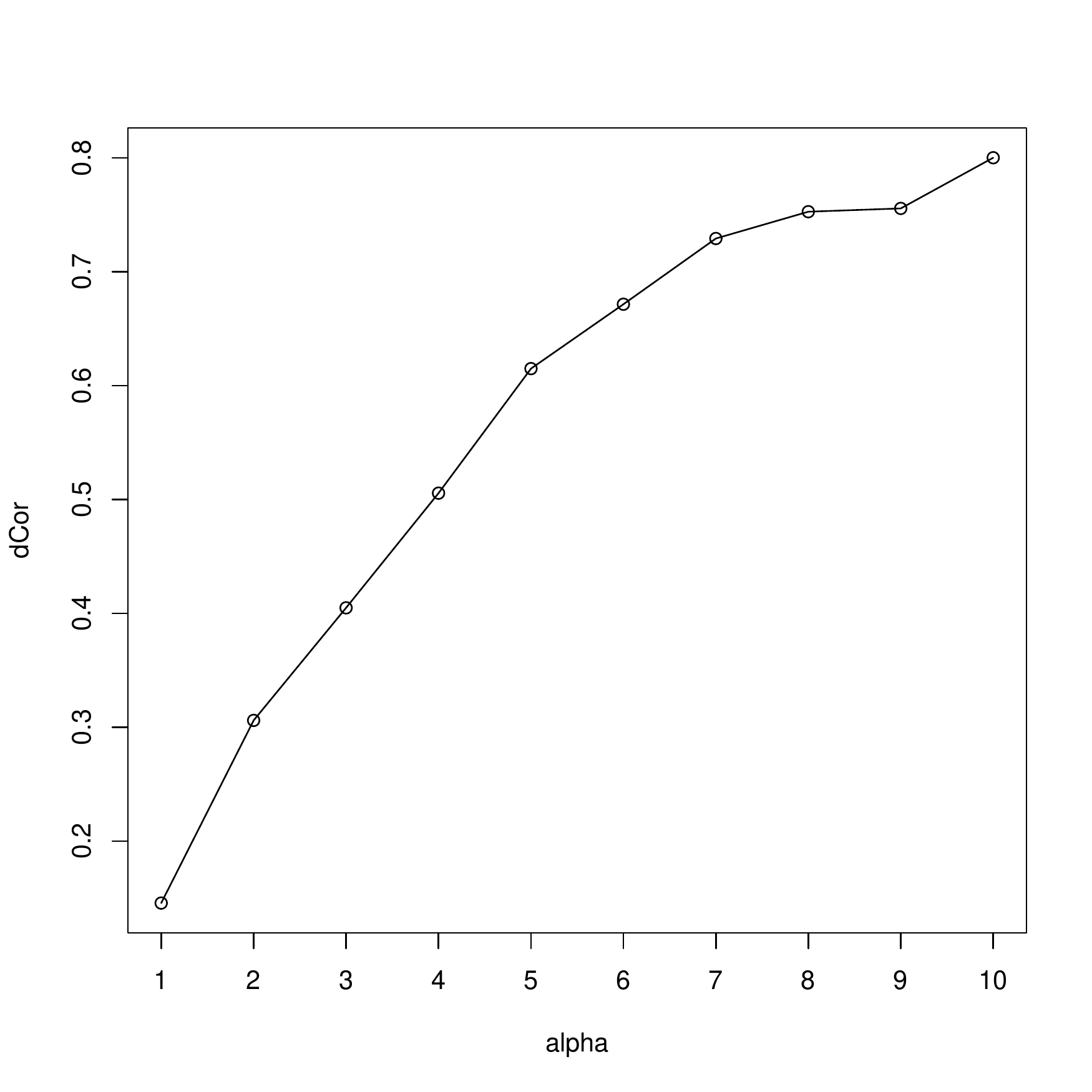}}
	\subfigure[mdm]{\includegraphics[width=0.245\linewidth]{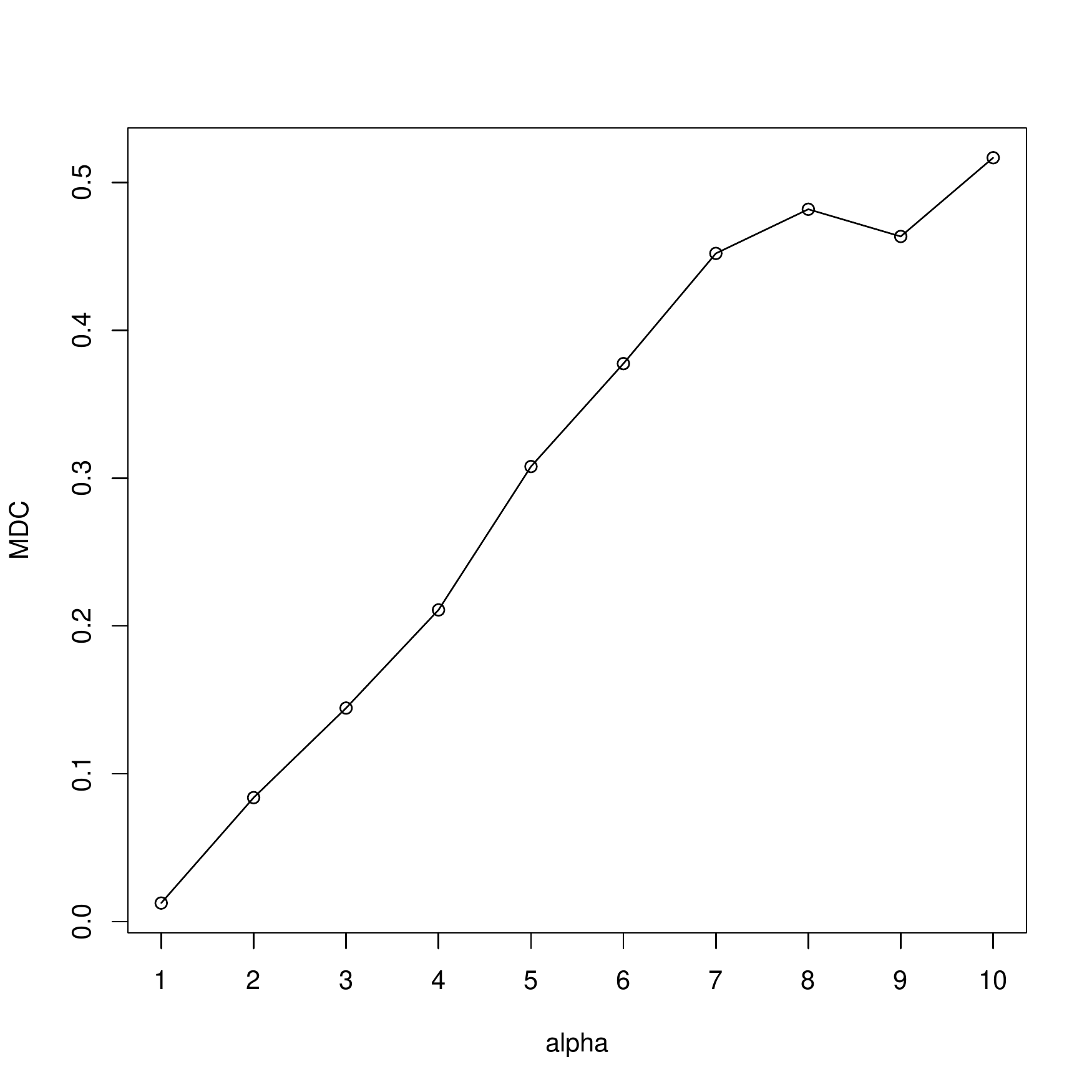}}
	\subfigure[dHSIC]{\includegraphics[width=0.245\linewidth]{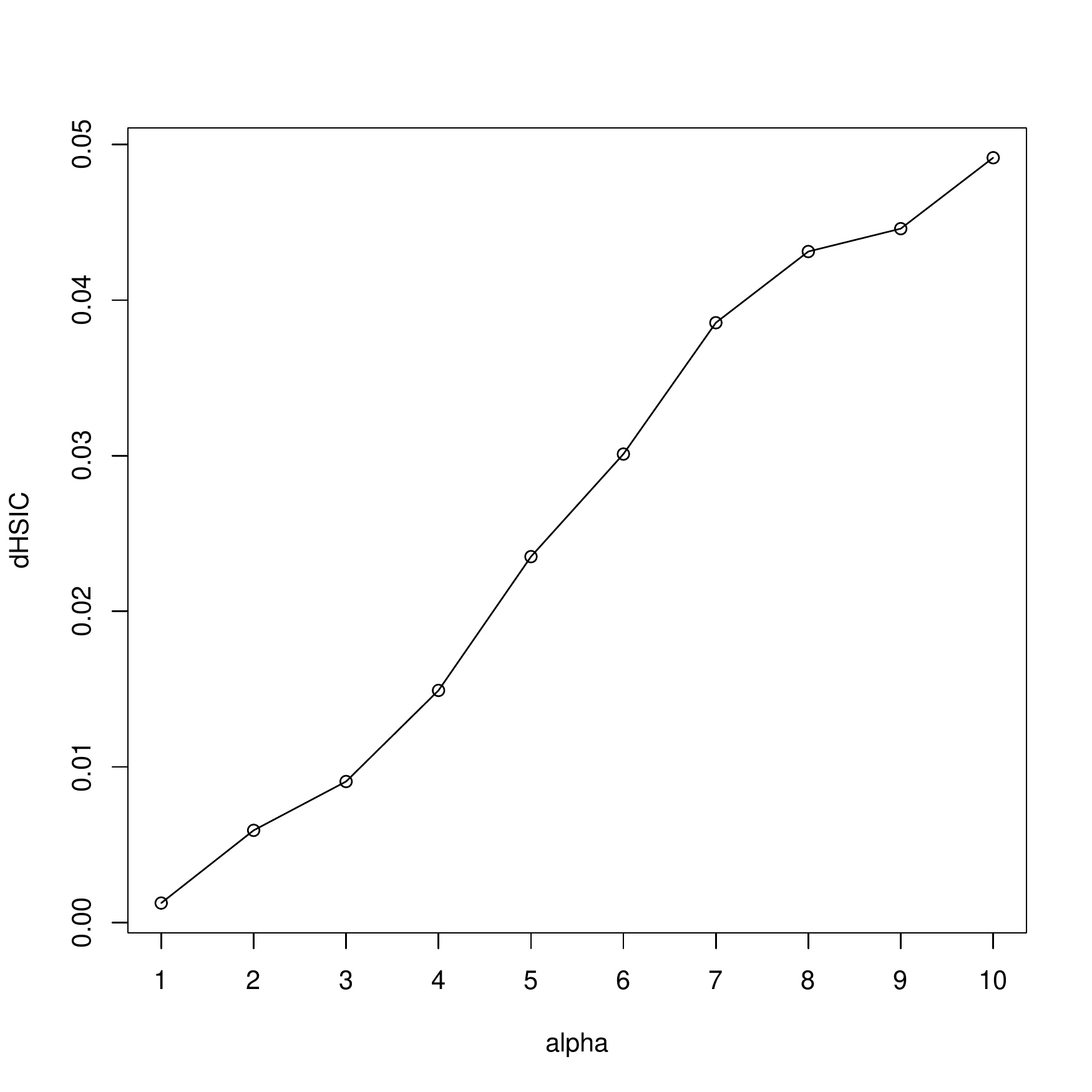}}
	\subfigure[NNS]{\includegraphics[width=0.245\linewidth]{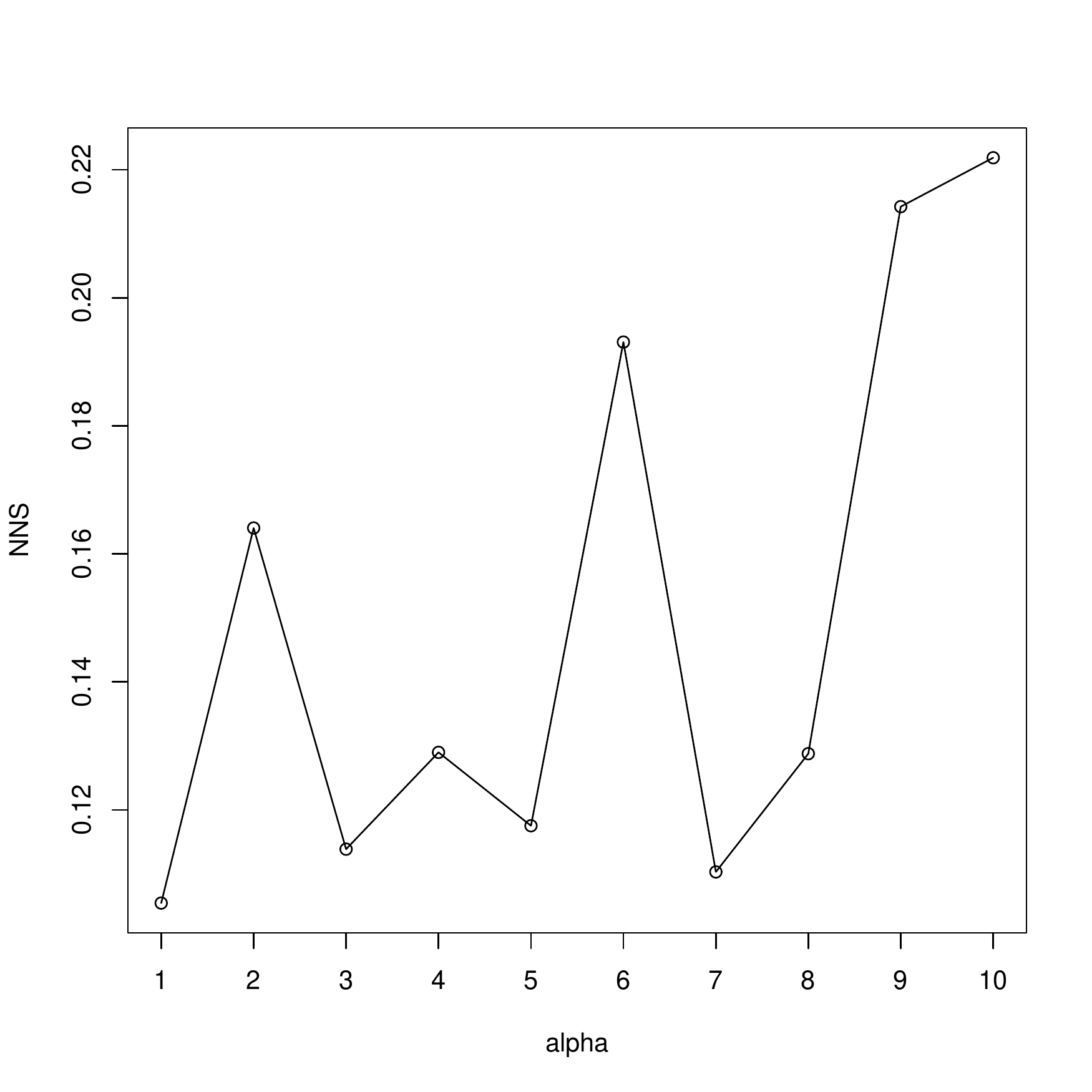}}
	\caption{Estimation of the independence measures from the simulated data of the bivariate Frank copula.}
	\label{fig:bifrank}
\end{figure}

\begin{figure}
	\centering
	\includegraphics[width=0.9\linewidth]{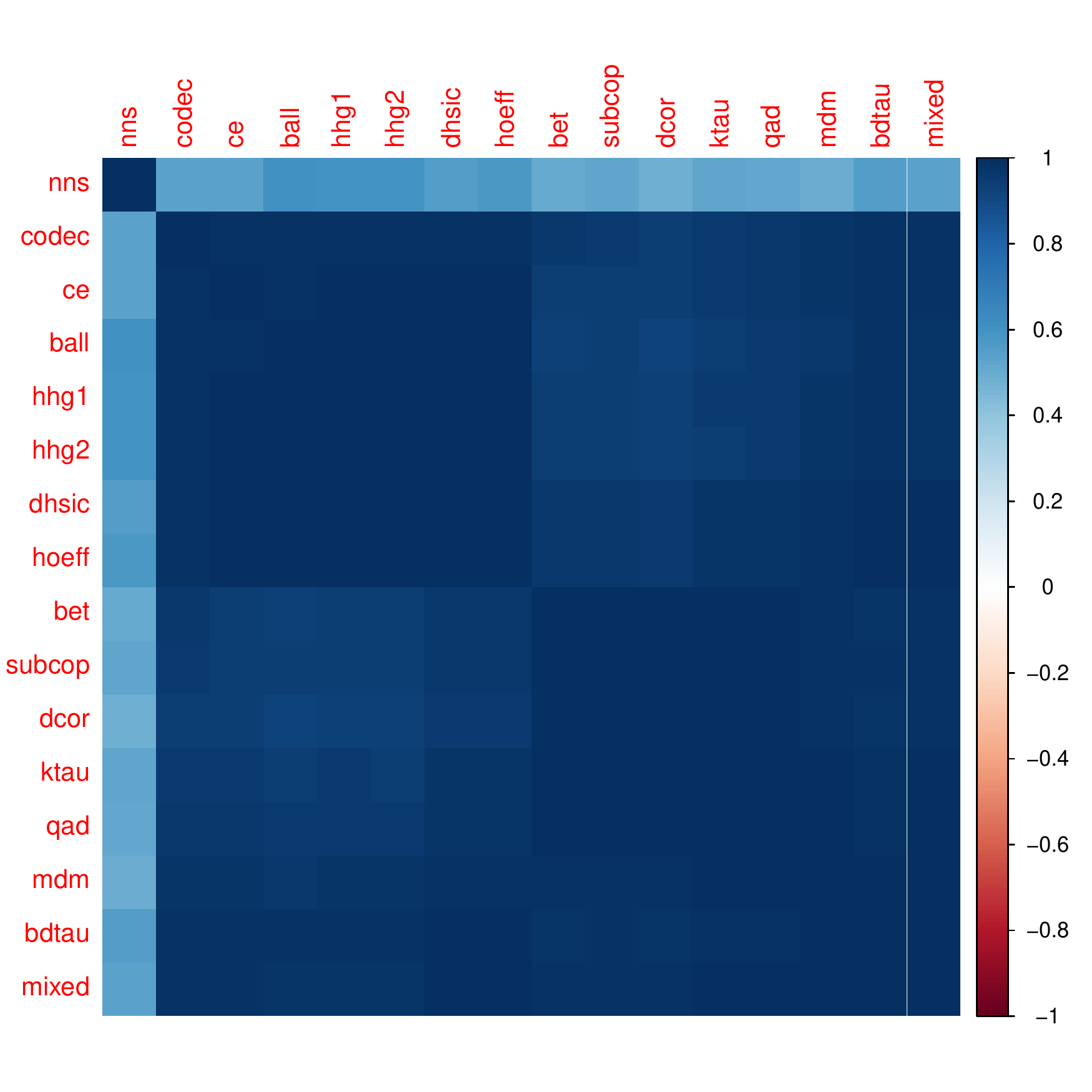}
	\caption{Correlation matrix of the independence measures estimated from the simulated data of the bivariate Frank copula.}
	\label{fig:bifrankcm}
\end{figure}

For the fifth experiment, the estimation result from the bivariate Frank copula is shown in Figure \ref{fig:bifrank} and the correlation matrix is shown in Figure \ref{fig:bifrankcm}. The measures are seperated into 3 groups according to the correlation matrix: \{NNS\}, \{CODEC, CE, Ball, HHG, dHSIC, Hoeff\}, and all the others.

\begin{figure}
	\subfigure[CE]{\includegraphics[width=0.3285\linewidth]{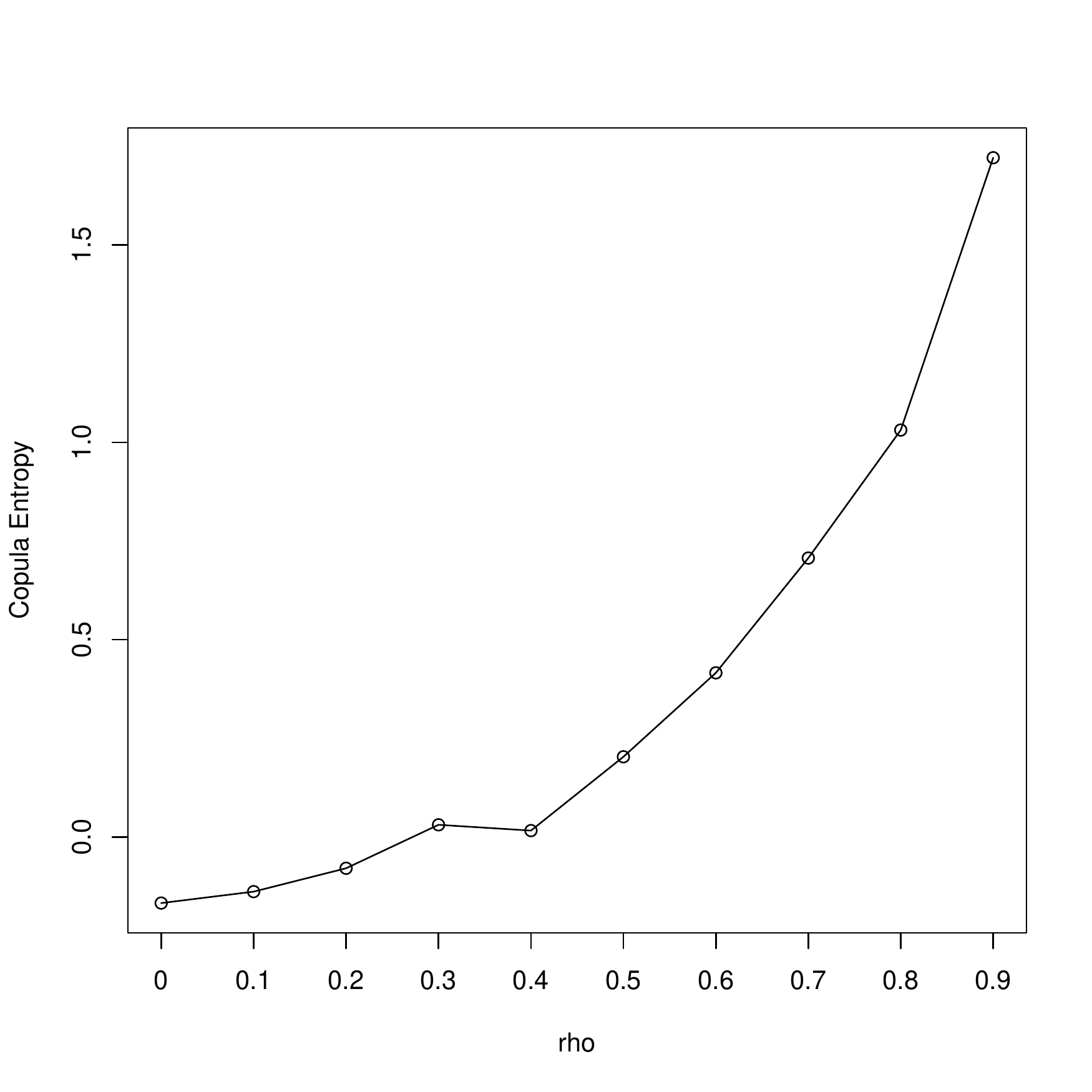}}
	\subfigure[BET]{\includegraphics[width=0.3285\linewidth]{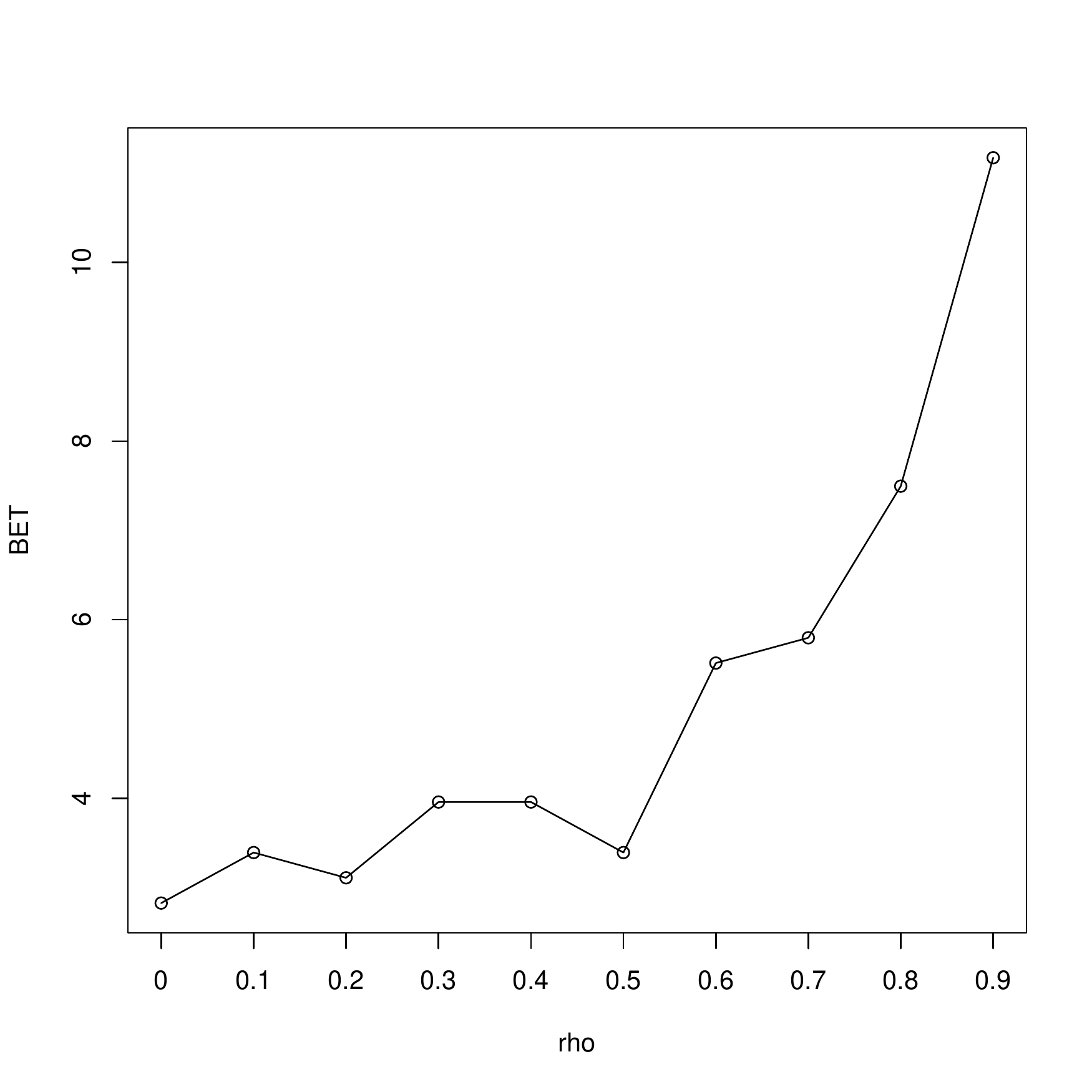}}
	\subfigure[mixed]{\includegraphics[width=0.3285\linewidth]{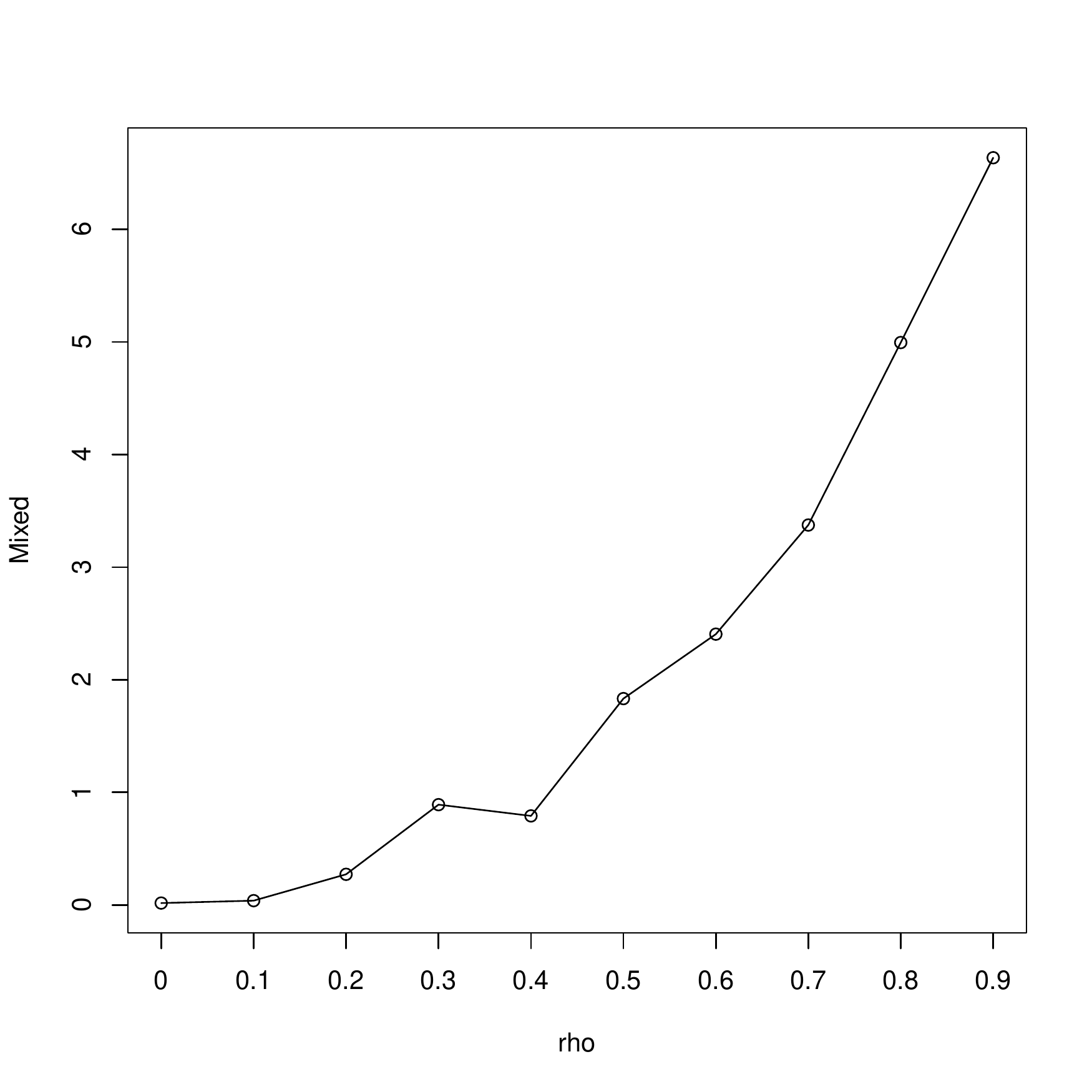}}
	\subfigure[subcop]{\includegraphics[width=0.3285\linewidth]{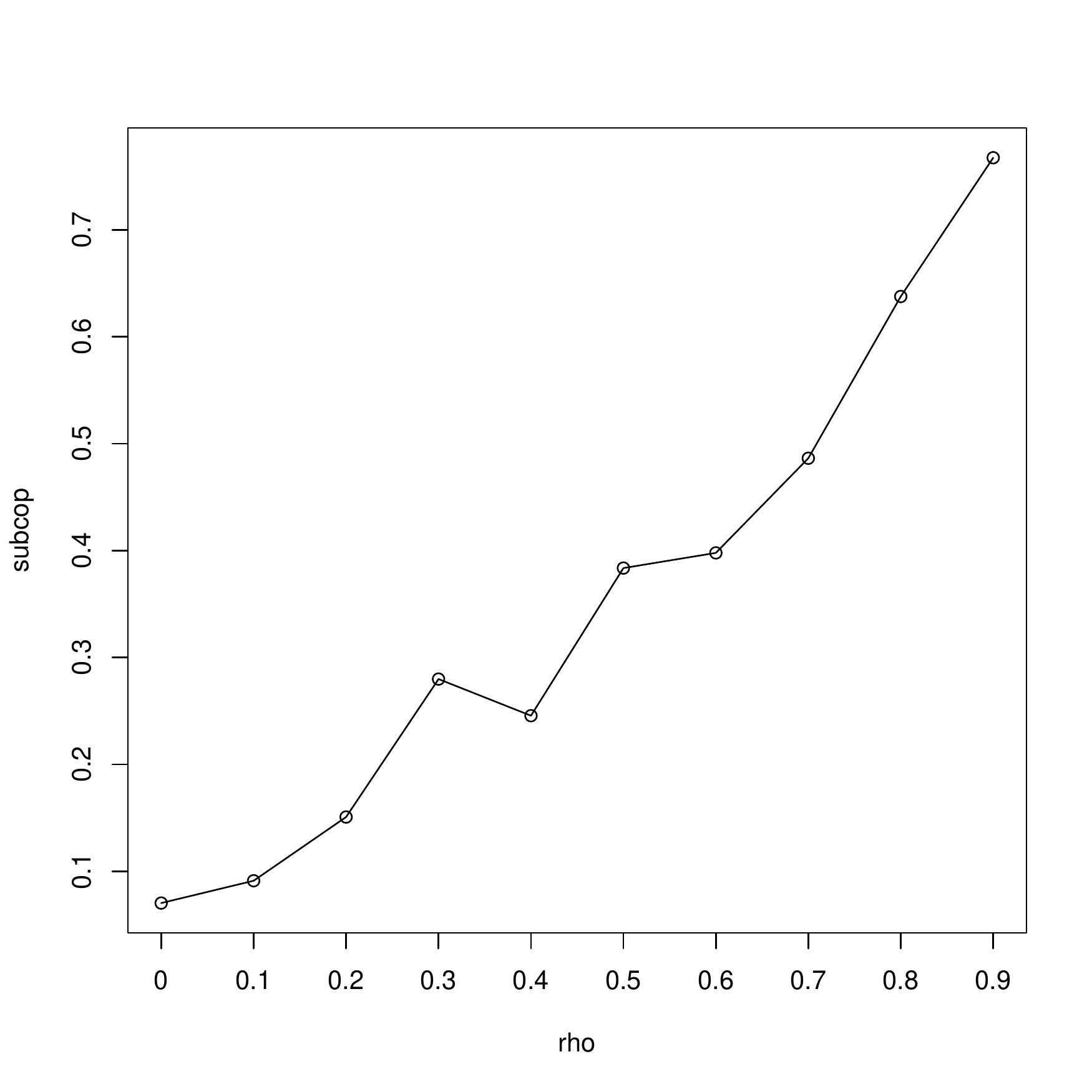}}
	\subfigure[dHSIC]{\includegraphics[width=0.3285\linewidth]{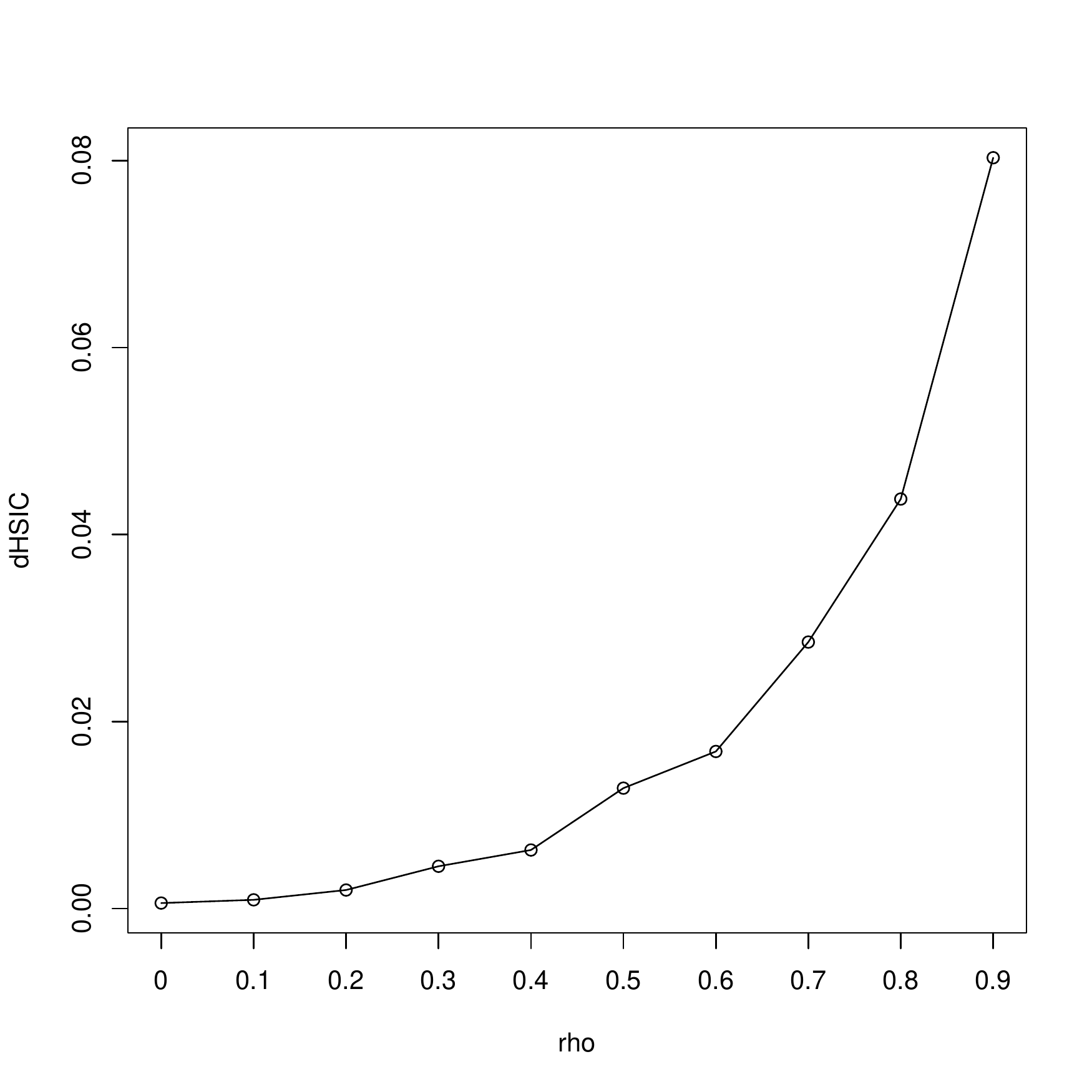}}
	\subfigure[NNS]{\includegraphics[width=0.3285\linewidth]{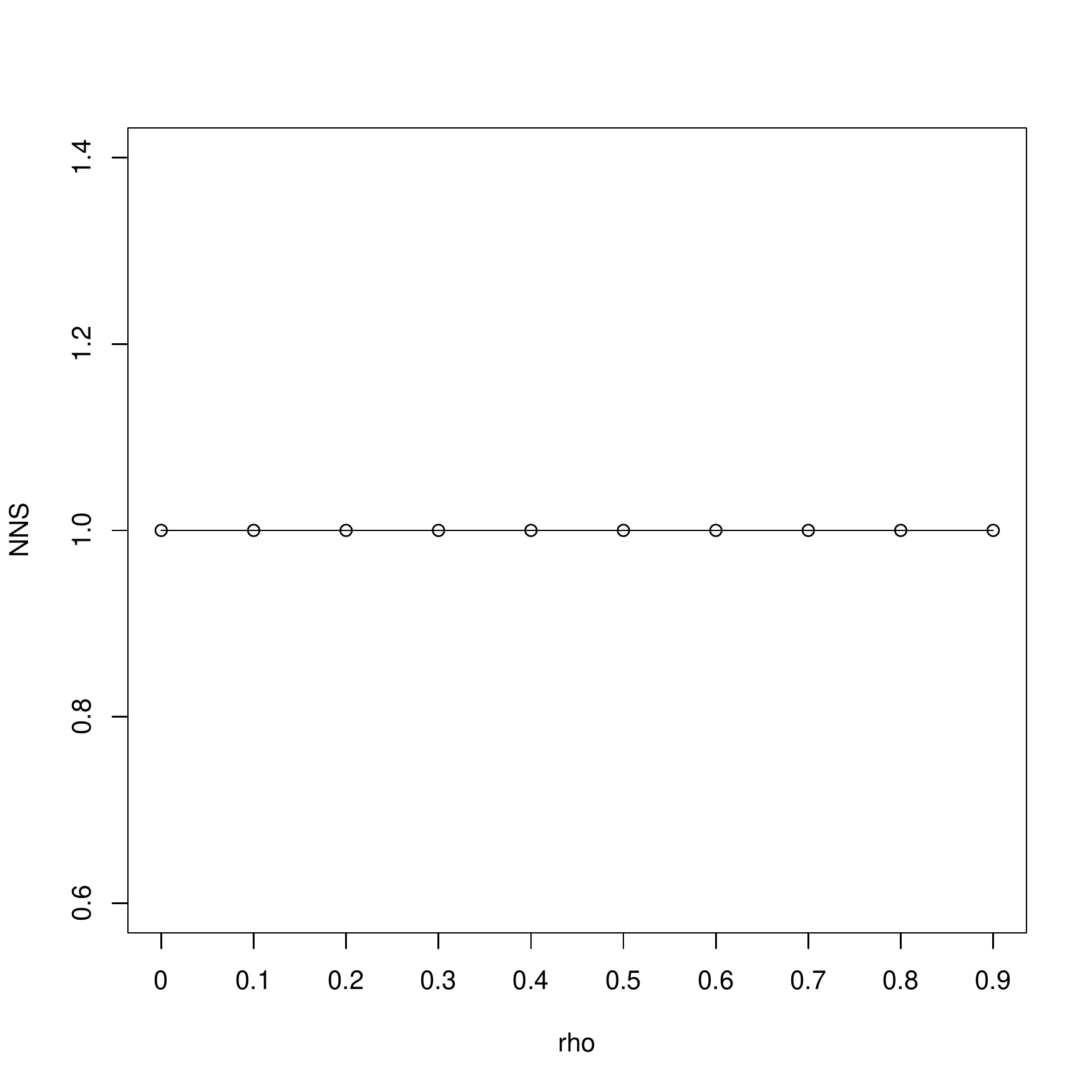}}
	\caption{Estimation of the multivariate independence measures from the simulated data of the trivariate normal distribution.}
	\label{fig:trinormal}
\end{figure}

\begin{figure}
	\centering
	\includegraphics[width=0.9\linewidth]{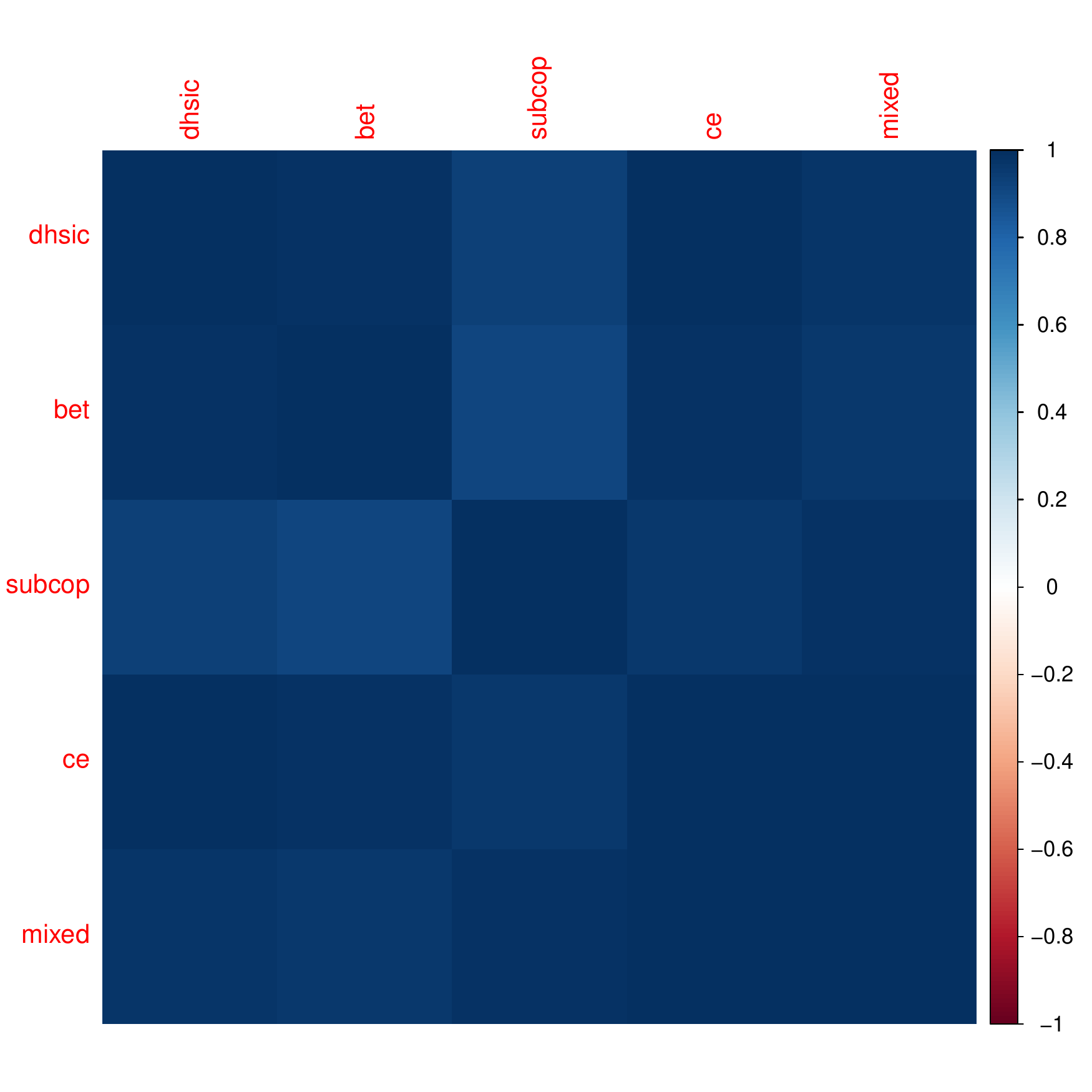}
	\caption{Correlation matrix of the independence measures estimated from the simulated data of the trivariate normal distribution.}
	\label{fig:trinormalcm}
\end{figure}

\paragraph{Multivariate normal distribution}
Only six measures are evaluated in the sixth and seventh experiments. The results for trivariate normal distribution are shown in Figure \ref{fig:trinormal}. There are 3 groups of measures according to the corresponding correlation matrix as shown in Figure \ref{fig:trinormalcm}: \{dHSIC, BET\}, \{subcop\}, and \{CE, mixed\}. The results of NNS are the contant 1 as $\rho$ increases, which means the dependence between three variables are considered as complete dependence by NNS. 

\begin{figure}
	\subfigure[CE]{\includegraphics[width=0.3285\linewidth]{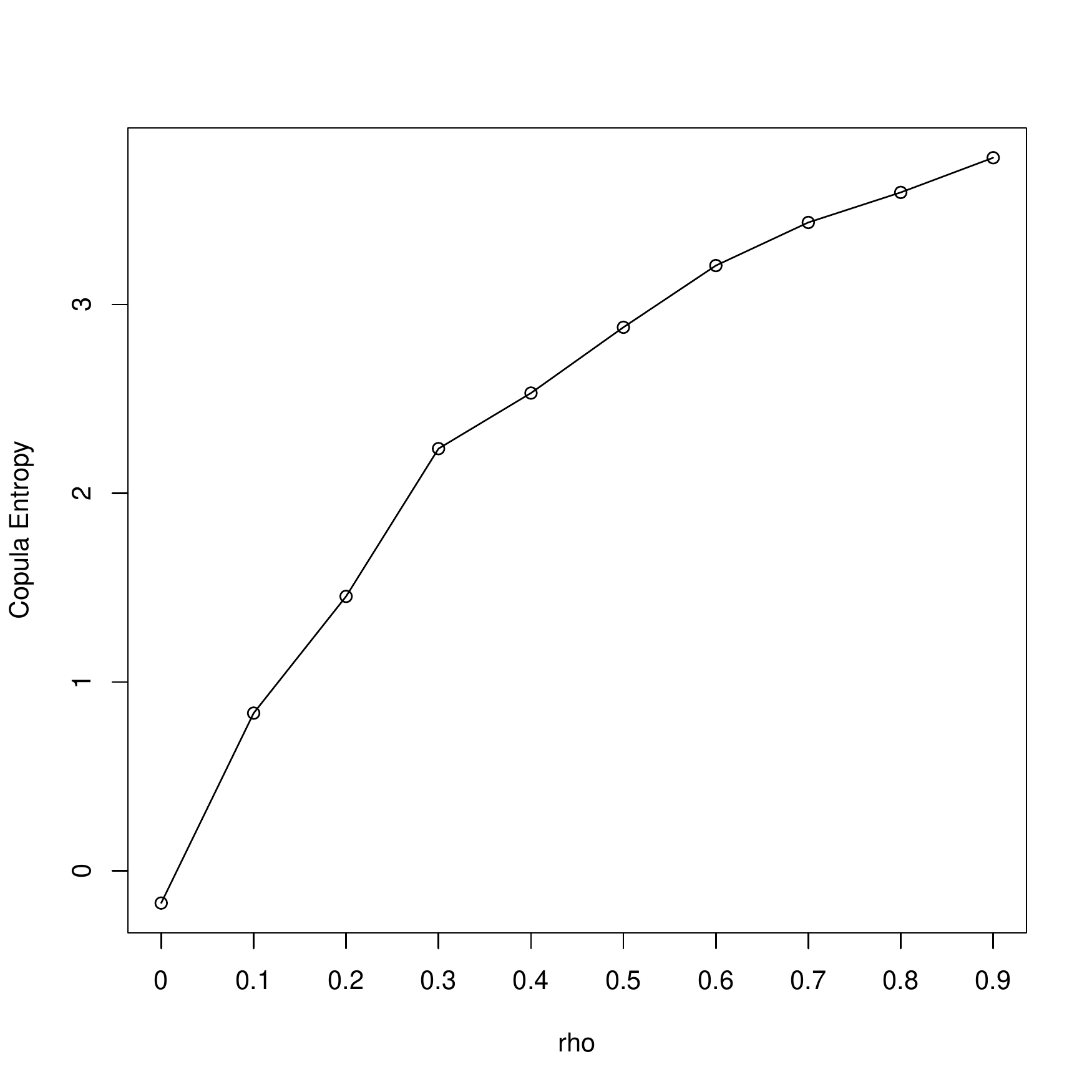}}
	\subfigure[BET]{\includegraphics[width=0.3285\linewidth]{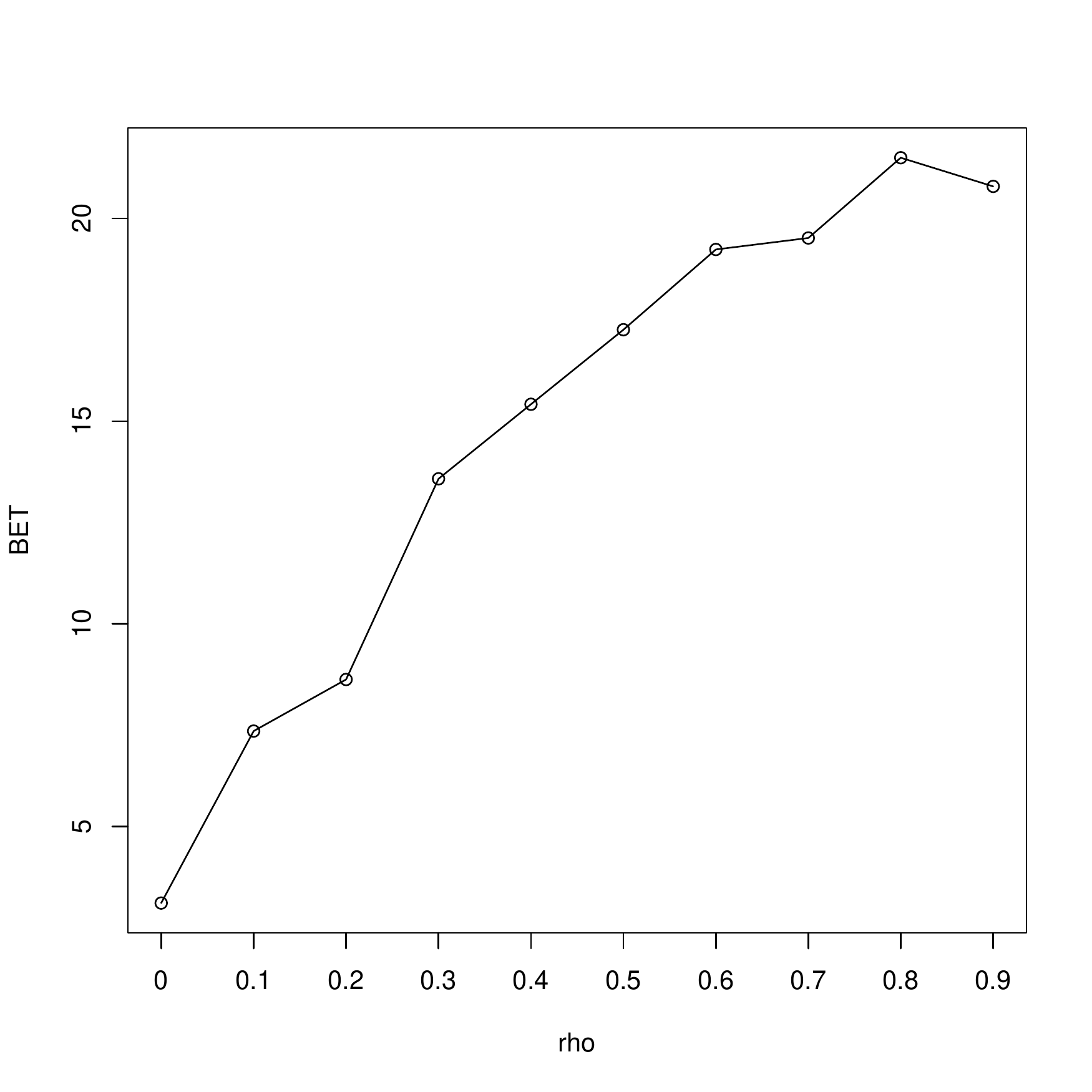}}
	\subfigure[mixed]{\includegraphics[width=0.3285\linewidth]{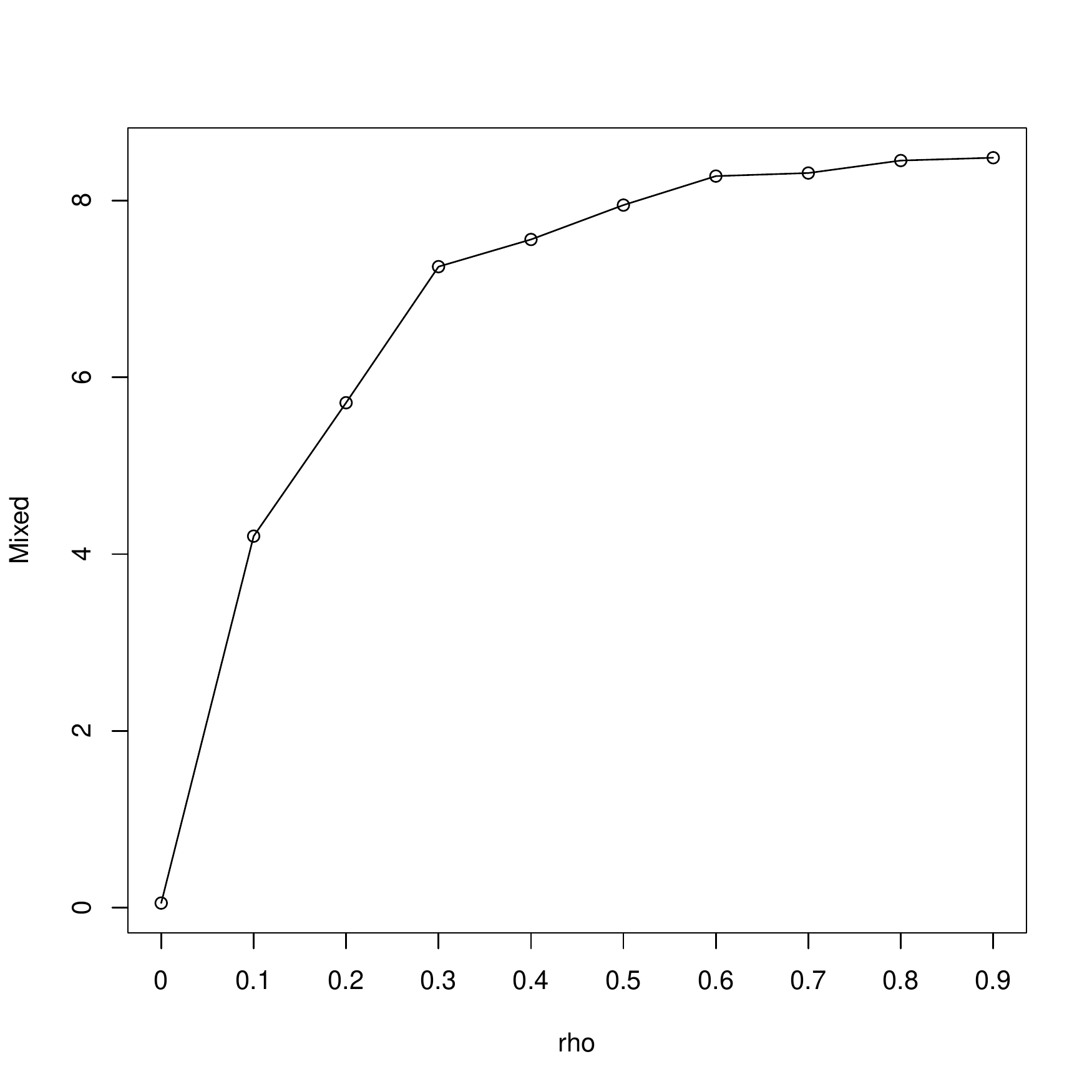}}
	\subfigure[subcop]{\includegraphics[width=0.3285\linewidth]{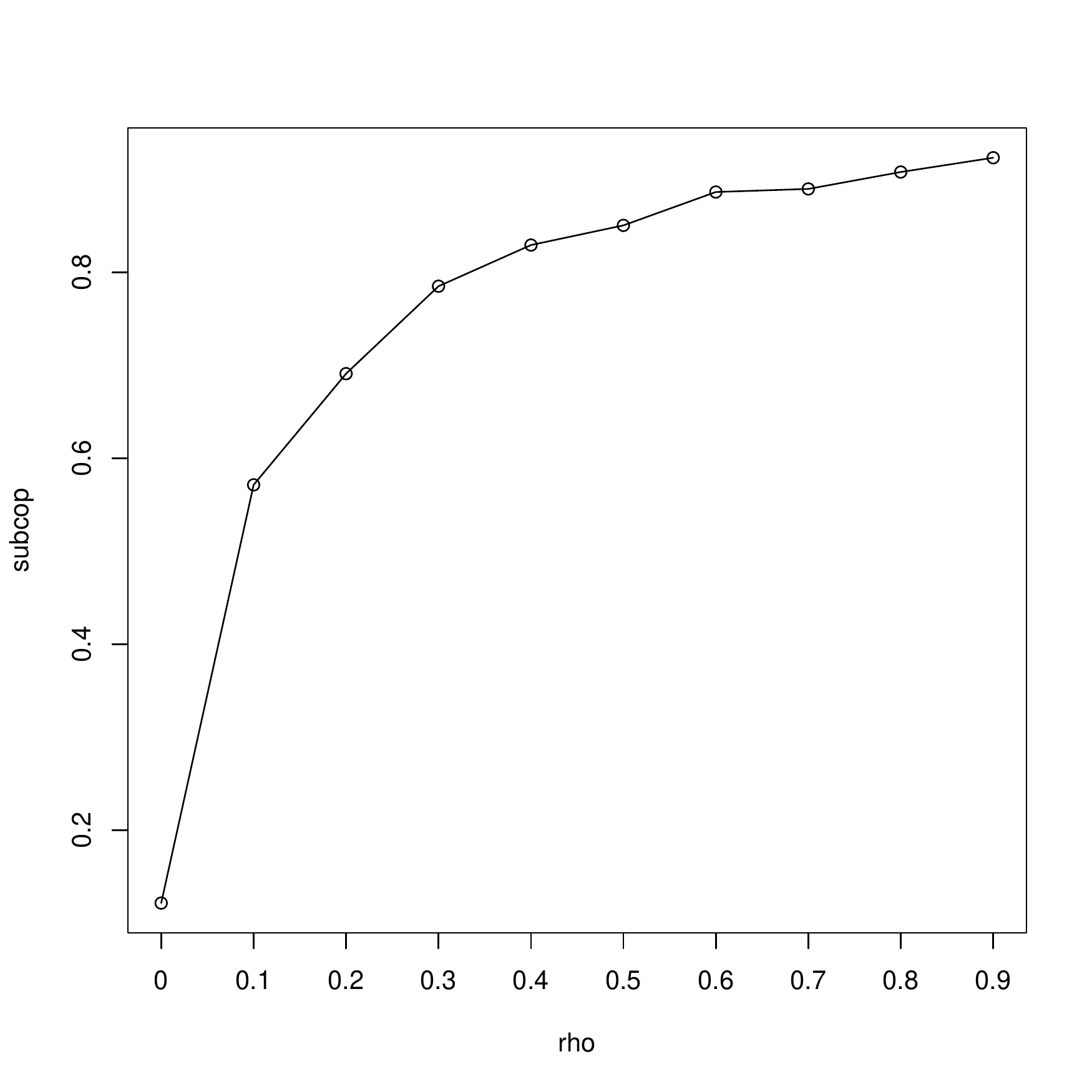}}
	\subfigure[dHSIC]{\includegraphics[width=0.3285\linewidth]{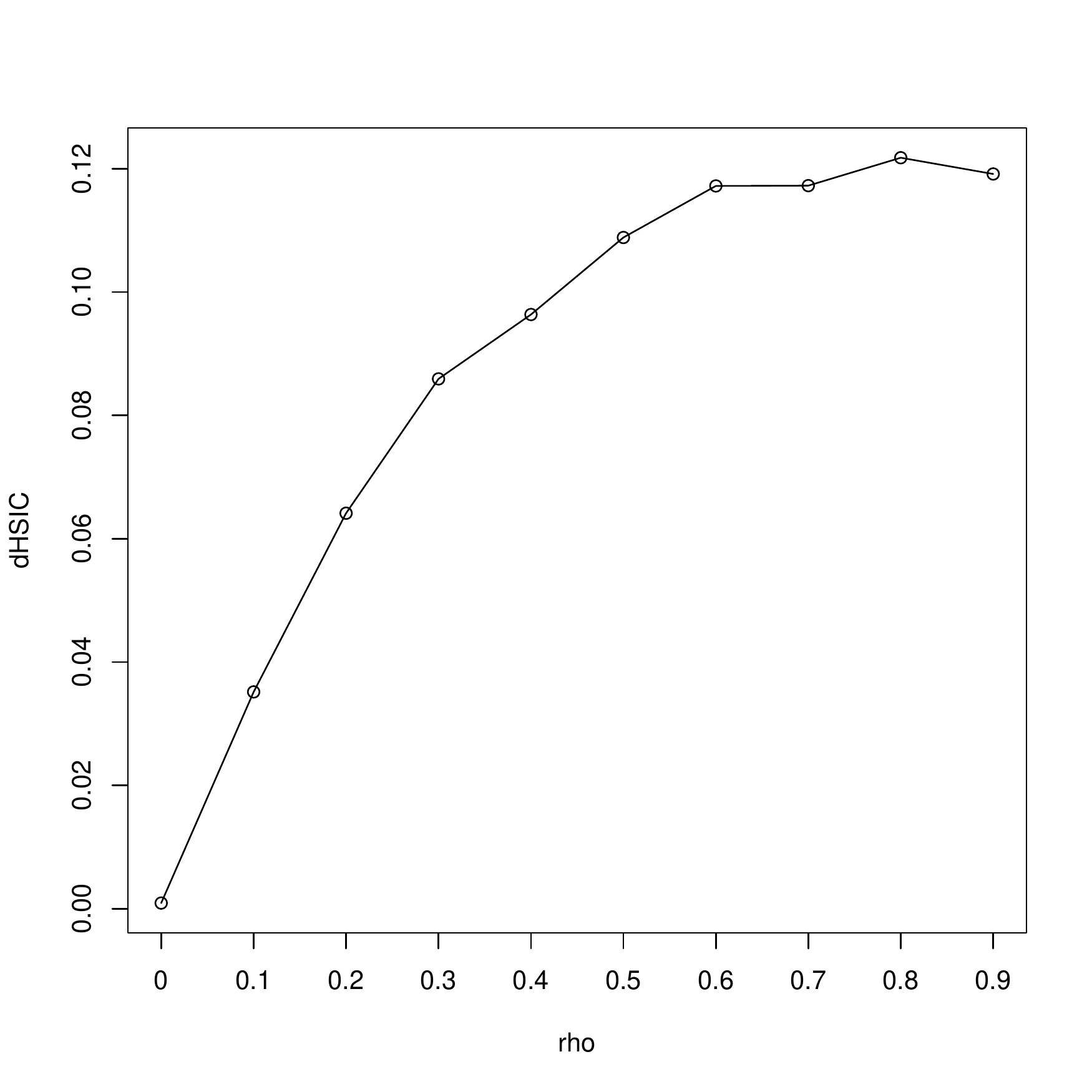}}
	\subfigure[NNS]{\includegraphics[width=0.3285\linewidth]{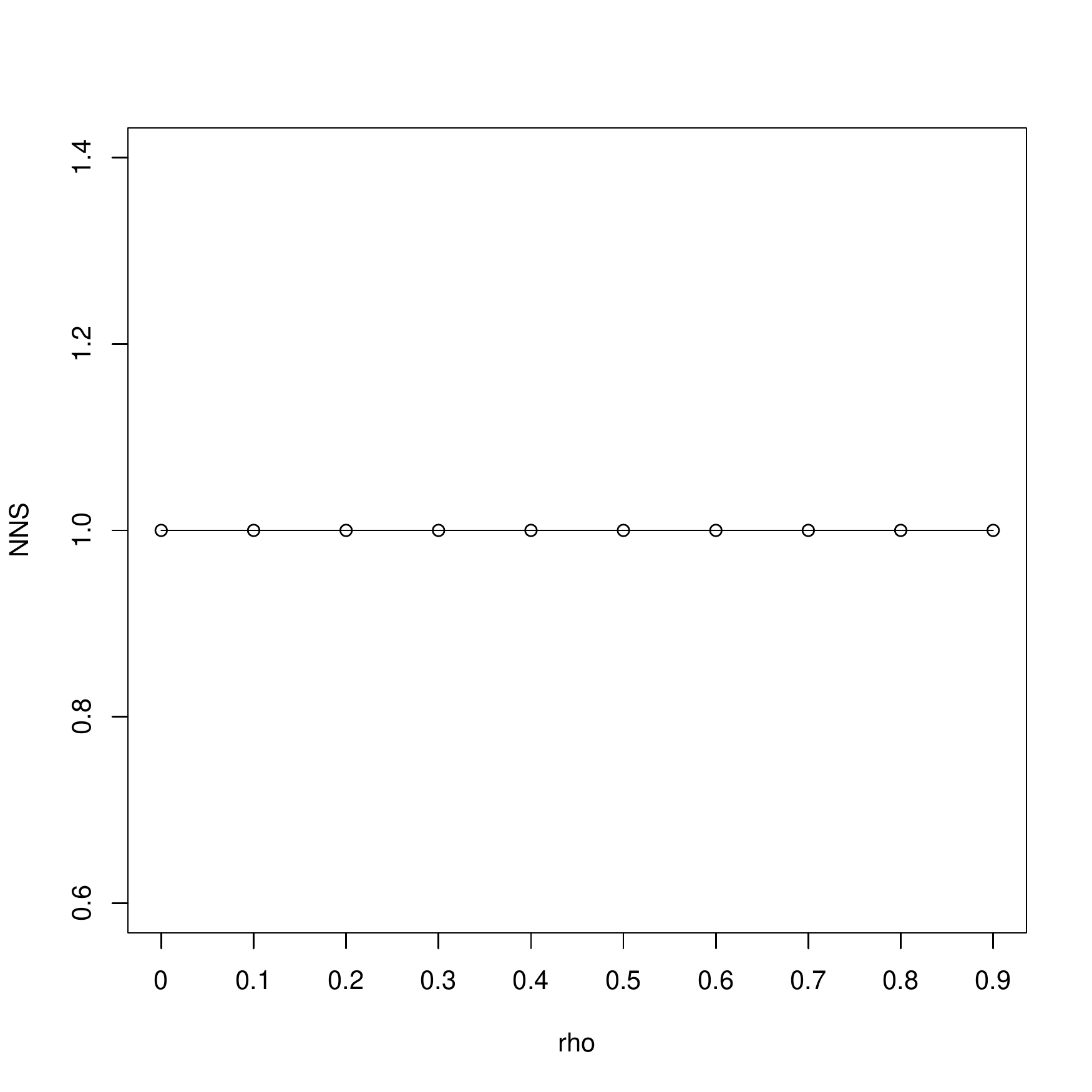}}
	\caption{Estimation of the multivariate independence measures from the simulated data of the trivariate Gumbel copula.}
	\label{fig:trigumbel}
\end{figure}

\begin{figure}
	\centering
	\includegraphics[width=0.9\linewidth]{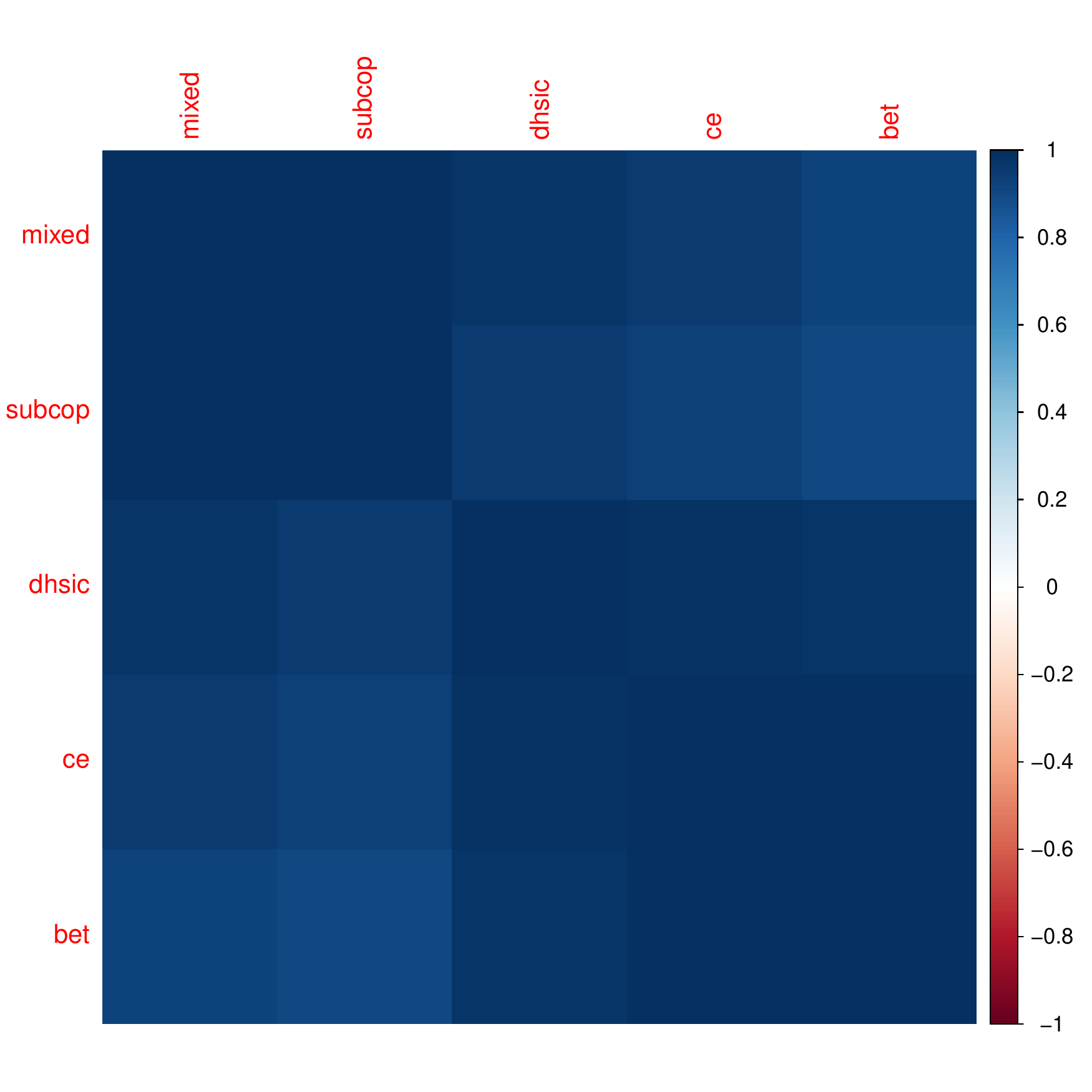}
	\caption{Correlation matrix of the independence measures estimated from the simulated data of the trivariate Gumble copula.}
	\label{fig:trigumbelcm}
\end{figure}

\paragraph{Multivariate Archimedean copula}
The results for trivariate Gumbel copula are shown in Figure \ref{fig:trigumbel}. There are 2 groups of measures according to the correlation matrix as shown in Figure \ref{fig:trigumbelcm}: \{mixed, subcop\} and \{dHSIC, CE, BET\}. the estimation of NNS is constant 1 again.

\begin{figure}
	\subfigure[CE]{\includegraphics[width=0.245\linewidth]{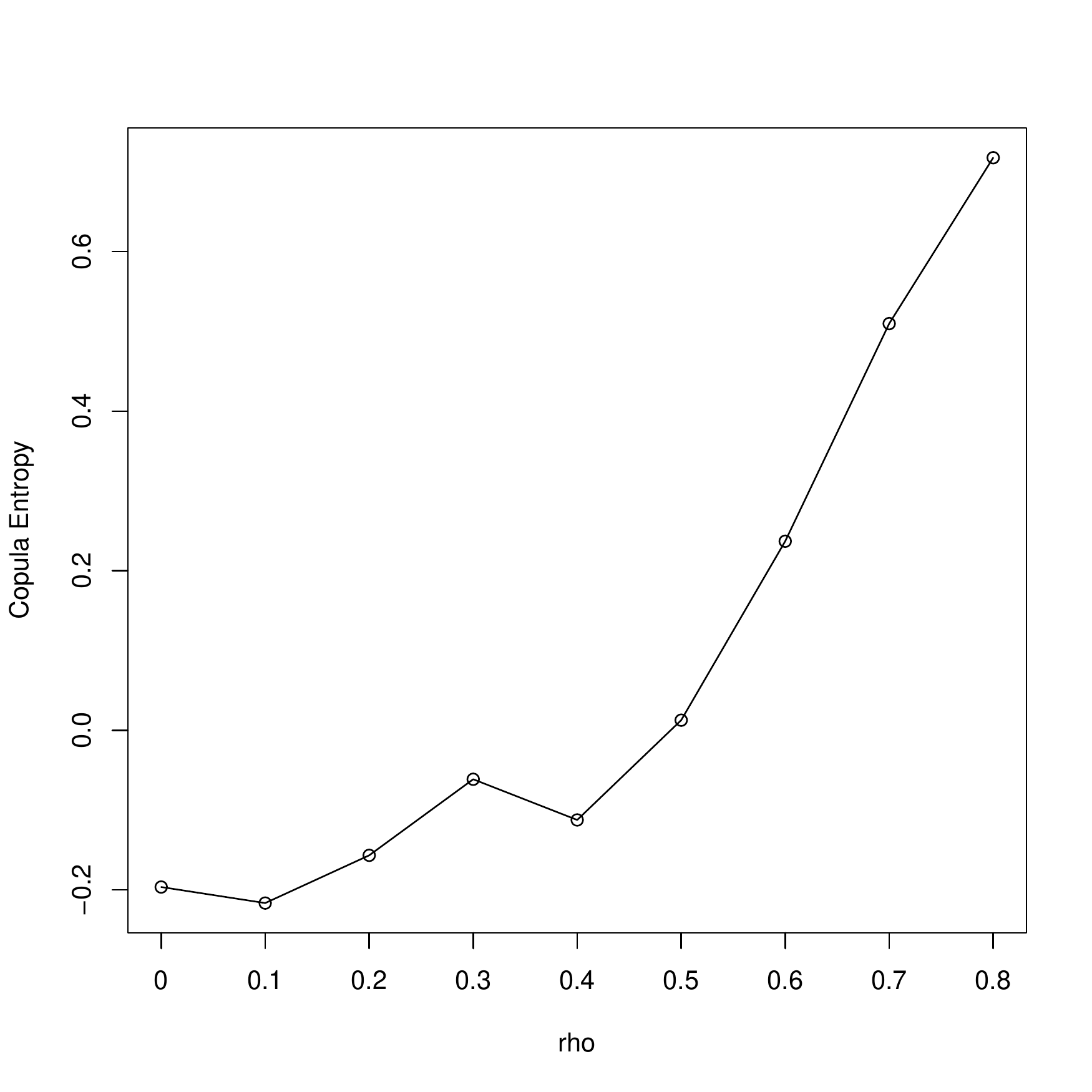}}
	\subfigure[HHG.chisq]{\includegraphics[width=0.245\linewidth]{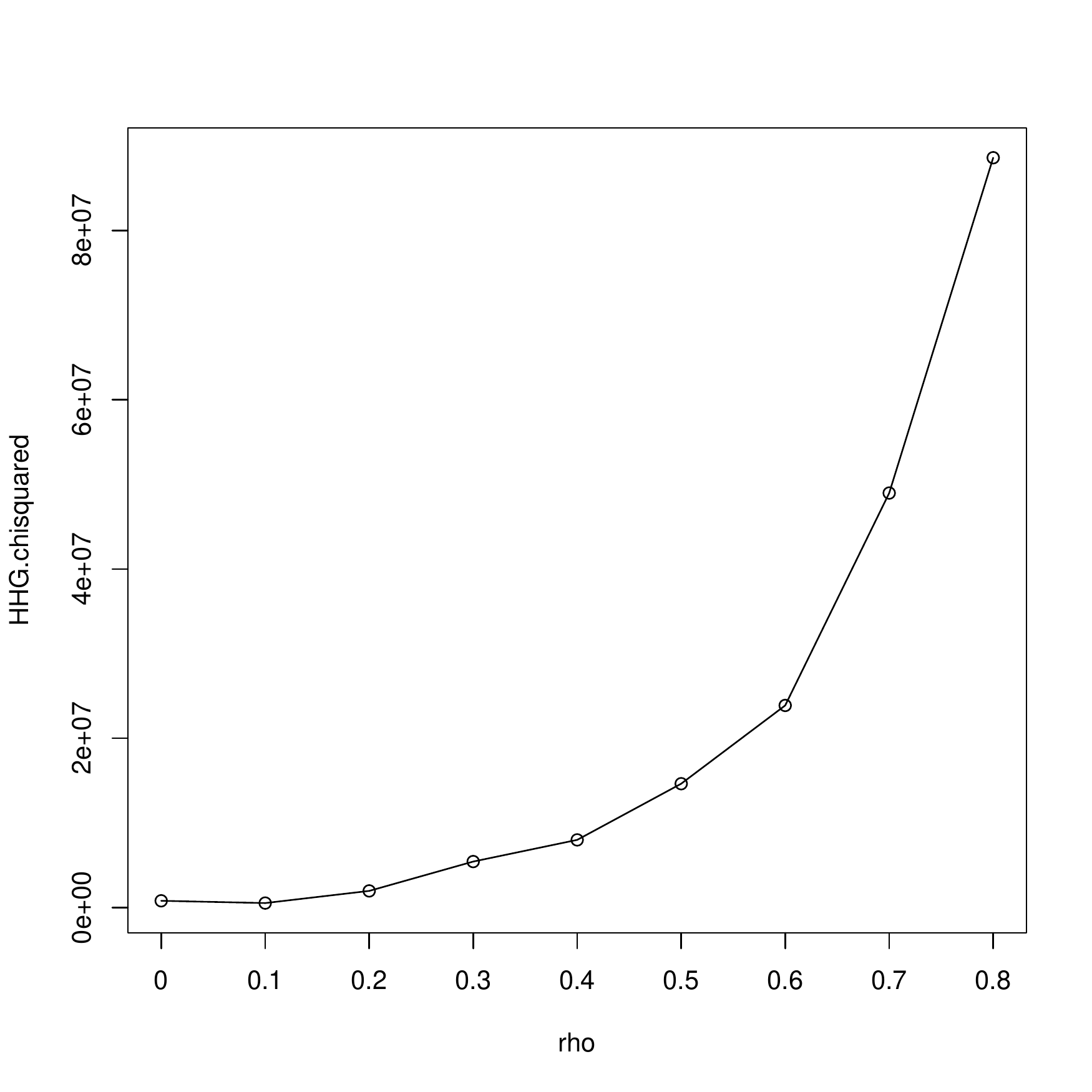}}
	\subfigure[HHG.lr]{\includegraphics[width=0.245\linewidth]{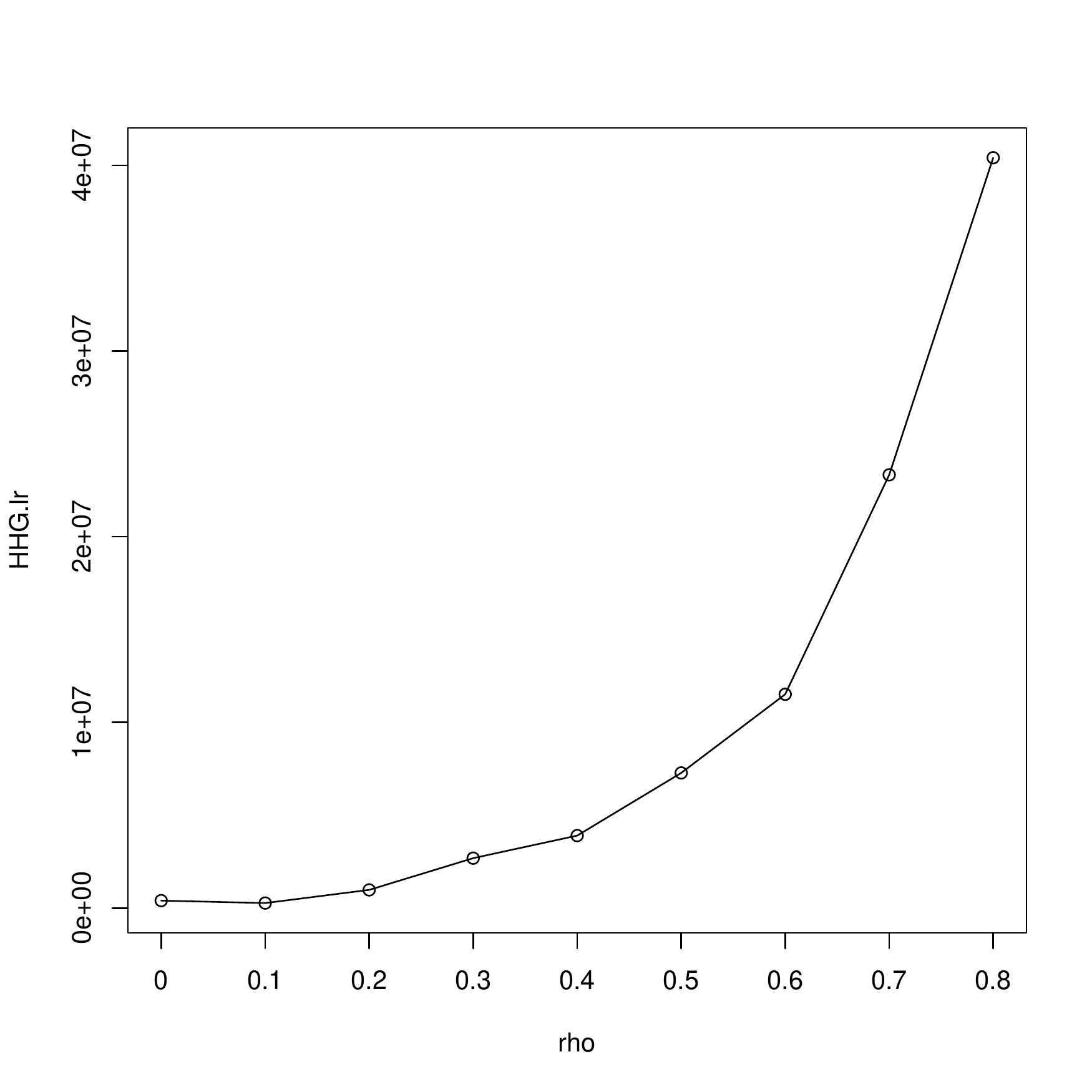}}
	\subfigure[Ball]{\includegraphics[width=0.245\linewidth]{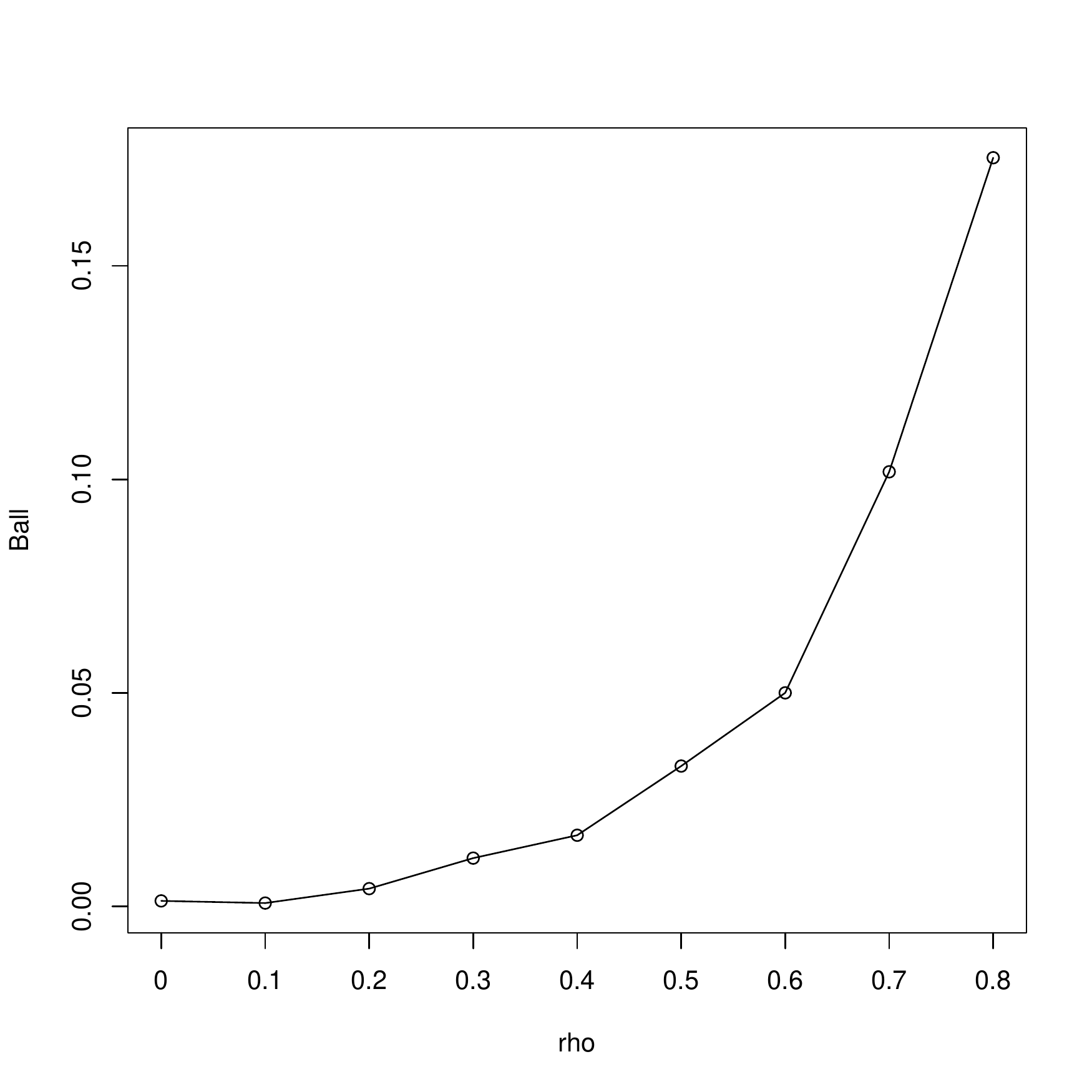}}
	\subfigure[BET]{\includegraphics[width=0.245\linewidth]{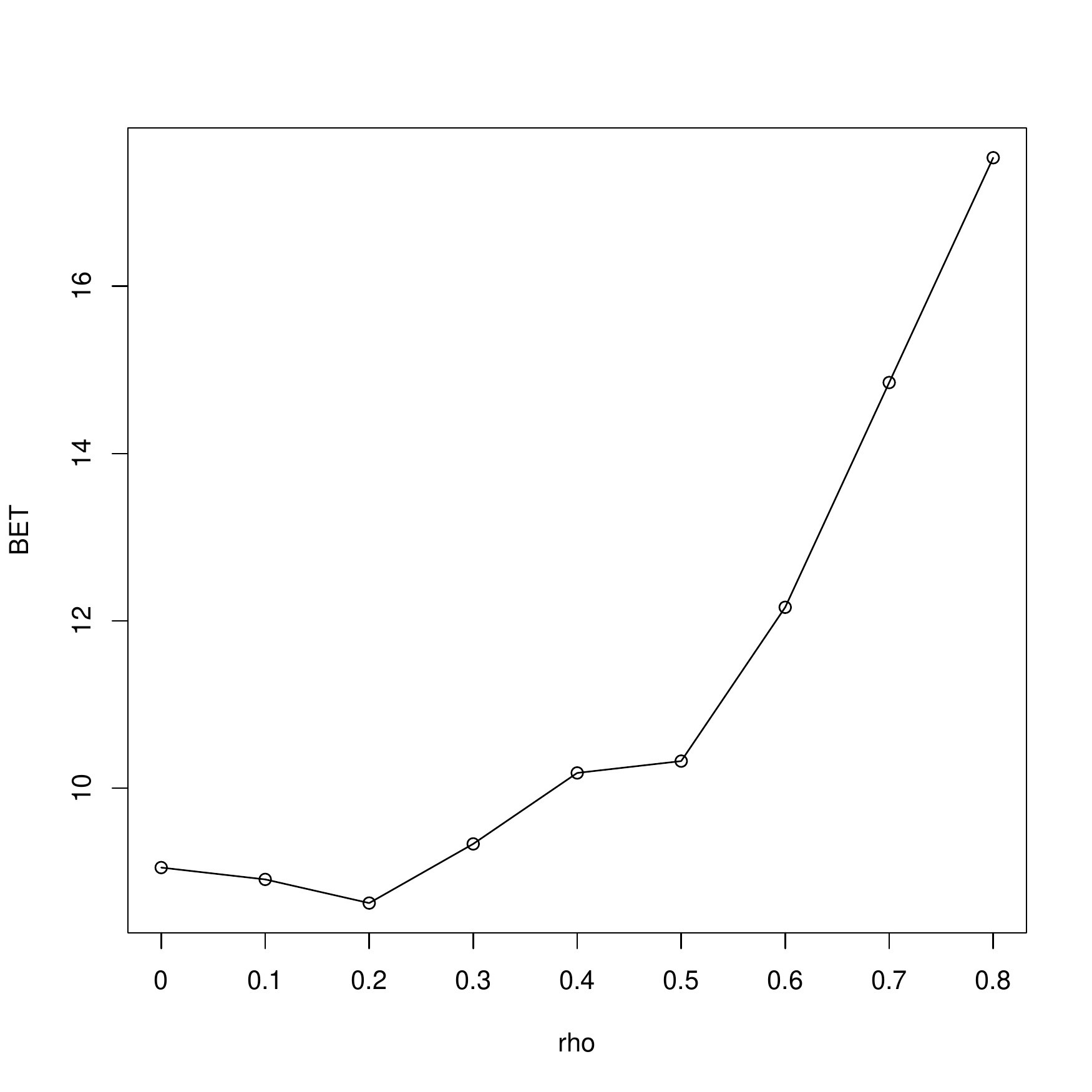}}
	\subfigure[QAD]{\includegraphics[width=0.245\linewidth]{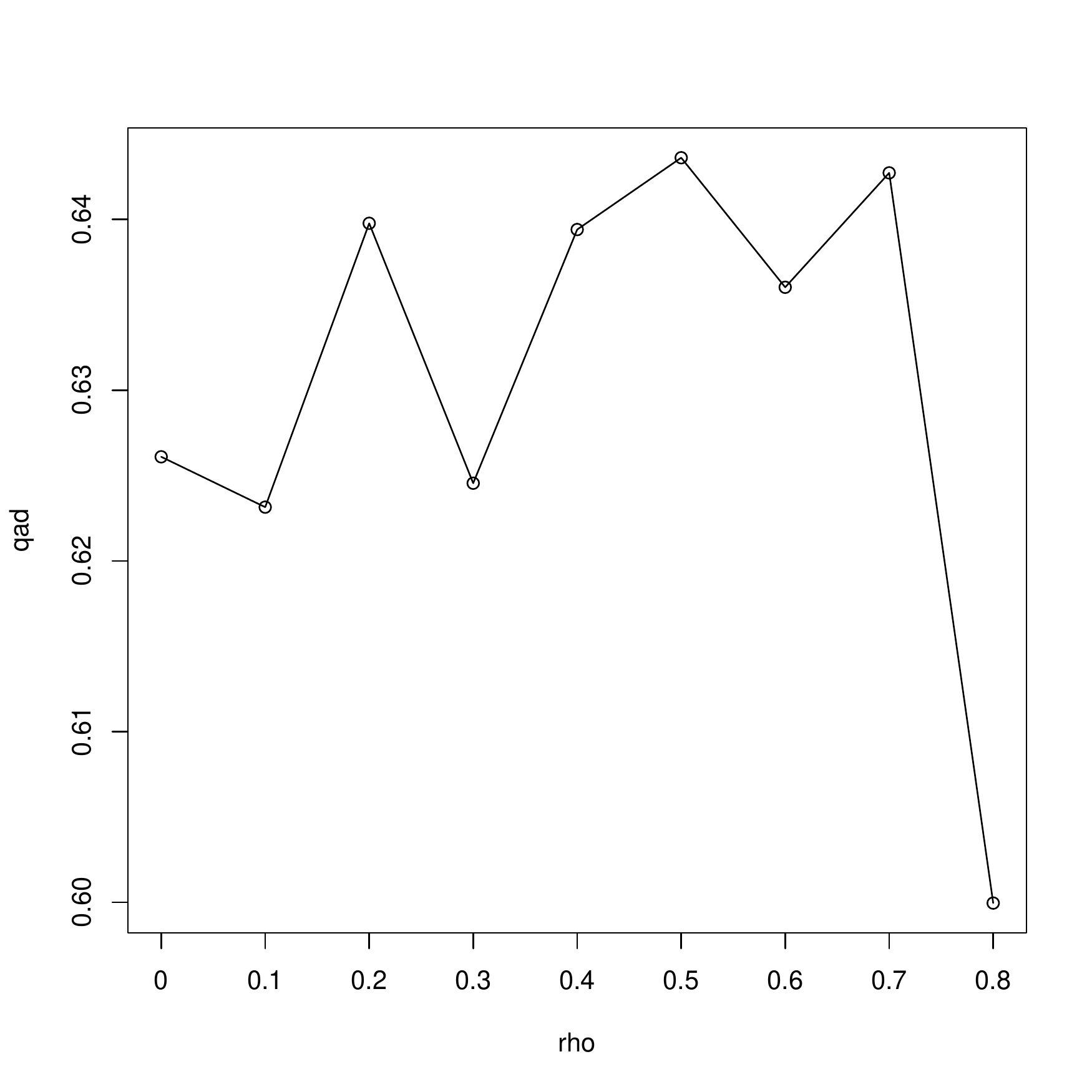}}
	\subfigure[dCor]{\includegraphics[width=0.245\linewidth]{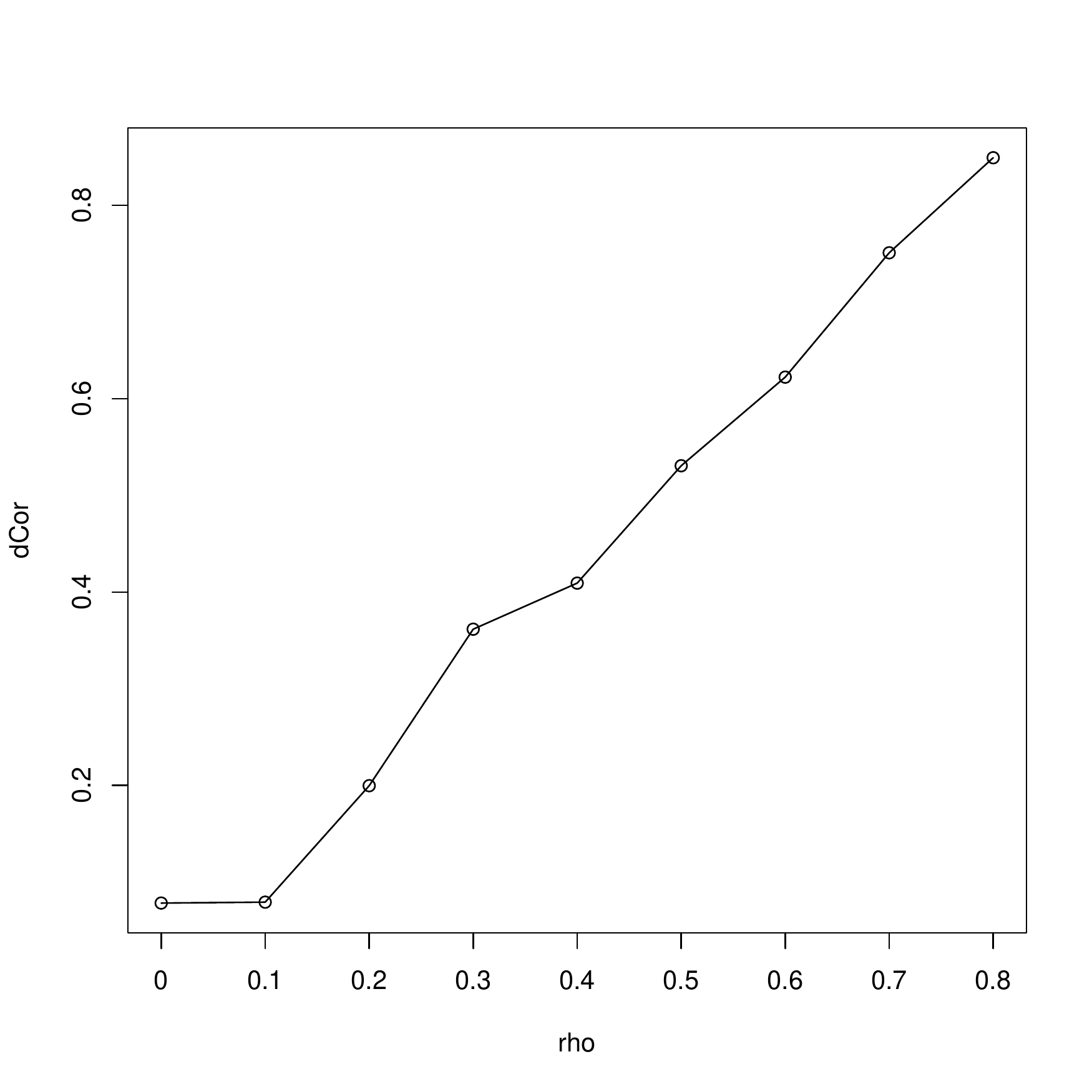}}
	\subfigure[mdm]{\includegraphics[width=0.245\linewidth]{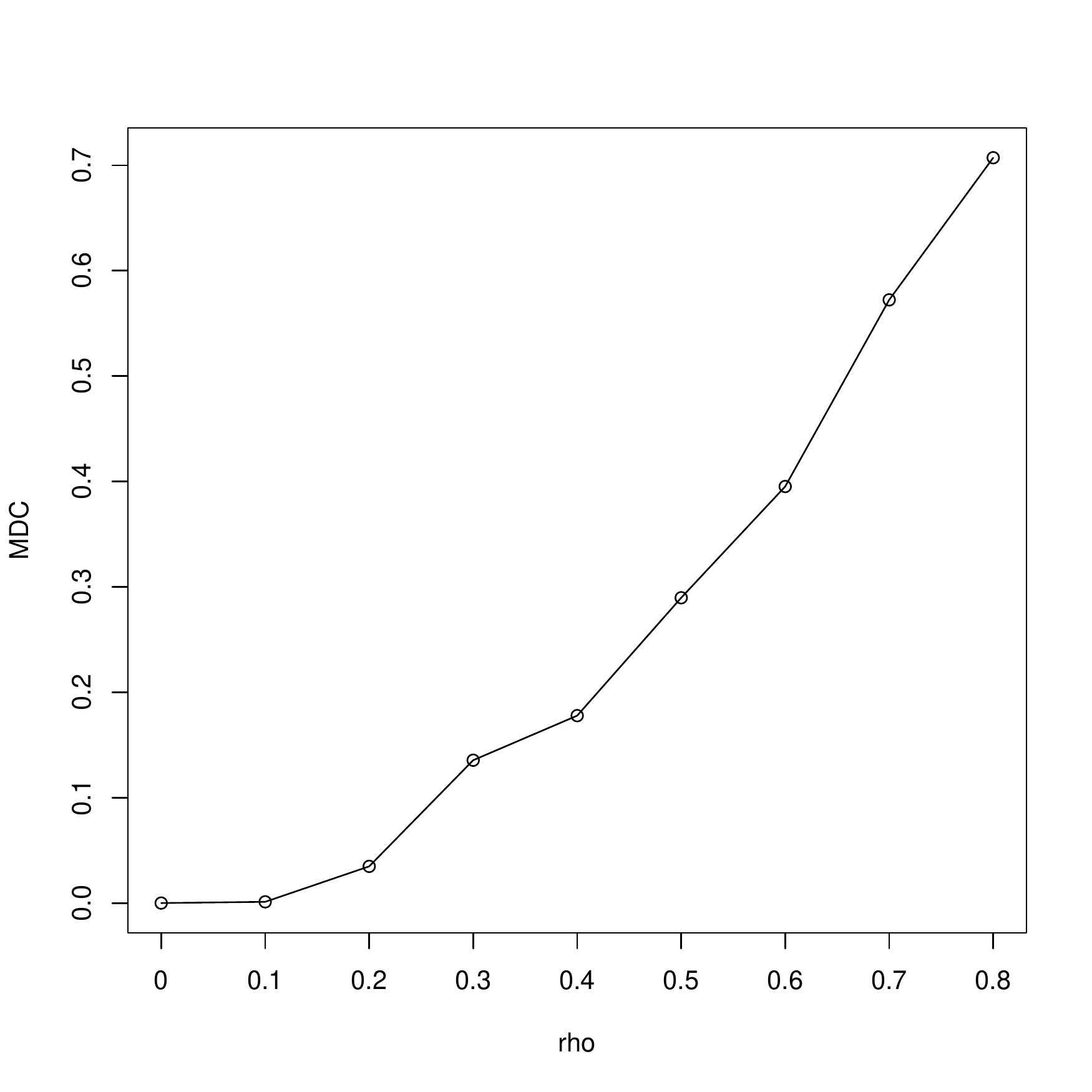}}
	\subfigure[dHSIC]{\includegraphics[width=0.245\linewidth]{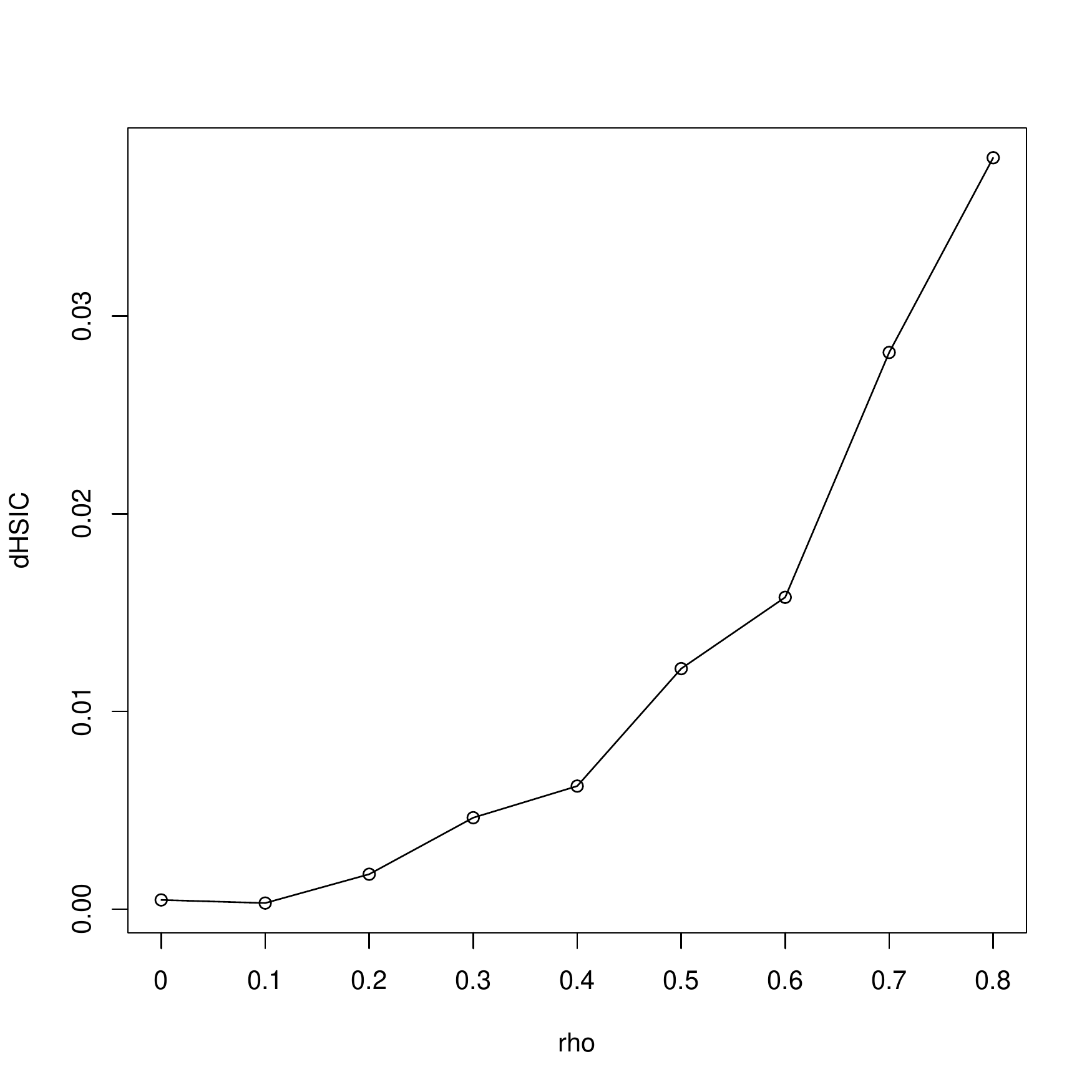}}
	\subfigure[NNS]{\includegraphics[width=0.245\linewidth]{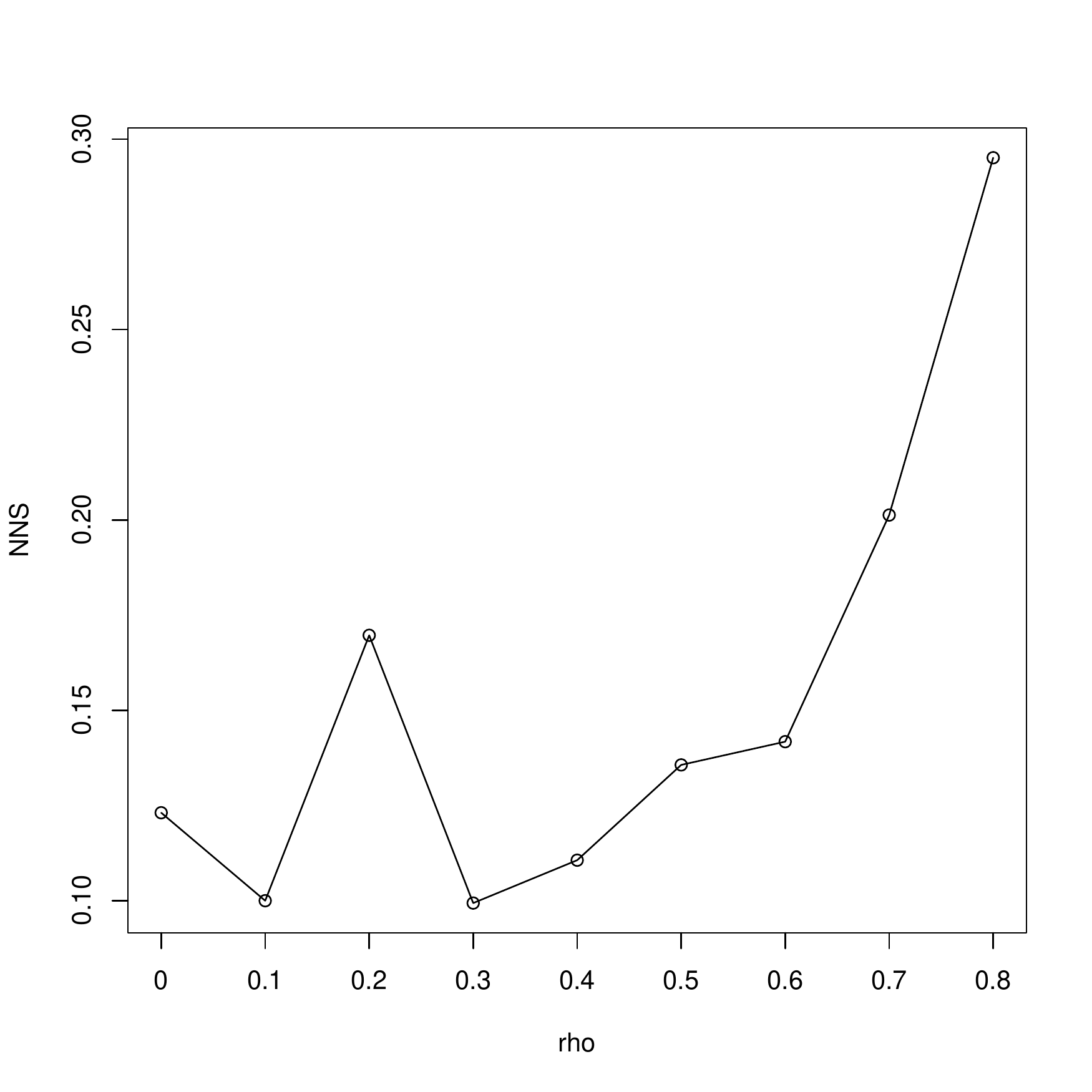}}
	\caption{Estimation of the independence measures between two random vectors from the simulated data of the quadvariate normal distribution.}
	\label{fig:quadnorm}
\end{figure}

\begin{figure}
	\centering
	\includegraphics[width=0.9\linewidth]{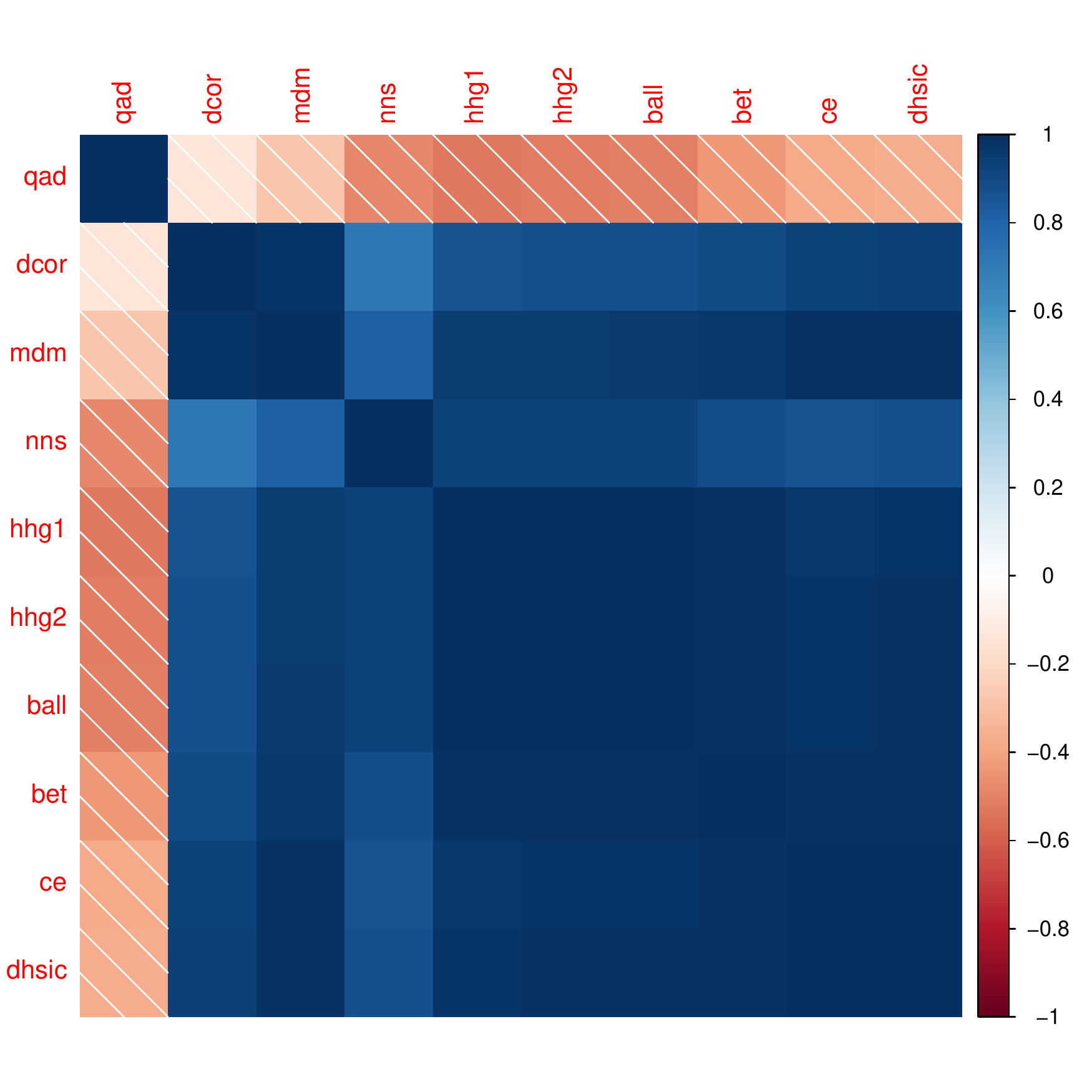}
	\caption{Correlation matrix of the independence measures between two random vectors estimated from the simulated data of the quadvariate normal distribution.}
	\label{fig:quadnormcm}
\end{figure}

\paragraph{Multivariate normal distribution for dependence between random vectors}
There are ten measures for independence between two random vectors in the eighth experiment. The estimation results is shown in Figure \ref{fig:quadnorm}. These measures are grouped into four parts according to the correlation matrix in Figure \ref{fig:quadnormcm}: \{QAD\}, \{dCor, MDM\}, \{NNS\}, and all the others. It can be learned from Figure \ref{fig:quadnorm} that QAD failed to estimate the measure strength. 

\subsubsection{Real data}
\begin{figure}
	\subfigure[CE]{\includegraphics[width=0.495\linewidth]{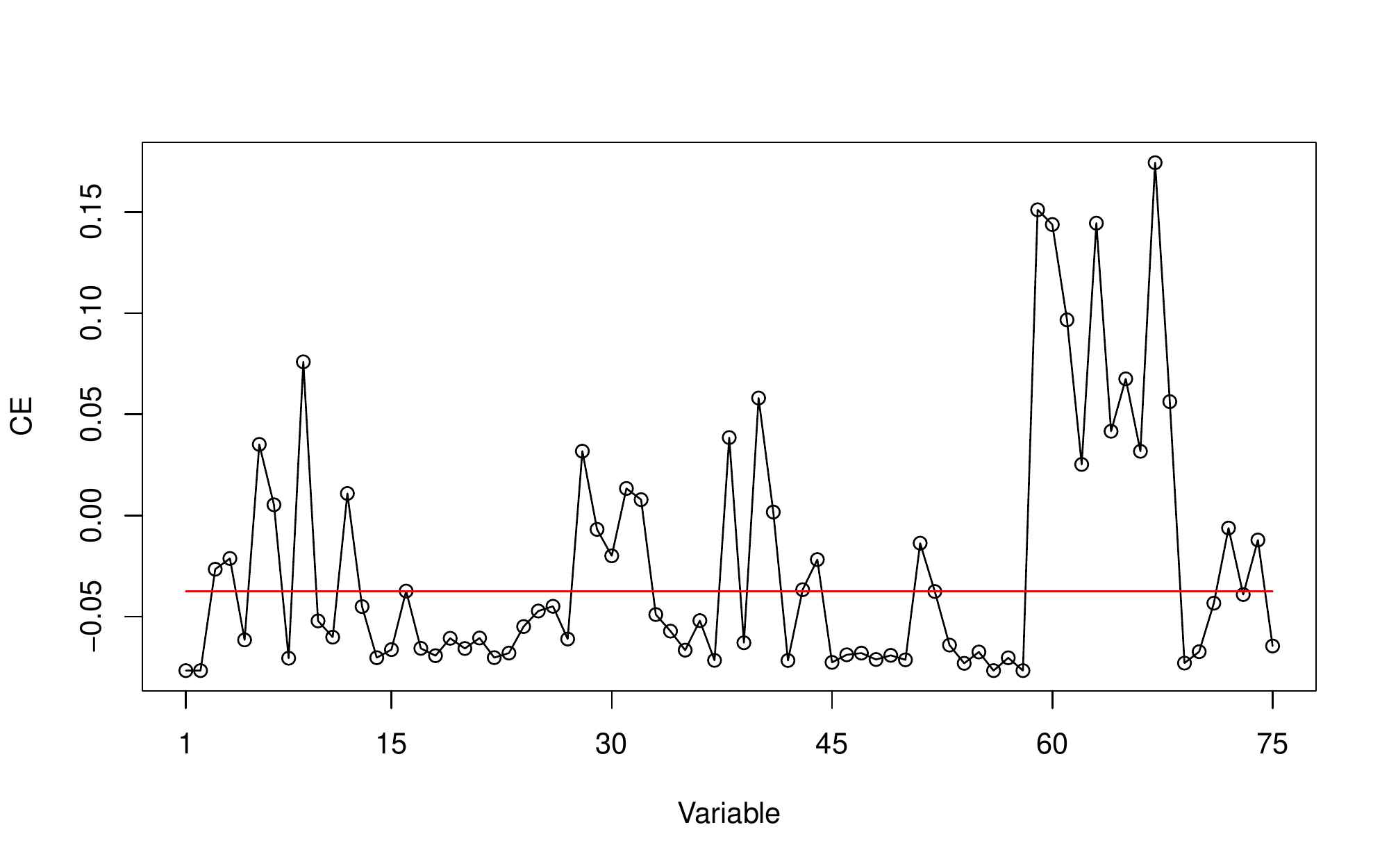}}
	\subfigure[Ktau]{\includegraphics[width=0.495\linewidth]{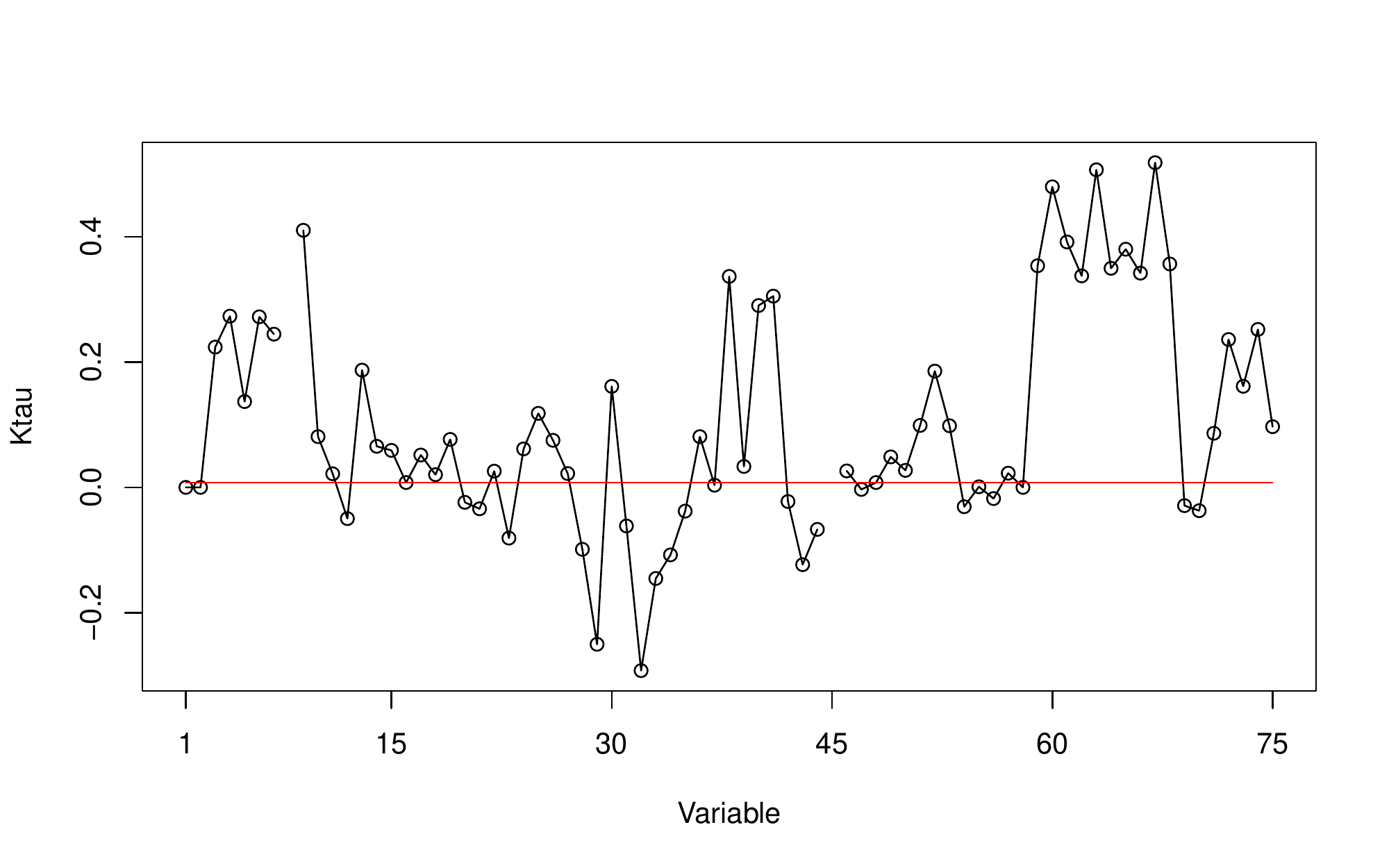}}
	\subfigure[Hoeff]{\includegraphics[width=0.495\linewidth]{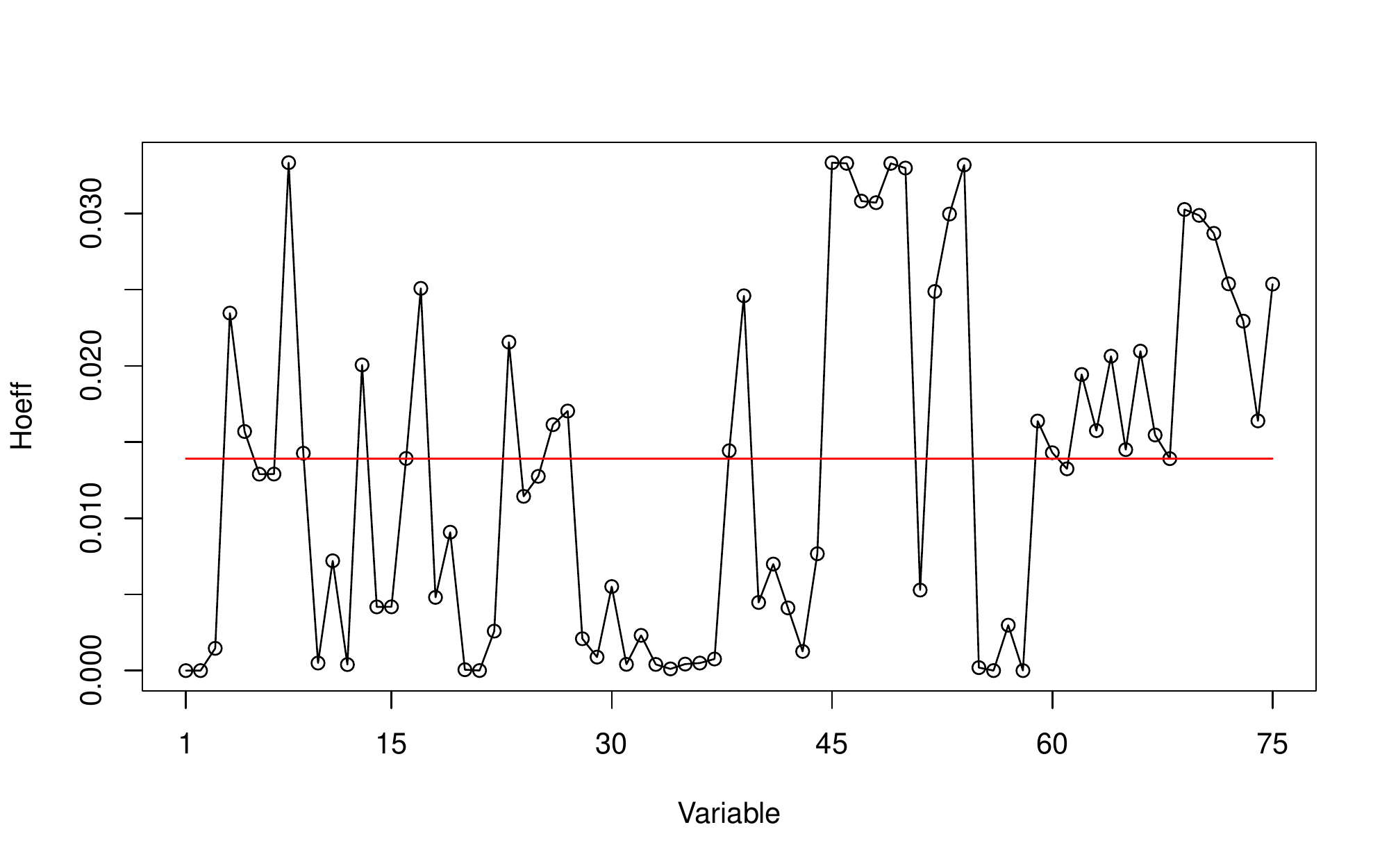}}
	\subfigure[BDtau]{\includegraphics[width=0.495\linewidth]{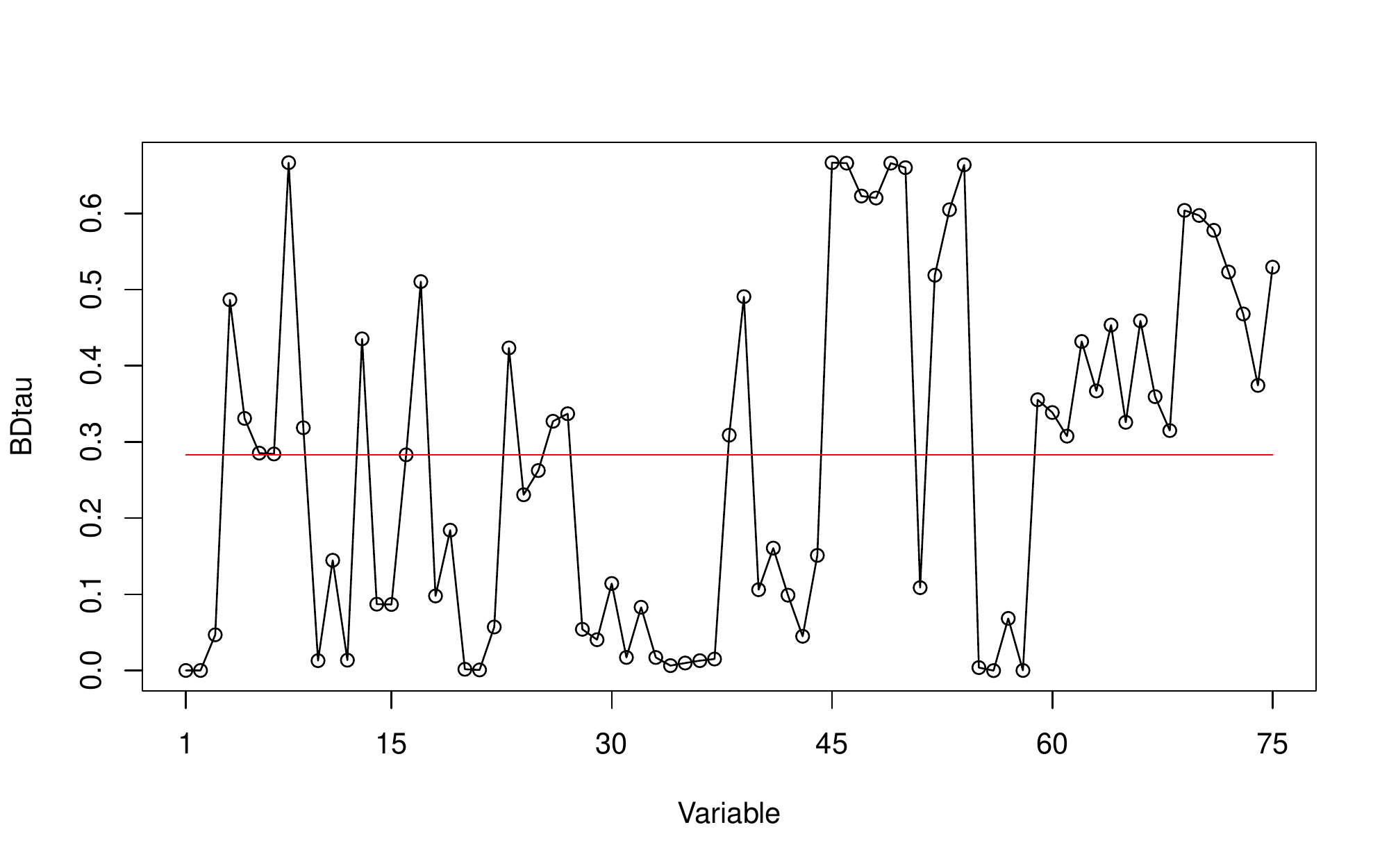}}
	\subfigure[HHG.chisq]{\includegraphics[width=0.495\linewidth]{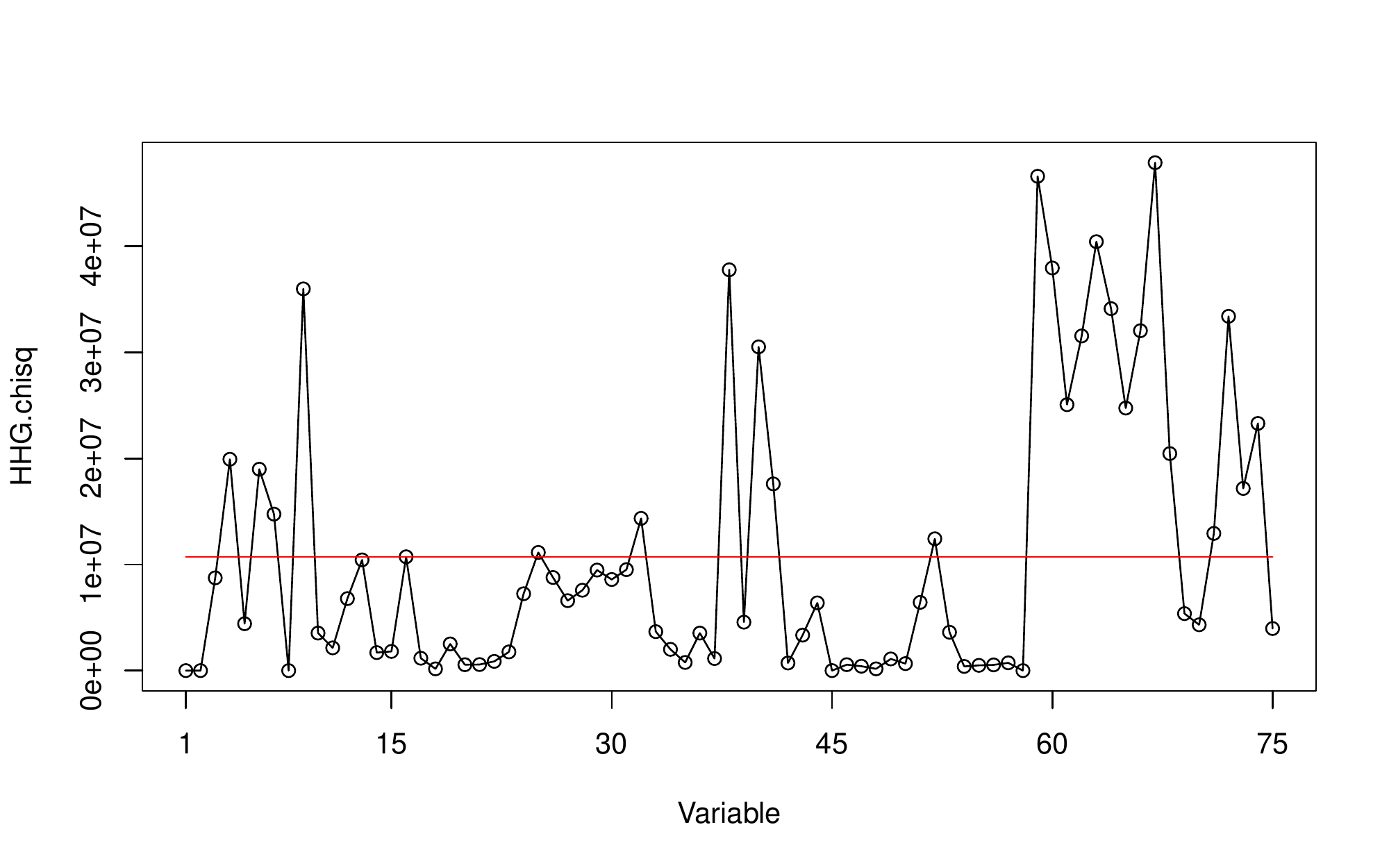}}
	\subfigure[HHG.lr]{\includegraphics[width=0.495\linewidth]{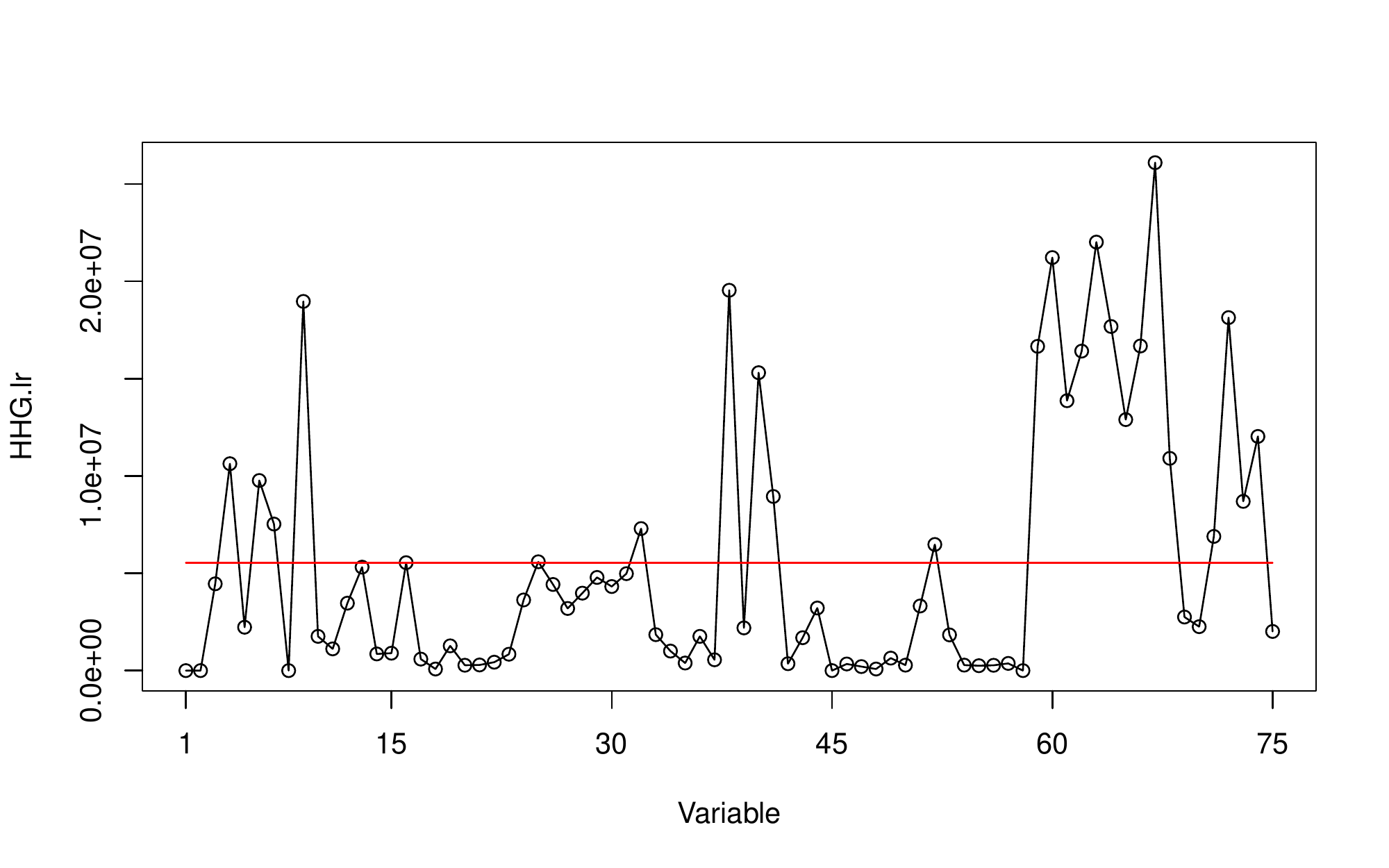}}
	\subfigure[Ball]{\includegraphics[width=0.495\linewidth]{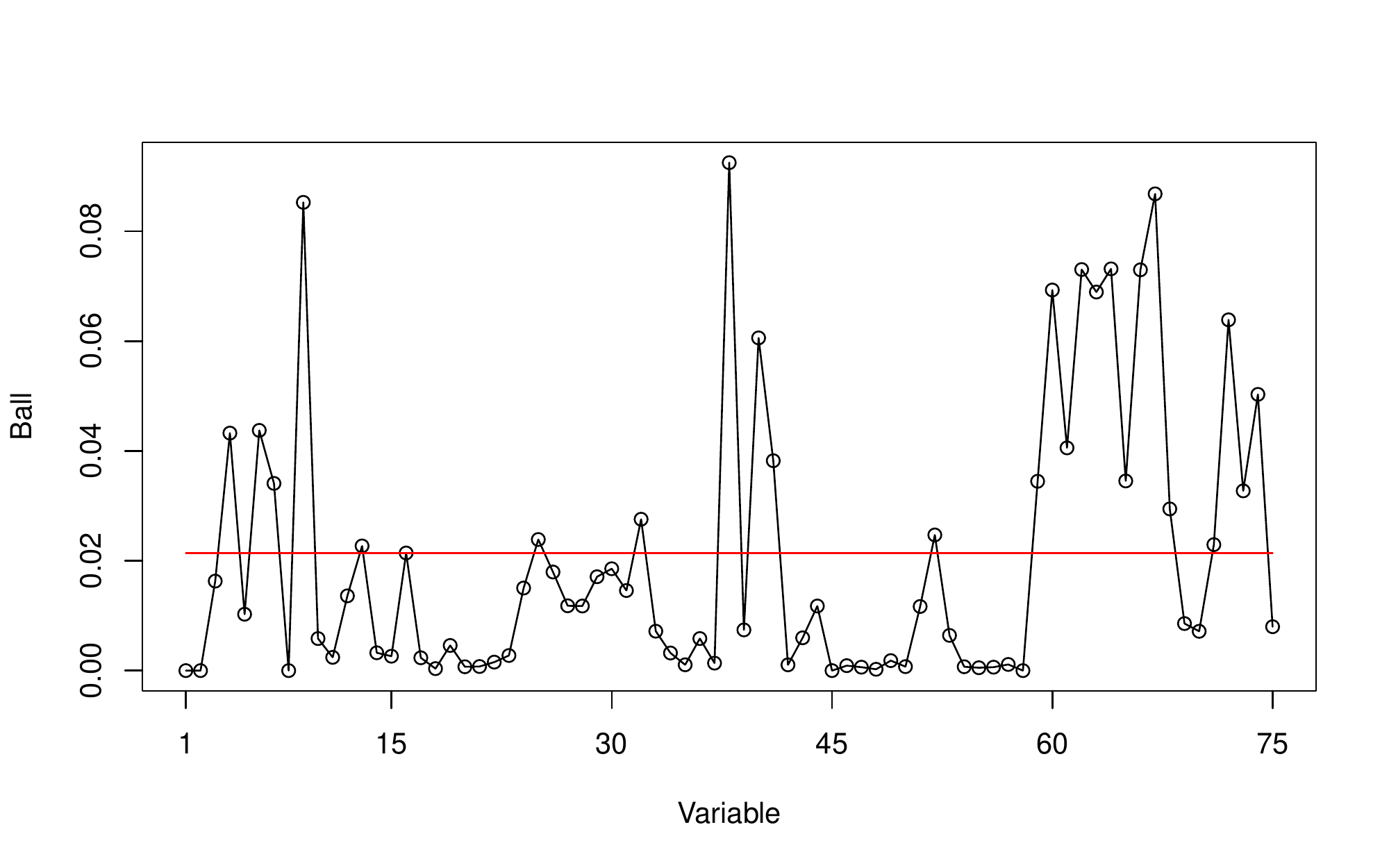}}
	\subfigure[BET]{\includegraphics[width=0.495\linewidth]{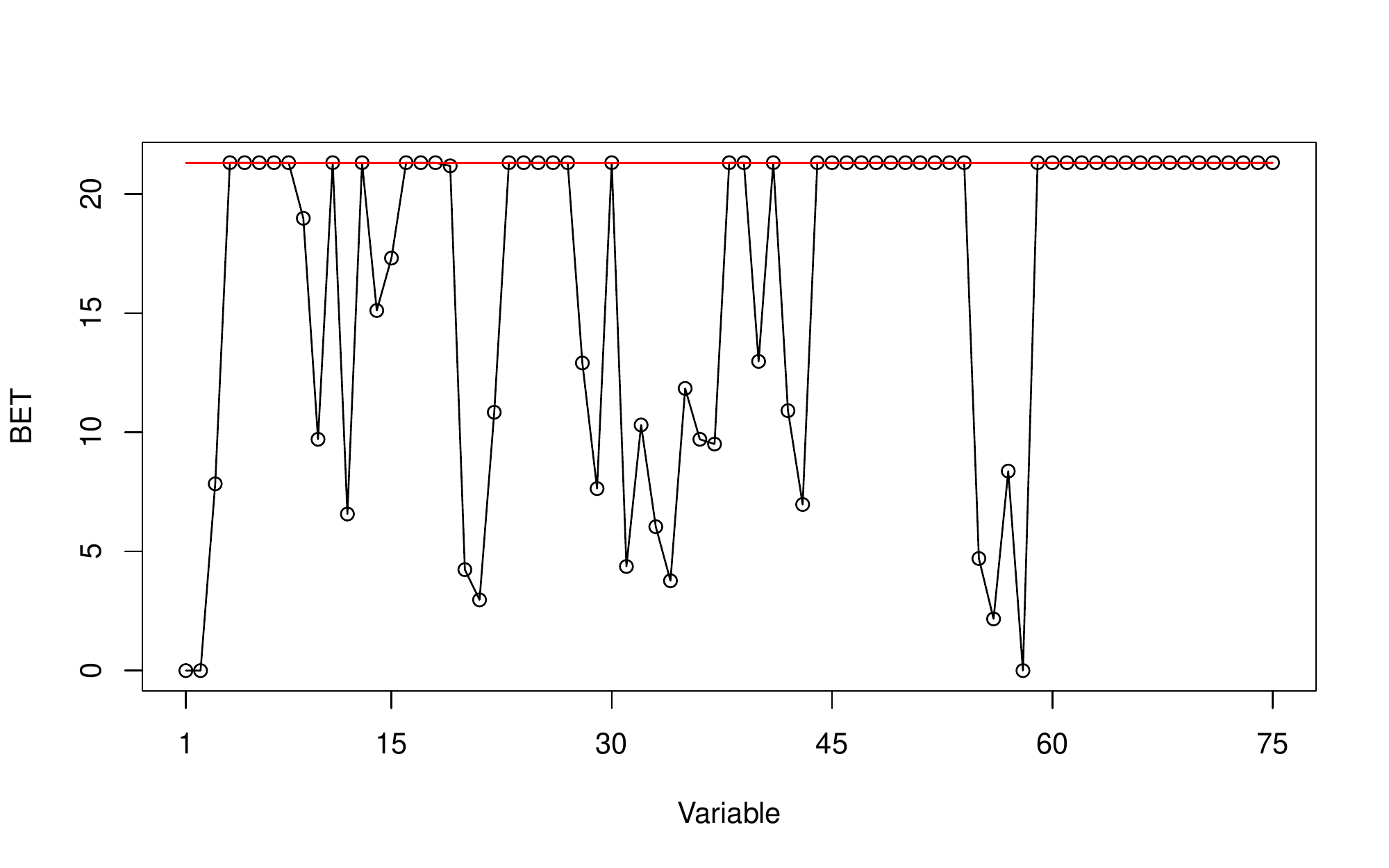}}
	\caption{Estimation of the independence measures between diagnosis and the other attributes in the UCI heart disease data.}
	\label{fig:heart}
\end{figure}

\paragraph{Heart disease data}
The estimation results of the 16 independence measures from the heart disease data are shown in Figure \ref{fig:heart}. The threshold for variable selection is the dependence between diagnosis and fbs(\#16), which is marked as a red line in all the subfigures. Based on the thresholds for each measure, the attributes that are useful for diagnosis were selected. Then they are compared to the set of the recommended attributes by professionals to derive the number of True Positive (TP) and False Positive (FP), as listed in Table \ref{tb:vs}. It can be learned from Table \ref{tb:vs} that CE selected 11 out of 13 recommended attributes at the cost of 7 FPs and that QAD, mixed, and MDM selected all the 13 attributes at the cost of high FP numbers.

The correlation matrix of the estimation of the 16 independence measures is shown in Figure \ref{fig:heartcm}. It can be learned from it that CE, dHSIC, mixed, dCor, MDM, Ball, and HHG presented relatively similar results according to the correlations between them. Particularly, they can be separated into three groups: \{CE, dHSIC, mixed\},\{dCor,MDM\}, and \{Ball,HHG\}.

\begin{table}
	\centering
	\caption{The selection results with the 16 independence measures on the UCI heart disease data.}
	\vskip2mm
	\begin{tabular}{l|c|c||l|c|c}
		\toprule
		\textbf{Measure}&TP&FP&\textbf{Measure}&TP&FP\\
		\midrule
		CE&11&7&dCor&9&9\\
		dHSIC&10&8&Ktau&10&23\\
		HHG.chisq&7&4&HHG.lr&7&4\\
		Hoeff&4&17&BDtau&4&19\\
		Ball&7&5&QAD&13&44\\
		BET&6&24&mixed&13&17\\
		mixed&9&14&MDM&13&18\\
		CODEC&3&4&NNS&3&11\\
		\bottomrule
	\end{tabular}
	\label{tb:vs}
\end{table}

\paragraph{Wine quality data}
For the wine quality data, we did not present all the estimation results. Instead, we normalized the estimation results of the measures between physicochemical and sensory attributes by setting the value of the results of the attribute "fixed acidity" as 0 and that of "alcohol" as 1. In this way, we can compare the estimation of the measures of each physicochemical attribute at a same scale and therefore can compare the ability of these measure to differentiate attributes. The normalized estimation results is shown in Figure \ref{fig:wine2}, from which it can learned clearly that Ktau, NNS, and CODEC present the results distinct from that of the other measures. This can also be confirmed by the correlation matrix between the measures in Figure \ref{fig:wine}. It can also be learned from Figure \ref{fig:wine2} that the estimation results of almost all the other measures for the attributes "density" and "residual sugar" have relatively high values than that for the other attributes. It can be further learned from Figure \ref{fig:wine2} that CE and BET are the two measures that presented higher relative strength for these two attributes than the other measures did.

\subsection{Conditional Independence Measures}

\subsubsection{Simulated data}
\paragraph{Trivariate normal distribution}
The estimation results of the 16 CI measures from the simulated data of trivariate normal distribution is shown in Figure \ref{fig:trinormci}. The results from FCIT, PCIT, CCIT, and wGCM are p-values while those of all the other measures are the statistics. 

It can be learned from Figure \ref{fig:trinormcicm} that PCIT and FCIT are different from all the others and that wGCM and CCIT are within a same group and all the remaining measures that decrease similarly as the $\rho_{xz}$ increases is in a group. 

RCoT as an approximation of KCI presented a result distinct from that of KCI. The three variants of the CMI estimator, including CE, CMI1, and CMI2, presents similar results. wGCM as a generalization of GCM presented a result different from that of GCM as the correlation between them indicates.

\paragraph{Trivariate normal copula}
The results of the 16 CI measures from the simulated data of trivariate normal copula is shown in Figure \ref{fig:trinormcopci} and Figure \ref{fig:trinormcopcicm}, which is very similar to that of the CI measures estimated from the simulated data of trivariate normal distribution.

\subsubsection{Real data}
\paragraph{Beijing air data}
The estimation results of the 16 CI measures for the CI between pressure and the future value of PM2.5 given the historical value of PM2.5 is shown in Figure \ref{fig:air}. 

There are three groups that can be identified according to the correlation matrix in Figure \ref{fig:aircm}: the first group includes CMI1, CMD, and FCIT; the second group includes KCI, KPC, pcop, GCM, and pcor, which increase linearly as time lag increases; CE, CDC and CODEC can be roughly put into the third group. The measures in the second group increase linearly as the time lag increases. The measures in the third group first increase sharply from 1 hour to 7 hours and then slowly at the remaining time as the time lag increases. 

The three variants of the CMI estimators presented the results different with each other. RCoT as the approximation of KCI presents very inconsistent estimates as contrast to KCI. wGCM as the generalization of GCM presents distinct results compared with GCM.

\section{Discussion}
\label{sec:discussion}
In this paper, we did experiments with both simulated data and real data to evaluating the measures in section \ref{sec:measures}. We want to understand: 1) the behavior of each measure in the cases of different dependence (linear or nonlinear, bivariate or multivariate, simulated or real cases); 2) the relationships between the measures in each case and in general; 3) which one is better in general.

\subsection{Independence Measures}
We did eight simulation experiments to evaluate the 16 independence measures. In each simulation, we let the control parameter ($\rho$ or $\alpha$) increases to generate a group of samples with increasing dependence between random variables. Here we first checked whether the estimated results of the measures reflect such monotonicity as shown in Table \ref{tb:correct}. It can be learned that all expect NNS can present right monotonicity in the first simulation experiments and CE, BET and dHSIC did well in all the eight simulation experiments.

\begin{table}
	\centering
	\caption{Monotonicity of the results of the independence measures in the eight simulation experiments.}
	\begin{tabular}{l|c|c|c|c|c|c|c|c}
		\toprule
		Measure&1&2&3&4&5&6&7&8\\
		\midrule
		CE&$\surd$ &$\surd$&$\surd$&$\surd$&$\surd$&$\surd$&$\surd$&$\surd$\\
		Ktau&$\surd$&$\surd$&$\surd$&$\surd$&$\surd$&&&\\
		Hoeff&$\surd$&$\surd$&$\surd$&$\surd$&$\surd$&&&\\
		BDtau&$\surd$&$\surd$&$\surd$&$\surd$&$\surd$&&&\\
		HHG.chisq&$\surd$&$\surd$&$\surd$&$\surd$&$\surd$&&&$\surd$\\
		HHG.lr&$\surd$&$\surd$&$\surd$&$\surd$&$\surd$&&&$\surd$\\
		Ball&$\surd$&$\surd$&$\surd$&$\surd$&$\surd$&&&$\surd$\\
		BET&$\surd$&$\surd$&$\surd$&$\surd$&$\surd$&$\surd$&$\surd$&$\surd$\\
		QAD&$\surd$&$\surd$&$\surd$&$\surd$&$\surd$&&&$\times$\\
		mixed&$\surd$&$\surd$&$\surd$&$\surd$&$\surd$&$\surd$&$\surd$&\\
		CODEC&$\surd$&$\surd$&$\surd$&$\surd$&$\surd$&&&\\
		subcop&$\surd$&$\surd$&$\surd$&$\surd$&$\surd$&$\surd$&$\surd$&\\
		dCor&$\surd$&$\surd$&$\surd$&$\surd$&$\surd$&&&$\surd$\\
		MDM&$\surd$&$\surd$&$\surd$&$\surd$&$\surd$&&&$\surd$\\
		dHSIC&$\surd$&$\surd$&$\surd$&$\surd$&$\surd$&$\surd$&$\surd$&$\surd$\\
		NNS&$\surd$&$\surd$&$\surd$&$\times$&$\times$&$\times$&$\times$&$\surd$\\
		\bottomrule
	\end{tabular}
\label{tb:correct}
\end{table}

From the first five simulated experiments, we can learn that all the independence measures presented satisfactory results, which means increase as the dependence strength increases, except NNS which did not present monotonically increasing dependence in the cases of Gumbel copula and Frank copula. In the case of bivariate normal distribution that simulate linear correlations, Ktau,dCor and subcop present linearly increasing results, and CE, Hoeff, Ball, HHG and dHSIC present strong nonlinearly increasing results, which means the former are linear measures while the latter are nonlinear measures approximately. In the case of bivariate normal copula associated with nonlinear exponential marginals, the situation is almost same. In the cases of three bivarite Archimedean copulas which simulate nonlinear dependence, CE, Hoeff, Ball and HHG presented linearly increasing results while Ktau, dCor and subcop presented strong nonlinear increasing results. In all the five simulations, the remaining measures presented weak nonlinear increasing results. These results suggest that the nonlinearity properties of the measures are different. From section \ref{sec:indep}, we know that Hoeff, Ball and HHG are all defined based on Hoeffding's principle which measures second order dependence and that both Ktau and subcop are defined first order dependence. The surprise is that dCor which is defined on second order dependence is also shown to has linearity.

Next, we study the relationship between the 16 independence measures based on the estimation results. In the simulation experiments, we derived the correlation matrix of the measures and conduct hierarchical clustering based on correlation. Based on clustering results, we can find that: 1) CE, Hoeff, Ball and HHG has similar behaviors in the simulations; 2) BDtau, CODEC, dHSIC tend to be clustered together which means similar behaviors; 3) Ktau, dCor and subcop presented similar results in the simulation experiments. As mentioned above, Hoeff, Ball and HHG are defined on Hoeffding's principle, which may lead to their similar behaviors. Ktau is a nonparametric linear correlation measure. The similarity between Ktau, dCor and subcop means the latter two can be considered as linear measures approximately. This is seemingly right for subcop, but contrary to the definition of dCor.

We did three simulations for multivariate cases. It is no surprise that most of the suitable measures works well. In the three simulations, CE, BET and dHSIC presented similar results according to correlations.

We can learn from the correlation matrix of the simulation experiments that the correlations between the estimation results of the measures are mostly strong since most of the measures presents the right monotonicity. However, in the real data case where the dependence relationships are complex, the correlations between the measures are relatively weaker than in the simulated data of which the distributions are governed by the simpler functions. It can be learned from Table \ref{tb:vs} that only CE, dHSIC and dCor can be considered as presenting good results with high TP and relatively low FP. The other measures, such as Ball, mixed, HHG, presented only moderate results. In the wine quality data, we compared the ability of the measures to identify the pysicochemical attributes related to wine quality. Figure \ref{fig:wine2} shows that after normalization, the strength of these measures are quite different. As mentioned before, CE and BET excelled to the others on the important attributes, such as "density" and "residual sugar". This implies that the ability of these measures will be differentiated in the complex real situations.

In summary, let us make a comparison between these measures in general. We can learn from Table \ref{tb:correct} that CE, BET and dHSIC did well on all the eight simulation experiments. In the real data experiments, CE, dHSIC and dCor work well on the heart disease data as shown in Table \ref{tb:vs} and CE and BET work well on the wine quality data according to Figure \ref{fig:wine2}. Based on these results, we can say CE is the best measures among these measures in terms of properties and performance. It is because that CE is a multivariate independence measure with rigorous definition, which give it several good properties that the others don't have, such as monotonically invariant, equivalent to correlation coefficient in Gaussian cases. It provides the unified theory of (in)dependence measure for any types of dependence since it is based on the copula theory. It is also because of the rank-based estimator of CE that is nonparametric and consistent.

\subsection{Conditional Independence Measures}
As introduced in section \ref{sec:ci}, the CI measures includes distance correlation based (CDC and CMD), the definitions based on \eqref{eq:eixyz} (CODEC and KPC), prediction based (FCIT, PCIT and CCIT), CMI based (CE, CMI1 and CMI2), partial correlation based (pcor, GCM, wGCM, KCI, RCoT and pcop). Kernel technique is used in the estimators of CDC, KPC, KCI and RCoT. However, the estimation results of the CI measures were not grouped in the same way as their definitions. All the measures, except prediction-based ones and wGCM, presents similar results in the simulation experiment with trivariate normal distribution and trivariate Gumbel copula.

In the experiments with the Beijing air data, the estimation results are more diverse than in the simulation experiments. This may be with many different reasons. For example, three CMI based estimators presented distinct results. This diversity is from the estimation methods. CE, CDC, and CODEC presented similar reasonable results. However, KCI, GCM, KPC, PCor and Pcop presented similar results. Since Pcor is a linear CI measure working under Gaussian assumptions, we can consider this groups of measures tending to be linear in these conditions. Considering the nonlinearity of the relationships between the factors in atmosphere system, we could say that CE, CDC, and CODEC are more preferable to the others in real practice. Moreover, CE is recommended for CI test due to its sound mathematical definition and consistent estimators.

\section{Conclusions}
\label{sec:conclusions}
In this paper, the 16 independence measures and 16 CI measures were evaluated with simulated and real data. For the independence measures, eight simulated data were generating from normal distribution, normal and Archimedean copula distribution to compare the ability of the measures in bivariate or multivariate, linear or nonlinear settings. Two datasets, including the heart disease data and the wine quality data, were used to test the true power of the measures in real conditions. For the CI measures, two simulated data with normal distribution and Gumbel copula, and one real data (the Beijing air data) were utilized to test the CI measures in prespecified linear or nonlinear settings and much complex real settings. Experimental results showed that most of the measures work well on the simulated data by presenting the right monotonicity of the simulations. However, both the independence measures and the CI measures differed considerably on much complex real data. Among the independence measures, only CE, dHSIC and dCor can be considered working well on the heart disease and CE and BET on the wine quality data with reference to domain knowledge. For the CI measures, we believe CE, CDE and CODEC presented much reasonable results than the others when considering the nonlinearity of the atmosphere system. We also found that the measures tend to be separated into groups based on the similar behaviors of them in each setting and in general. According to the performance in the experiments, we recommend CE as a good unified framework for both independence and CI measure. The reason that CE excelled to the others is due to its rigorous distribution-free definition and consistent nonparametric estimator.

\bibliographystyle{unsrt}
\bibliography{bench}

\addtocounter{figure}{-1}

\begin{figure}
	\addtocounter{subfigure}{8}
	\subfigure[QAD]{\includegraphics[width=0.495\linewidth]{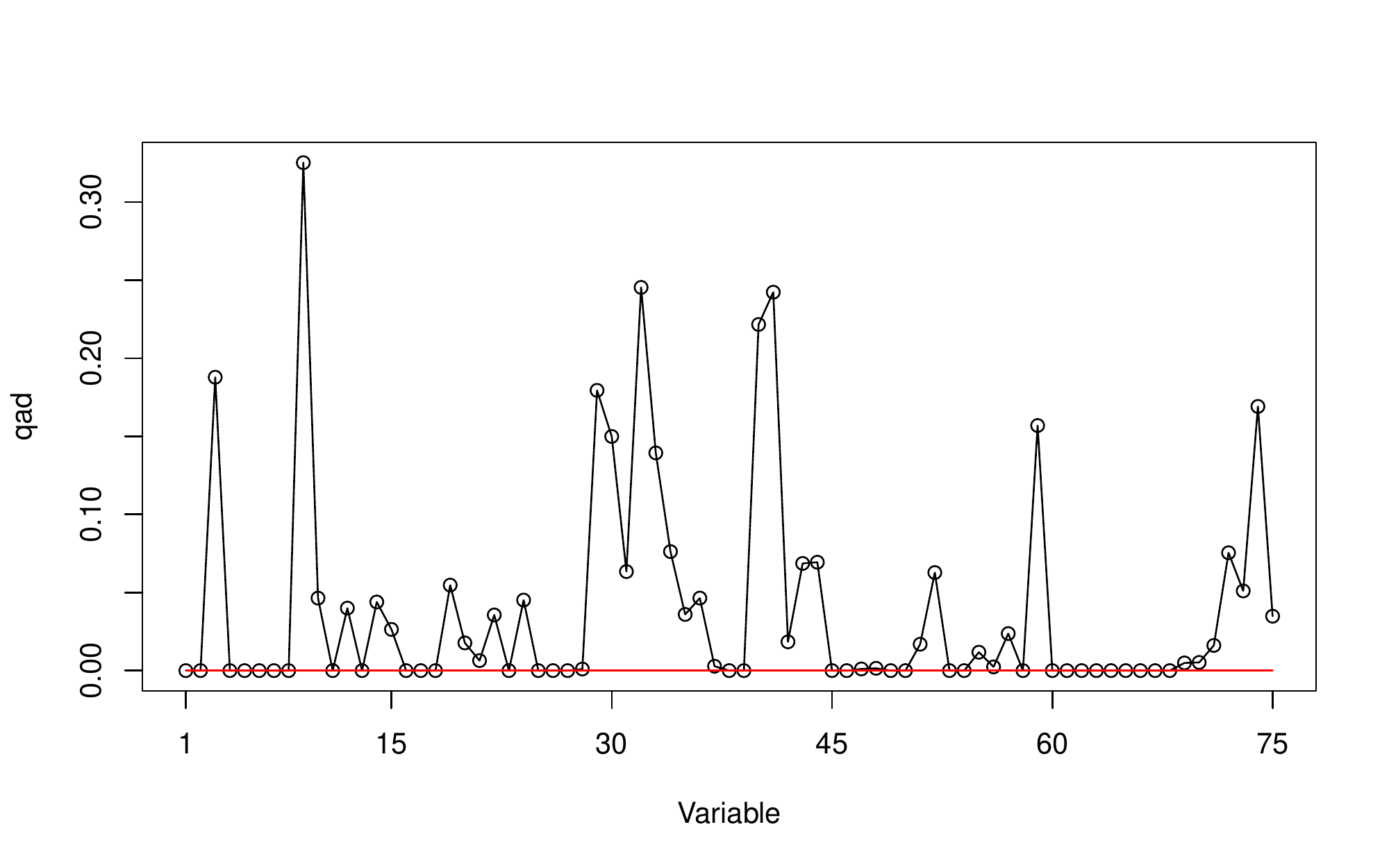}}
	\subfigure[mixed]{\includegraphics[width=0.495\linewidth]{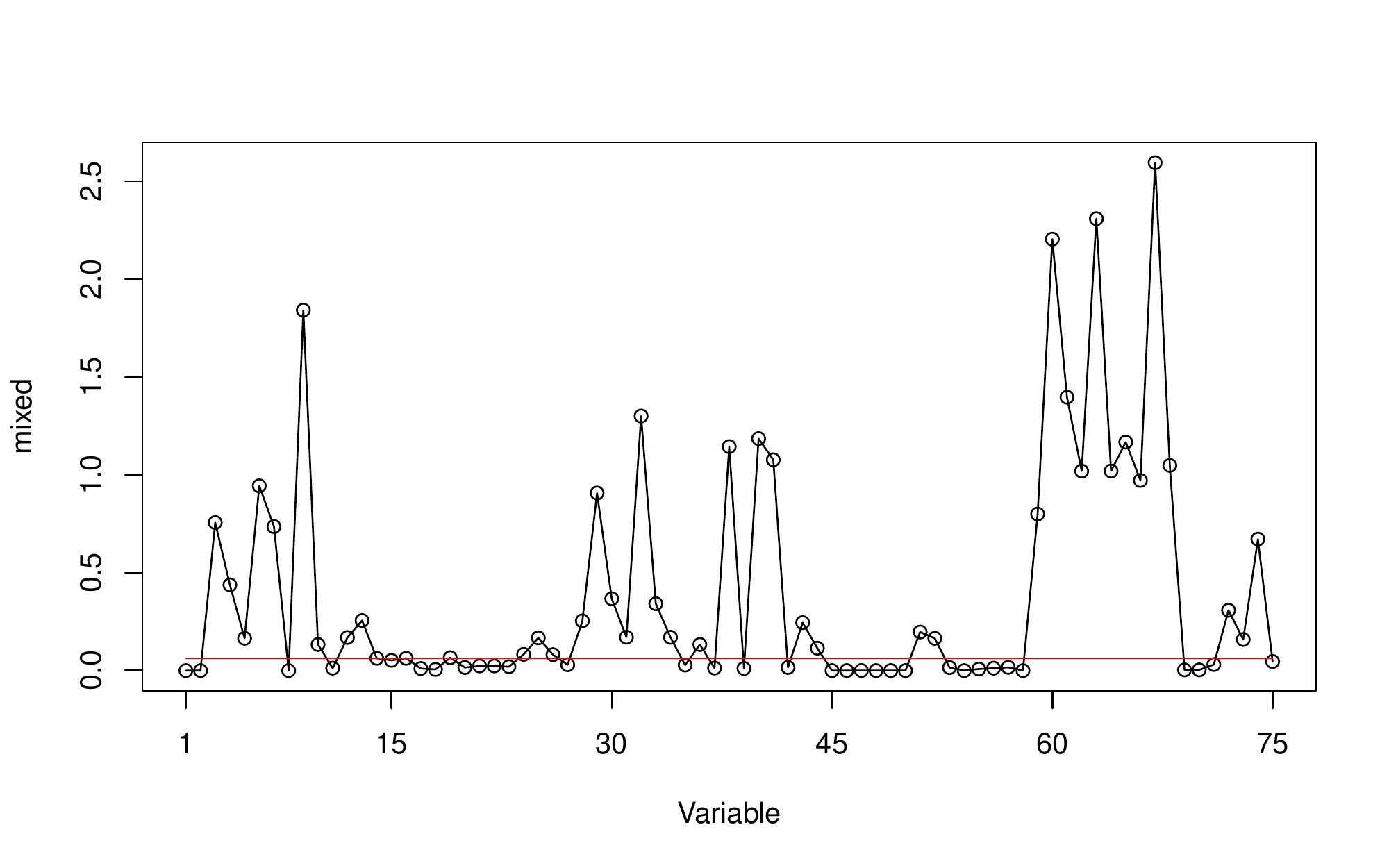}}
	\subfigure[CODEC]{\includegraphics[width=0.495\linewidth]{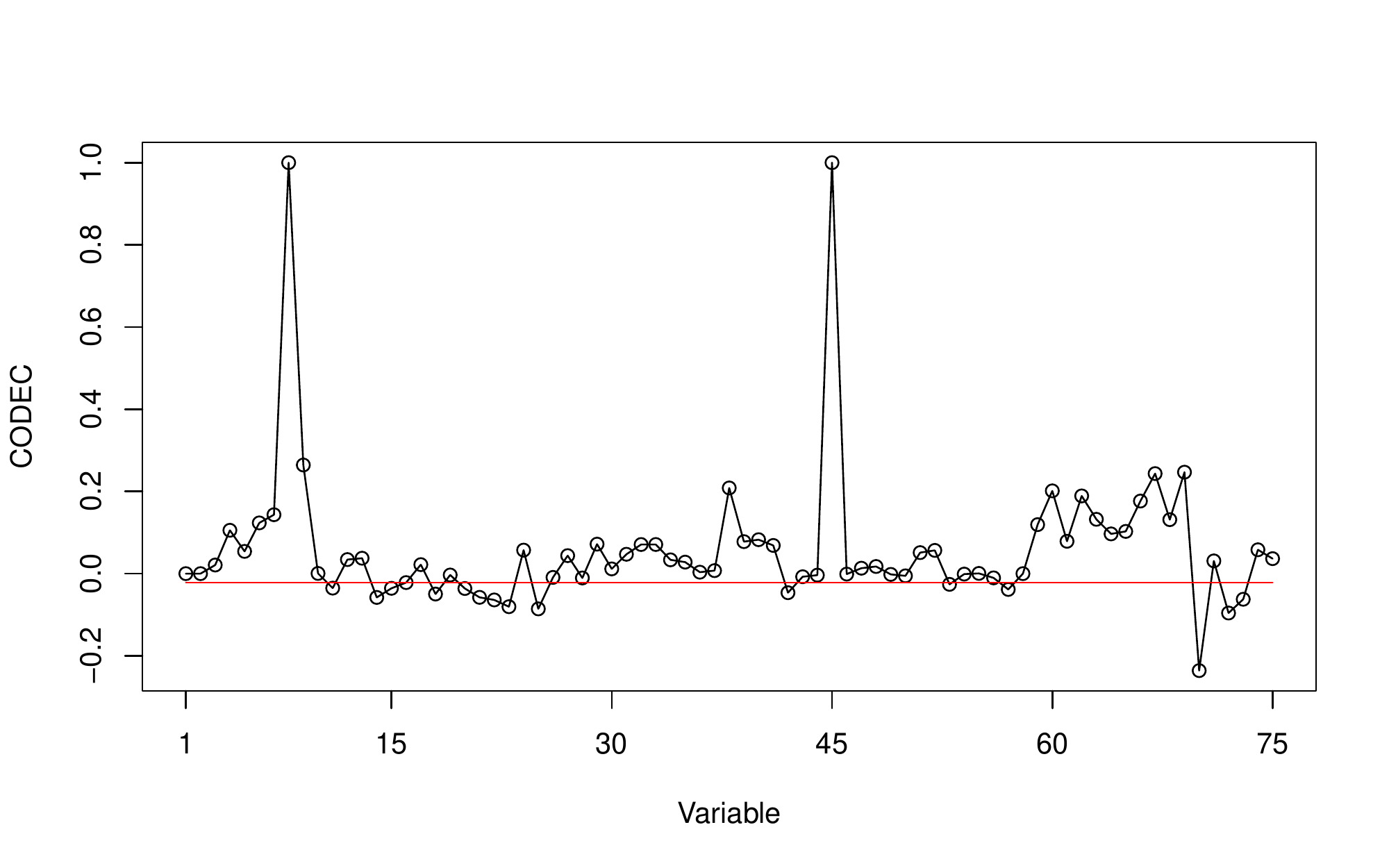}}
	\subfigure[subcop]{\includegraphics[width=0.495\linewidth]{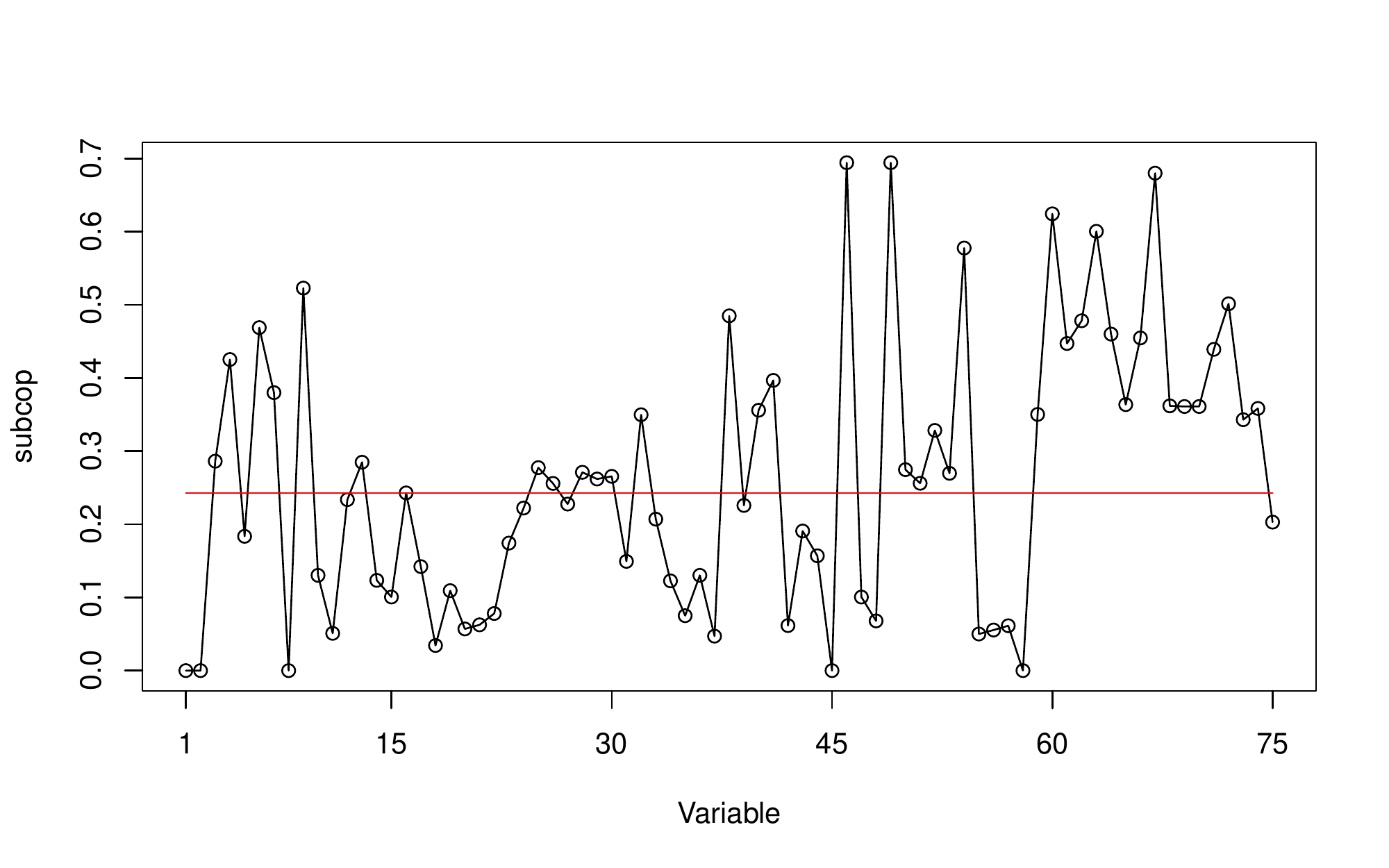}}
	\subfigure[dCor]{\includegraphics[width=0.495\linewidth]{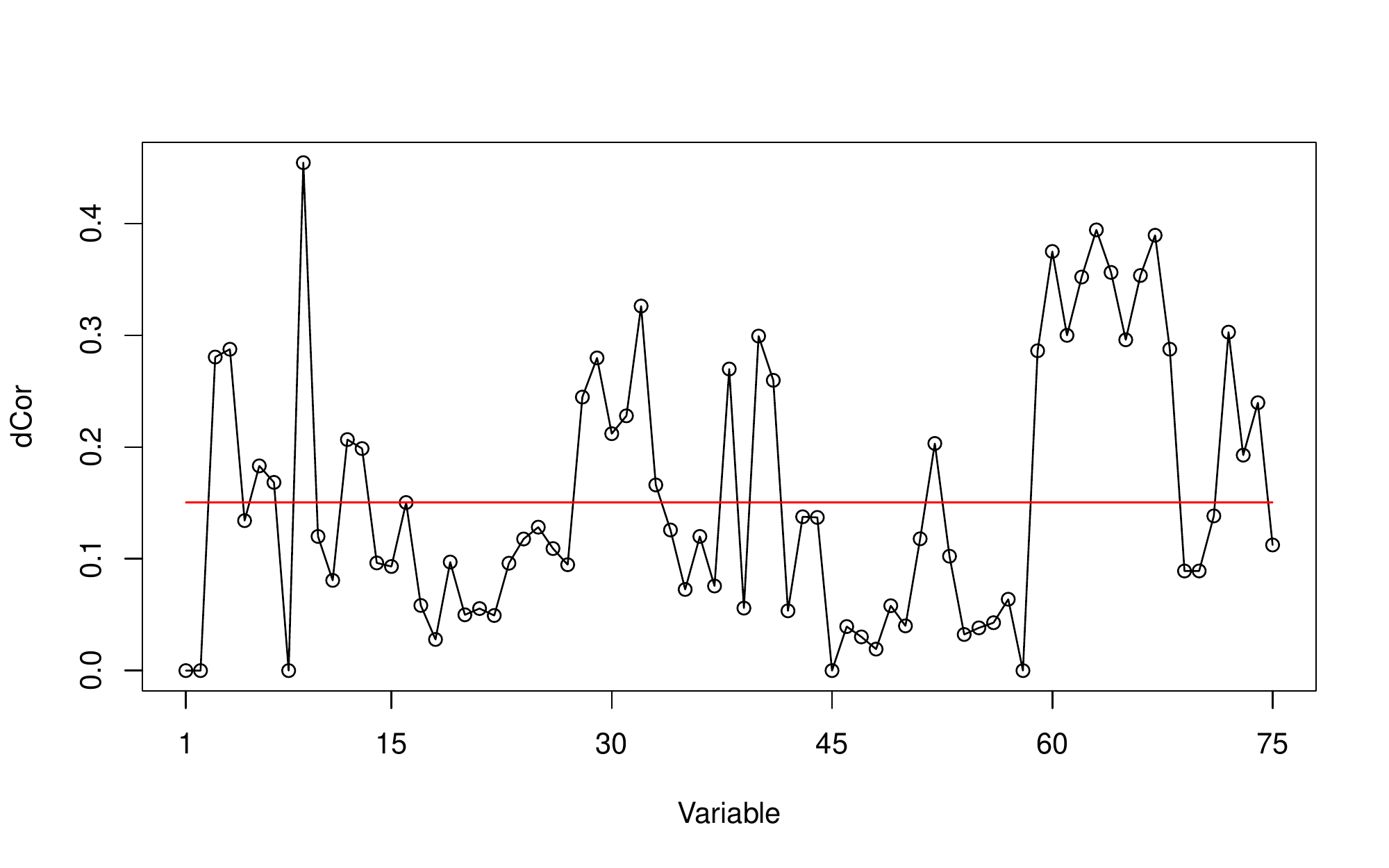}}
	\subfigure[mdm]{\includegraphics[width=0.495\linewidth]{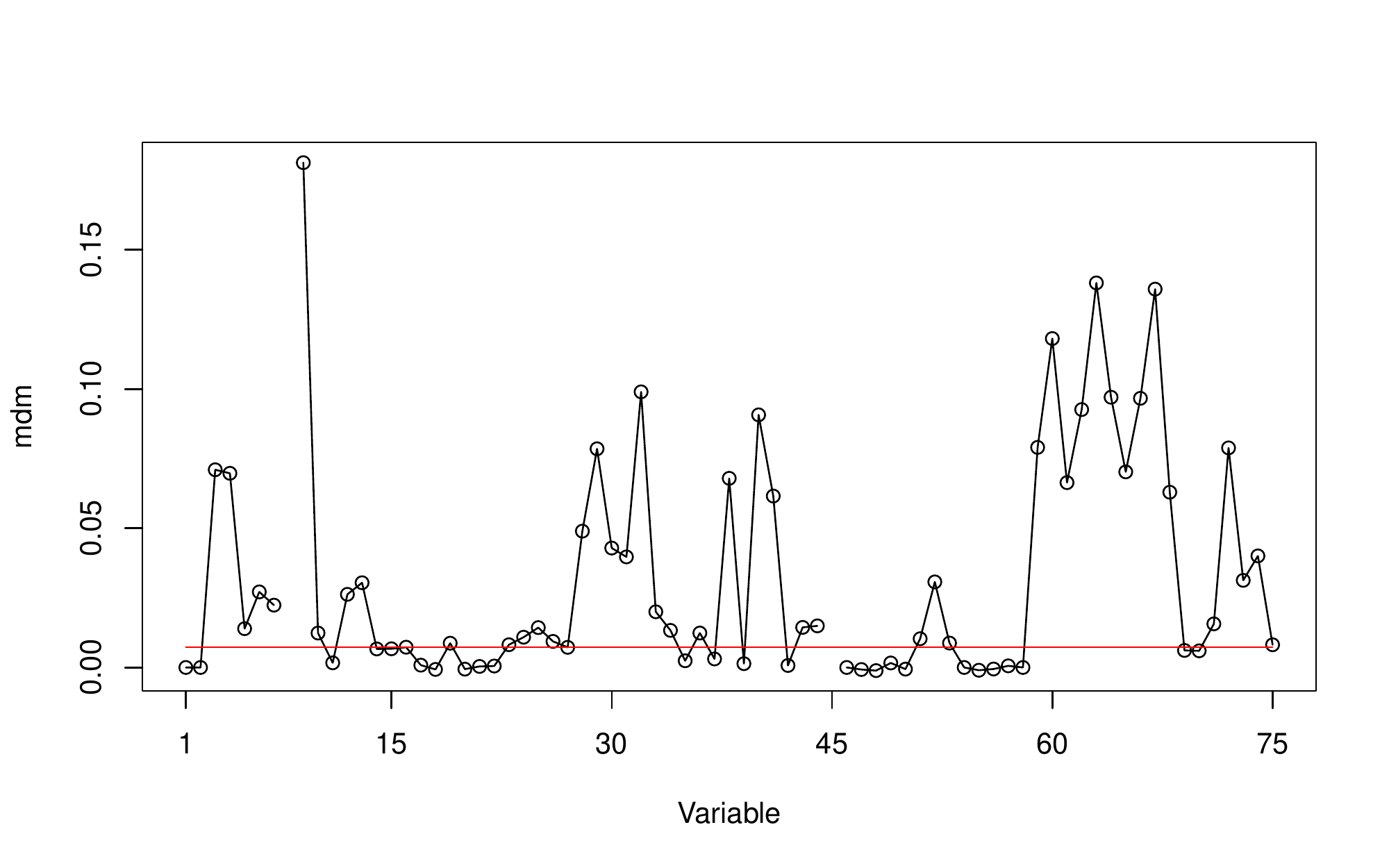}}
	\subfigure[dHSIC]{\includegraphics[width=0.495\linewidth]{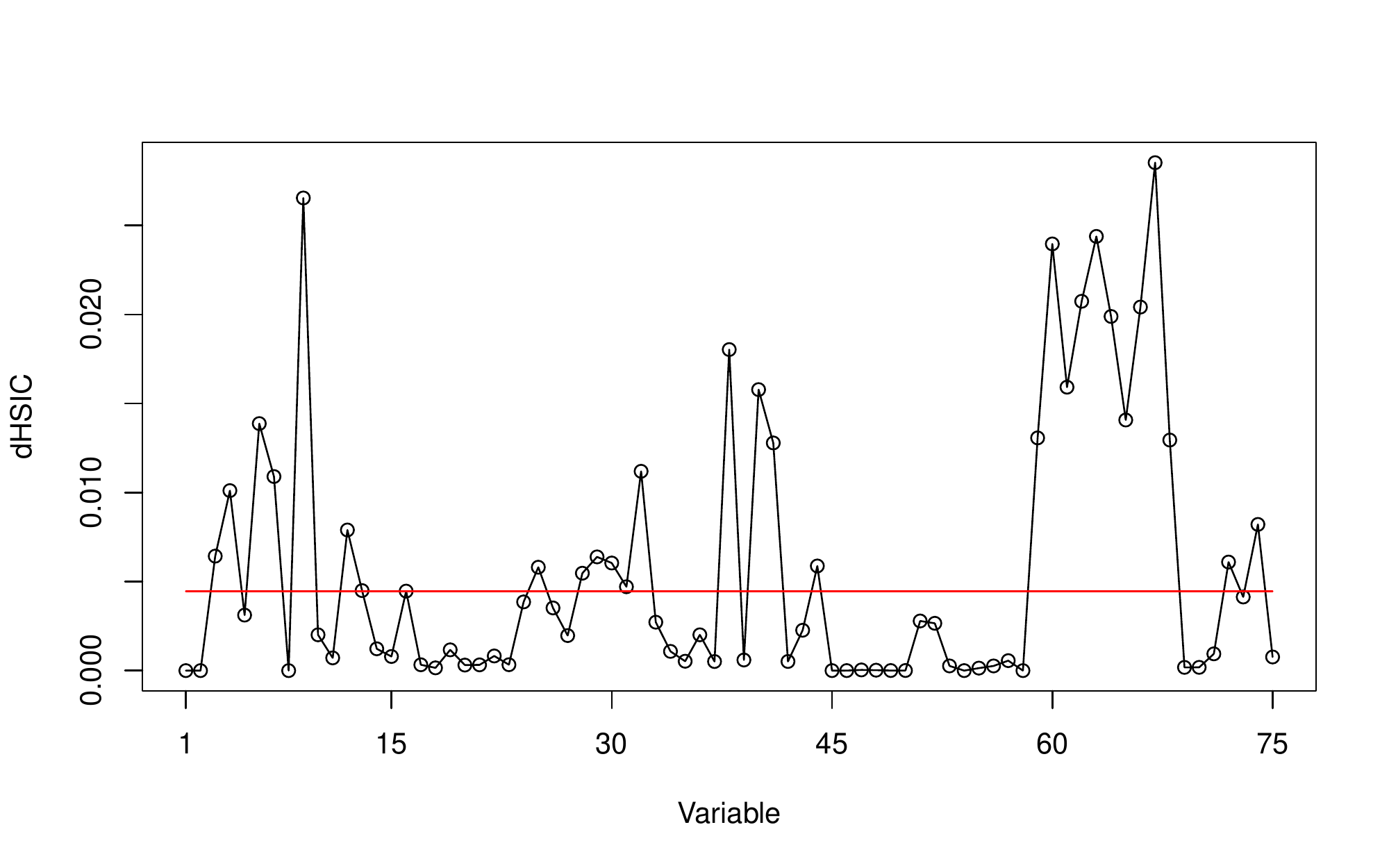}}
	\subfigure[NNS]{\includegraphics[width=0.495\linewidth]{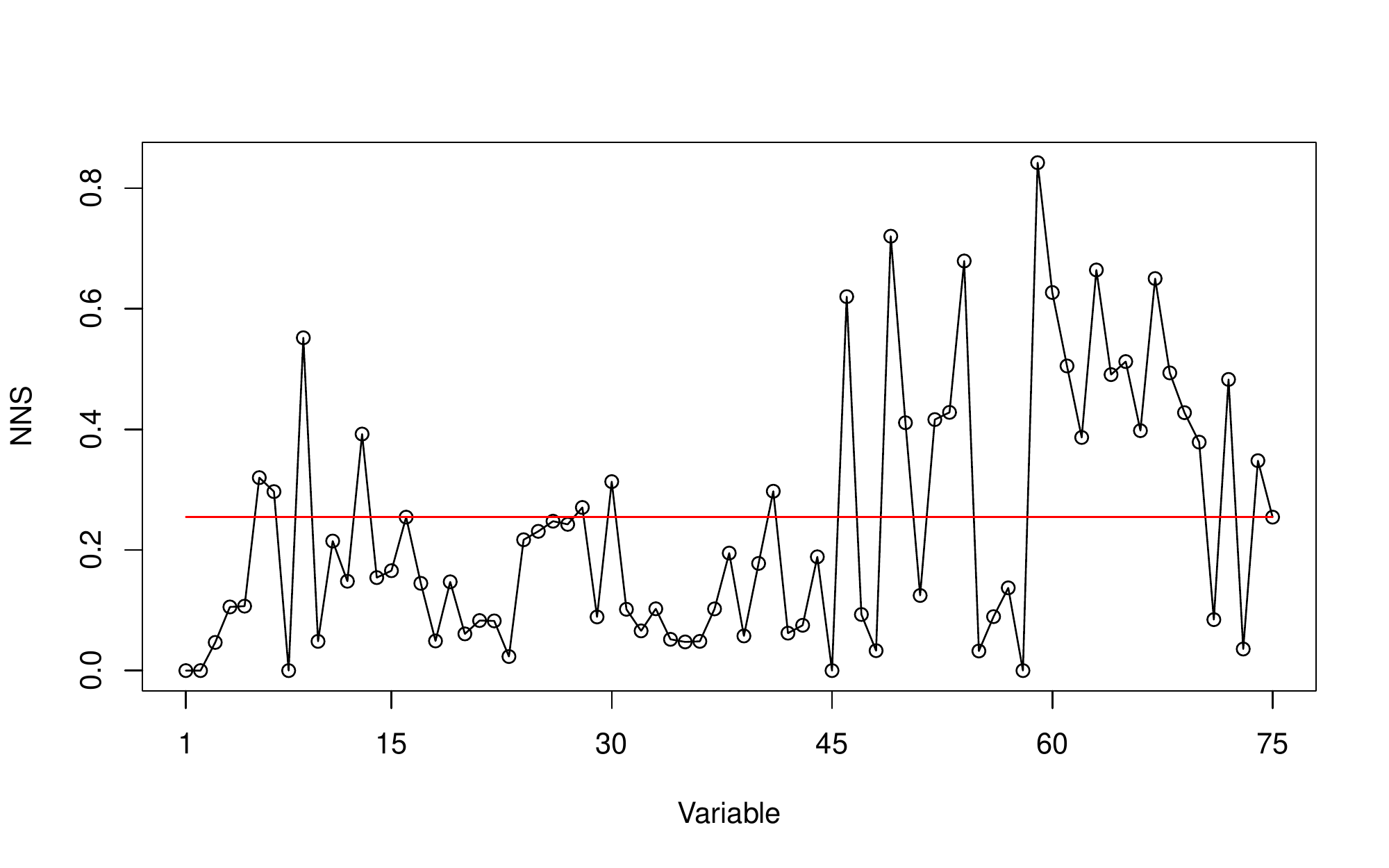}}
	\caption{Estimation of the independence measures between diagnosis and the other attributes in the UCI heart disease data.}
\end{figure}

\begin{figure}
	\centering
	\includegraphics[width=0.9\linewidth]{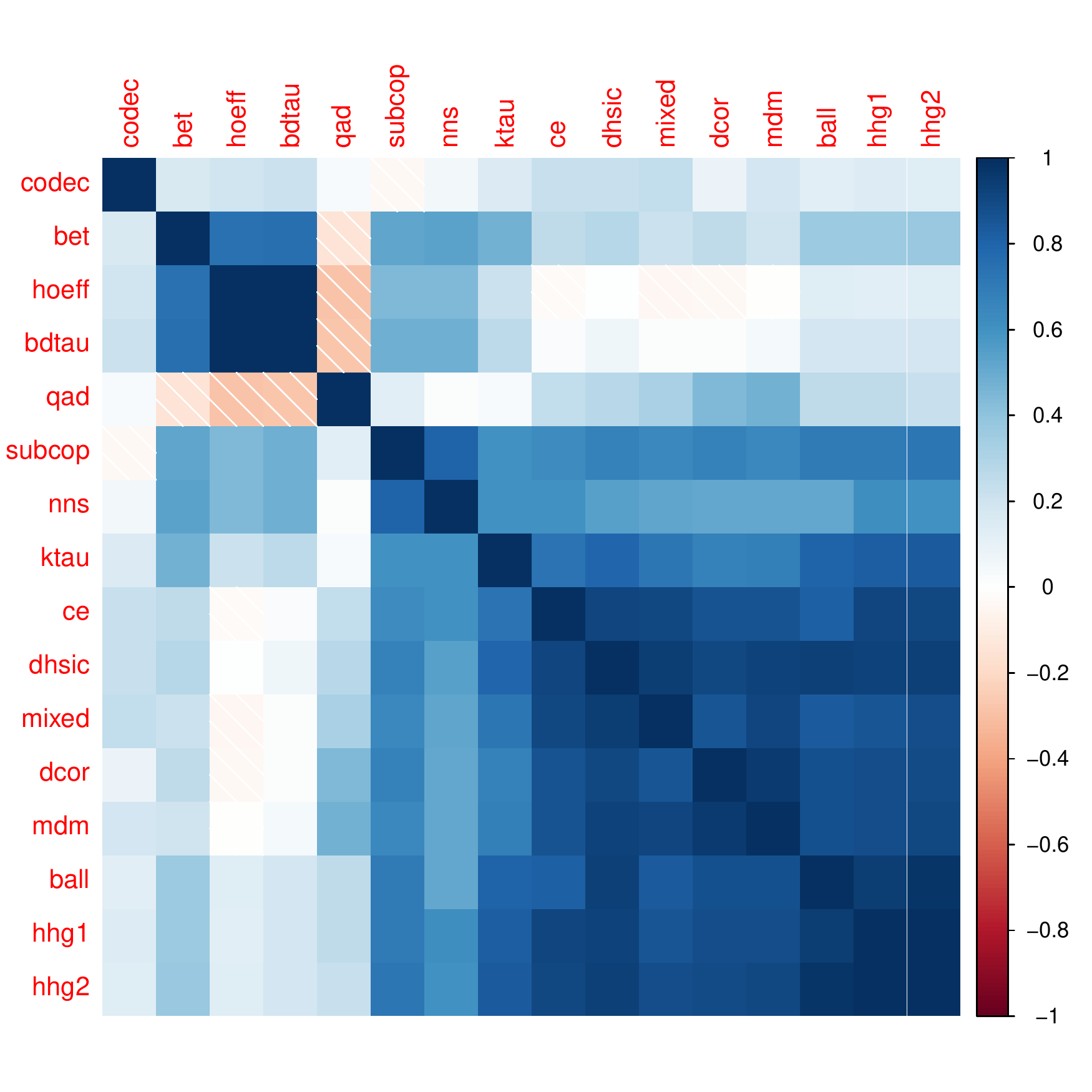}
	\caption{Correlation matrix of the independence measures estimated from the UCI heart disease data.}
	\label{fig:heartcm}
\end{figure}

\begin{figure}
	\centering
	\includegraphics[width=0.9\linewidth]{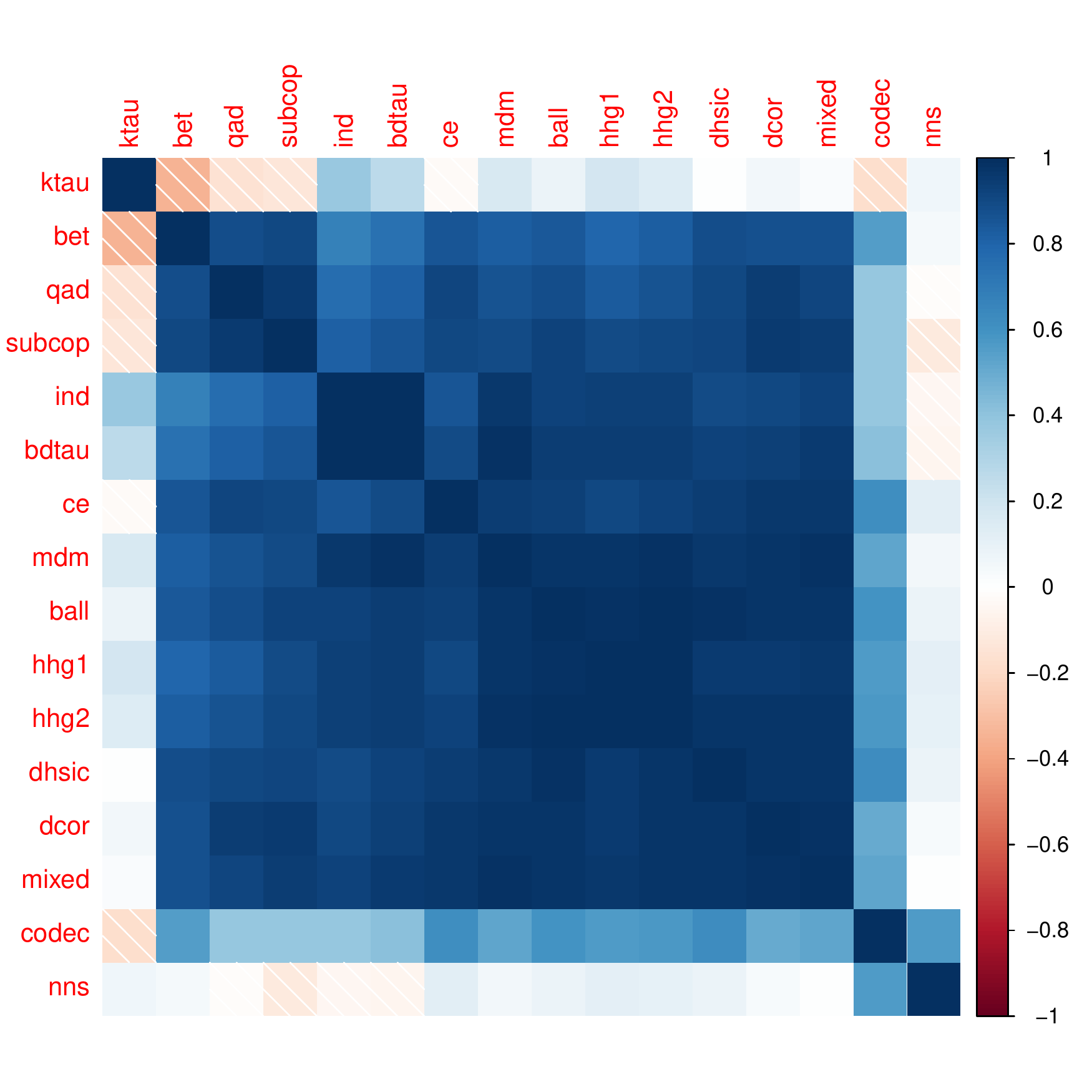}
	\caption{Correlation matrix of the independence measures estimated from the UCI wine quality data.}
	\label{fig:wine}
\end{figure}

\begin{figure}
	\centering
	\includegraphics[width=0.9\linewidth]{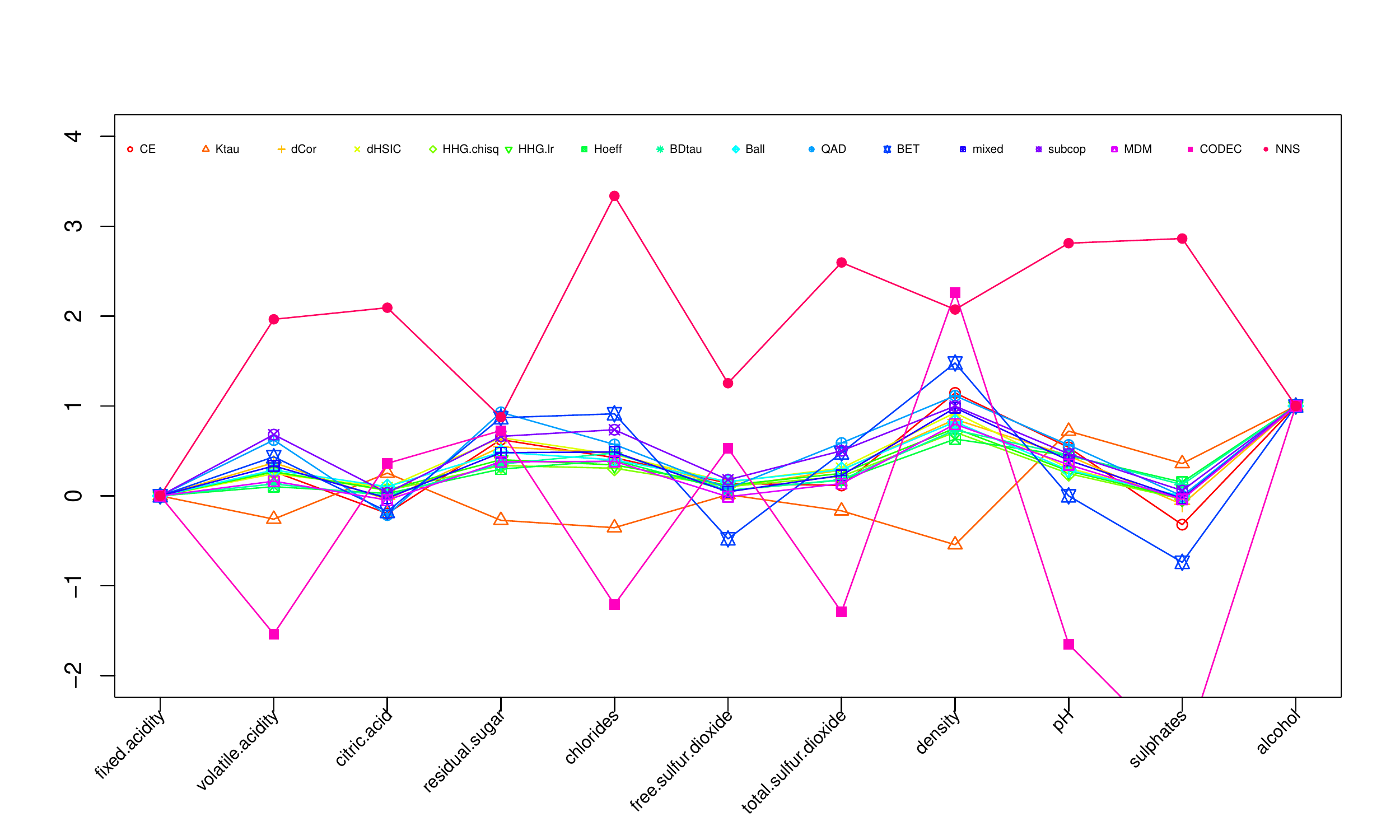}
	\caption{The normalized independence measures estimated from the UCI wine quality data.}
	\label{fig:wine2}
\end{figure}

\begin{figure}
	\subfigure[CE]{\includegraphics[width=0.245\linewidth]{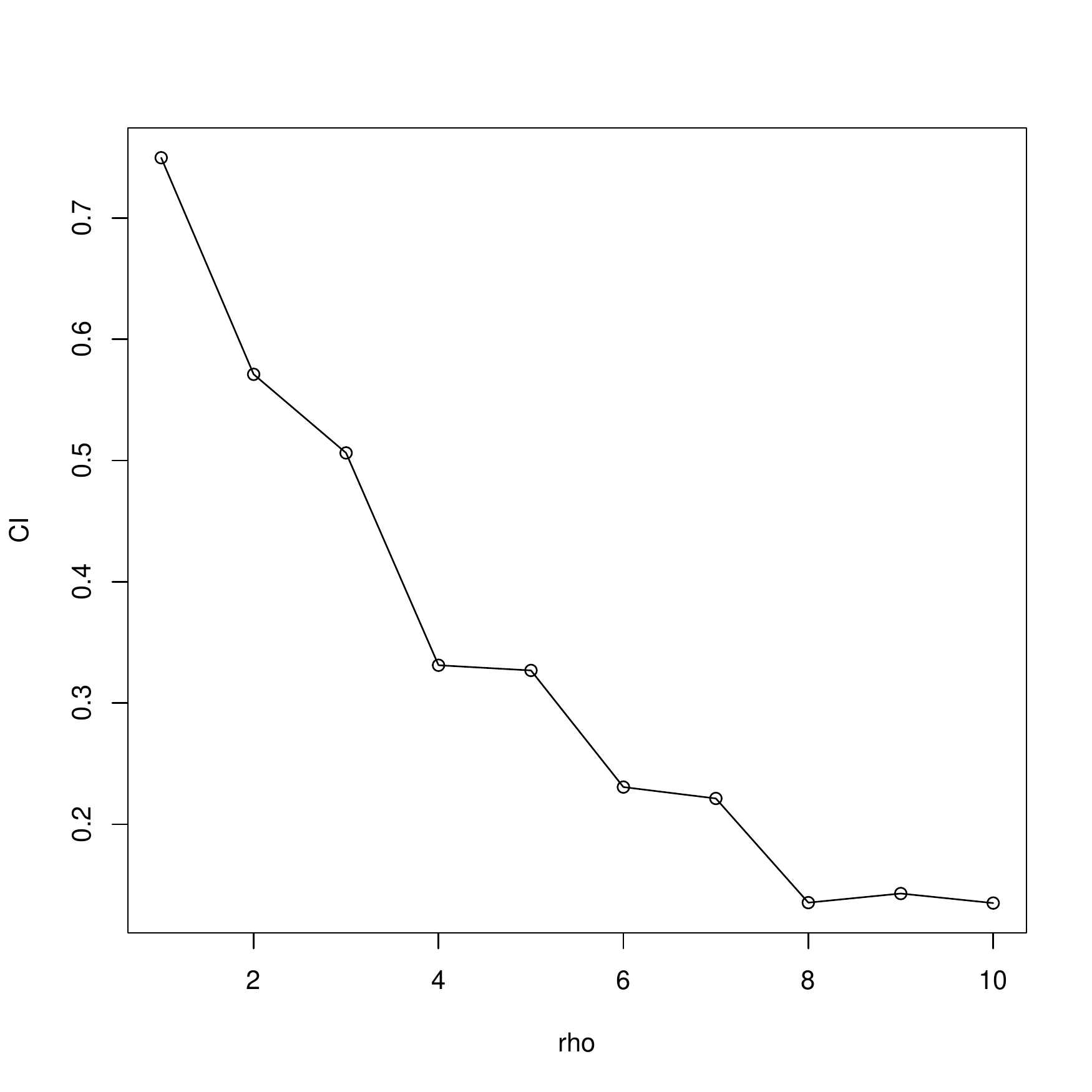}}
	\subfigure[KCI]{\includegraphics[width=0.245\linewidth]{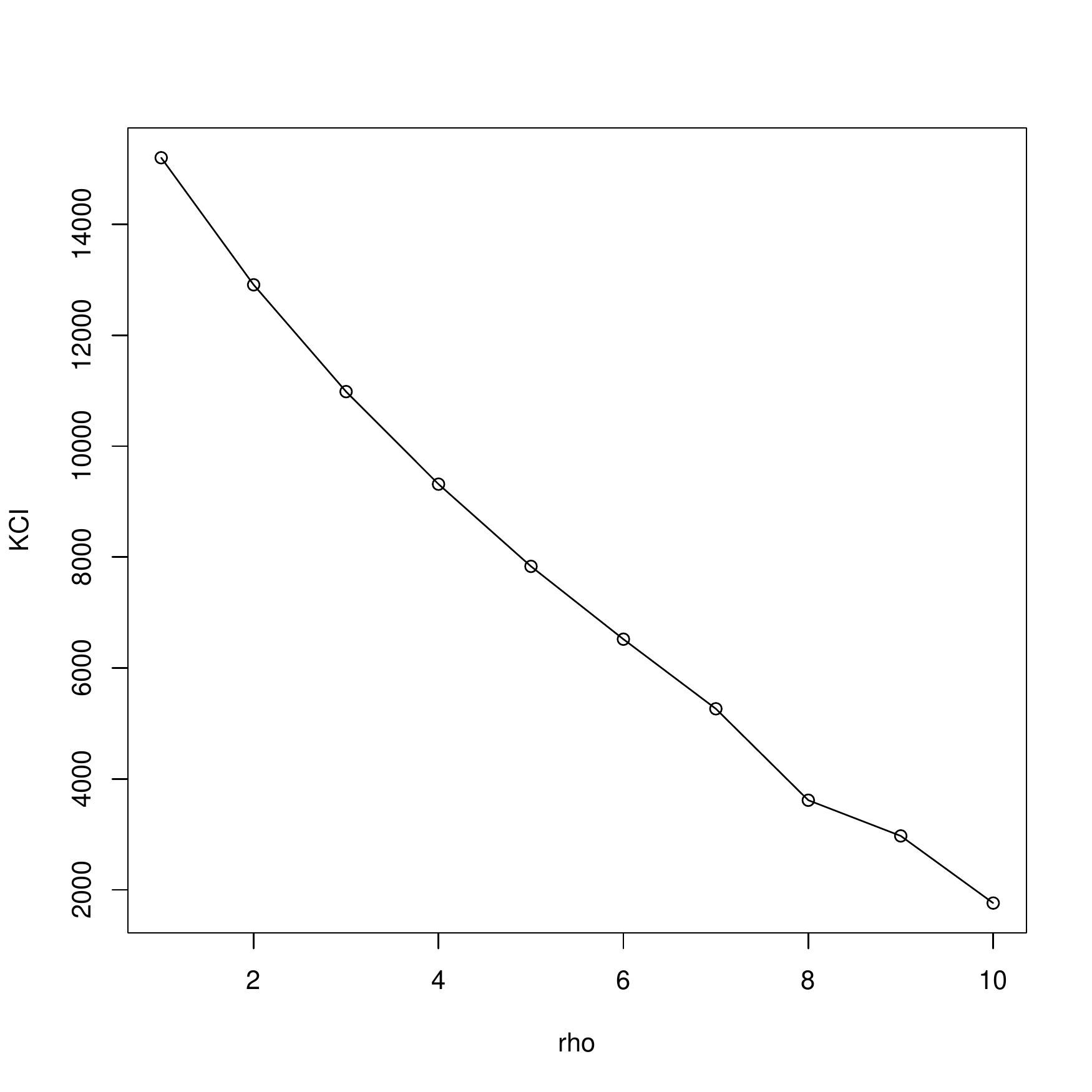}}
	\subfigure[RCoT]{\includegraphics[width=0.245\linewidth]{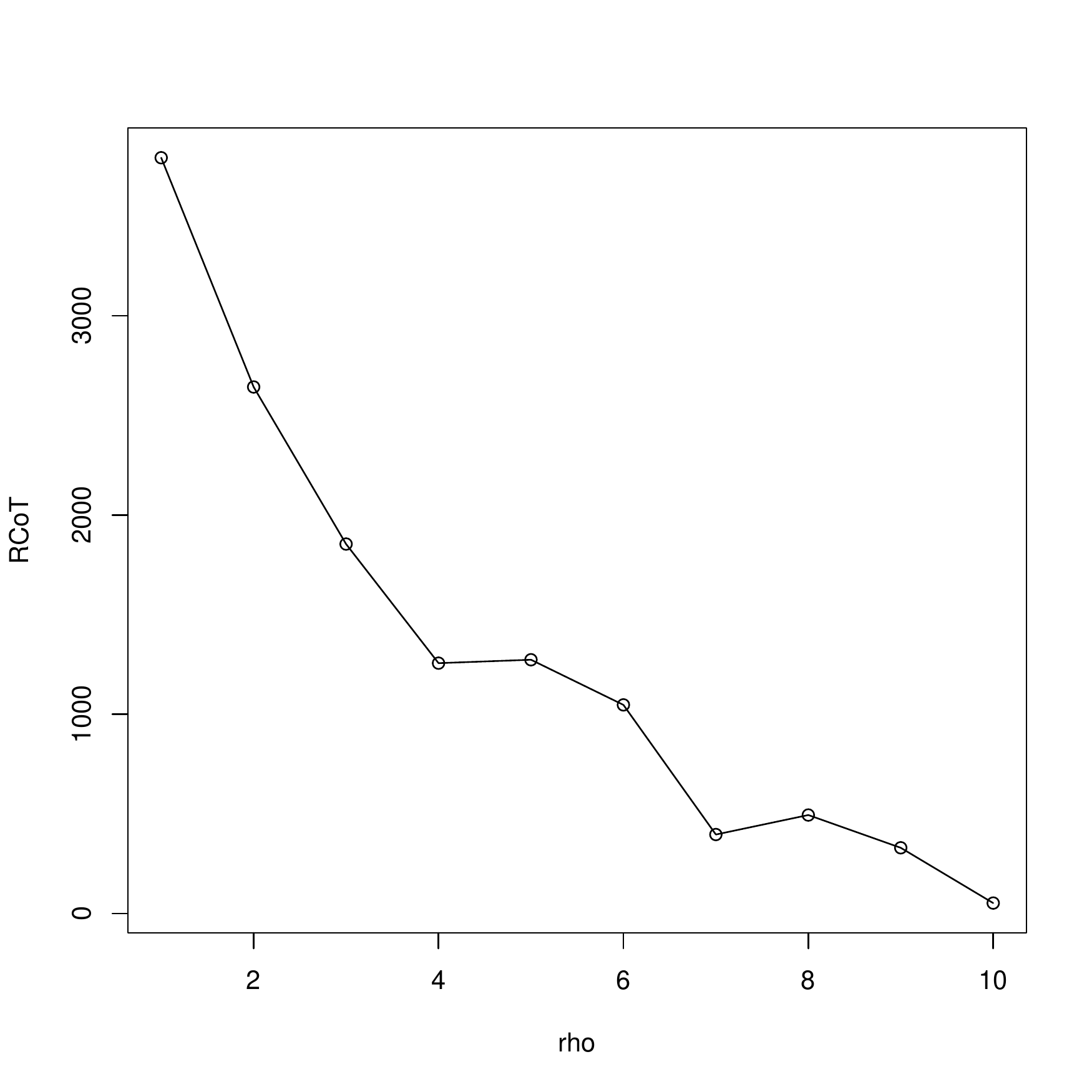}}
	\subfigure[CDC]{\includegraphics[width=0.245\linewidth]{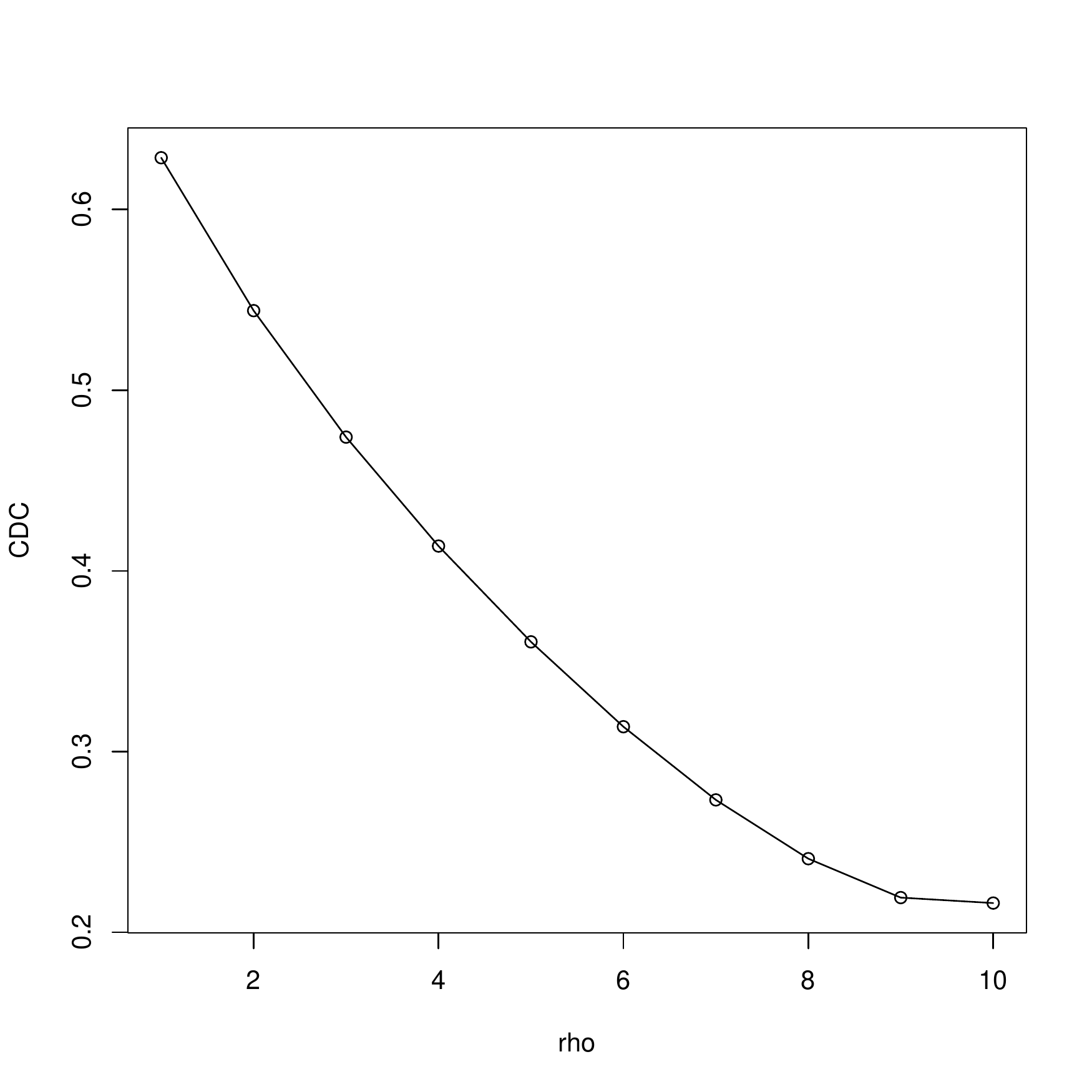}}
	\subfigure[GCM]{\includegraphics[width=0.245\linewidth]{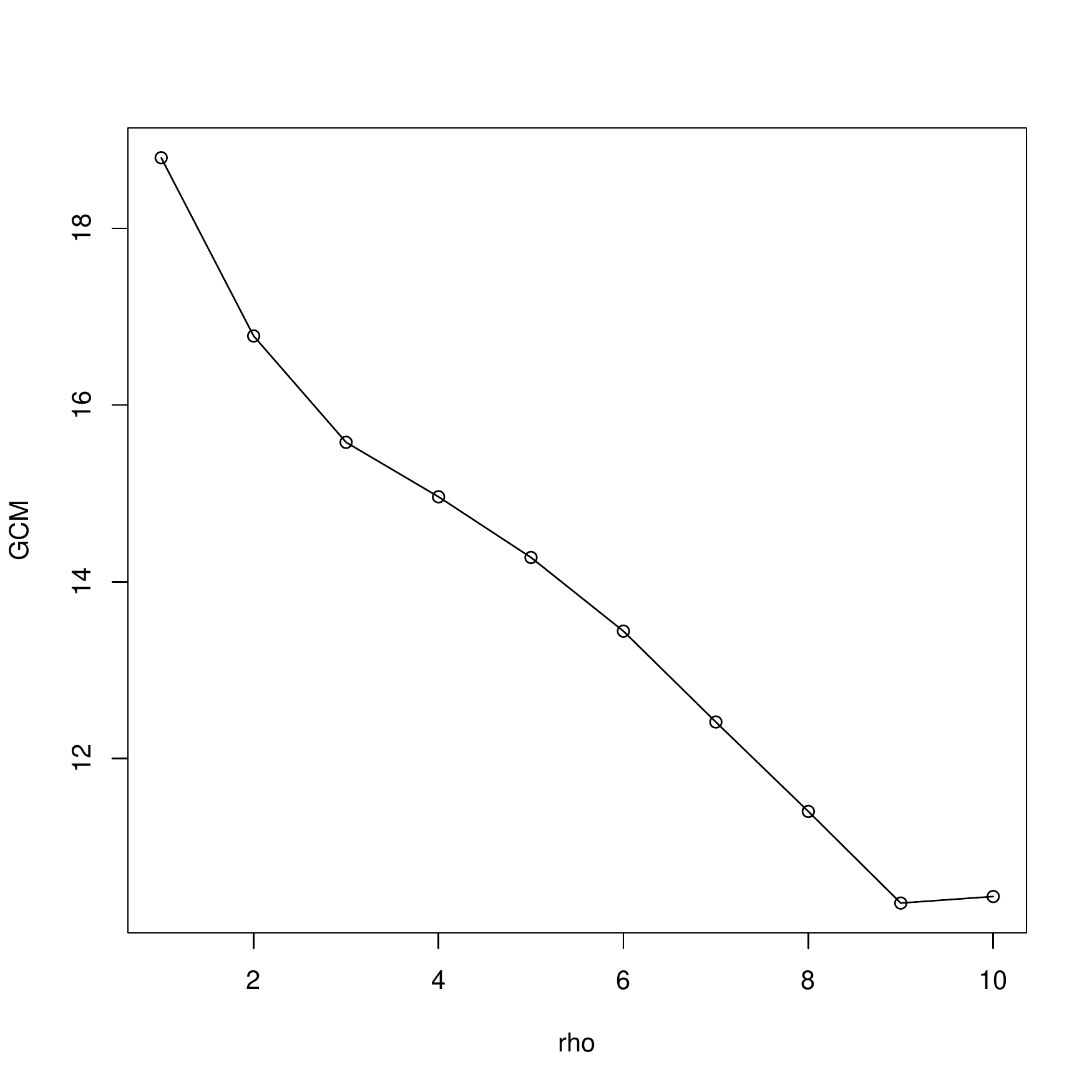}}
	\subfigure[wGCM]{\includegraphics[width=0.245\linewidth]{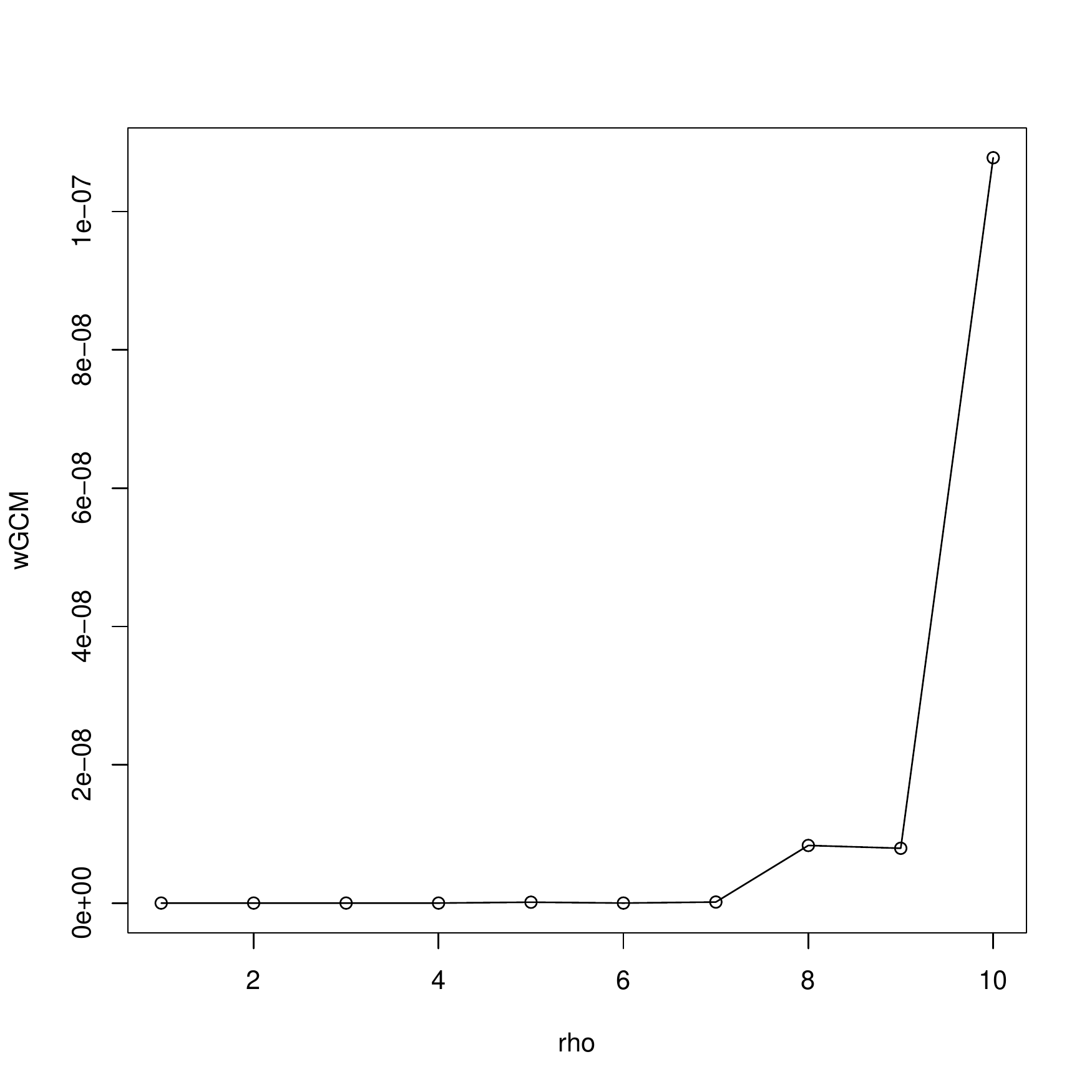}}
	\subfigure[CODEC]{\includegraphics[width=0.245\linewidth]{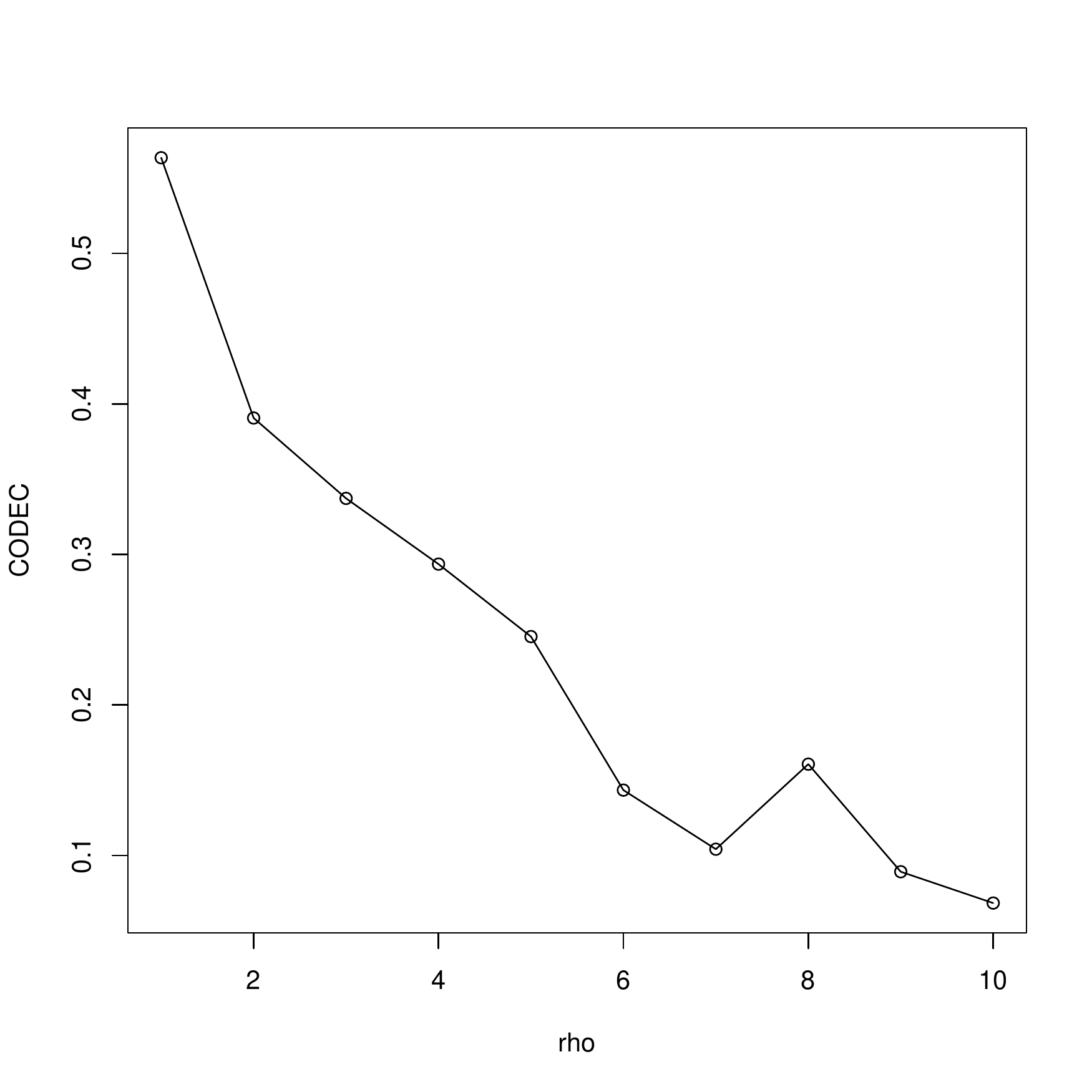}}
	\subfigure[KPC]{\includegraphics[width=0.245\linewidth]{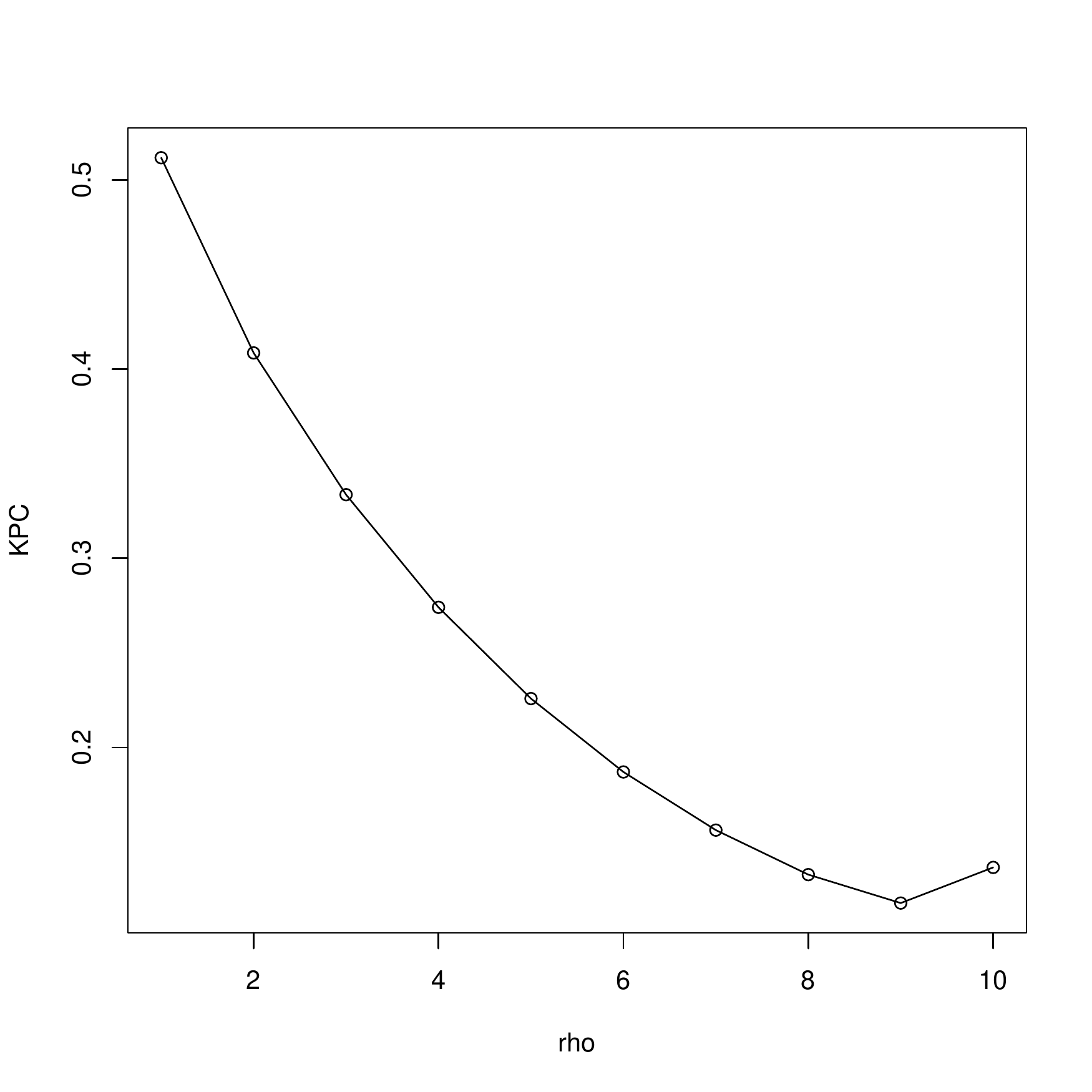}}
	\subfigure[PCor]{\includegraphics[width=0.245\linewidth]{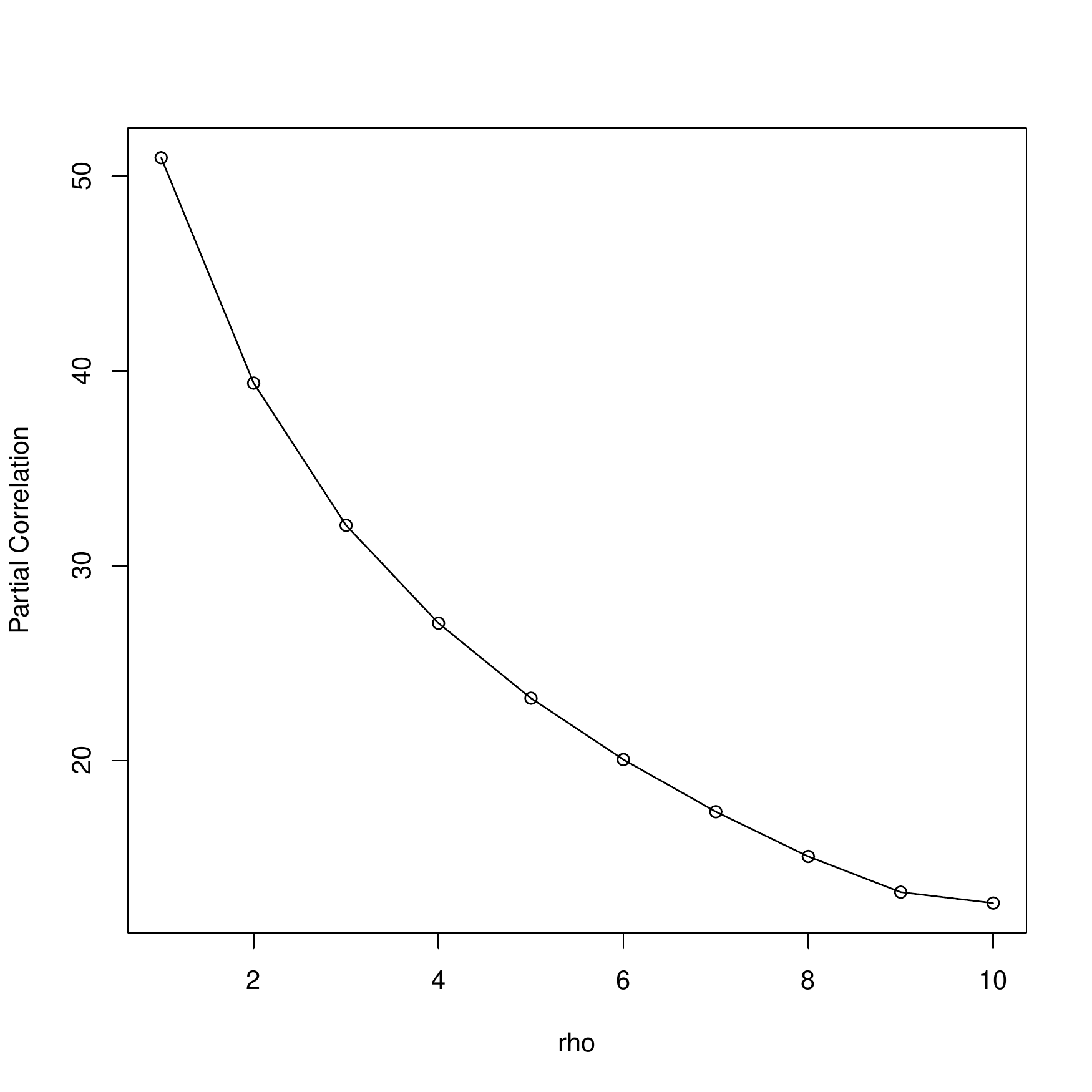}}
	\subfigure[CMDM]{\includegraphics[width=0.245\linewidth]{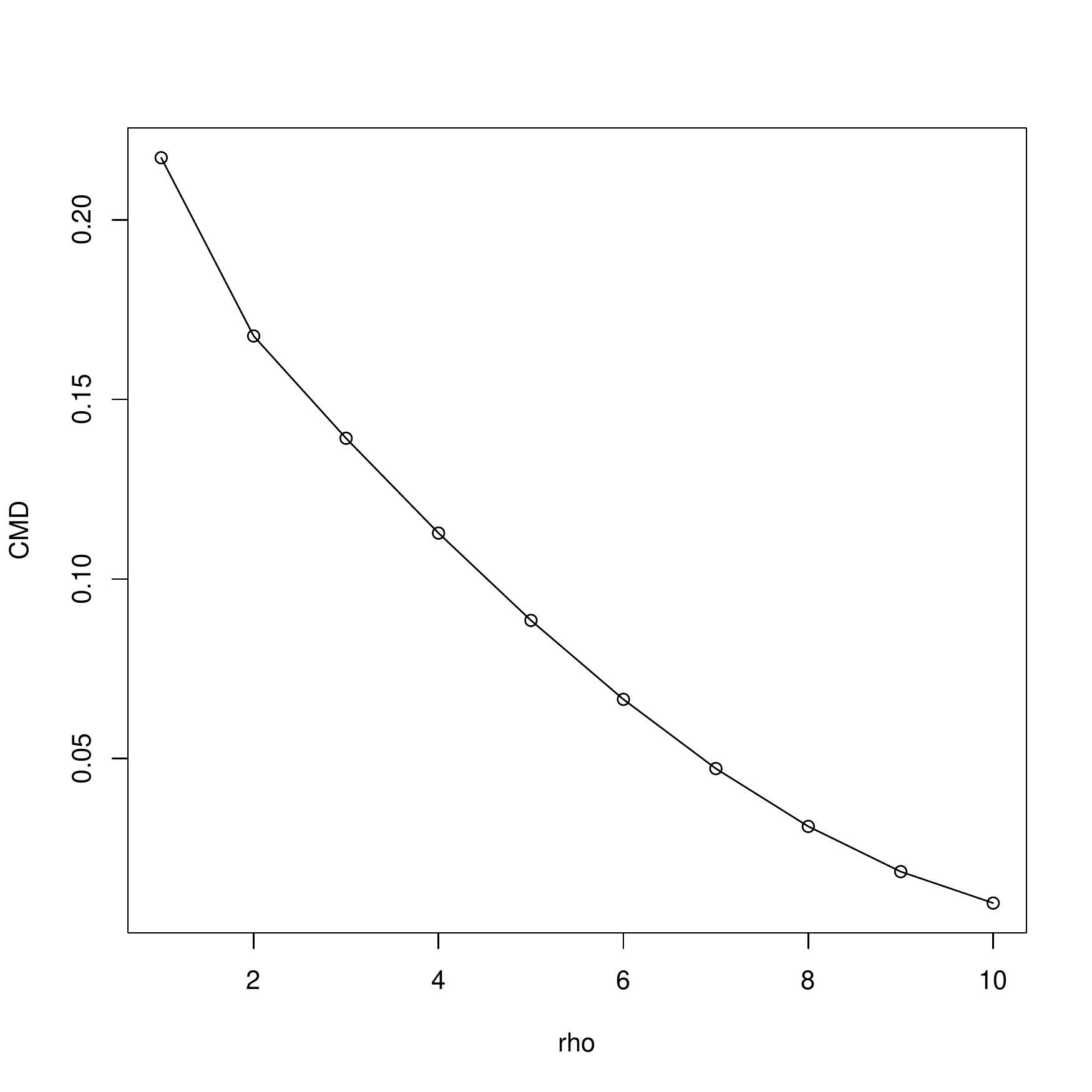}}
	\subfigure[PCop]{\includegraphics[width=0.245\linewidth]{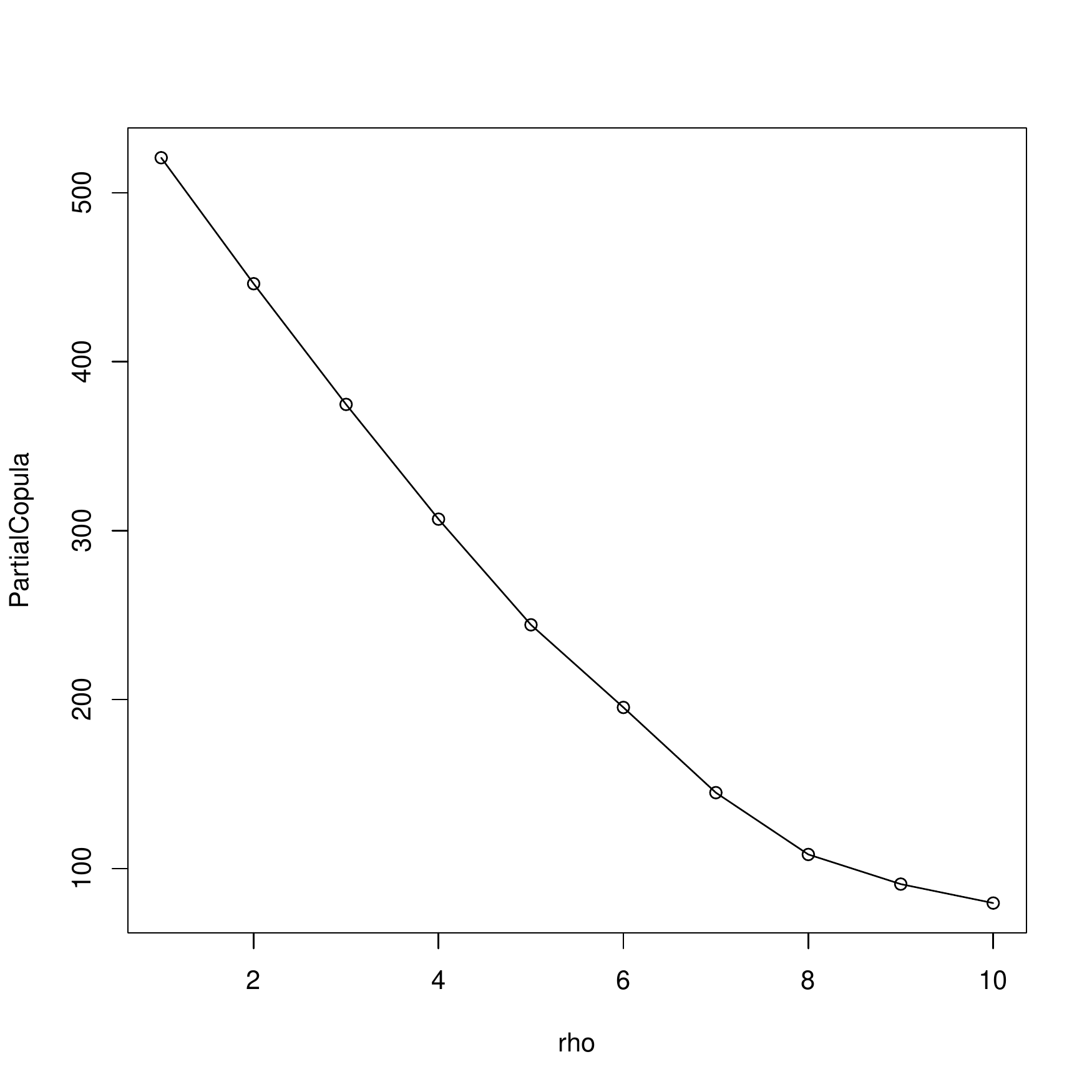}}
	\subfigure[CMI1]{\includegraphics[width=0.245\linewidth]{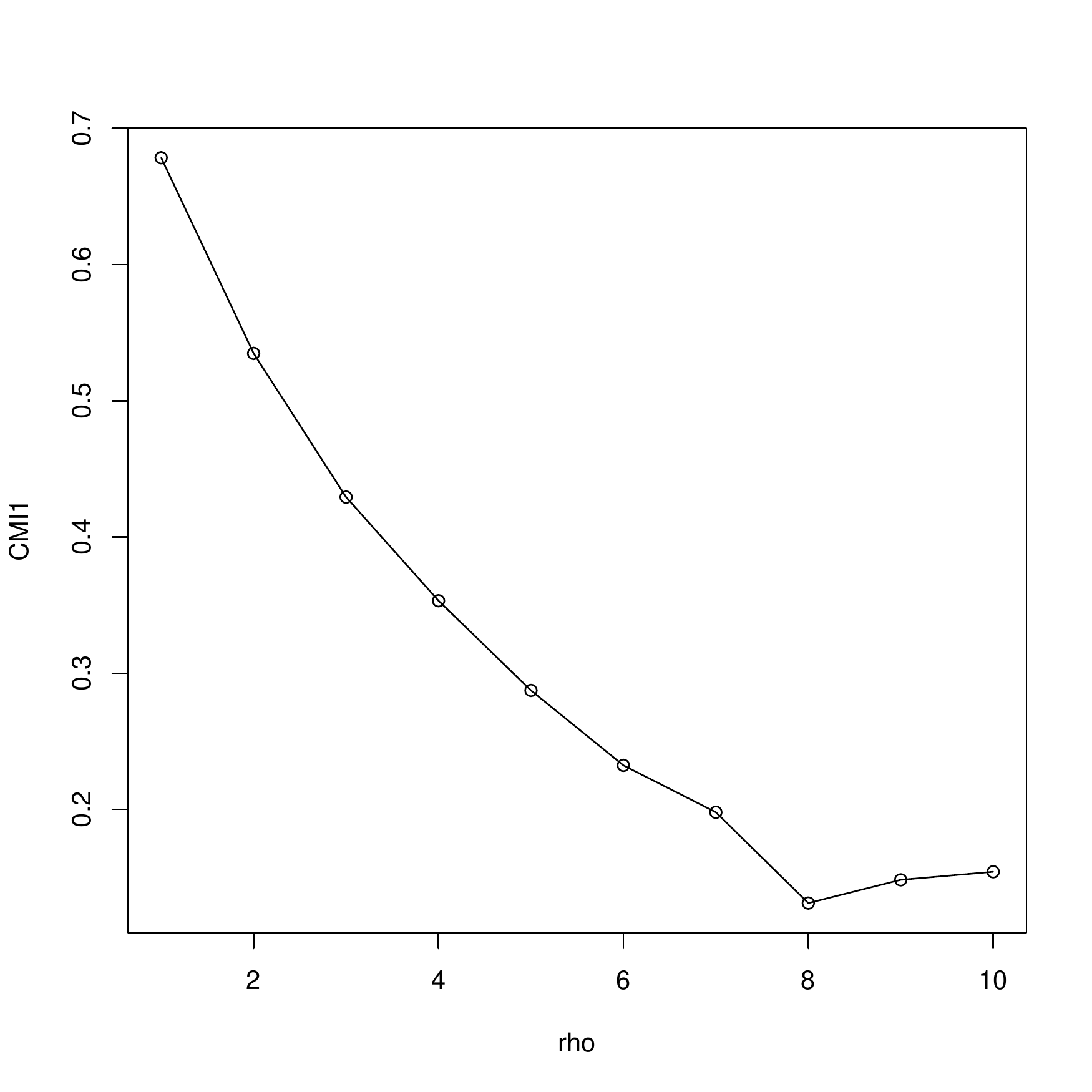}}
	\subfigure[CMI2]{\includegraphics[width=0.245\linewidth]{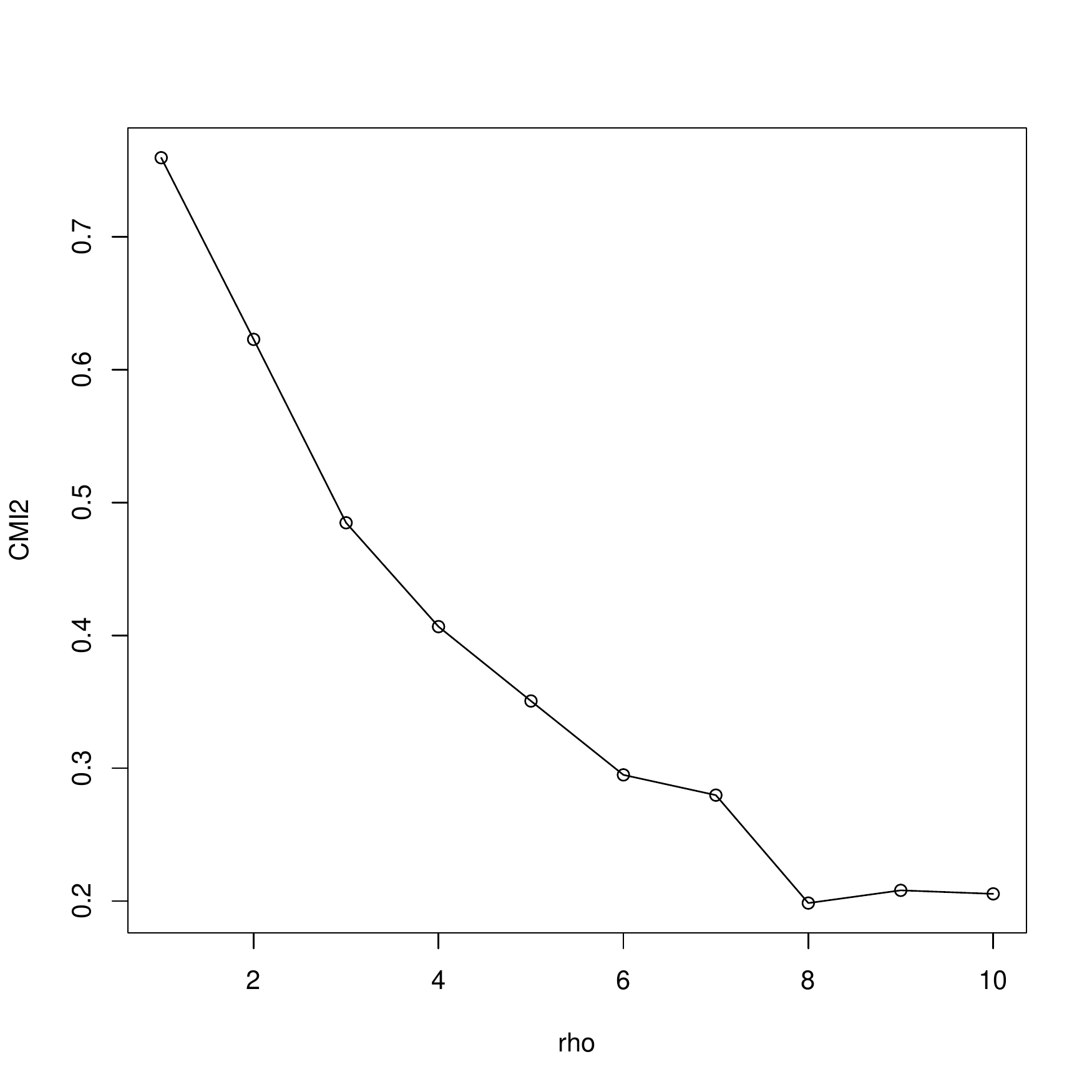}}
	\subfigure[FCIT]{\includegraphics[width=0.245\linewidth]{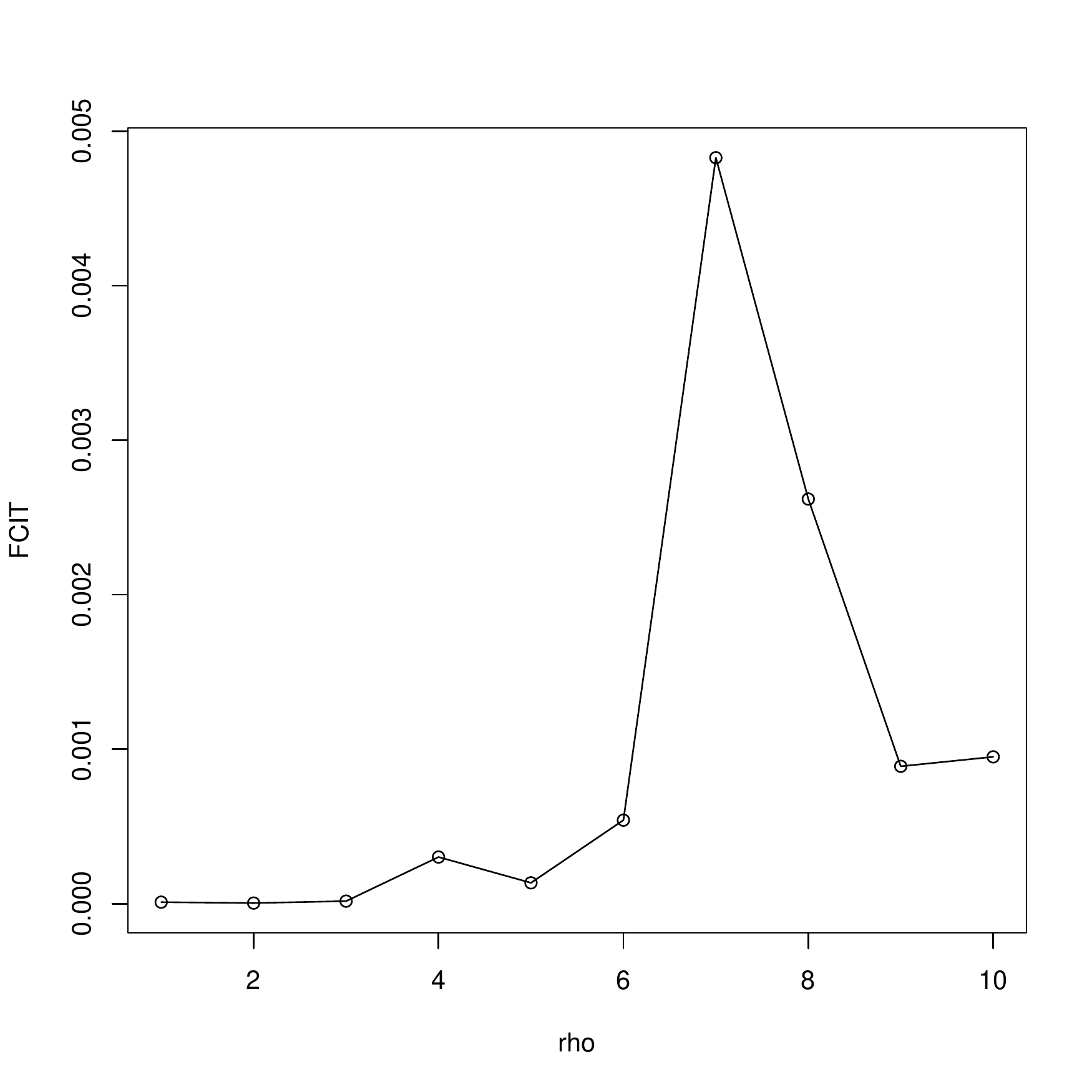}}
	\subfigure[CCIT]{\includegraphics[width=0.245\linewidth]{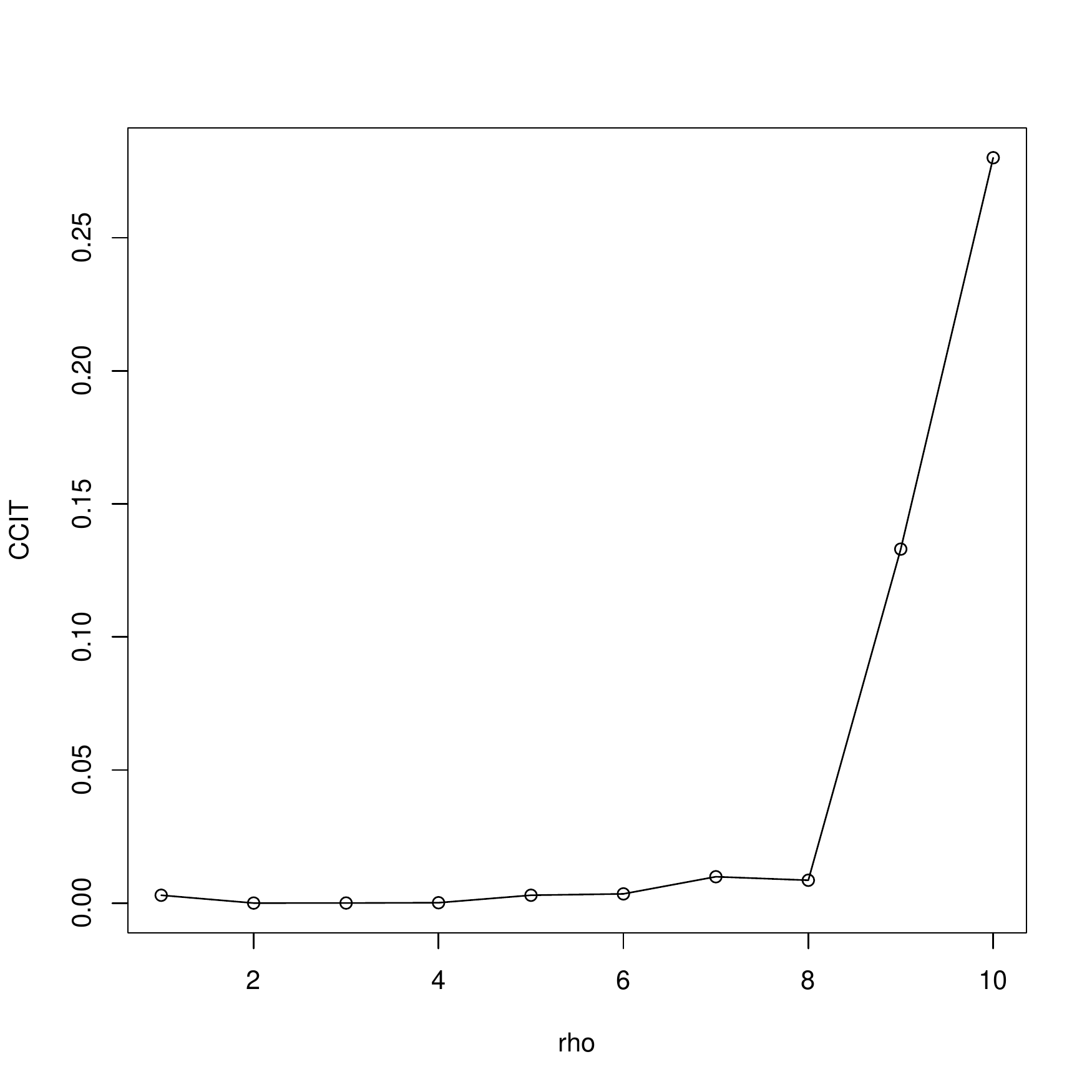}}
	\subfigure[PCIT]{\includegraphics[width=0.245\linewidth]{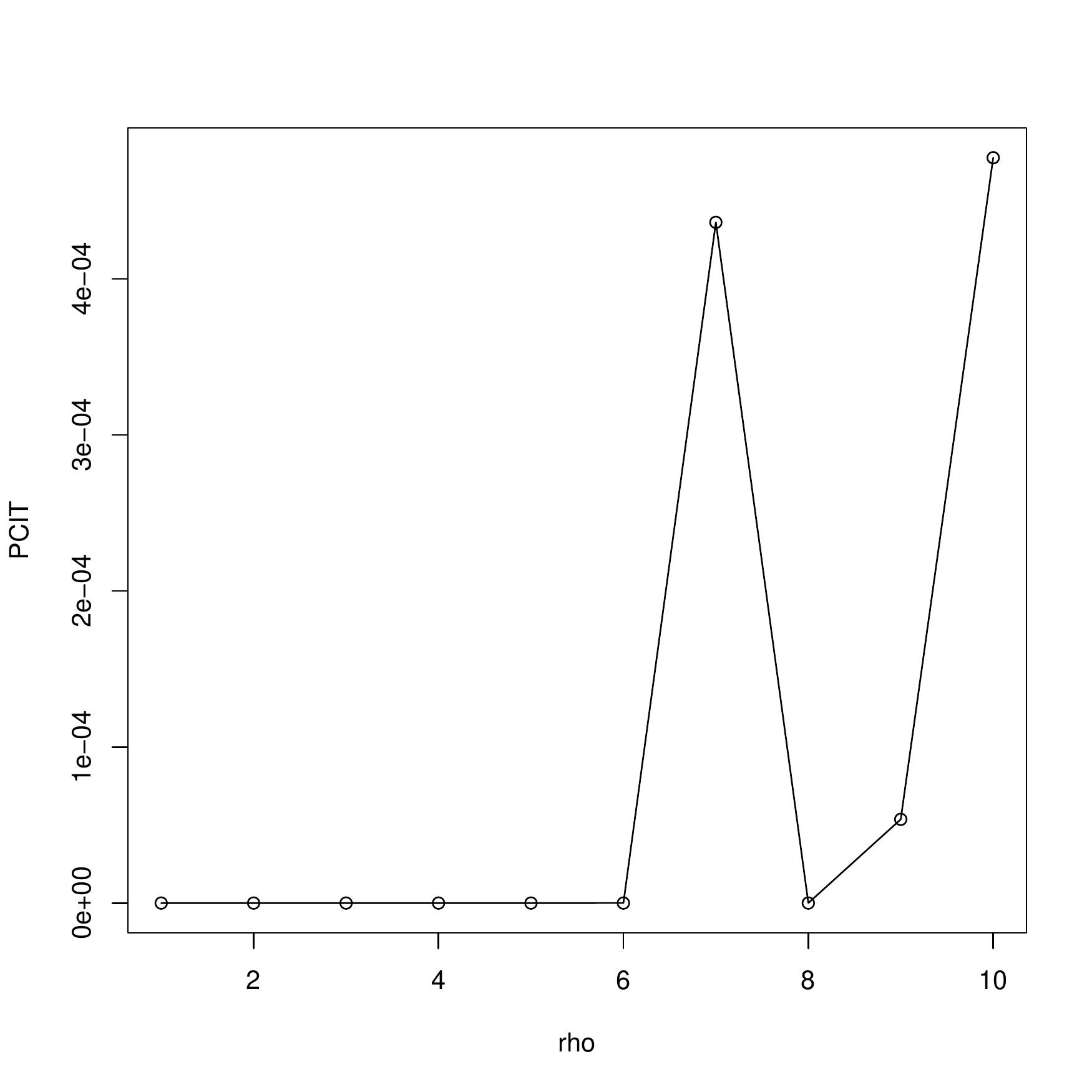}}
	\caption{Estimation of the CI measures from the simulated data of the trivariate normal distribution.}
	\label{fig:trinormci}
\end{figure}

\begin{figure}
	\centering
	\includegraphics[width=0.9\linewidth]{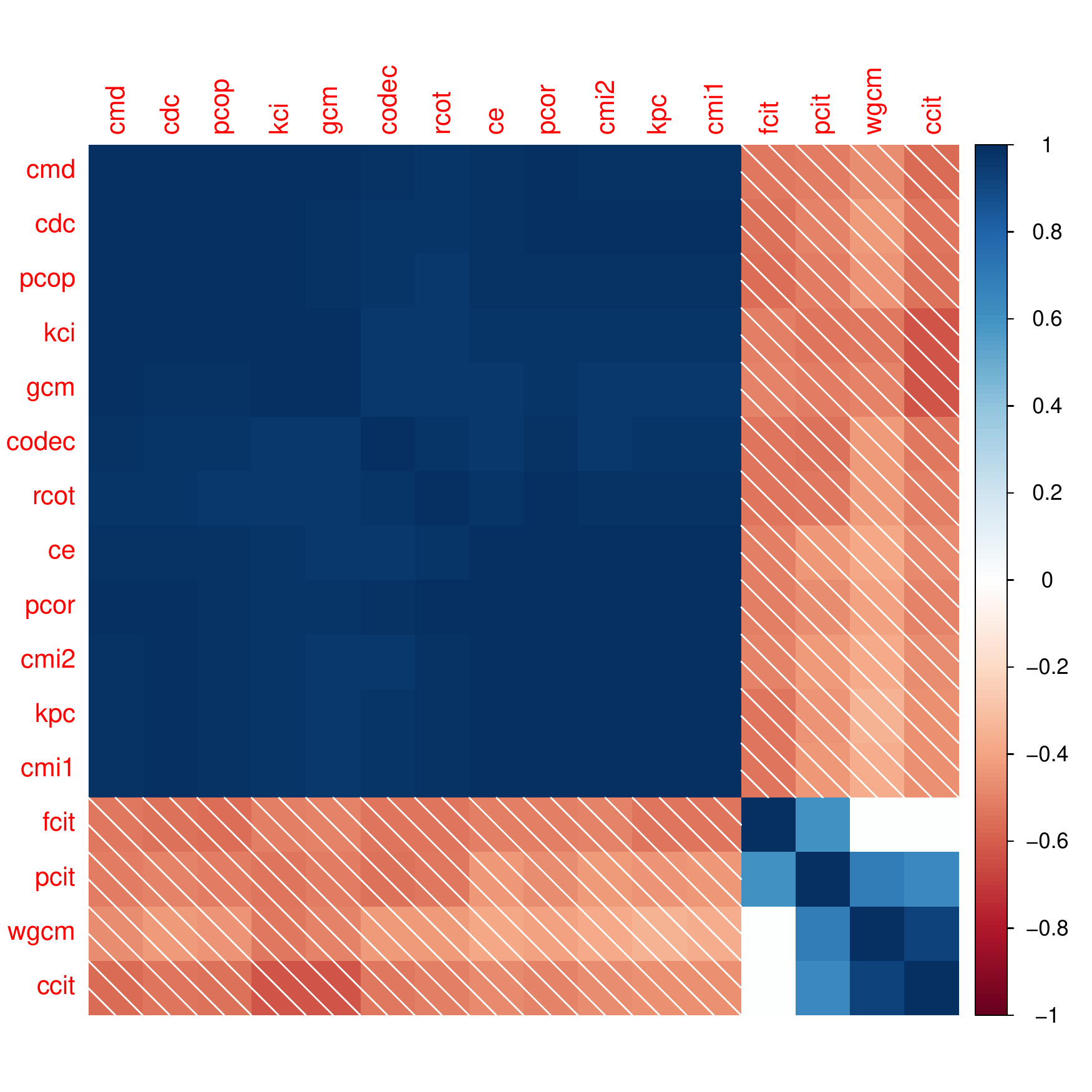}
	\caption{Correlation matrix of the CI measures estimated from the simulated data of the trivariate normal distribution.}
	\label{fig:trinormcicm}
\end{figure}

\begin{figure}
	\subfigure[CE]{\includegraphics[width=0.245\linewidth]{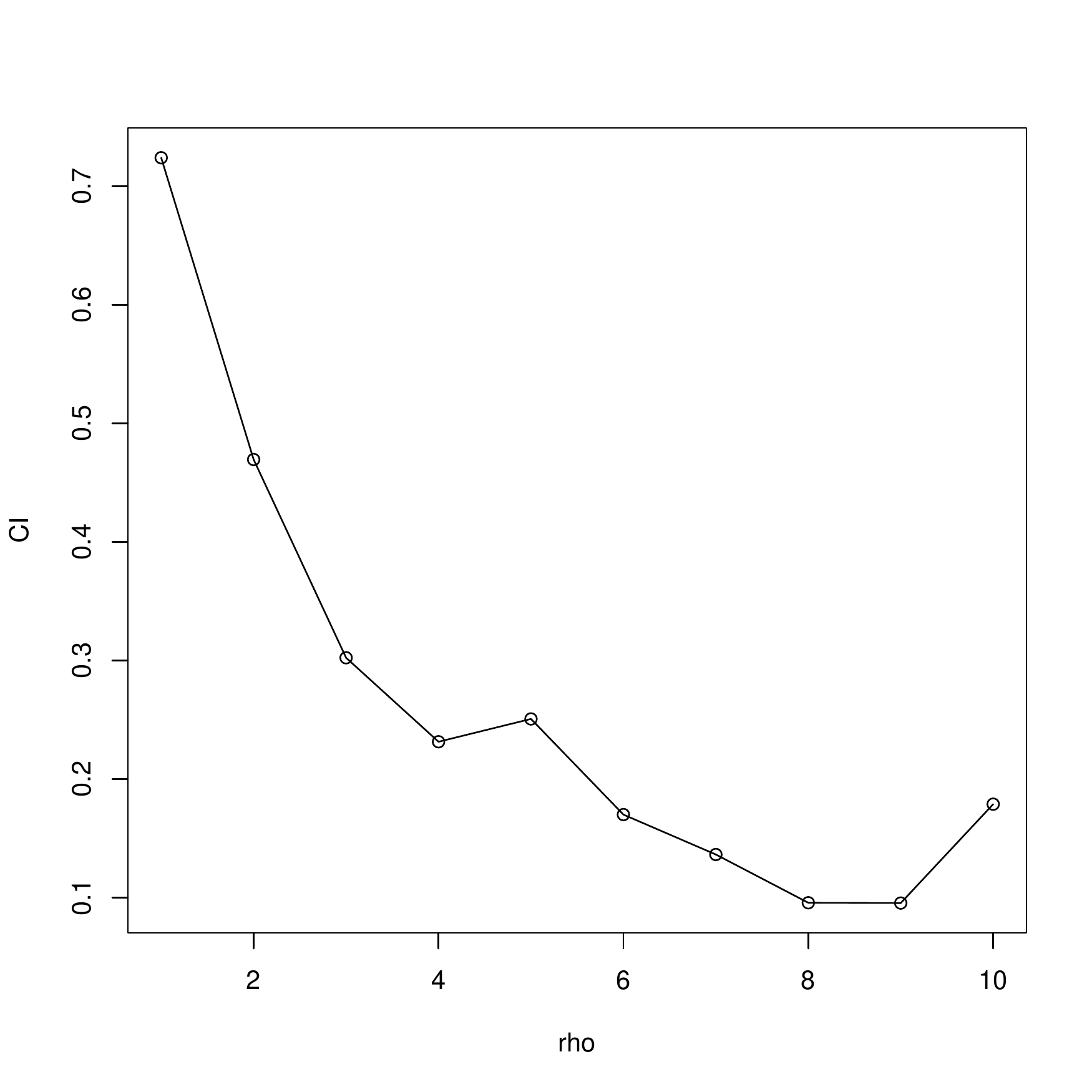}}
	\subfigure[KCI]{\includegraphics[width=0.245\linewidth]{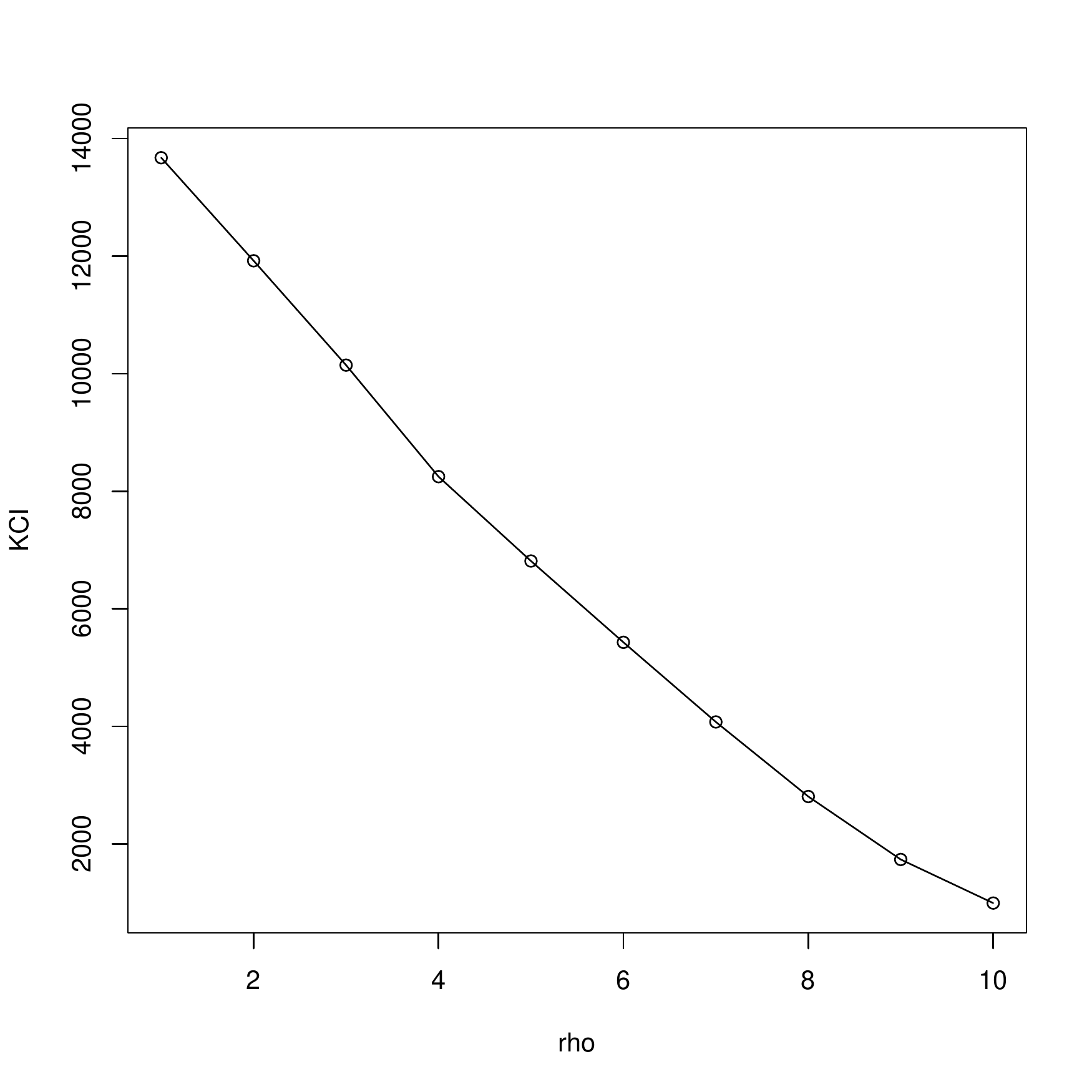}}
	\subfigure[RCoT]{\includegraphics[width=0.245\linewidth]{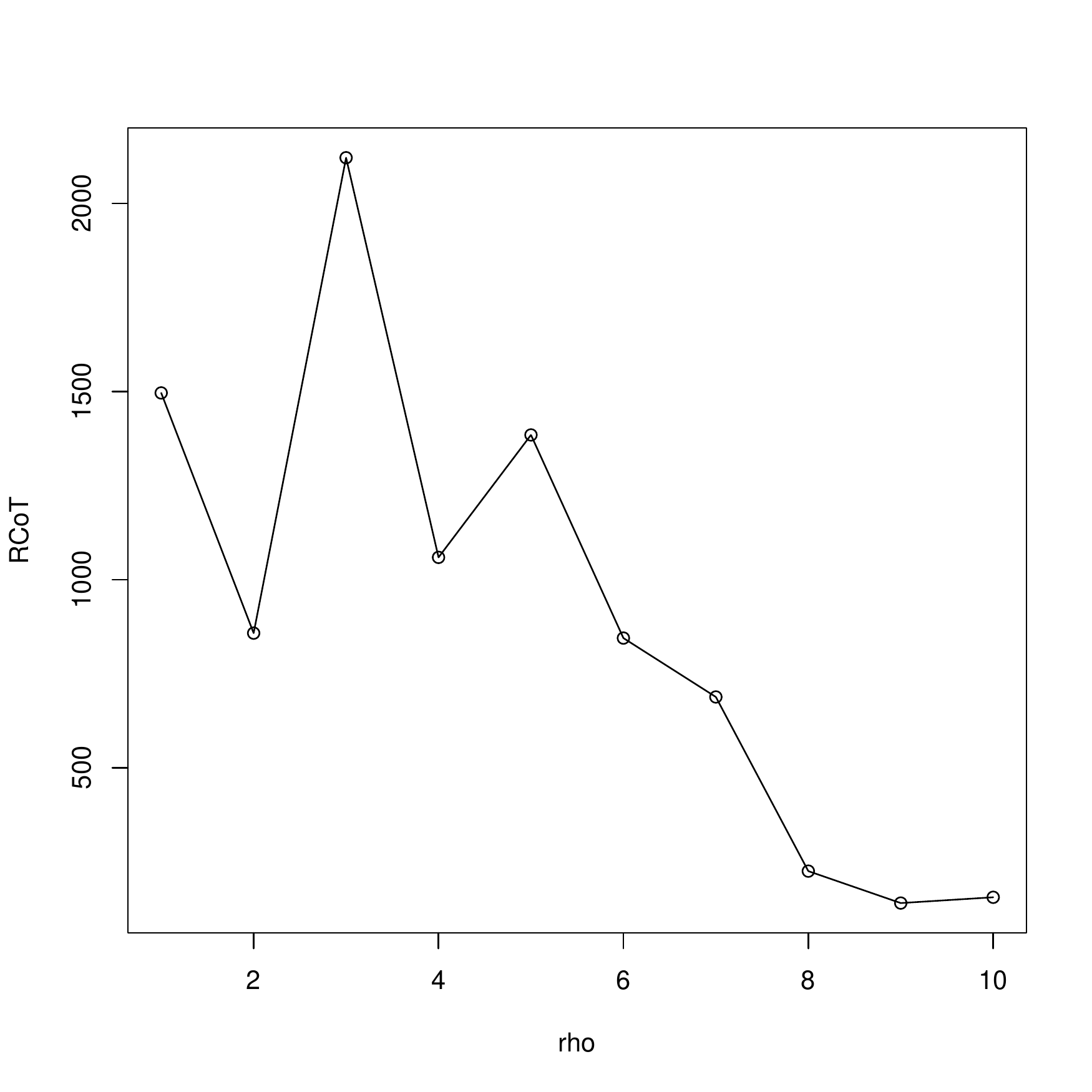}}
	\subfigure[CDC]{\includegraphics[width=0.245\linewidth]{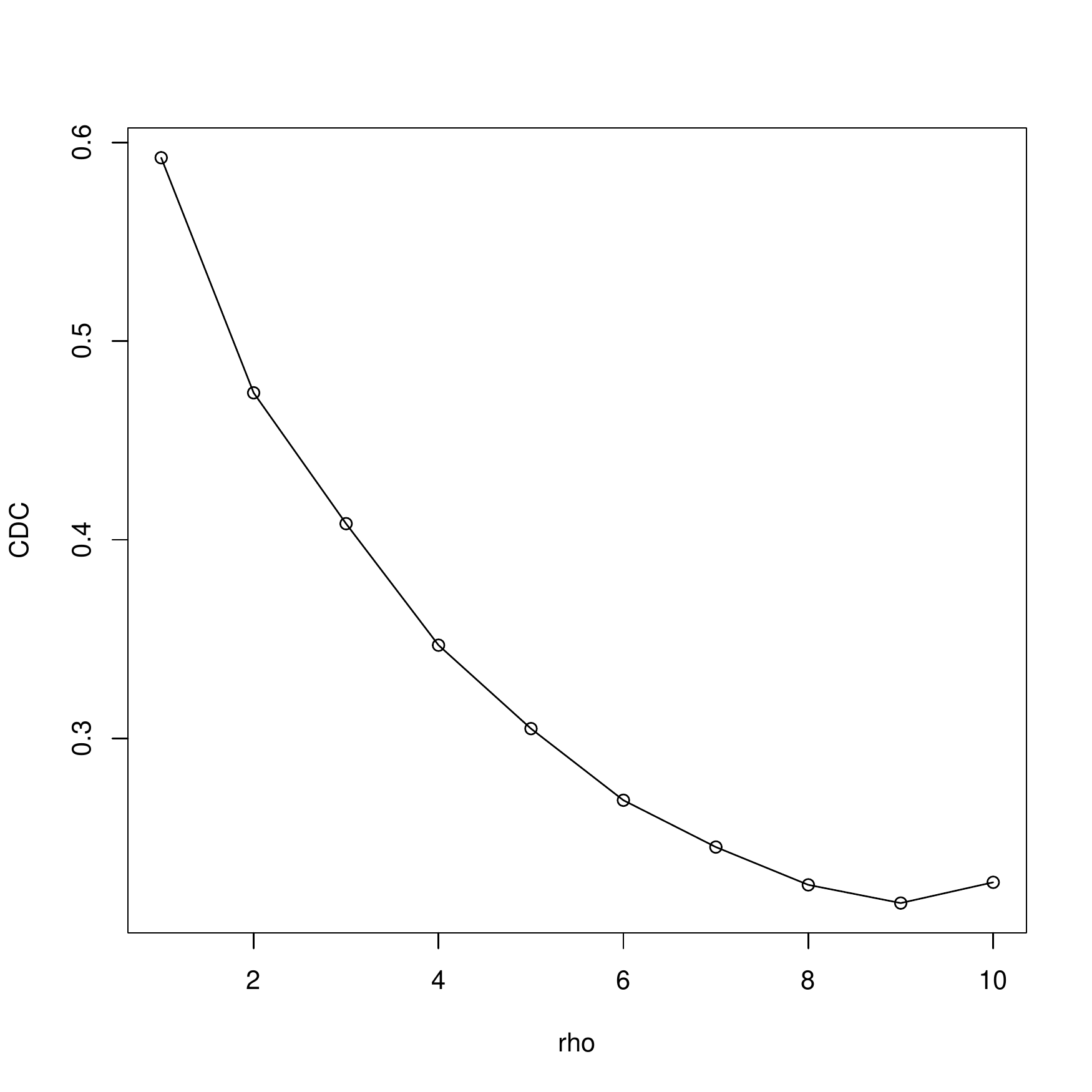}}
	\subfigure[GCM]{\includegraphics[width=0.245\linewidth]{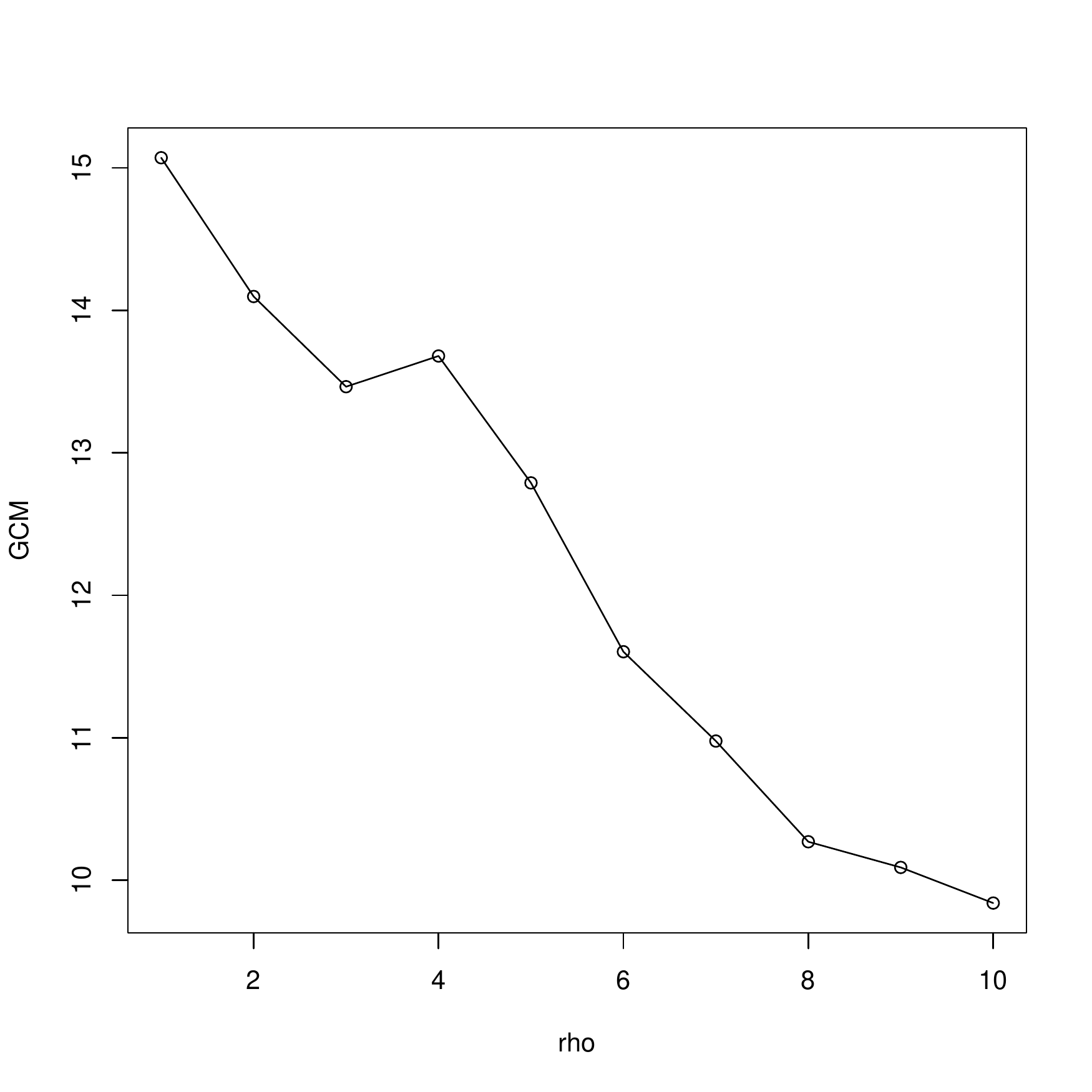}}
	\subfigure[wGCM]{\includegraphics[width=0.245\linewidth]{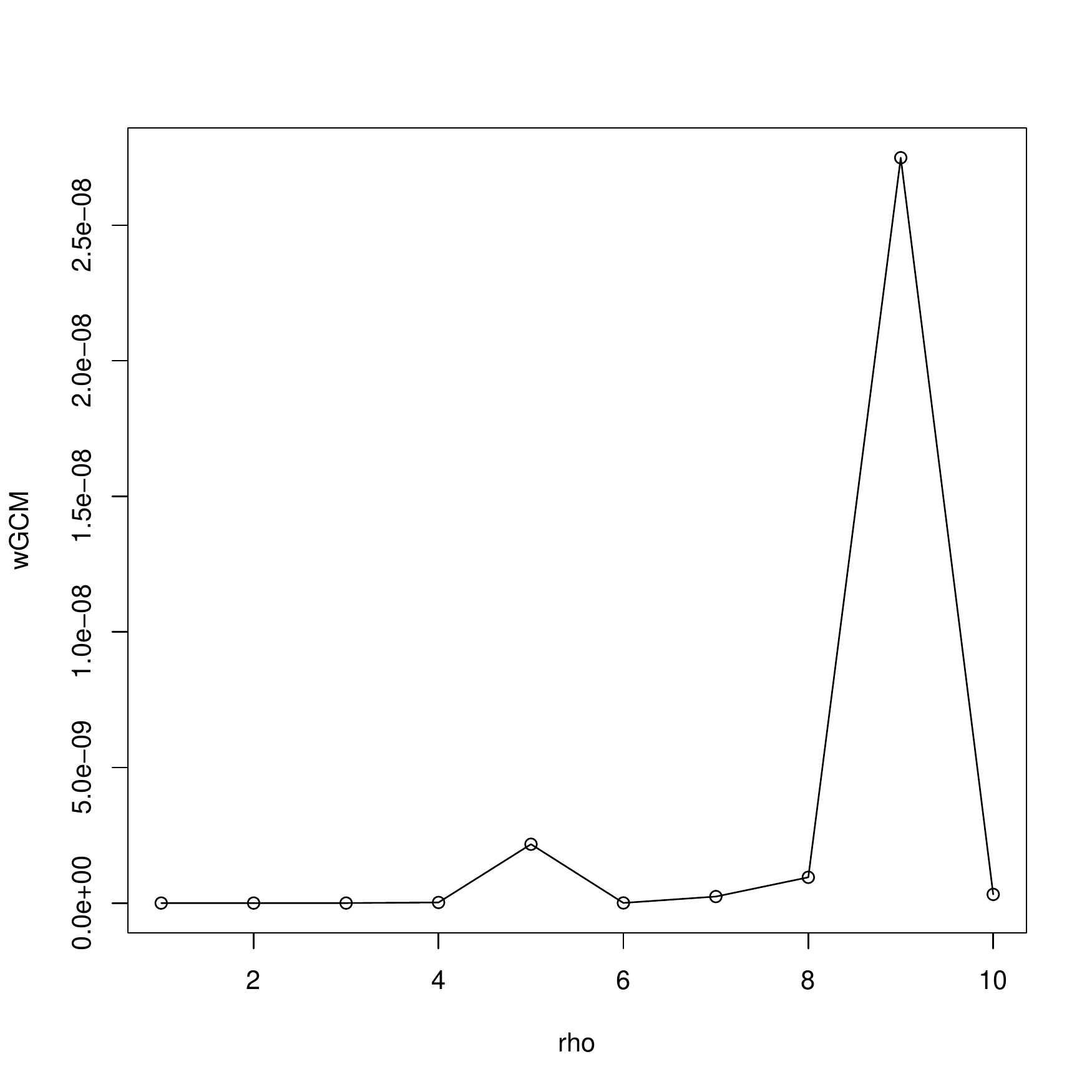}}
	\subfigure[CODEC]{\includegraphics[width=0.245\linewidth]{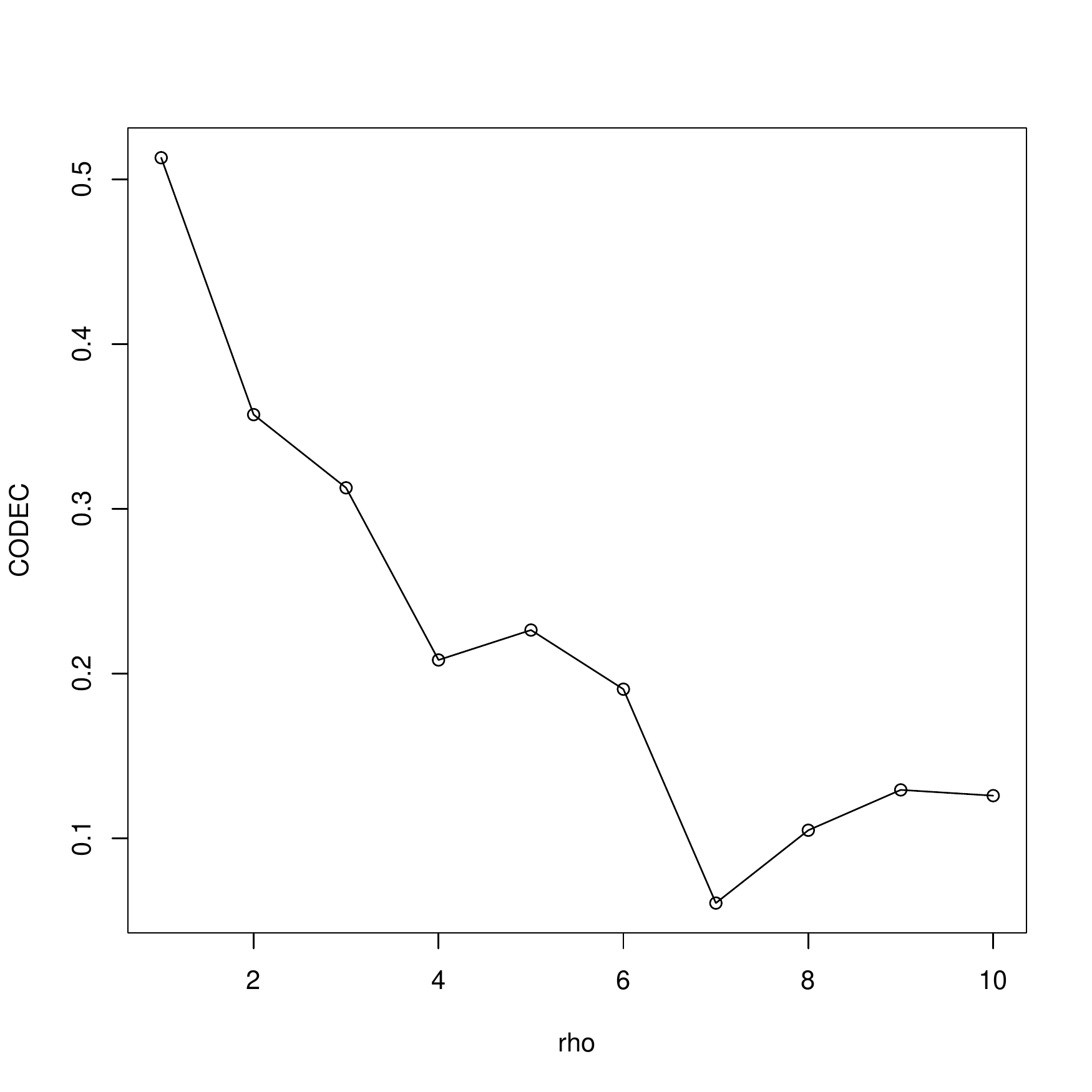}}
	\subfigure[KPC]{\includegraphics[width=0.245\linewidth]{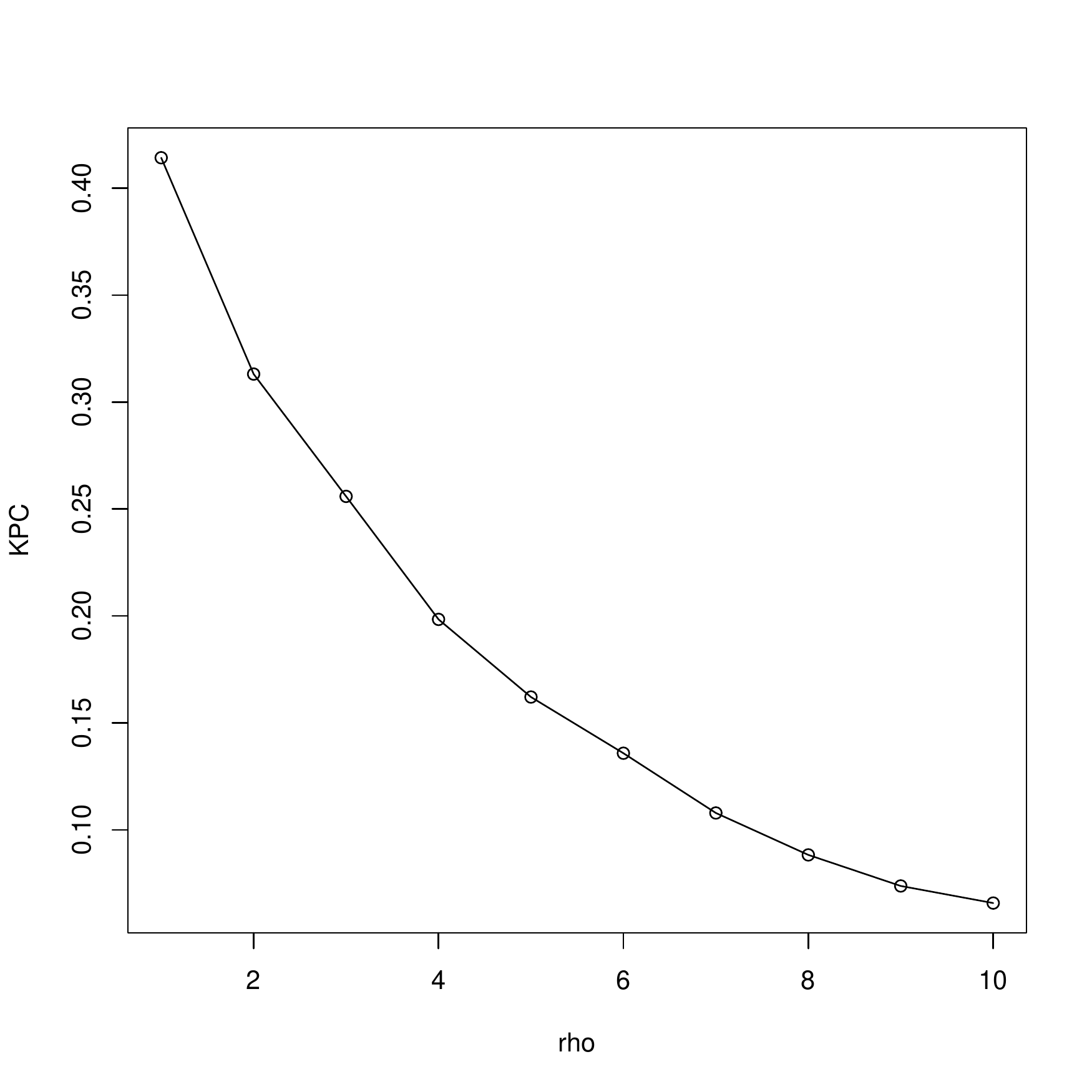}}
	\subfigure[PCor]{\includegraphics[width=0.245\linewidth]{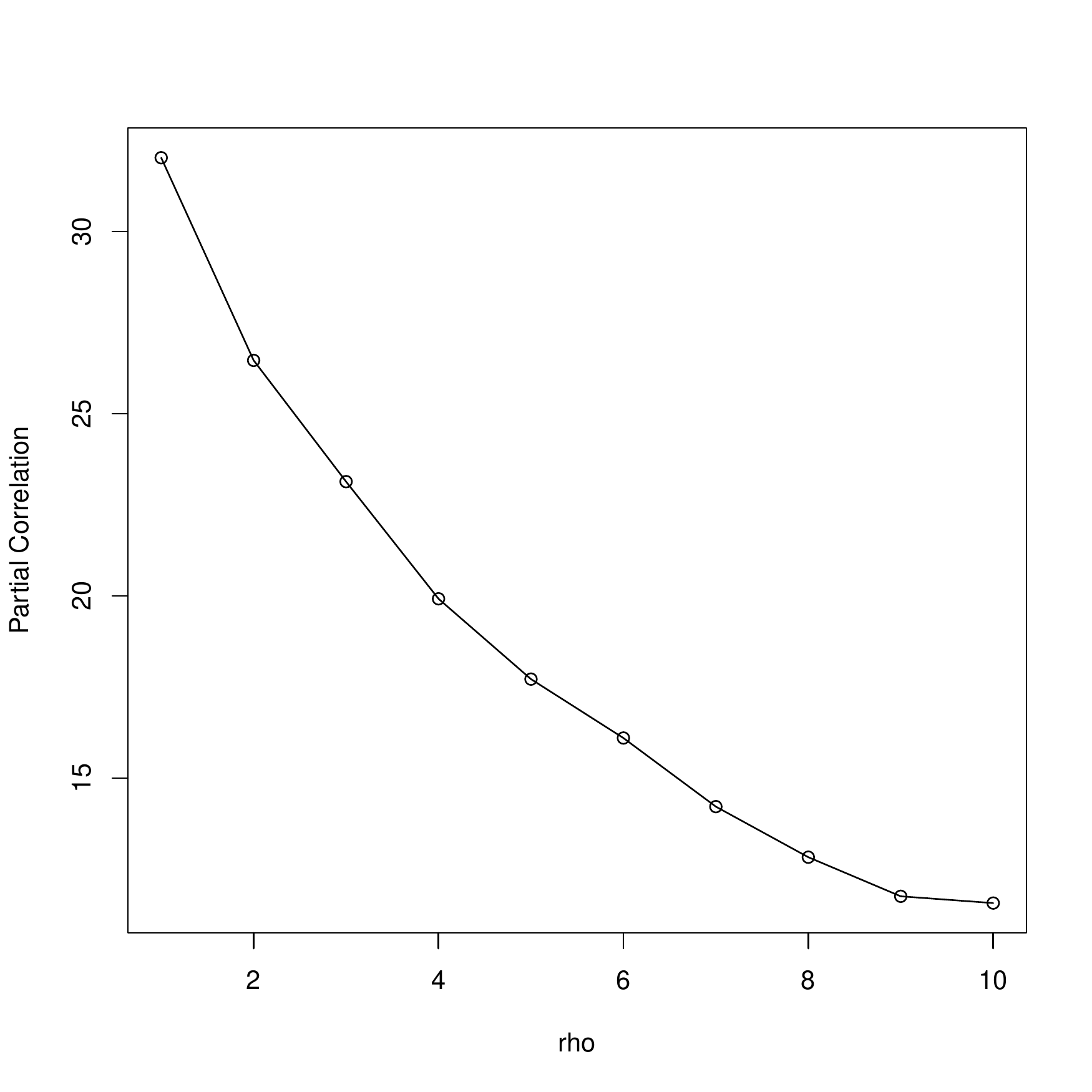}}
	\subfigure[CMDM]{\includegraphics[width=0.245\linewidth]{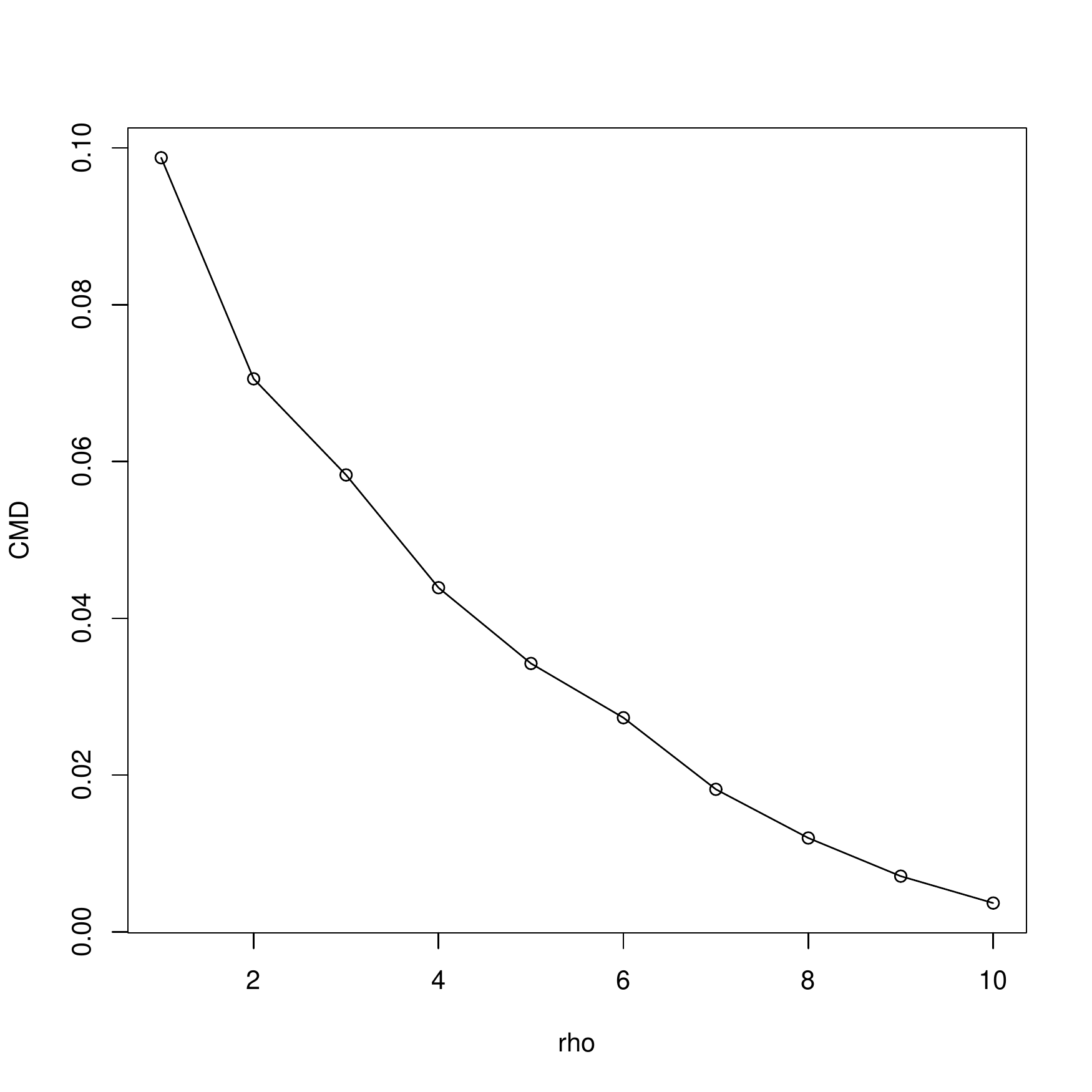}}
	\subfigure[PCop]{\includegraphics[width=0.245\linewidth]{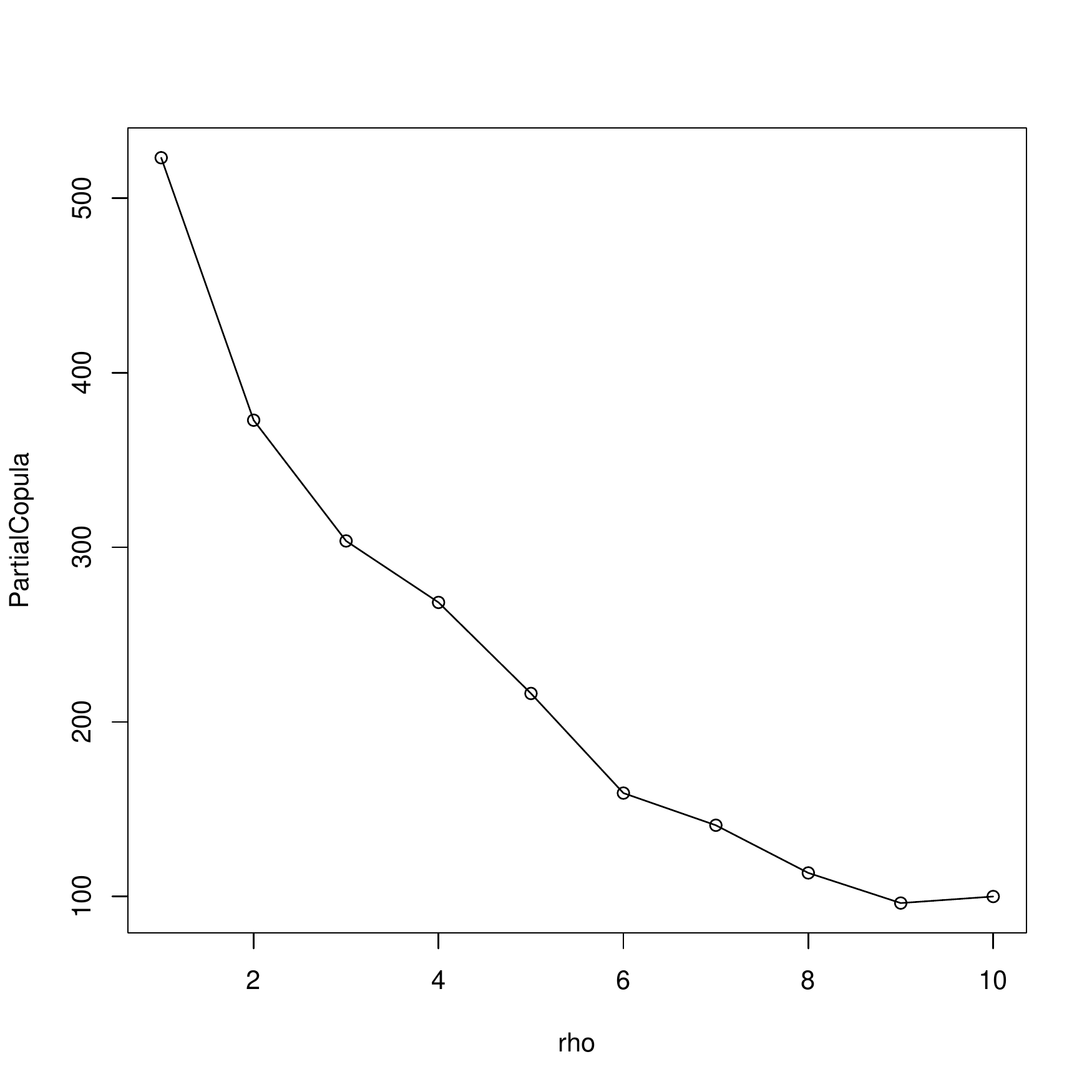}}
	\subfigure[CMI1]{\includegraphics[width=0.245\linewidth]{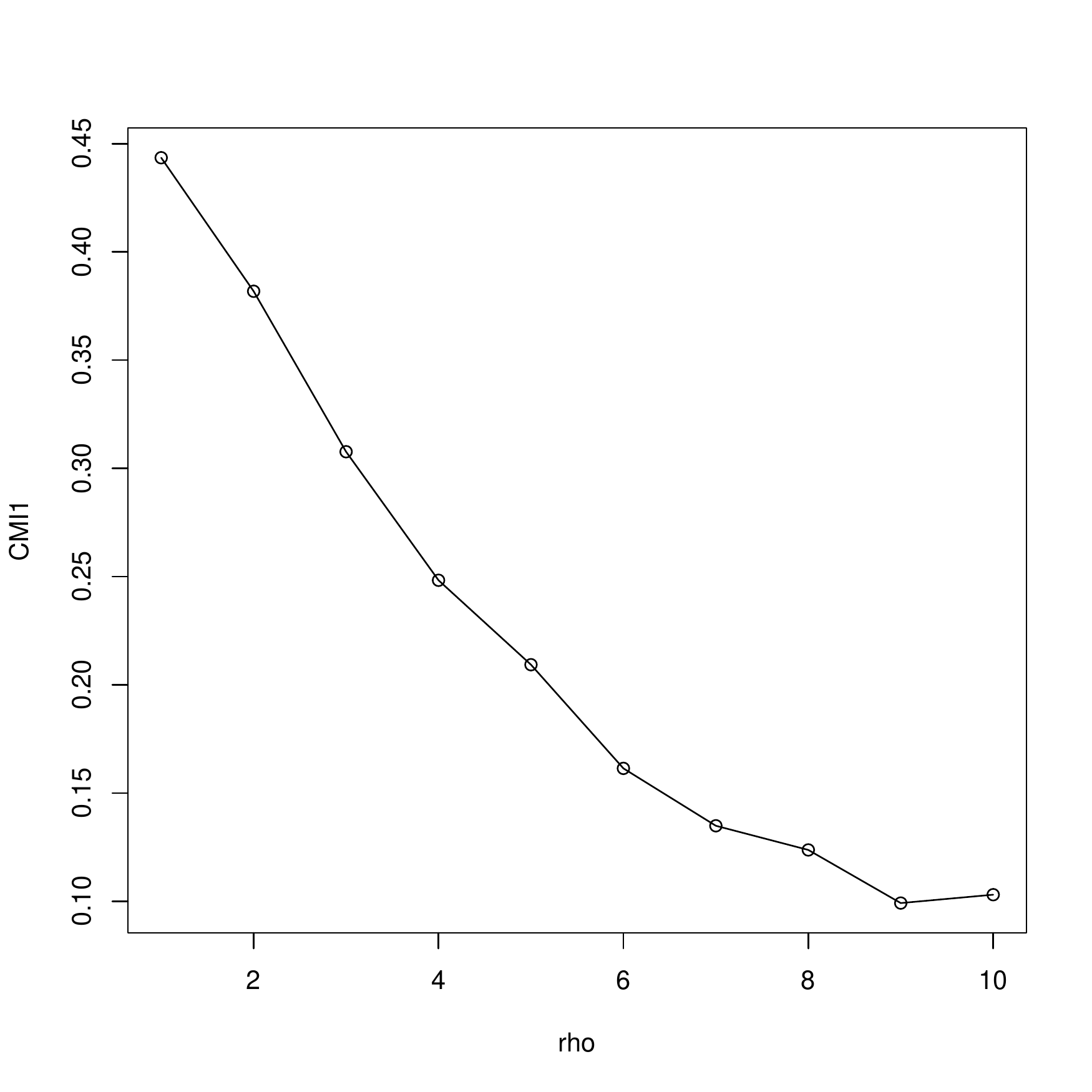}}
	\subfigure[CMI2]{\includegraphics[width=0.245\linewidth]{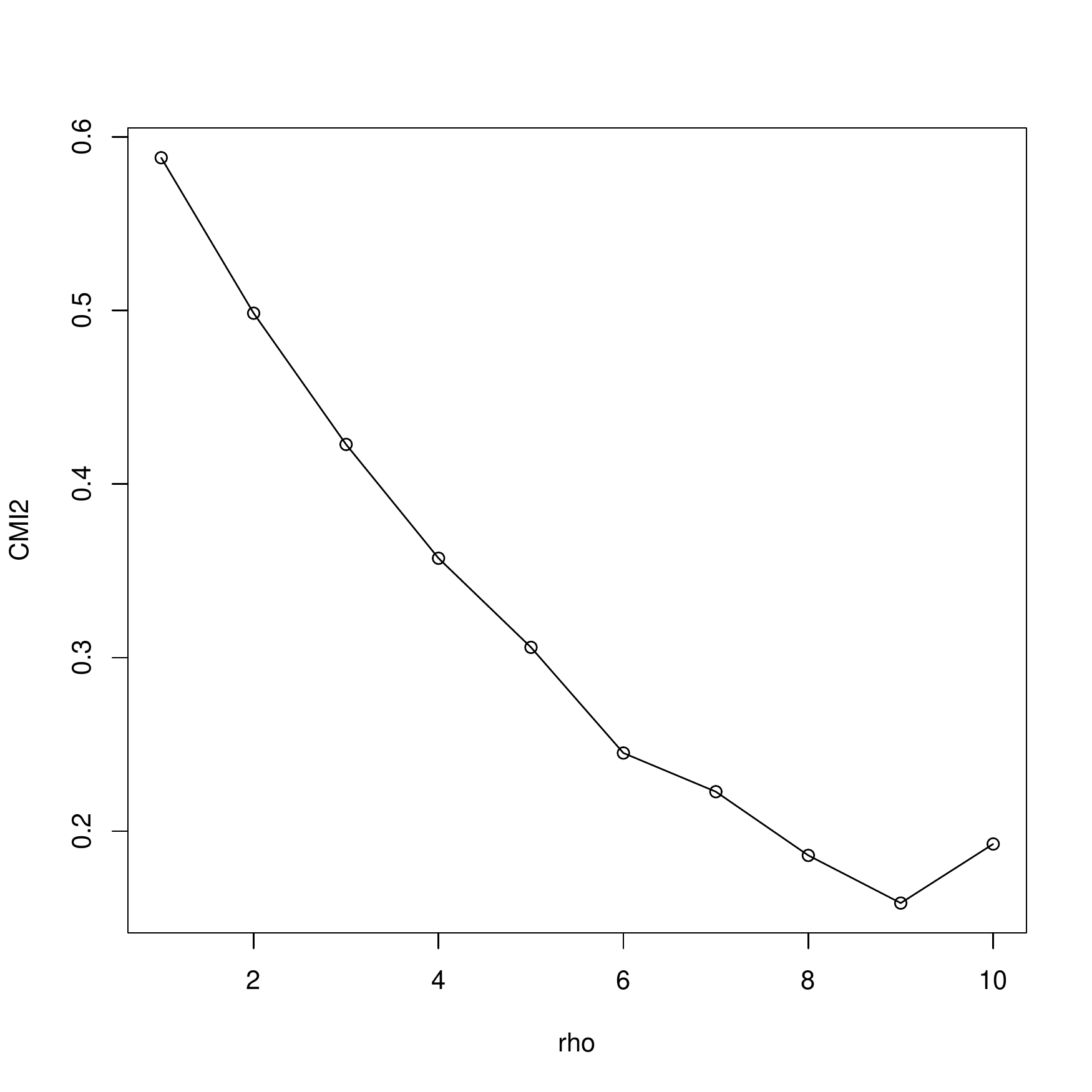}}
	\subfigure[FCIT]{\includegraphics[width=0.245\linewidth]{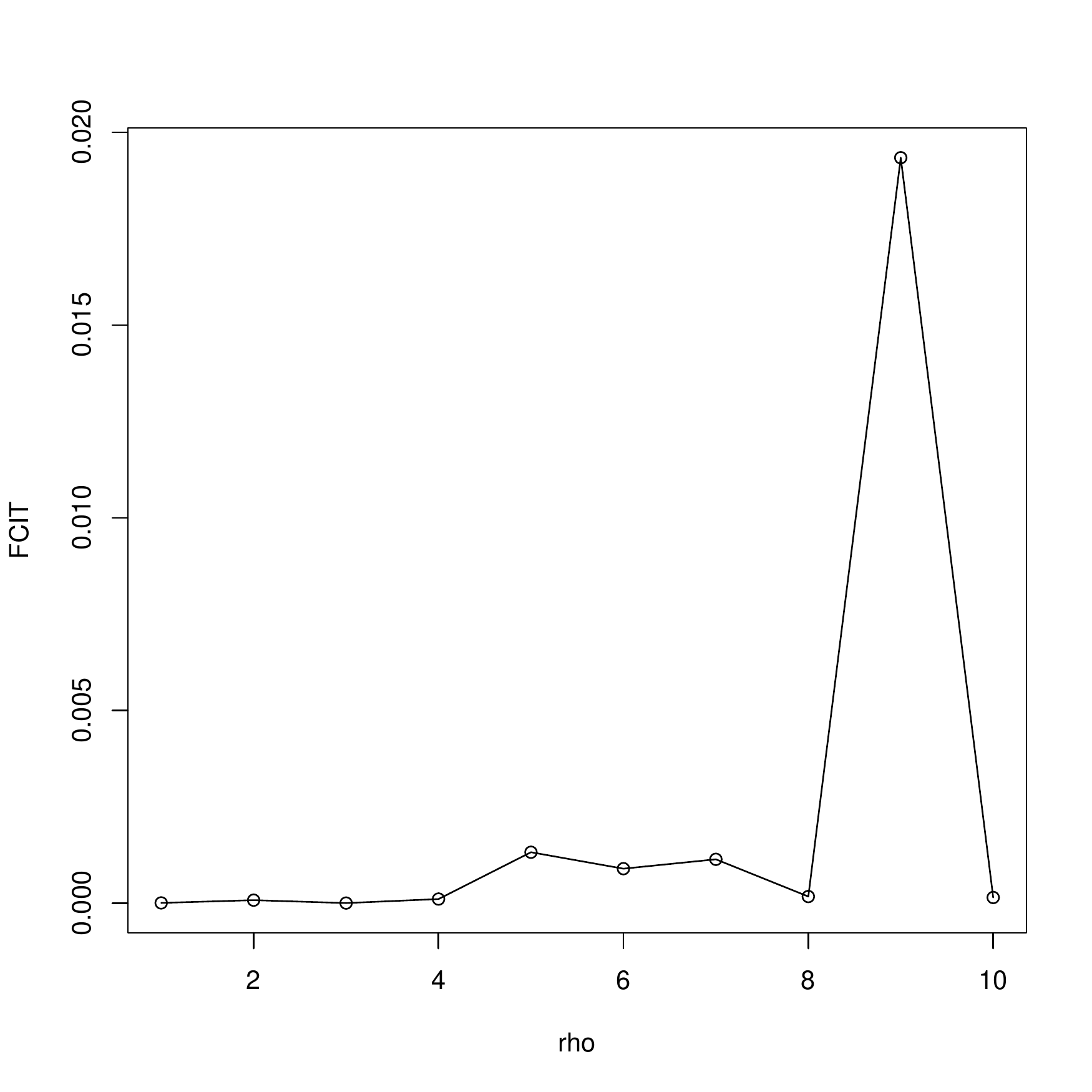}}
	\subfigure[CCIT]{\includegraphics[width=0.245\linewidth]{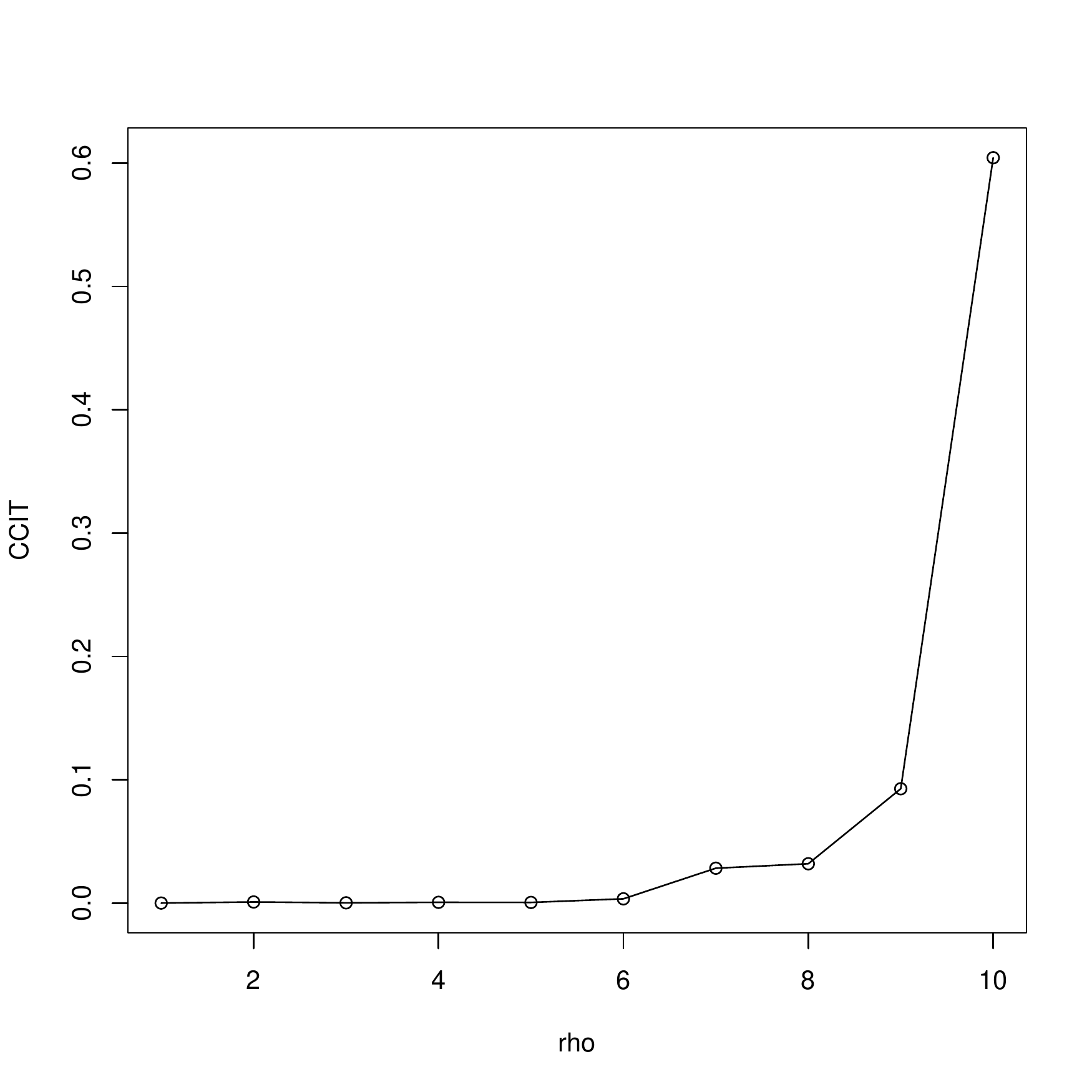}}
	\subfigure[PCIT]{\includegraphics[width=0.245\linewidth]{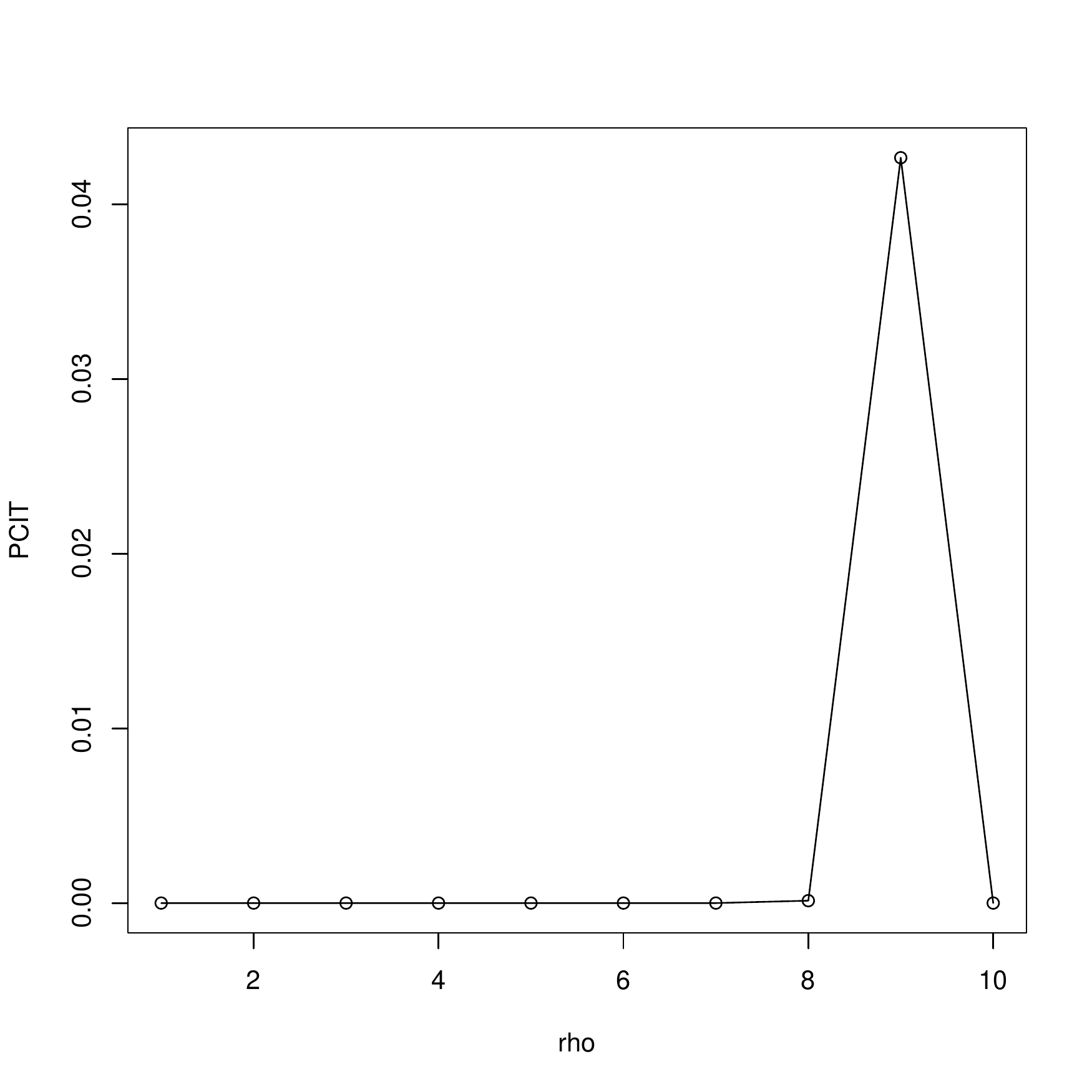}}
	\caption{Estimation of the CI measures from the simulated data of the trivariate normal copula.}
	\label{fig:trinormcopci}
\end{figure}

\begin{figure}
	\centering
	\includegraphics[width=0.9\linewidth]{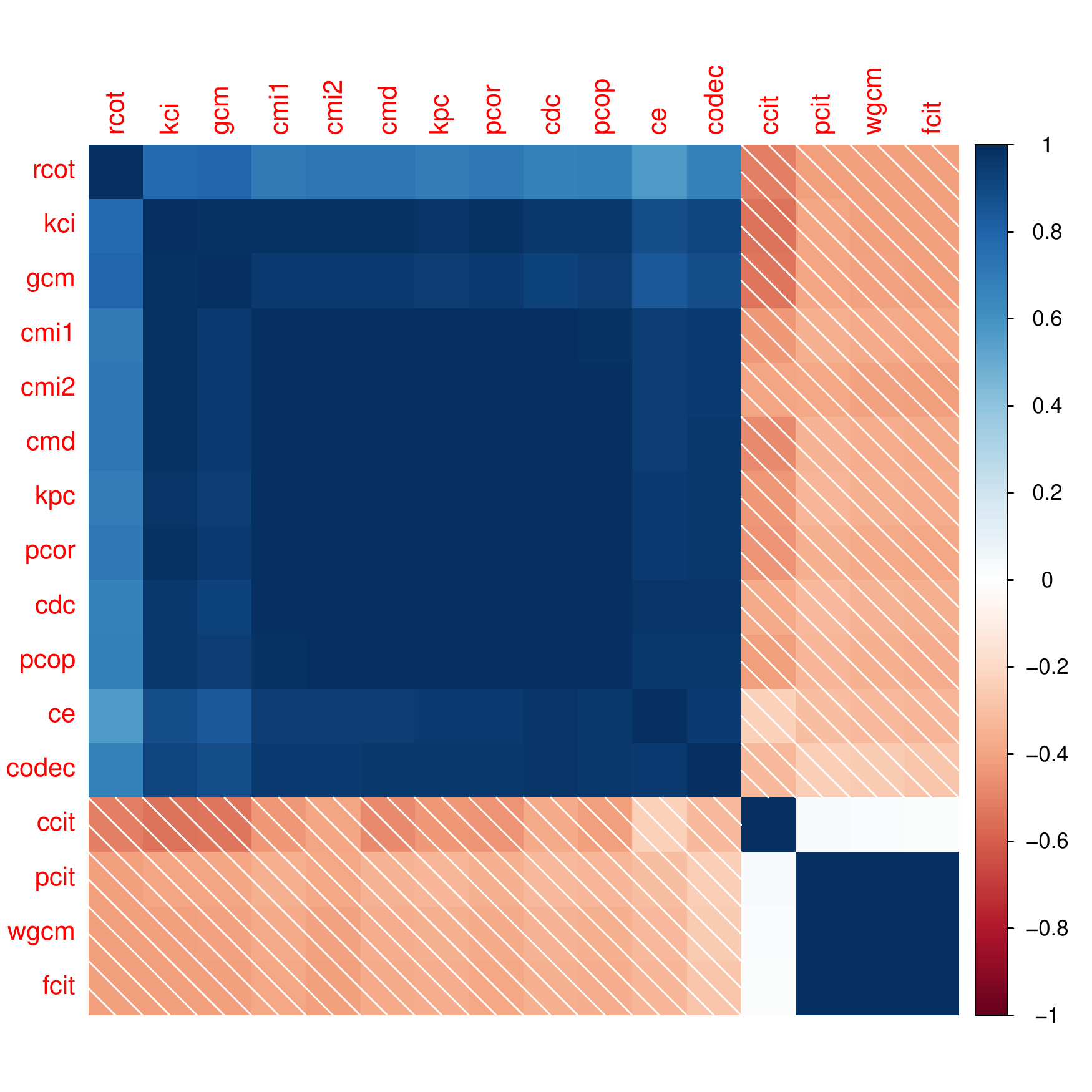}
	\caption{Correlation matrix of the CI measures estimated from the simulated data of the trivariate normal copula.}
	\label{fig:trinormcopcicm}
\end{figure}

\begin{figure}
	\subfigure[CE]{\includegraphics[width=0.245\linewidth]{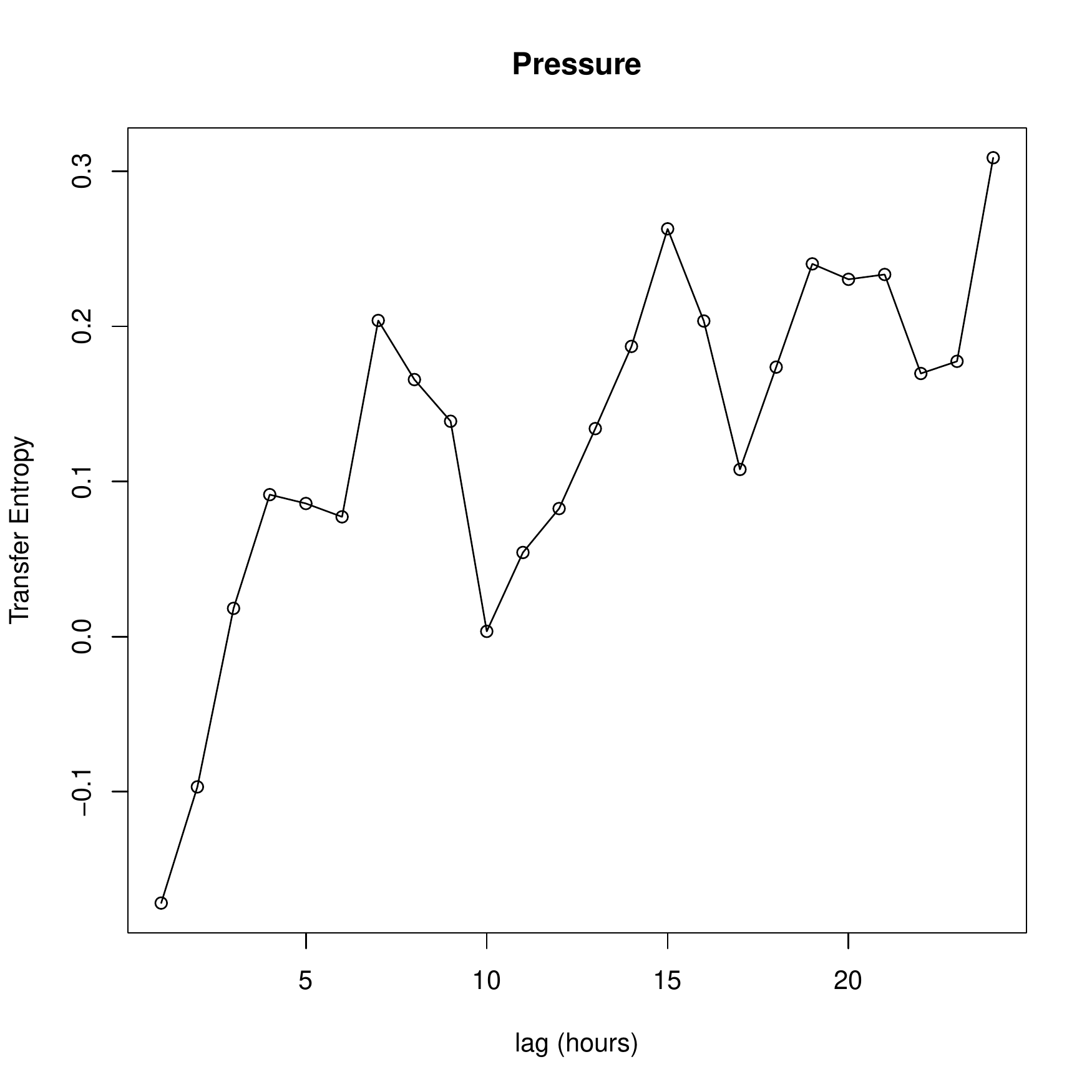}}
	\subfigure[KCI]{\includegraphics[width=0.245\linewidth]{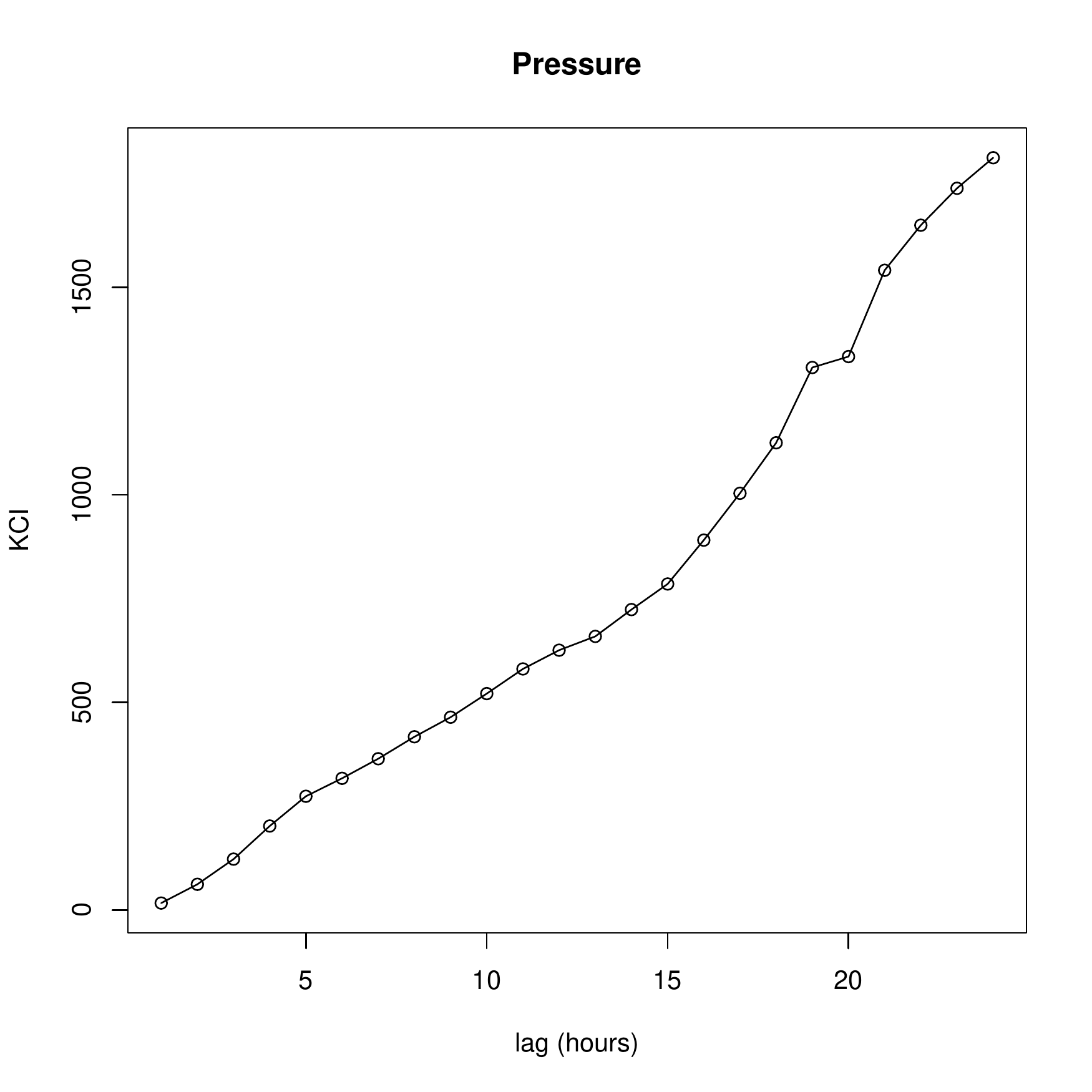}}
	\subfigure[RCoT]{\includegraphics[width=0.245\linewidth]{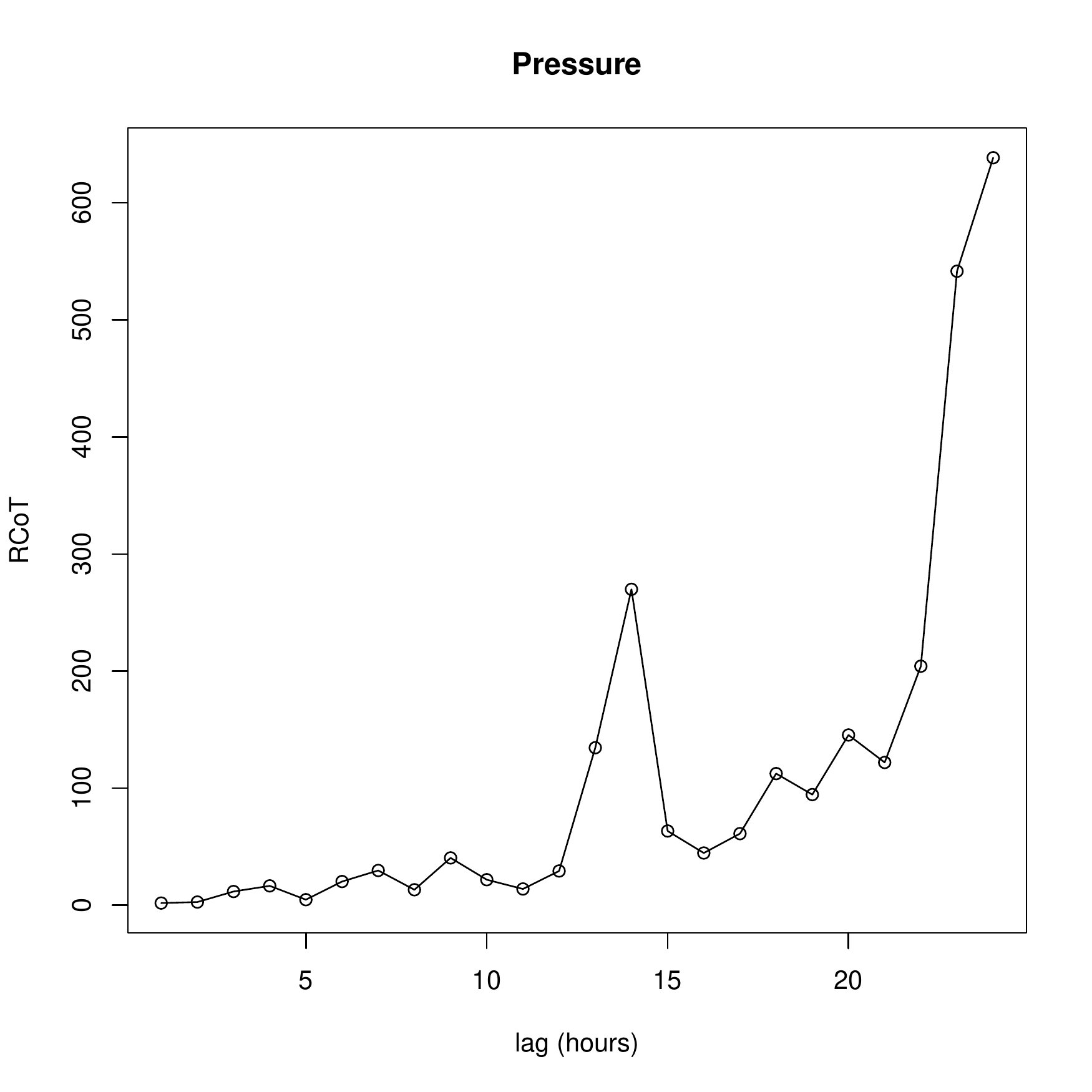}}
	\subfigure[CDC]{\includegraphics[width=0.245\linewidth]{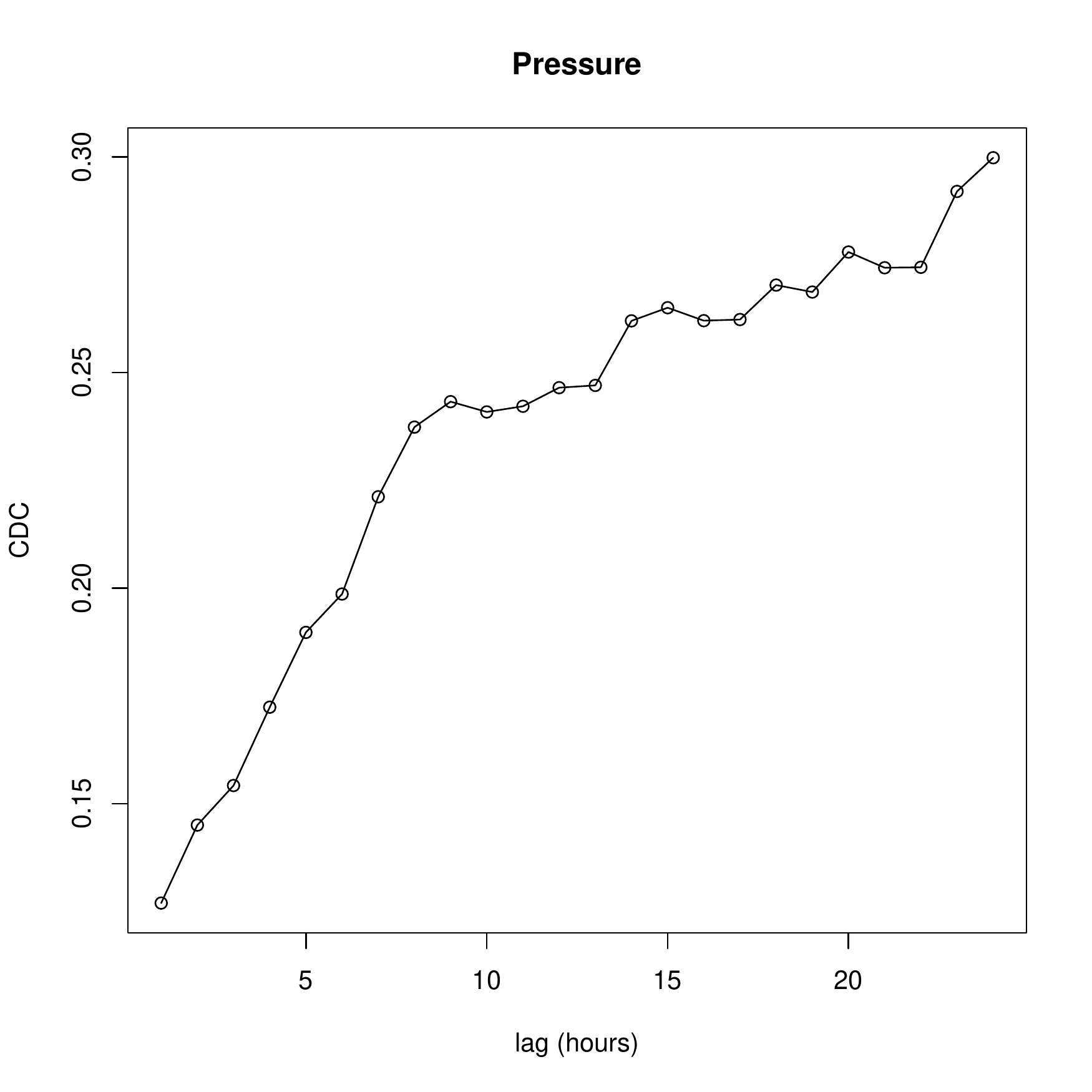}}
	\subfigure[GCM]{\includegraphics[width=0.245\linewidth]{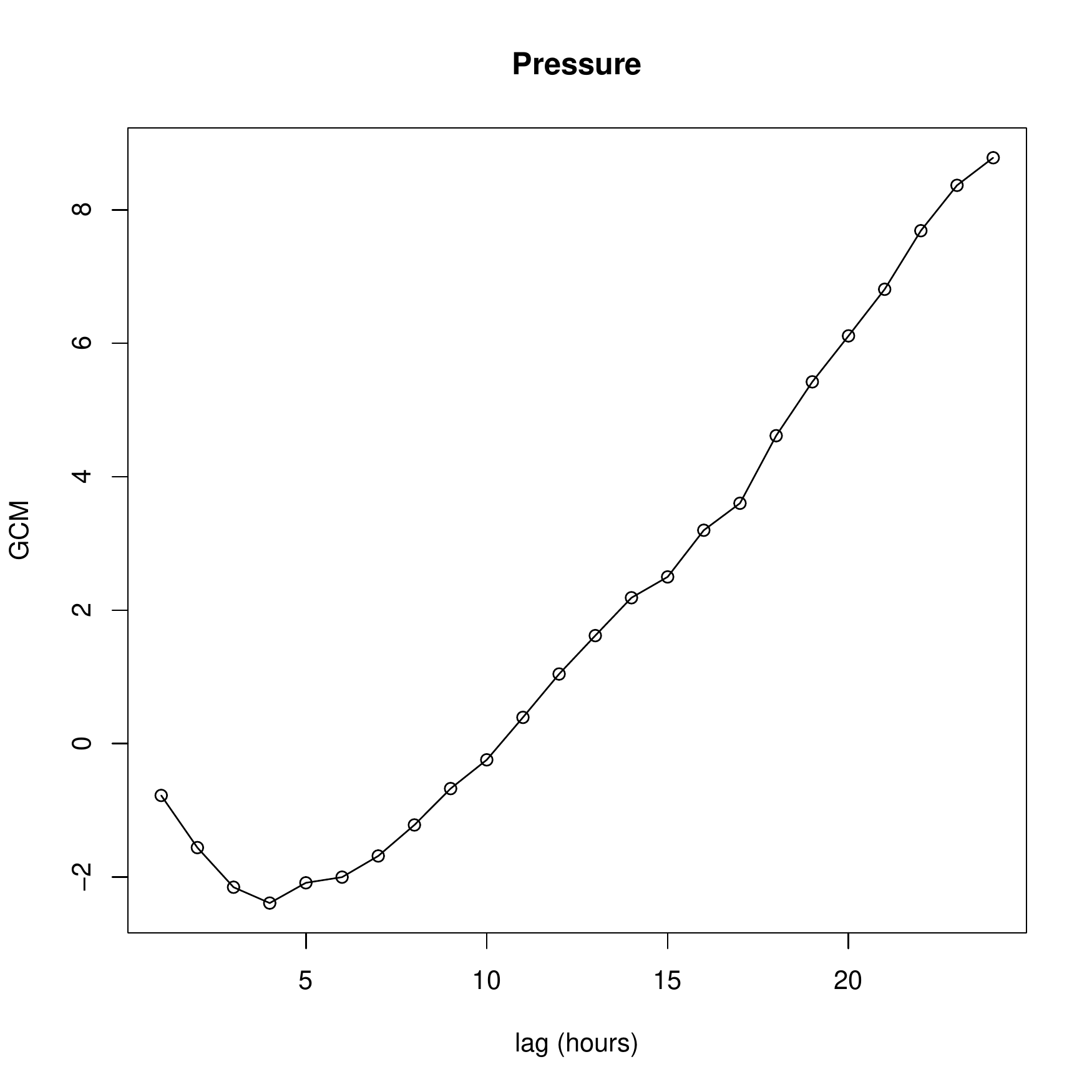}}
	\subfigure[wGCM]{\includegraphics[width=0.245\linewidth]{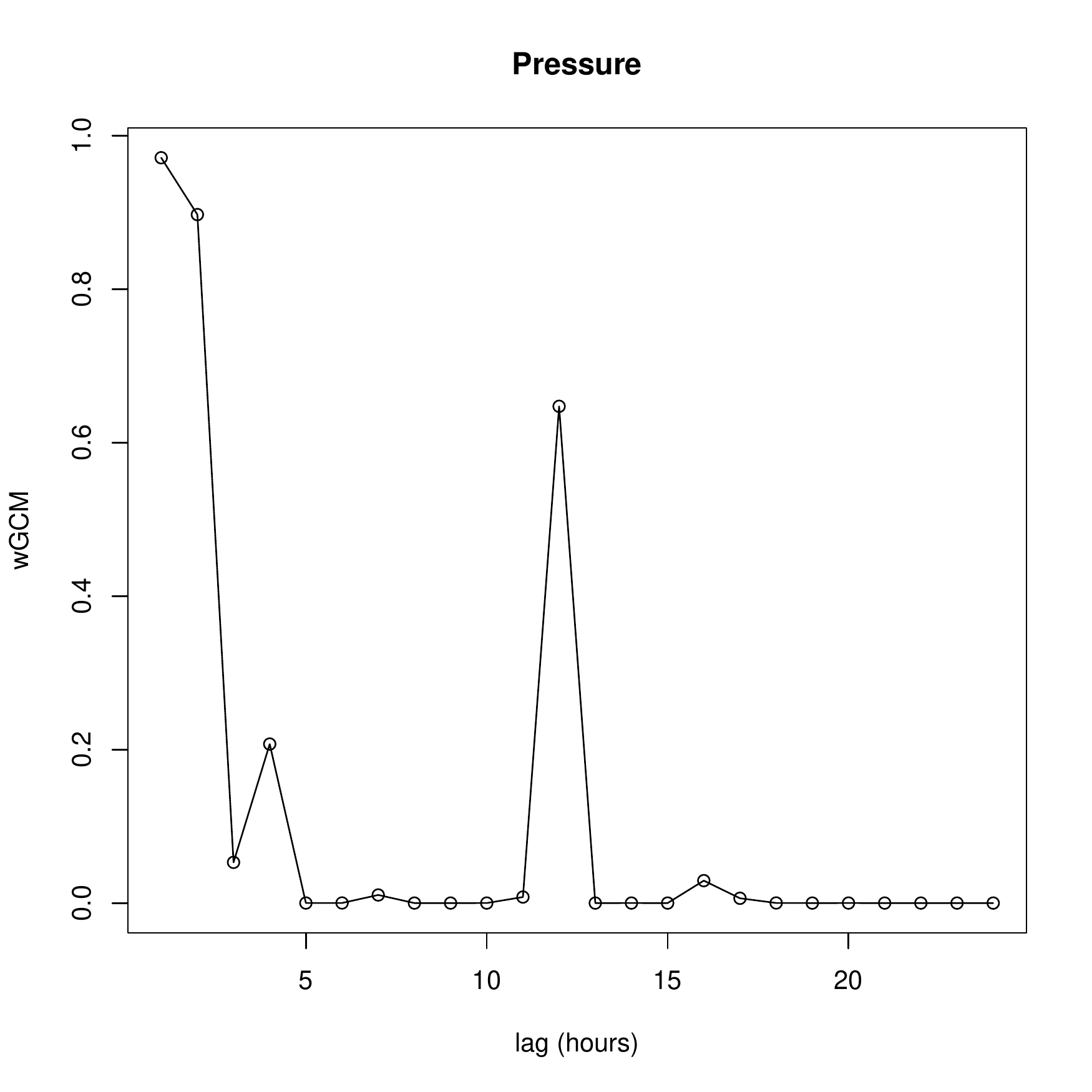}}
	\subfigure[CODEC]{\includegraphics[width=0.245\linewidth]{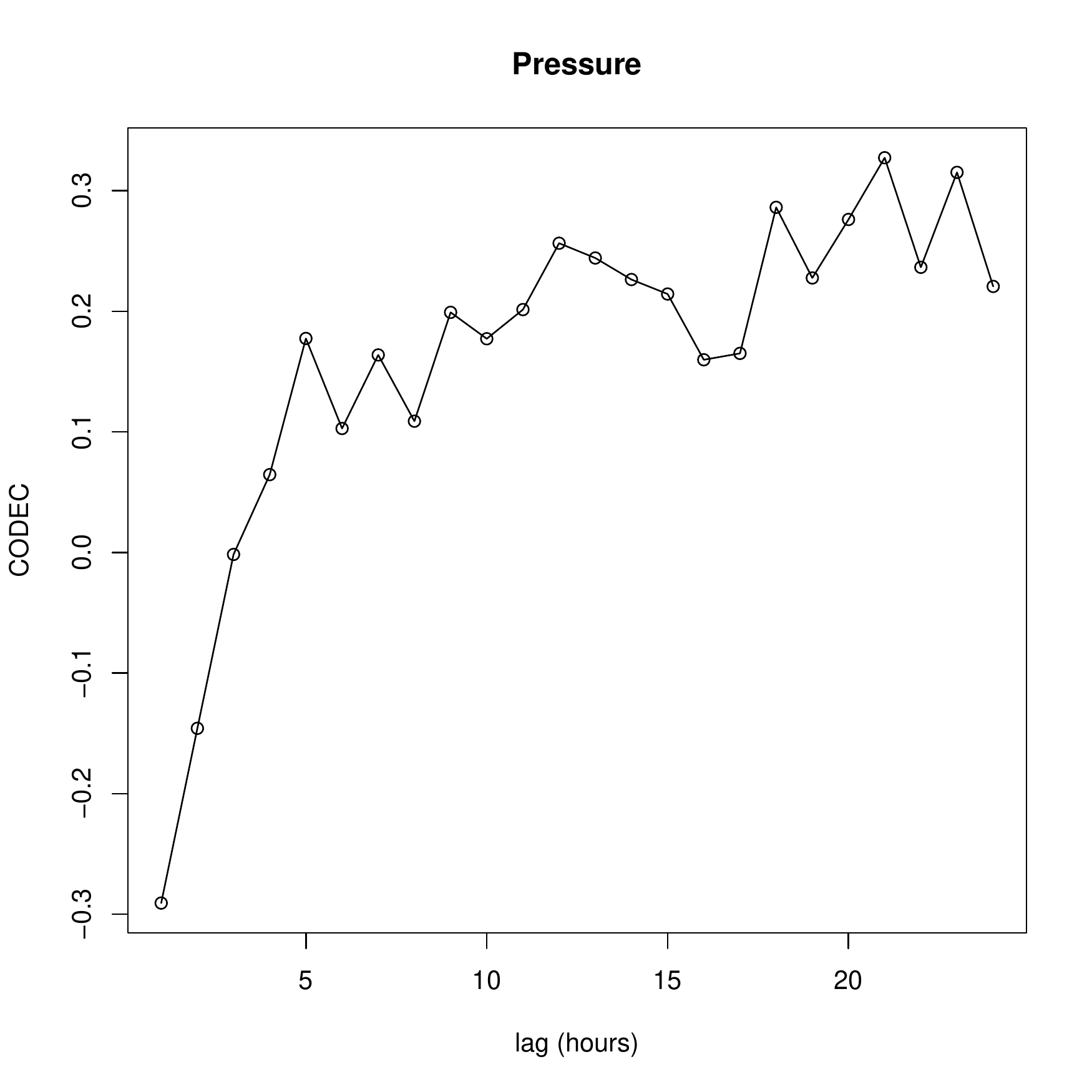}}
	\subfigure[KPC]{\includegraphics[width=0.245\linewidth]{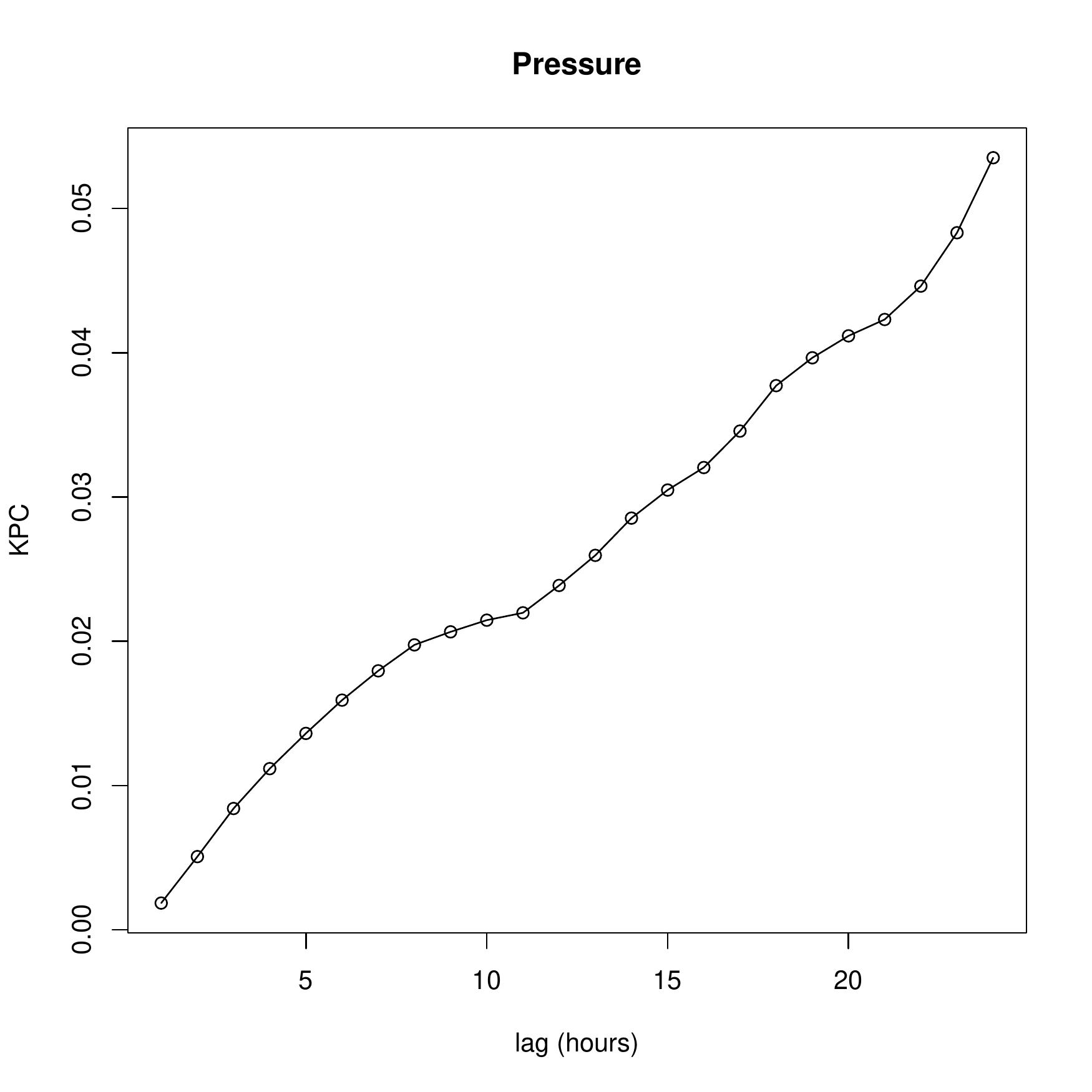}}
	\subfigure[PCor]{\includegraphics[width=0.245\linewidth]{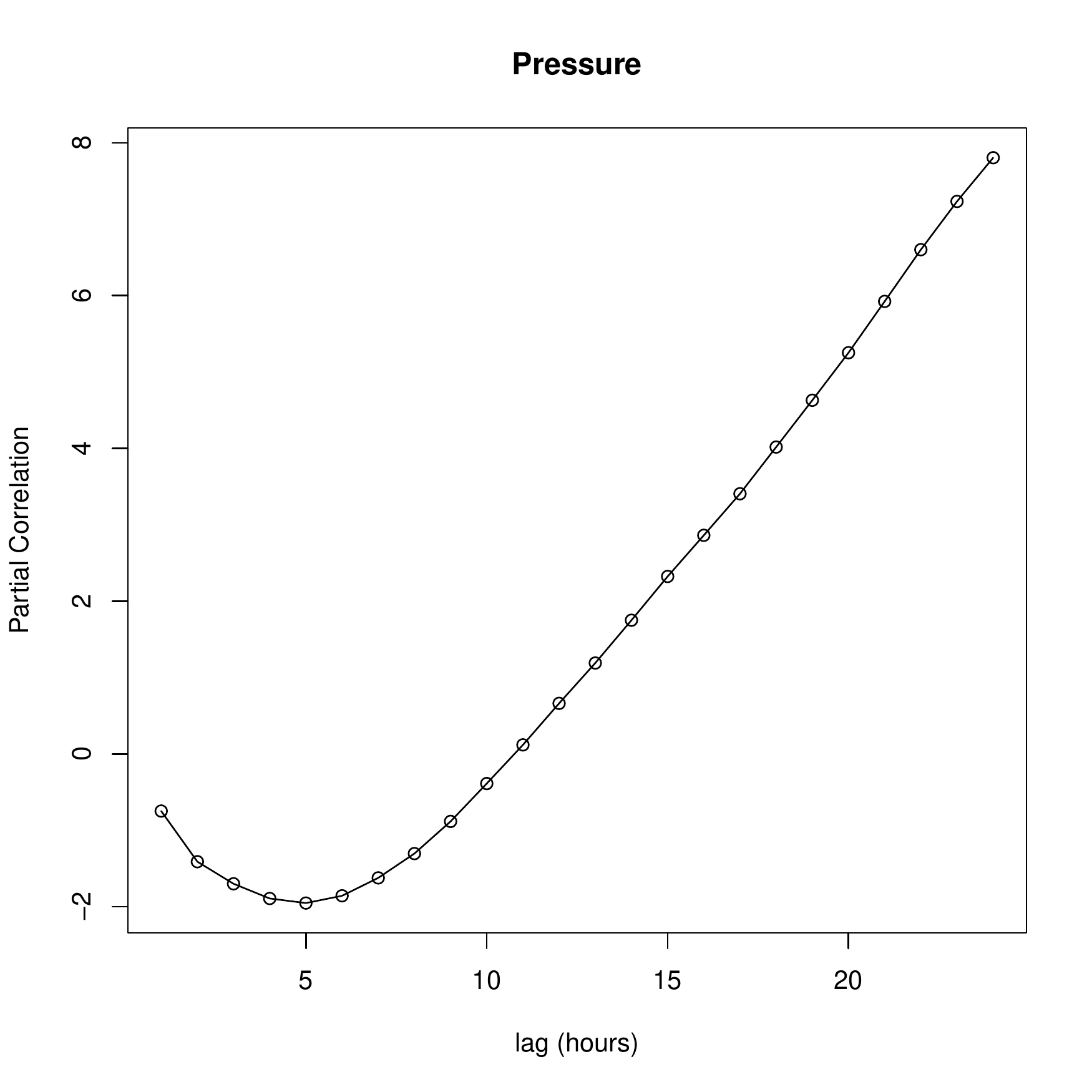}}
	\subfigure[CMDM]{\includegraphics[width=0.245\linewidth]{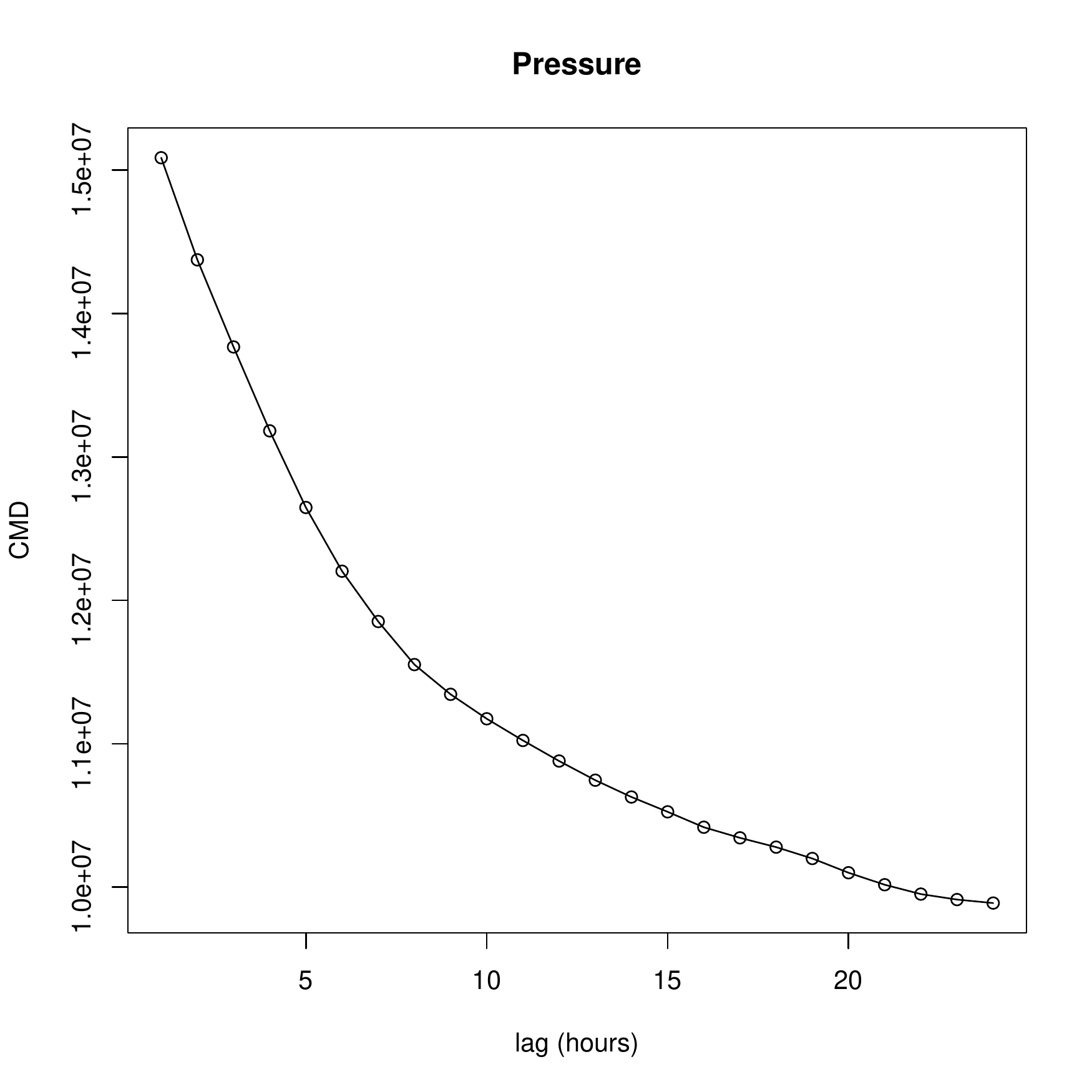}}
	\subfigure[PCop]{\includegraphics[width=0.245\linewidth]{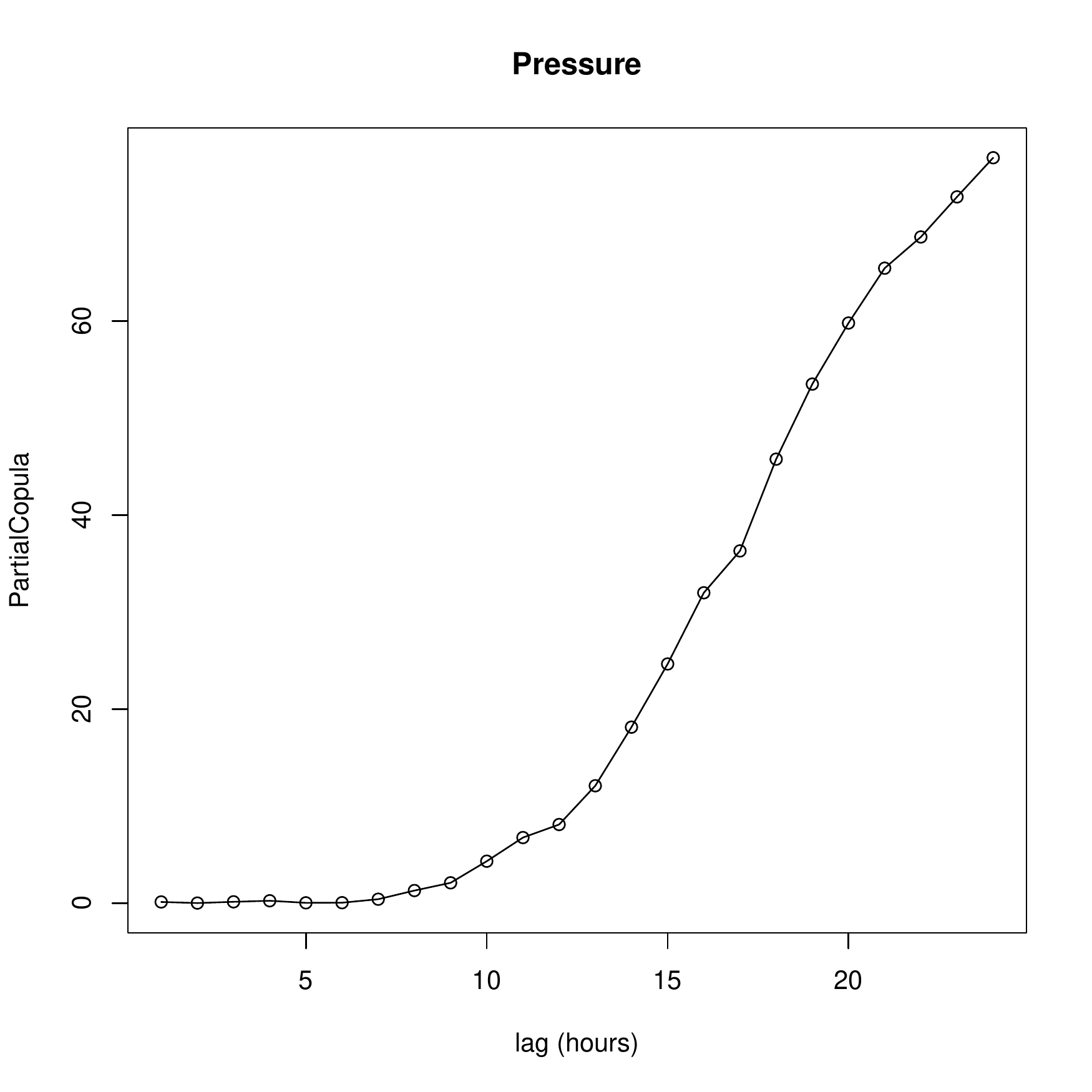}}
	\subfigure[CMI1]{\includegraphics[width=0.245\linewidth]{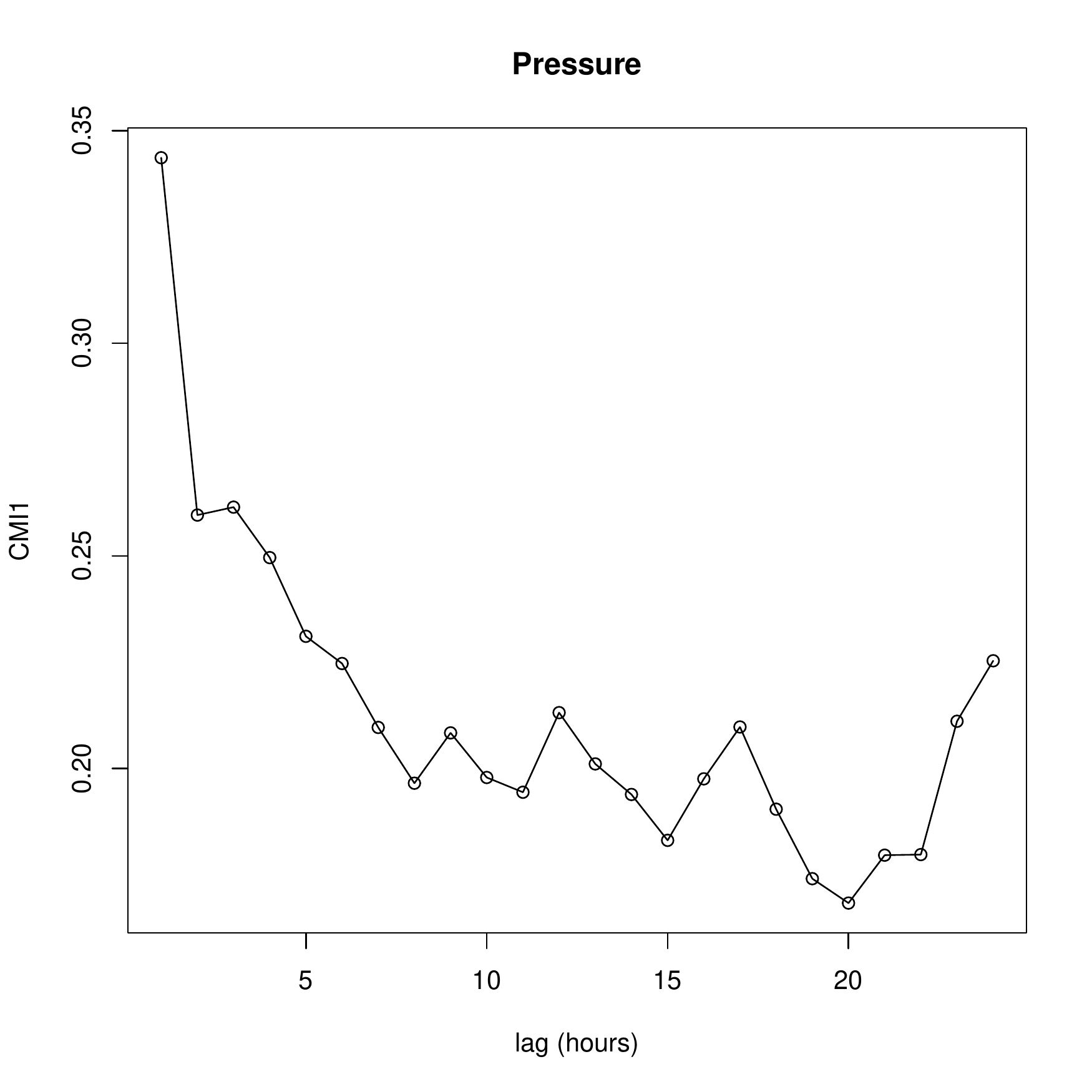}}
	\subfigure[CMI2]{\includegraphics[width=0.245\linewidth]{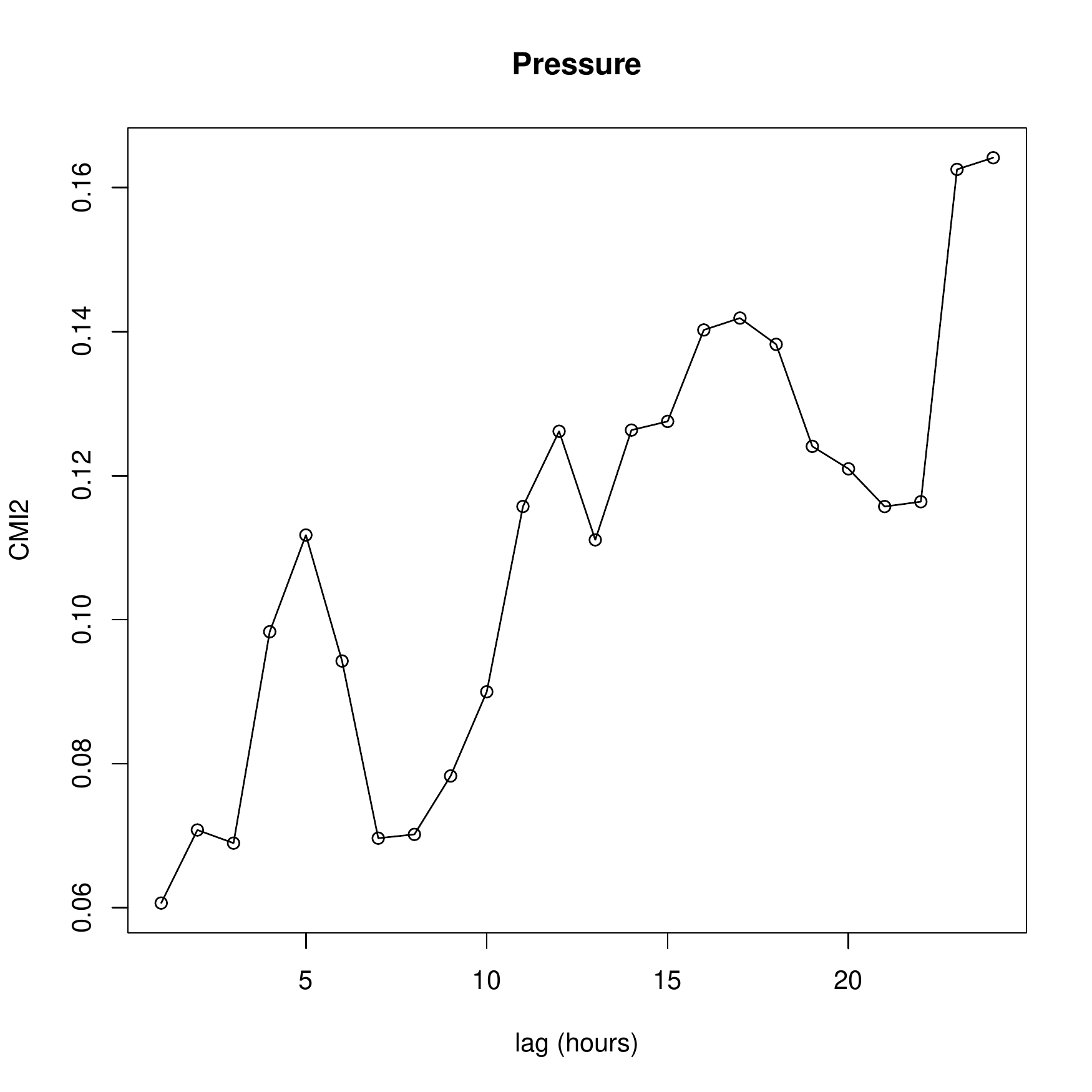}}
	\subfigure[FCIT]{\includegraphics[width=0.245\linewidth]{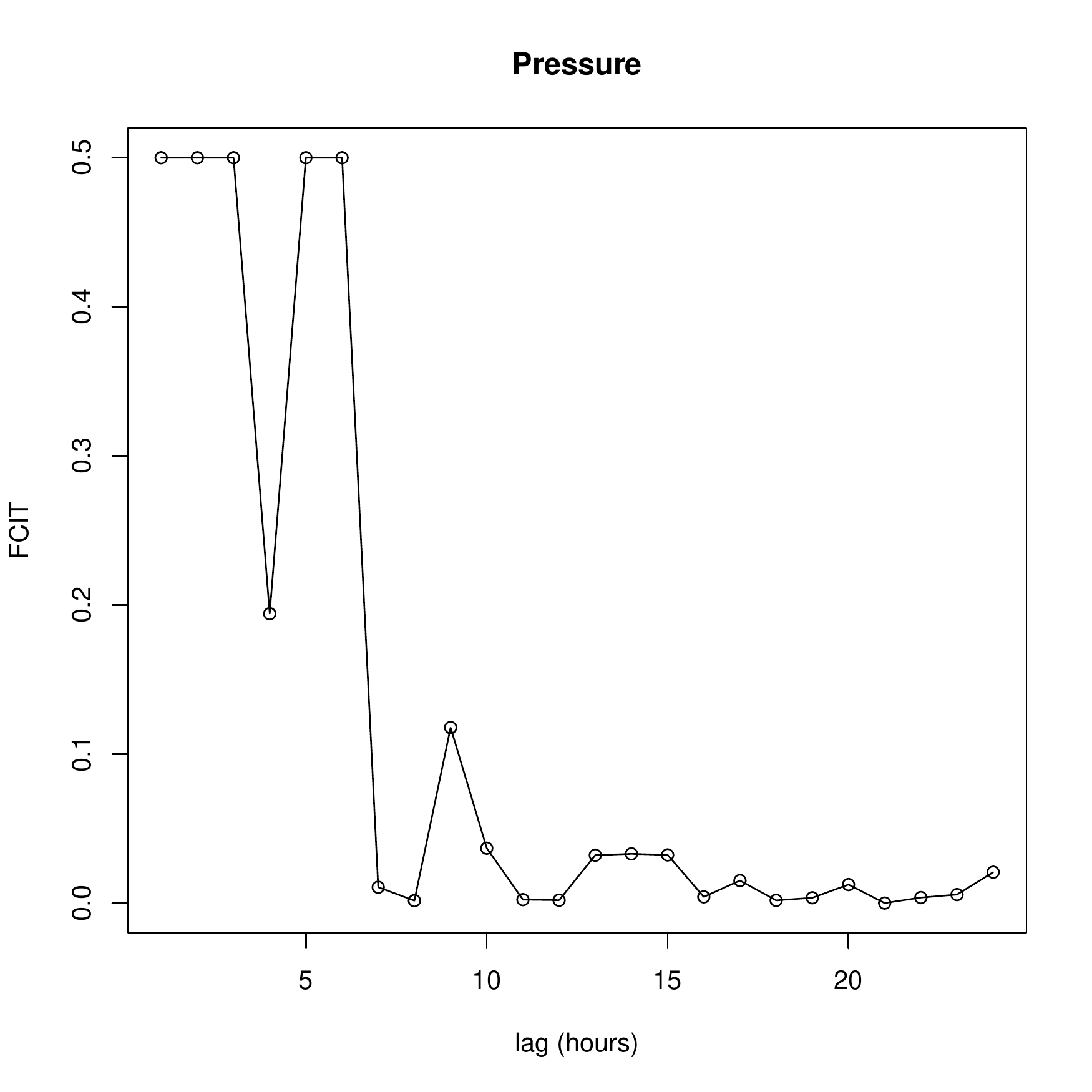}}
	\subfigure[CCIT]{\includegraphics[width=0.245\linewidth]{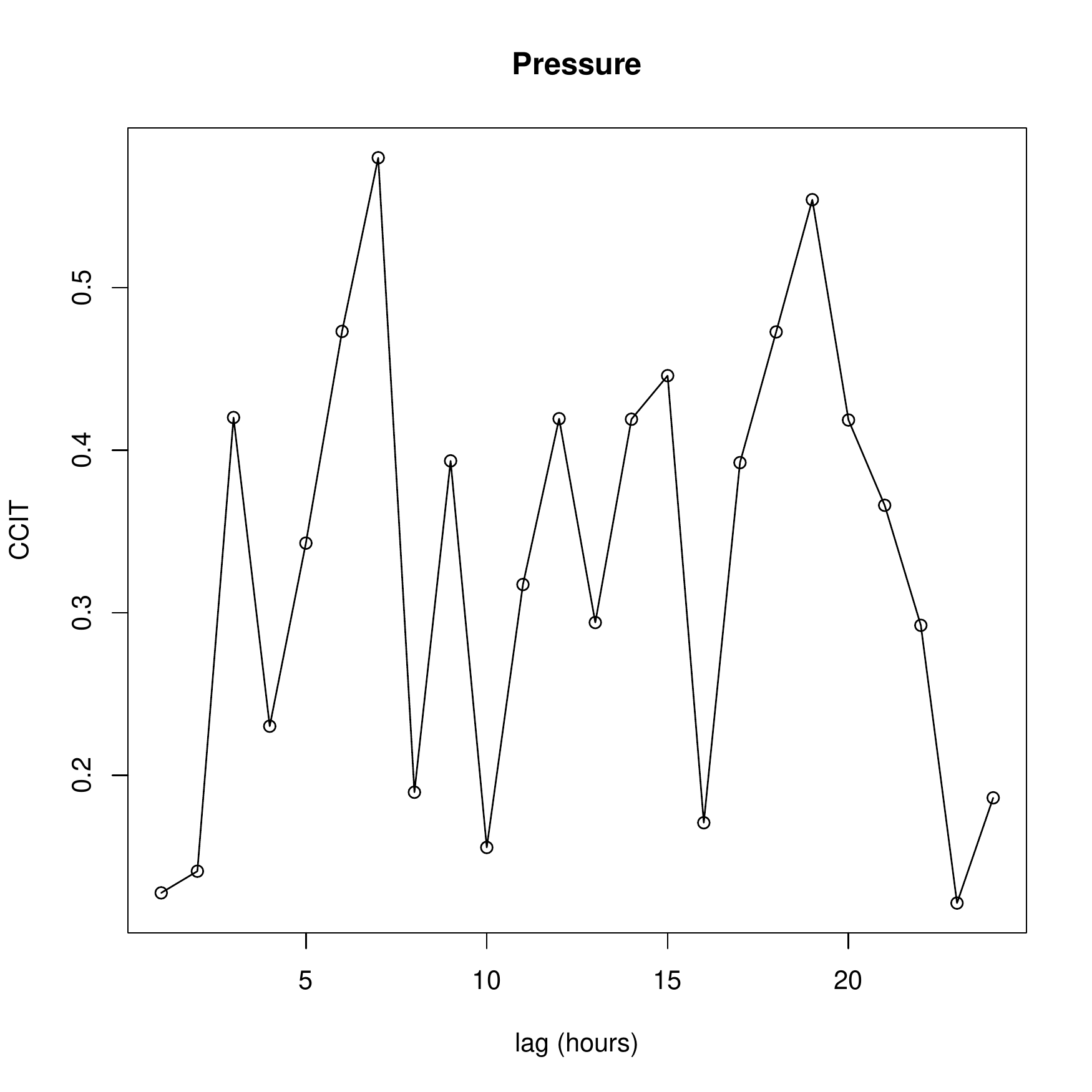}}
	\subfigure[PCIT]{\includegraphics[width=0.245\linewidth]{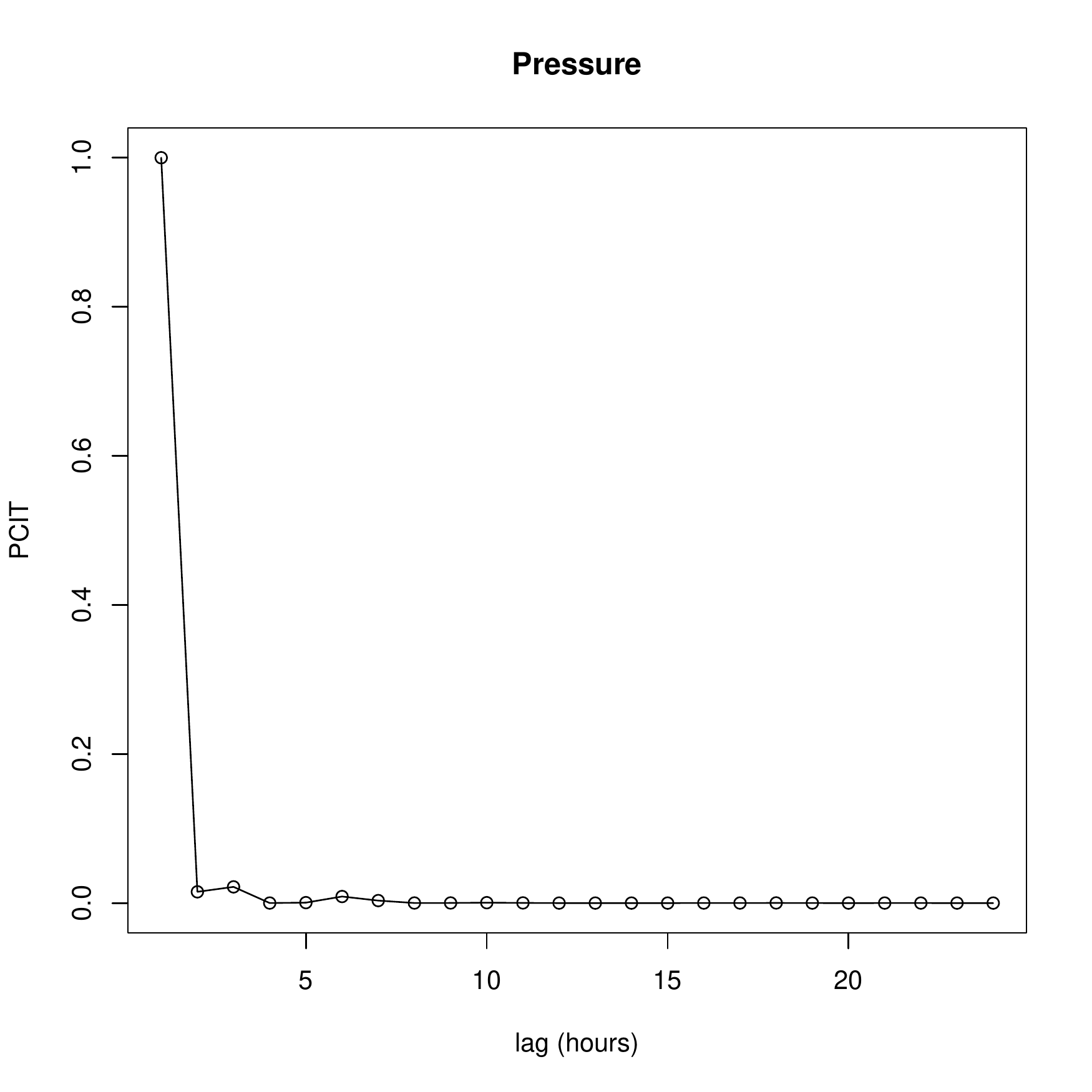}}
	\caption{Estimation of the CI measures between PM2.5 and pressure in the UCI Beijing air data.}
	\label{fig:air}
\end{figure}

\begin{figure}
	\centering
	\includegraphics[width=0.9\linewidth]{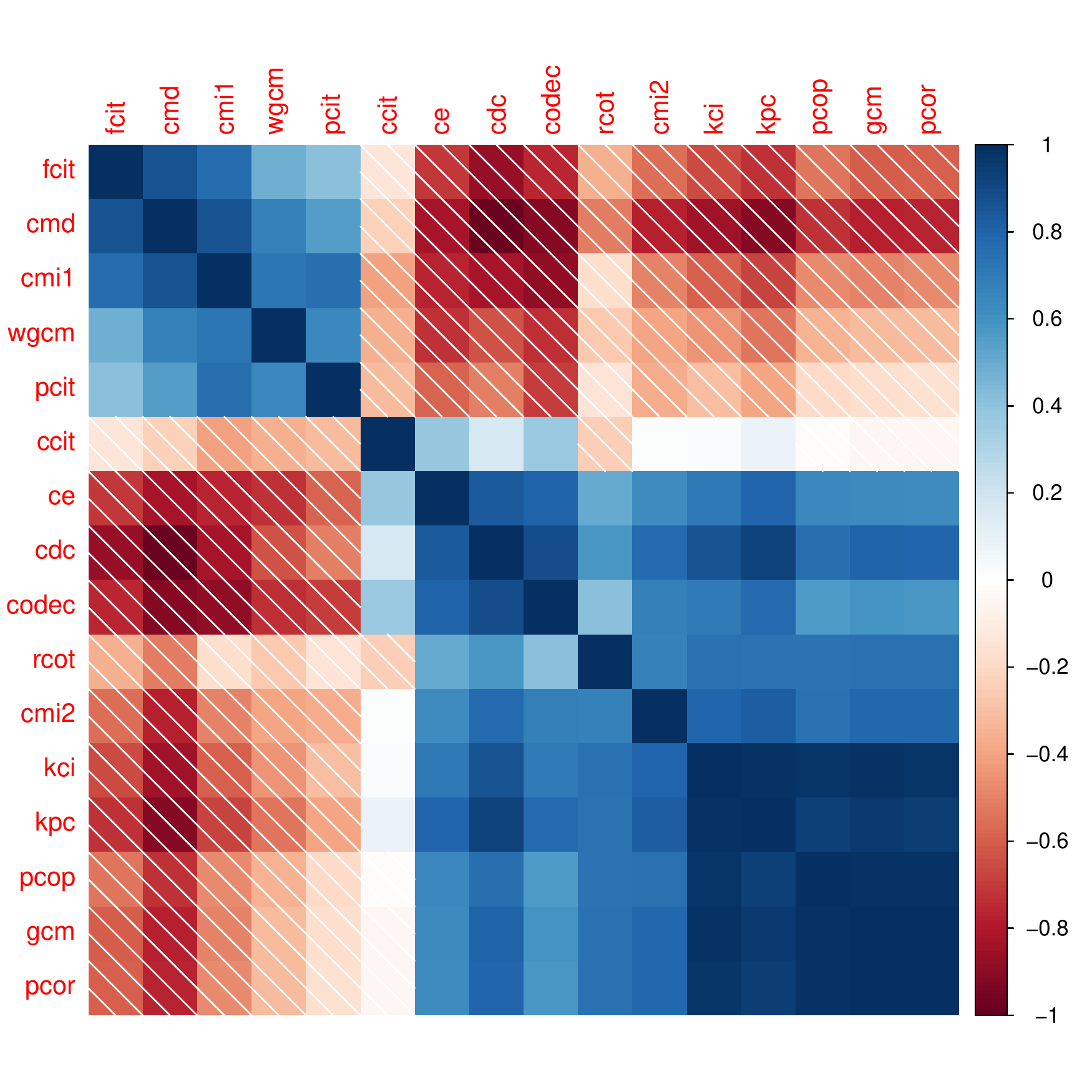}
	\caption{Correlation matrix of the CI measures estimated from the UCI Beijing air data.}
	\label{fig:aircm}
\end{figure}

\end{document}